%% file: main.tex
\documentclass[a4paper,11pt]{article}
\usepackage{jheppub} %
\usepackage{lmodern}
\usepackage{amsmath}
\usepackage{amssymb}
\usepackage{bm}
\usepackage{mathtools}
\usepackage{tikz}
\usepackage{tikz-cd}
\usetikzlibrary{scopes}
\usetikzlibrary{backgrounds}
\usetikzlibrary{arrows,automata}
\usetikzlibrary{patterns}
\usepackage{physics}
\usepackage{verbatim}
\usepackage{placeins}

\DeclareMathOperator{\dvol}{dvol}

\DeclareMathOperator{\Eext}{ext}

\usepackage{hyperref}
\usepackage{graphicx}
\usepackage{subcaption}
\usepackage{slashed}

\title{\boldmath Random matrix product state models of gravitationally prepared states}

\author[a,b]{Sunghoon Jung,}
\author[a]{Sungjung Kim,}
\author[a,c,d]{Jiwoo Park}
\author[a]{and Seokhyeon Song}
\affiliation[a]{Center for Theoretical Physics, Department of Physics and Astronomy, Seoul National University, Seoul 08826, Korea}
\affiliation[b]{Astronomy Research Center, Seoul National University, Seoul 08826, Korea}
\affiliation[c]{Department of Physics, Princeton University, Princeton, NJ 08544, U.S.A.}
\affiliation[d]{Department of Physics, Massachusetts Institute of Technology, Cambridge, MA 02139, U.S.A.}

\emailAdd{sunghoonj@snu.ac.kr}
\emailAdd{sjkimphya@gmail.com}
\emailAdd{jp7247@princeton.edu}
\emailAdd{hypercube256@snu.ac.kr}

\abstract{Gravitationally prepared states are quantum field theoretic states prepared by gravitational path integrals with spatial boundaries that have fixed boundary conditions for gravity but not for matter fields. They can be interpreted as quantum field theoretic states of closed universes encoding quantum gravitational effects of the past. We propose a method of modeling gravitationally prepared states in two dimensions with random matrix product states (RMPS). Such RMPS models allow us to compute quantities which include contributions of higher topologies and replica geometries to all orders. The bra-ket wormhole phase transition is a crucial physical property of gravitationally prepared states. We show that a RMPS model shows behavior qualitatively similar to the bra-ket wormhole phase transition if it satisfies the \textit{spectral gapping property}, which we define. We define a class of models called $\mathrm{O}(k)$ models satisfying this property. A novel advantage of RMPS models is that they allow us to compute effects corresponding to \textit{off-shell wormholes}, i.e., wormhole topologies without semiclassical solutions. In particular, numerical results in RMPS models suggest that off-shell wormholes lead to nonzero long-distance correlators in gravitationally prepared states. We also define RMPS models in continuous space, and discuss implications for studying de Sitter gravitationally prepared states.}

\keywords{2D Gravity, Cosmological models, AdS-CFT Correspondence, de Sitter space}
\arxivnumber{2512.11966}

\usepackage{etoolbox}
\makeatletter
\patchcmd{\l@section}{1.0em}{0.35em}{}{}
\makeatother
\let\originaltableofcontents\tableofcontents
\renewcommand{\tableofcontents}{\begingroup\small\originaltableofcontents\endgroup}
\renewcommand{\afterTocRuleSpace}{\clearpage}

\begin{document}
\maketitle
\setlength{\parskip}{0.8em}
\setlength{\emergencystretch}{3em}

\flushbottom

\section{Introduction}

One of the key questions motivating the study of quantum gravity is that of the early universe, when the universe was very small and dense, and thus quantum gravity effects were important. Theoretically, expanding universes (such as inflationary universes) are interesting objects to study as they underwent a period of evolution in the past where extreme quantum gravity effects occurred, i.e. large fluctuations in the spacetime geometry and its topology. In such a period of evolution, the universe cannot be described semiclassically. However, in the far future (such as after reheating for an inflationary universe), such quantum gravity effects become relatively weak, and we can describe the state of the universe semiclassically in terms of a QFT state on a classical background geometry. However, if we believe information to be conserved, this QFT state must encode information about the universe's non-semiclassical past and all its exotic quantum gravity effects. From here we get an idea of the quantum state of an expanding universe in the far future as a QFT state $\ket{\psi}$ of the matter fields of the universe that encodes information about a quantum gravitational past evolution.

This QFT state $\ket{\psi}$ is modeled naturally in the following way \cite{chen_bra-ket_2021, teresi_islands_2022}. We consider a $d$-dimensional gravitational path integral \cite{gibbons_action_1977}  with a $d-1$-dimensional spacelike boundary $\Sigma$. On $\Sigma$, we fix boundary conditions for gravity. However, we do not fix boundary conditions for the matter fields of the system, and instead view the gravitational path integral as a functional of the values of the matter fields $\varphi_i$ on $\Sigma$. Interpreting this functional as a wavefunctional, we can define a QFT state $\ket{\psi}$ of the matter fields on $\Sigma$. In other words, we can write $\ket{\psi}$ as
\begin{equation}
    \ket{\psi} = \int Dg|_{\partial g = \Sigma} \prod_i D\varphi_i  \exp \left( -S\left[g, \varphi_i \right] \right)  \ket{\varphi_{i, \Sigma} }
\end{equation}
where $S\left[g, \varphi_i \right]$ is the action of gravity coupled to the matter fields, and $\ket{\varphi_{i, \Sigma} }$ are eigenstates of values of the matter fields $\varphi_i$ on $\Sigma$.

We call $\ket{\psi}$ a gravitationally prepared state. It is a QFT state encoding information about quantum gravitational histories. In the cosmological context, we can interpret the geometries $g$ as different possible histories of the universe's geometry, and the surface $\Sigma$ as a cutoff surface in the far future where gravity is weak enough that we assume gravity to be classical from that point onward. For inflationary universes \cite{guth_inflationary_1981, linde_eternally_1986}, a natural interpretation of the surface $\Sigma$ would be the reheating surface where inflation ends.

An interesting aspect of gravitationally prepared states is that they encode nonperturbative quantum gravity effects. In particular, they include the effects of higher topologies and replica effects in the gravitational path integral. Higher topologies are contributions with different topologies than the dominant saddlepoint geometry. Replica effects are the contributions of geometries connecting multiple copies of the gravitationally prepared state. They are of interest to quantum gravity due to their involvement in purifying the Page curve of Hawking radiation at late times \cite{almheiri_replica_2020, penington_replica_2020} and their relationship to quantum chaos \cite{saad_semiclassical_2019, chen_comments_2023}. A more thorough account of replica effects, including their explicit constructions, has been studied in \cite{Geng:2024xpj, Bao:2025plr, Geng:2025efs}.

Recently, there has been an interest in studying the higher topology contributions and replica effects of two dimensional gravitational path integrals using random matrices \cite{saad_jt_2019, witten_matrix_2020, maxfield_path_2021, jafferis_jt_2023}. The idea is that the gravitational path integral computes an ensemble-averaged quantity, where the ensemble is a probability distribution over random Hermitian matrices. Such methods allow us to compute quantities which include higher topology and replica effects to all orders, which cannot be done with just semiclassical gravity.

In this paper, we apply this idea to two dimensional gravitationally prepared states. In particular, we construct random matrix product states modeling two dimensional gravitationally prepared states. Such models allow us to numerically compute quantities which include higher topology effects and replica effects to all orders. They generalize the study of random matrix models of pure gravity to gravity with matter fields, where we can compute things such as correlators and entropies of spatial subregions.

Therefore, by constructing random matrix product state models of gravitationally prepared states, we get a new toolbox for studying nonperturbative effects in quantum gravity, with a much richer set of physical effects that we can study and calculate. In particular, we can compute quantities interpreted as the contributions of \textit{off-shell wormholes}, i.e. wormhole topologies without semiclassical solutions, in the gravitational path integral. We will study some particularly illuminating examples of such effects, such as the contribution of off-shell wormholes to the inner product $\braket{\psi}{\psi}$ and long-distance correlators, in this paper.

\paragraph{Organization of the paper.} In section~\ref{sec:gps}, we review features of two-dimensional gravitationally prepared states using the AdS JT+CFT model studied in \cite{chen_bra-ket_2021}. We discuss two important physical effects that can be computed from the semiclassical approximation: the bra-ket wormhole phase transition and entropy bounding.

In section~\ref{sec:rmps}, we motivate the program of modeling 2D gravitationally prepared states with random matrix product states based on three arguments --- the topological expansion behavior of random matrices, the entropy bounding behavior of matrix product states, and a holographic argument. We then identify a condition for an RMPS model to satisfy in order to be a model with well-behaved semiclassical geometries, which we call the \textit{spectral gapping condition}. We then present a broad class of models --- the $\mathrm{O}(k)$ models --- for which the spectral gapping condition holds. We analytically and numerically compute the physical properties of the $\mathrm{O}(k)$ models.

In section~\ref{sec:offshell}, we compute higher-topology RMPS corrections and discuss their conjectural interpretation as off-shell-wormhole effects. RMPS models provide a novel setup where these effects can be computed. In particular, we compute the contributions of off-shell wormholes to the inner product and to correlators of local operators.

In section~\ref{sec:crmps}, we discuss an RMPS model in continuous space. The continuous model satisfies all the properties of gravitationally prepared states that we checked for the discrete model, with the added benefit that it gives physical results for all real-valued distances.

In section~\ref{sec:dsrmps}, we discuss gravitationally prepared states in two dimensional de Sitter space and related issues in relation to RMPS models.

In section~\ref{sec:discussion}, we summarize our work and discuss avenues for future study.

\section{Gravitationally prepared states}

\label{sec:gps}

\subsection{Motivation and definition}

\label{ssec:gpsdefinition}

We consider the gravitationally prepared state $\ket{\psi}$ given by
\begin{equation}
    \ket{\psi} = \int Dg|_{\partial g = \Sigma} \prod_i D\varphi_i  \exp \left( -S\left[g, \varphi_i \right] \right)  \ket{\varphi_{i, \Sigma} } .
\end{equation}
We can say that
\begin{equation}
    \ket{\psi}  \in \mathcal{H}_{\text{QFT}, \Sigma}
\end{equation}
where $\mathcal{H}_{\text{QFT}, \Sigma}$ is the Hilbert space of the matter fields $\varphi_i$ on $\Sigma$.

To interpret the geometries $g$ as histories of an expanding universe, it may seem reasonable to take them to be Lorentzian geometries of signature $(1,d-1)$. However, formally, we need not restrict ourselves to Lorentzian geometries. The gravitational path integral may be better understood if the geometries are Euclidean. Furthermore, states prepared by Euclidean time evolution have been studied before in different contexts, such as the Hartle-Hawking no-boundary state for quantum gravity \cite{hartle_wave_1983} and the Bunch-Davies state for de Sitter cosmology \cite{bunch_quantum_1997}. Thus we will work with Euclidean geometries $g$, and take $S$ to be the Euclidean action. We note that in theories which behave well under analytical continuation, we may be able to analytically continue the geometries $g$ to be Lorentzian geometries or hybrids of Lorentzian and Euclidean geometries \cite{hertog_holographic_2012, maldacena_two_2021}. We also touch upon this issue further in section~\ref{sec:dsrmps}. Such geometries may have more realistic interpretations as real-time histories of an expanding universe.

\paragraph{The AdS JT+CFT model.} The AdS JT+CFT model is a model of a gravitationally prepared state in $d=2$ dimensions. It was studied extensively in \cite{chen_bra-ket_2021}. In this section we will reorganize their results. We consider $\text{AdS}_2$ JT gravity (i.e. JT gravity with a negative cosmological constant $\Lambda = - 1$) \cite{jackiw_lower_1985, teitelboim_gravitation_1983, engelsoy_investigation_2016, maldacena_conformal_2016} coupled to conformal matter fields $\varphi_i$. In other words, the action is given by
\begin{equation}
    S_{\text{JT+CFT}}\left[g, \phi, \varphi_i\right] = - \phi_0 \chi(g) - \frac{1}{4\pi} \int d^2x \sqrt{g}   \phi (R + 2) - \frac{1}{2\pi} \int_{\partial g} dx \sqrt{h} \phi K + S_{\text{CFT}} [\varphi_i, g] .
\end{equation}
Here $\chi(g)$ is the Euler characteristic of $g$, $h$ is the induced metric on $\partial g$, and $K$ is the extrinsic curvature of $\partial g$.

In the AdS JT+CFT model, the gravitational degrees of freedom are $g$ and $\phi$. Thus we need to fix boundary conditions for $g$ and $\phi$. This is given by
\begin{equation}
    ds^2 = \frac{dz^2+ dx^2 }{z^2}, \quad  \phi = \frac{\phi_r}{z} \quad \text{as }z \rightarrow 0
\end{equation}
along with periodicity conditions for $x$:
\begin{equation}
    x \approx x + L.
\end{equation}
We understand the boundary as being placed at $z = \epsilon$ where $\epsilon$ is infinitesimal. We can understand $L$ as the normalized length of the boundary, and $\phi_r$ as the normalized dilaton field value at the boundary. Thus, boundary conditions are given by a set $(L, \phi_r)$. However, it turns out that the model has a symmetry: the value of the path integral remains unchanged under the transformation
\begin{equation}
    (L, \phi_r) \rightarrow (\alpha L, \alpha \phi_r) \quad (\alpha >0)
\end{equation}
which allows us to keep $\phi_r$ fixed in our discussion.

We can now define the AdS JT+CFT gravitationally prepared state
\begin{equation}
    \ket{\psi} = \int Dg  D\phi  |_{(L, \phi_r)} \prod_i D\varphi_i  \exp \left( -S_{\text{JT+CFT}}\left[g, \phi, \varphi_i\right] \right)  \ket{\varphi_{i, \Sigma} }
\end{equation}
where $(L, \phi_r)$ indicate the corresponding boundary conditions for $g, \phi$.

We remark that for any 2D gravity theory, there is a topological expansion in terms of the number of boundaries and genus of a given geometry $g$. Given a geometry with $n$ boundaries and genus $g$, the Euler characteristic is given by $\chi(g) = (2-2g-n)$. If the action of the gravity theory contains the topological term $-\phi_0 \chi(g)$, the contribution of the geometry in the gravitational path integral is weighted by $e^{\phi_0 (2-2g-n)}$. Relative to the topology of $n$ disconnected disks, this geometry is suppressed by a factor of $e^{-2 \phi_0 (g+n-1)}$. We call this topological suppression. We say that a geometry is of higher topology if it has a higher topological suppression factor. Note that we can understand topological suppression in 2D gravity intuitively by saying that creating a wormhole costs a factor of $e^{-2\phi_0}$, as a wormhole connecting two disconnected geometries increases $n$ by 1 and a wormhole connecting a connected geometry to itself increases $g$ by 1.

Since the fluctuation of spacetime topology is an exotic quantum gravity effect, we expect topological suppression to be strong in a physical model of quantum gravity. Furthermore, the parameter $\phi_0$ is equal to the entropy of a black hole in AdS JT gravity \cite{maldacena_conformal_2016, almheiri_entropy_2019}, which physically should be a large quantity. Thus, we will take $\phi_0 \gg 1$.

\subsection{Bra-ket wormhole phase transition}

\label{ssec:gpsbkwh}

In the JT+CFT model, we consider the inner product $\braket{\psi}{\psi}$. We remark that computing the inner product amounts to imposing transparent boundary conditions for the matter fields between the boundaries of the bra and ket, as
\begin{equation}
    \braket{\varphi^\prime_{i, \Sigma} } {\varphi_{i, \Sigma} } = \delta[\varphi^\prime_{i, \Sigma} - \varphi_{i, \Sigma} ].
\end{equation}
We can compute the gravitational path integral semiclassically by looking for saddlepoint solutions, i.e. geometries minimizing the effective action which is given by
\begin{equation}
    S_{\text{eff}}[g,g^\prime] = - \ln \left( \int   \prod_i D\varphi_i  \exp \left( -S_{\text{JT+CFT}} \left[g, \varphi_i \right]  \right)  \right).
\end{equation}
In JT+CFT this is given by the sum of the pure JT gravity action with the effective action of the conformal matter fields. Finding a saddlepoint solution amounts to finding a geometry which satisfies the semiclassical field equations of JT gravity coupled to conformal matter:
\begin{equation}
	R = - 2, \quad (g_{ab} \nabla^2 - \nabla_a \nabla_b ) \phi - g_{ab} \phi = 2 \pi \left< T^{\text{CFT}}_{ab} \right>.
\end{equation}
These equations can be obtained by varying $S_{\text{eff}}$ by $\phi$ and $g_{\text{ab}}$. Useful formulae for computing $\left< T^{\text{CFT}}_{ab} \right>$ are given in appendix~\ref{sec:cft}.

To lowest order in the topological expansion, we have the contribution of geometries that are topologically equivalent to two disconnected disks capping off the bra and ket boundaries. We find such a saddlepoint geometry, where we have the following geometry for each disk:
\begin{equation}
	ds^2 = \frac{dz^2 + dx^2}{\frac{L^2}{4\pi^2} \sinh^2 \frac{2\pi z}{L}}, \quad \phi = \frac{c}{12} + \frac{\phi_r}{\frac{L}{2\pi} \tanh \frac{2\pi z}{L}}, \quad x = x + L.
\end{equation}
Here $c$ is the central charge of the matter CFT. We call this geometry the double disk geometry. We note that the disk geometry prepares the vacuum state of the matter CFT; i.e. the contribution of this geometry to the density matrix $\ket{\psi} \bra{\psi}$ is proportional to the vacuum density matrix $\ket{0_{\text{CFT},L}} \bra{0_{\text{CFT},L}}$ where $\ket{0_{\text{CFT},L}} $ is the vacuum state of the matter CFT on a circle of length $L$.

The effective action of the conformal matter fields contains a divergent term proportional to and only depending on $L$ and the total number of boundaries. We remove this term by suitably rescaling our definition of $\ket{\psi}$. We see that the double disk geometry has the following effective action:
\begin{equation}
	S_{\text{eff, disk}} = -2 \phi_0 - \frac{2\pi \phi_r }{L} + \frac{c}{6} + \frac{c}{3} \ln 2 .
\end{equation}
This is equal to the sum of the topological term, the boundary action of JT gravity, and the CFT effective action.

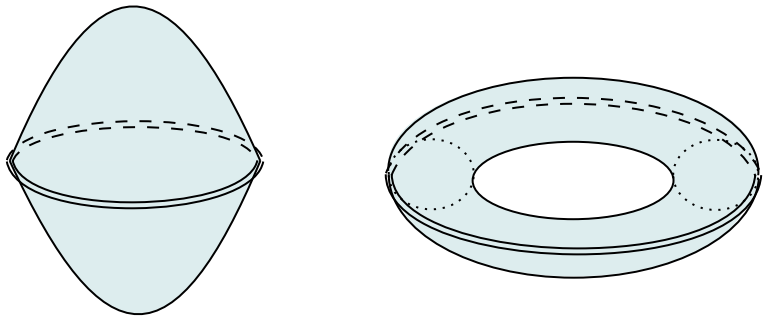
\begin{figure}[t]
	\begin{center}
    \input{figures/tikz/diskbkwh}
	\end{center}
	\caption{Picture of the disc and BKWH geometries in $\braket{\psi}{\psi}$. The meeting boundaries of the bra and the ket are denoted by a double line.}
    \label{fig:diskbkwh}
\end{figure}

However, there is another saddlepoint geometry of higher order in the topological expansion. When $L \gg 1$, we find a saddlepoint geometry which is topologically equivalent to an annulus connecting the two boundaries. It is given by
\begin{equation}
	ds^2 = \frac{dz^2 + dx^2}{\left( \frac{\beta}{\pi} \sin \frac{\pi  z}{\beta} \right)^2}, \quad
	\phi = \frac{c}{3} + \frac{c}{4} \frac{\frac{\pi}{2} - \frac{\pi  z}{\beta} }{\tan  \frac{\pi  z}{\beta} } ,\quad 0 < z <  \beta, \quad x = x+ L,
\end{equation}
where the periodicity in the $z$ direction is given by
\begin{equation}
	\beta = \frac{8 \phi_r}{c } .
\end{equation}
This geometry is a Euclidean wormhole connecting the bra and ket; thus we call it a bra-ket wormhole. We note that, due to the periodicity in Euclidean time, this geometry prepares a thermal state for the matter CFT; i.e. the contribution of this geometry to the density matrix $\ket{\psi} \bra{\psi}$ is proportional to the thermal density matrix $e^{-\beta H_{\text{CFT}, L}}$ where $H$ is the Hamiltonian of the matter CFT on a circle of length $L$. A picture of both the double disk and bra-ket wormhole geometries is given in figure~\ref{fig:diskbkwh}.

We remark that in the bra-ket wormhole geometry, the wormhole is sourced by the matter CFT's Casimir energy in the periodic $z$ direction. Thus, the inner product between the bra and the ket is essential, as it gives transparent boundary conditions for the matter CFT between the boundaries of the bra and the ket that allow periodicity of the matter CFT in the $z$ direction and thus the necessary Casimir energy. This mechanism is essentially equivalent to the generation of traversable wormholes from boundary interactions \cite{maldacena_eternal_2018, gao_traversable_2017}.

The effective action of the bra-ket wormhole geometry is given by
\begin{equation}
    S_{\text{eff, BKWH}} = - \frac{\pi c^2 L}{128 \phi_r} .
\end{equation}
Comparing with the effective action of the disk geometry, we see that $S_{\text{eff, BKWH}} < S_{\text{eff, disk}} $ when $L > L_{\text{crit}}$, where
\begin{equation}
    L_{\text{crit}} = \frac{256 \phi_0 \phi_r}{\pi c^2}.
\end{equation}
Thus, for $L > L_{\text{crit}}$, the bra-ket wormhole geometry dominates over the disk geometry in the gravitational path integral.

We see that the AdS JT+CFT prepared state has two distinct physical phases \cite{chen_bra-ket_2021, milekhin_bra-ket_2022}: the disk dominant phase at $L< L_{\text{crit}}$, where the inner product is dominated by the disk geometry, and the bra-ket wormhole dominant phase at $L> L_{\text{crit}}$, where the inner product is dominated by the bra-ket wormhole geometry.
We call the phase transition at $L= L_{\text{crit}}$ the bra-ket wormhole phase transition.

In the disk dominant phase, correlators of CFT field operators in the gravitationally prepared state are approximately given by
\begin{equation}
    \frac{\bra{\psi}O(x)O(0)\ket{\psi}}{\braket{\psi}{\psi}} \approx \bra{0_{\text{CFT},L}}O(x)O(0)\ket{0_{\text{CFT},L}}.
\end{equation}
In the bra-ket wormhole dominant phase, the same correlators are approximately given by
\begin{equation}
    \frac{\bra{\psi}O(x)O(0)\ket{\psi}}{\braket{\psi}{\psi}}  \approx \frac{\Tr \left( e^{-\beta H_{\text{CFT}, L}} O(x)O(0) \right)}{\Tr \left( e^{-\beta H_{\text{CFT}, L}} \right)}.
\end{equation}
Let $O(x)$ be a primary operator with conformal dimension $\Delta$. Then we have, for $|x| \gg 1$,
\begin{equation}
\label{eq:corrdepgps}
    \frac{\bra{\psi}O(x)O(0)\ket{\psi}}{\braket{\psi}{\psi}} \sim \begin{cases}
        |x|^{-2\Delta} & L < L_{\text{crit}}\\
        \exp(- \frac{ \pi c \Delta}{4 \phi_r} |x|) & L > L_{\text{crit}}.
    \end{cases}
\end{equation}
Thus the decaying behavior of correlators switches from a power law to exponential decay at the bra-ket wormhole phase transition.

We remark that, when $L>L_{\text{crit}}$, correlators and other observables in the state $\ket{\psi}$ may look like those observed in a mixed state, that is, the thermal state prepared by the bra-ket wormhole. However, $\ket{\psi}$ is still a pure state, although it is likely a highly disordered one such that it mimics the behavior of a mixed state.

The bra-ket wormhole is in fact very crucial to gravitationally prepared states. In \cite{chen_bra-ket_2021} it was argued that, assuming that the semiclassical approximation holds, the bra-ket wormhole should exist in order for the gravitationally prepared state to satisfy strong subadditivity of entropies of subsystems. Their argument extends to higher dimensions, and thus we expect bra-ket wormholes to exist generally for gravitationally prepared states of any dimension.

As an aside, one may wonder if, in the quantity $\braket{\psi}{\psi}^2$, one also has a similar ``bra-bra-ket-ket wormhole" saddlepoint geometry, where each bra or ket is connected to the other bra or ket by an annulus, or a ``bra-ket-bra-ket wormhole" saddlepoint geometry, where each bra is connected by an annulus to a ket in the other inner product. In both cases, it is easy to see that the Weyl contribution to the CFT stress tensor cancels out with the part of the CFT stress tensor coming from the Casimir effect, making it impossible for such saddlepoint geometries to exist. Thus we can expect that in quantities of the type $\braket{\psi}{\psi}^n$, the only saddlepoint geometries are collections of disks and bra-ket wormholes.

\subsection{Entropy bounding}

\label{ssec:gpsentropy}

We know that the parameter $\phi_0$ is equal to the black hole entropy in AdS JT gravity. Thus, it is an entropic parameter. In the context of cosmology, $\phi_0$ is related to the de Sitter entropy \cite{gibbons_cosmological_1977, arkani-hamed_measure_2007, Shaghoulian_2022} of dS JT gravity. The de Sitter entropy is defined as the log of the gravitational path integral on a sphere. For dS JT gravity, we can understand this as the log of the gravitational path integral over geometries topologically equivalent to the 2- sphere $S^2$.
\begin{equation}
    S_{\text{dS}} = \ln \left( \int_{\text{topologically }S^2} Dg D\phi \exp (-S[g])\right).
\end{equation}
Here $S[g]$ is the action of pure JT gravity with positive cosmological constant $\Lambda = 1$. Integrating out $\phi$ we see that the sole contribution to the path integral is that of the 2-sphere, and so
\begin{equation}
    S_{\text{dS}} = 2\phi_0 .
\end{equation}
Thus, $\phi_0$ is equal to half the de Sitter entropy of dS JT gravity.

Following the operational interpretation of \cite{arkani-hamed_measure_2007}, we may expect that, as a finite but large entropy value, $\phi_0$ enforces a bound on some entropy observables in the AdS JT+CFT prepared state. Indeed, we see that it does: the entanglement entropy of a connected spatial subregion, i.e. an interval, is bounded by
\begin{equation}
    S_{\text{max}} = 2 \phi_0 + C,
\end{equation}
where $C$ is a $\mathcal{O}(\phi_0^0)$ constant. More generally, for the union of $m$ mutually disconnected intervals, we see that the entanglement entropy is bounded by $m S_{\text{max}}$. In other words, for a general spatial subregion $R$, we have
\begin{equation}
    S[R] \leq \frac{|\partial R|}{2} S_{\text{max}},
\end{equation}
where $|\partial R|$ is the cardinality of $\partial R$ as a set of points. We remark that this bound takes the form of an area law in one spatial dimension, as $|\partial R|$ is a measure of the ``area" of $\partial R$. This is a surprising result, for in either the vacuum state or a thermal state of a 2D CFT, the entropy of an interval goes to infinity as the length of the interval goes to infinity.

Furthermore, we see that the $n$-Renyi entropy is also similarly bounded,
\begin{equation}
    S_n[R] \leq \frac{|\partial R|}{2} S_{\text{max}, n},
\end{equation}
where $S_{\text{max}, n} = 2 \phi_0 + C_n$ for some constant $C_n$.

\paragraph{Entanglement entropy bounding.} We will show the entanglement entropy bounding result given above. We will do this by computing the entanglement entropy of a spatial subregion using the island formula \cite{almheiri_page_2020, hartman_islands_2020}, which arises from quantum corrections to the Ryu-Takayanagi proposal \cite{ryu_holographic_2006, hubeny_covariant_2007, faulkner_quantum_2013, engelhardt_quantum_2015} and was originally studied in the context of the black hole information paradox \cite{almheiri_entropy_2019, almheiri_replica_2020, penington_replica_2020, penington_entanglement_2020}. In our context, the island formula involves finding an island region $I$ in the dominant saddlepoint geometry of $\braket{\psi}{\psi}$. The island formula is given by the minimum extremum value of the generalized entropy $S_{\text{gen}}$ of $R \cup I$, where $I$ can be any one-dimensional subset of the dominant saddlepoint geometry.
\begin{equation}
    S[R] = \min \Eext \left( S_{\text{gen}}[R \cup I] \right),
\end{equation}
\begin{equation}
    S_{\text{gen}}[R \cup I] = S_{\text{CFT}}[R \cup I] + \sum_{p \in \partial I} \left( \phi_0 +\phi_r(p) \right) .
\end{equation}
Here $S_{\text{CFT}}$ is the CFT entropy, and $\sum_{p \in \partial I} \left( \phi_0 +\phi_r(p) \right)$ is the analogue of the gravitational area term in JT gravity. Useful formulae for computing the CFT entropy are given in appendix~\ref{sec:cft}.

One possible choice of $I$ is the empty set, in which case
\begin{equation}
     S_{\text{gen}}[R \cup I] = S_{\text{CFT}}[R].
\end{equation}
We call this the islandless entropy. We note that $S_{\text{gen}}[R \cup I]$ contains a large $O(\phi_0)$ term,
\begin{equation}
    S_{\text{gen}}[R \cup I] \supset |\partial I| \phi_0.
\end{equation}
Thus, such choices of $I$ are suppressed in the island formula.

In the disk dominant phase, $L < L_{\text{crit}} \sim \mathcal{O}(\phi^0)$, and on the disk geometry, the CFT entropy of an interval is at most $\frac{c}{3} \ln \frac{L}{\pi} $. Thus, in the disk dominant phase, the CFT entropy of an interval is always $\mathcal{O}(\ln \phi_0)$ or lower. Furthermore, there are no nontrivial islands on the disk with $|\partial I| = 0$, and thus the islandless entropy always dominates. The entropy bound is satisfied without the need for nontrivial islands.

In the bra-ket wormhole dominant phase, nontrivial island configurations can dominate over the islandless entropy. Below we will continue to work in the bra-ket wormhole dominant phase.

We remark that in the bra-ket wormhole dominant phase, $I$ can be the entire width of the bra-ket wormhole, i.e. the subset $W$ defined by $z = \frac{\beta}{2}$. In this case, $I$ is nontrivial, but $|\partial I| = 0$. Generally, we have the following property: given a choice of island $I$, we have
\begin{equation}
    S_{\text{gen}}[I \cup R] = S_{\text{gen}} [I^c \cup R^c],
\end{equation}
where $I^c$ is a one-dimensional subset of the bra-ket wormhole geometry such that $I \cup I^c$ is homologous to $W$. For example, when $I = W$, we have
\begin{equation}
    S_{\text{gen}}[W \cup R] = S_{\text{CFT}} [ R^c].
\end{equation}
From the above property, we can see that the island formula agrees with the fact that the entanglement entropy of $R$ must be equal to that of its complement $R^c$, i.e. $S[R] = S[R^c]$.

Since short intervals increase $|\partial R|$ by $2$ but only increase $S[R]$ by $\mathcal{O}(\phi_0^0)$, it is enough to consider the case where each interval is long, i.e. $\gg 1$. By $S[R] = S[R^c]$, we can also assume the spacing between neighboring intervals to be large. We will work under this assumption below.

We consider the general case of $R$ consisting of $m$ disjoint intervals. We denote the intervals as $x_{2i-1} < x < x_{2i} $ with $i = 1 ... m$. Then, we find an extremal choice of $I$ given by the $n$ line segments
\begin{equation}
    z = \frac{\beta}{2} , \quad x_{2i-1} <x< x_{2i} .
\end{equation}
The orientation of $I$ relative to $R$ is chosen so that $I$ purifies the CFT entropy of $R$. We call this the maximally disjoint island configuration. We see that
\begin{equation}
	S_{\text{gen}}[I \cup R] = m S_{\text{max}},
\end{equation}
where
\begin{equation}
    S_{\text{max}} = 2 \phi_0 +  \frac{c}{2} + \frac{c}{3} \ln \frac{\beta}{\pi}.
\end{equation}
Thus, we see that $S[R] \leq  m S_{\text{max}}$.

The CFT entropy of a large interval of length $d$ on the bra-ket wormhole geometry is given by
\begin{equation}
    S_{\text{CFT}}[R] = \frac{\pi c d }{ 3\beta}.
\end{equation}
Thus the entropy bound is saturated when $d$ exceeds a critical interval length $d_{\text{crit}}$, given by
\begin{equation}
\label{eq:entropydcrit}
    d_{\text{crit}} = \frac{ 3\beta}{\pi c } S_{\text{max}} .
\end{equation}
We see that $d_{\text{crit}}$ is $O(\phi_0)$.

The maximally disjoint island configuration is the extremal island configuration with the largest number of disjoint intervals. The entropy bound is saturated if and only if the maximally disjoint island configuration is dominant. This occurs if and only if the lengths of all intervals and all spacings between neighboring intervals are longer than $d_{\text{crit}}$.

\begin{figure}[t]
    \begin{center}

    \input{figures/tikz/island}
    \end{center}
    \caption{Picture of a spatial subregion $R$ given by the union of two intervals $R_1$, $R_2$ and the dominant choice of island $I$ given by the union of two line segments $I_1$, $I_2$ atop the bra-ket wormhole geometry of the AdS JT+CFT prepared state. We assume the lengths of $R_1$, $R_2$ and the distances between them are larger than $d_{\text{crit}}$.}
    \label{fig:island}
\end{figure}
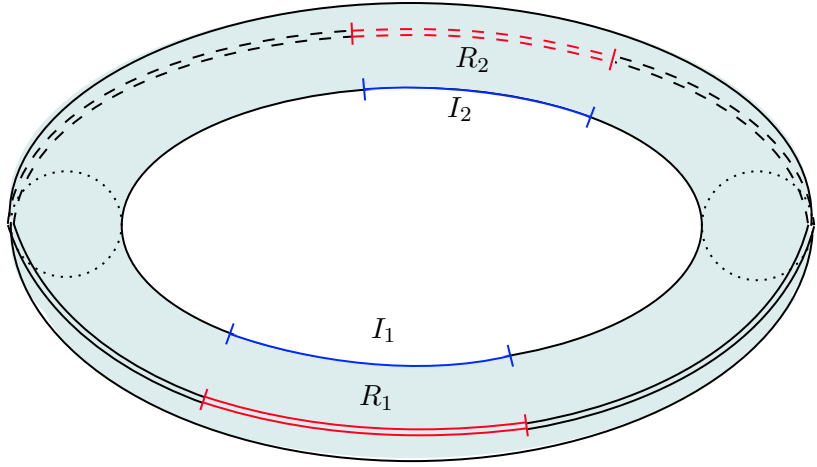

As an example, we consider the case where $R$ consists of $2$ intervals: $0 < x < d$ and $d +\ell < x < d + \ell + d^\prime $. We denote $ \ell^\prime = L -d -\ell - d^\prime $. The extremal choices for $I$ in this case are:
\begin{equation}
    \begin{cases}
        z = \frac{\beta}{2} , \quad 0 < x < d \quad \text{and} \quad z = \frac{\beta}{2} , \quad d +\ell < x < d + \ell + d^\prime  & (1)\\
        z = \frac{\beta}{2} , \quad 0 < x  < d + \ell + d^\prime   & (2)\\
        z = \frac{\beta}{2} , \quad  0 < x < d \text{ or }  d +\ell < x \leq L   & (3)\\
        z = \frac{\beta}{2} , \quad 0 < x  < d &(4)\\
        z = \frac{\beta}{2} , \quad d+\ell < x  < d + \ell + d^\prime& (5)\\
        z = \frac{\beta}{2} &(6)\\
        \varnothing & (7)
    \end{cases}
\end{equation}
(1) is the maximally disjoint island configuration, and (2), (3), (4), (5) are non-maximally disjoint island configurations. (6) is the $I = W$ case. We have
\begin{equation}
    S_{\text{gen}}[I \cup R] = \begin{cases}
        2 S_{\text{max}} & (1)\\
         S_{\text{max}} + \frac{\pi c \ell }{ 3\beta}  & (2)\\
         S_{\text{max}} + \frac{\pi c \ell^\prime }{ 3\beta}  & (3)\\
         S_{\text{max}} + \frac{\pi c d^\prime }{ 3\beta} &(4)\\
         S_{\text{max}} + \frac{\pi c d }{ 3\beta}&(5)\\
         \frac{\pi c (\ell + \ell^\prime ) }{ 3\beta} & (6)\\
         \frac{\pi c (d + d^\prime ) }{ 3\beta} & (7).
    \end{cases}
\end{equation}
Thus we have
\begin{equation}
	S [R] =\min \left(  \frac{\pi c \min(d+ d^\prime, \ell + \ell^\prime ) }{ 3\beta}  , \quad  S_{\text{max}} + \frac{\pi c \min(d, d^\prime, \ell,\ell^\prime) }{ 3\beta},  \quad 2S_{\text{max}} \right) ,
\end{equation}
and so the entropy bound is saturated when $d, d^\prime, \ell,\ell^\prime > d_{\text{crit}}$. A figure of the dominant choice of islands in this case is given in figure~\ref{fig:island}.

Finally, we note on how the island formula is derived. We follow the logic of \cite{almheiri_replica_2020, penington_replica_2020}. Assuming the replica trick, the entanglement entropy of $R$ can be computed as a limit of the $n$-Renyi entropy of $R$,
\begin{equation}
    S[R] = \lim_{n \rightarrow 1} S_n [R].
\end{equation}
The $n$-Renyi entropy of $R$ is given by
\begin{equation}
    S_n[R]= \frac{n \ln \Tr \rho_R - \ln \Tr (\rho_R^n)}{n-1} ,
\end{equation}
where $\rho_R = \Tr_{R^c} \ket{\psi} \bra{\psi}$ is the unnormalized density matrix of $R$. The existence of nontrivial islands can be argued from the existence of semiclassical connected $2n$-boundary geometry contributions in $\Tr (\rho_R^n)$. These geometries connect together $n$ bra-ket wormholes in a way that essentially creates an island-like region in the bra-ket wormhole geometry, purifying the CFT entropy of $R$. We call these geometries $n$-Renyi replica wormholes.

We remark that, because the entropy bound for a spatial subregion in the AdS JT+CFT prepared state comes from the effect of $n$-Renyi replica wormholes, which are higher topology contributions in the gravitational path integral, the entropy bound is a nonperturbative quantum gravitational effect.

\paragraph{$n$-Renyi entropy bounding.} We shall now show that $n$-Renyi replica wormholes imply a bound for $S_n[R]$ in the same form as $S[R]$, namely
\begin{equation}
    S_n[R] \leq \frac{|\partial R|}{2} S_{\text{max}, n},
\end{equation}
where $S_{\text{max}, n} = 2 \phi_0 + C_n$ for some constant $C_n$.

To see this, we consider the case where $R$ consists of $m$ disjoint intervals. In $\Tr (\rho_R^n)$, there are $n$ bra boundaries $\Sigma_{\text{bra}, i}$ and $n$ ket boundaries $\Sigma_{\text{ket}, j}$, for $j = 1, ..., n$. In $R^c$, the matter CFT on $\Sigma_{\text{bra}, j}$ is identified with the matter CFT on $\Sigma_{\text{ket}, j}$. In $R$,  the matter CFT on $\Sigma_{\text{bra}, j}$ is identified with the matter CFT on $\Sigma_{\text{ket}, j+1}$, where we take $\Sigma_{\text{ket}, n+1} = \Sigma_{\text{ket}, 1}$.

We consider the following $n$-Renyi replica wormhole geometry. In $R$, far from $\partial R$, we locally connect $\Sigma_{\text{bra}, j}$ with $\Sigma_{\text{ket}, j+1}$ with the bra-ket wormhole geometry
\begin{equation}
	ds^2 = \frac{dz^2 + dx^2}{\left( \frac{\beta}{\pi} \sin \frac{\pi  z}{\beta} \right)^2}, \quad
	\phi = \frac{c}{3} + \frac{c}{4} \frac{\frac{\pi}{2} - \frac{\pi  z}{\beta} }{\tan  \frac{\pi  z}{\beta} } ,\quad 0 < z <  \beta,
\end{equation}
while in $R^c$, far from $\partial R$, we locally connect $\Sigma_{\text{bra}, j}$ with $\Sigma_{\text{ket}, j}$ with the bra-ket wormhole geometry. Near $\partial R$, we need some interpolating geometry. For our purposes we don't need to know the metric of the interpolating geometry exactly. We call this the maximally disjoint $n$-Renyi replica wormhole geometry.

\begin{figure}[t]
    \begin{center}
    \input{figures/tikz/nrenyiwh}
    \end{center}
    \caption{Picture of the maximally disjoint $2$-Renyi replica wormhole geometry related to the spatial subregion $R = R_1 \cup R_2$ in the AdS JT+CFT prepared state. When the lengths of $R_1$, $R_2$ and the distances between them are larger than $d_{\text{crit},n}$, this geometry becomes the dominant contribution to $\Tr (\rho_R^n)$.}
    \label{fig:nrenyiwh}
\end{figure}
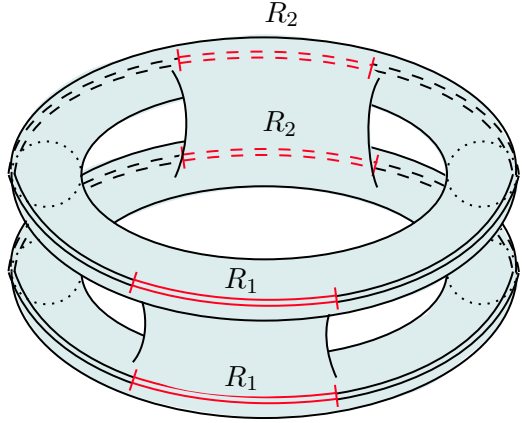

By its construction, the effective action of this geometry is equal to the effective action of $n$ bra-ket wormholes with additional terms localized at $\partial R$, i.e. near the endpoints of the intervals. Thus, we see that the maximally disjoint $n$-Renyi replica wormhole geometry contributes
\begin{equation}
   (\text{$n$-Renyi replica wormhole geometry}) \approx (D_n)^m e^{-2(n-1)m \phi_0} \braket{\psi}{\psi}^n.
\end{equation}
Here $D_n$ is a $\mathcal{O}(\phi_0^0)$ constant, and the factor of $e^{-2(n-1) m \phi_0}$ is the topological suppression of the $n$-Renyi replica wormhole relative to the geometry of $n$ bra-ket wormholes. As a result, we get
\begin{equation}
    S_n[R]= \frac{n \ln \Tr \rho_R - \ln \Tr (\rho_R^n)}{n-1} \leq m S_{\text{max},n},
\end{equation}
where
\begin{equation}
    S_{\text{max},n}= 2\phi_0 - \frac{\ln D_n}{n-1}.
\end{equation}
This gives us the entropy bound we stated above.

In the absence of $n$-Renyi replica wormholes, the Renyi entropy of a large interval $d$ is equal to that of the CFT $n$-Renyi entropy on the bra-ket wormhole geometry, given by
\begin{equation}
    S_{\text{CFT}, n}[R] = \frac{\pi (n+1) c d }{ 6n \beta},
\end{equation}
plus an $\mathcal{O}(\phi_0^0)$ constant $B_n$ coming from deformations of the bra-ket wormhole geometry near $\partial R$. Thus the entropy bound is saturated when $d$ exceeds a critical interval length $d_{\text{crit}, n}$, given by
\begin{equation}
    d_{\text{crit},n} = \frac{ 6n \beta}{\pi (n+1) c } (S_{\text{max}, n} - B_n) .
\end{equation}
We see that $d_{\text{crit}, n}$ is $\mathcal{O}(\phi_0)$.

The maximally disjoint $n$-Renyi replica wormhole geometry is the $n$-Renyi replica wormhole geometry with the largest Euler characteristic. The entropy bound is saturated if and only if the maximally disjoint island configuration is dominant. This occurs if and only if the lengths of all intervals and all spacings between neighboring intervals are longer than $d_{\text{crit}, n}$.

\subsection{A factorization puzzle}

\label{ssec:gpsfactorization}

We remark that in the AdS JT+CFT gravitational prepared state, both the bra-ket wormhole phase transition and entropy bounding are due to replica effects, i.e. the contribution of wormholes connecting different copies of the state $\ket{\psi}$. However, these replica effects are related to a factorization puzzle \cite{harlow_wormholes_2016, guica_construction_2017}: it seems that, apparently, $\ket{\psi}$ does not factorize:
\begin{equation}
    (\ket{\psi} \ket{\psi}) \neq (\ket{\psi})(\ket{\psi}) ?
\end{equation}
because replica effects are seemingly included in the left hand side of the equation but not in the right hand side.

Naively, one may suspect that, on the basis of factorization, replica effects must be ruled out. However, as we will show below, replica effects need not be ruled out to restore factorization. Furthermore, replica effects have proven to be necessary in the study of quantum gravity. One such context is the quantum chaos of black holes. To reproduce the ramp behavior of the spectral form factor of a black hole at late times, one needs to consider a double cone geometry, which is a replica effect \cite{saad_semiclassical_2019, chen_comments_2023}. Another context is black hole evaporation. In a holographic setting, we can compute the entanglement entropy of Hawking radiation, and obtaining the exact value, which agrees with information conservation, crucially depends on the inclusion of replica effects \cite{almheiri_replica_2020, penington_replica_2020}.

In many of the situations where replica effects are relevant, such as the cases given above, replica effects are related to quantum chaos or disordered systems \cite{cotler_black_2016, cotler_chaos_2017, witten_matrix_2020, maxfield_path_2021, jafferis_jt_2023}. Furthermore, ensemble models provide a natural setting in which factorization can be restored in the existence of replica effects \cite{cotler_chaos_2017, witten_matrix_2020, maxfield_path_2021, jafferis_jt_2023, hernandez-cuenca_wormholes_2024}.

Motivated by this, we propose a resolution to the factorization problem of gravitationally prepared states by considering random state models. In a random state model, given the Hilbert space $\mathcal{H}_{\text{QFT}}$ of states of the matter field on $\Sigma$, we define a specific ensemble of kets in $\mathcal{H}_{\text{QFT}}$, and take $\ket{\psi}$ to be a typical selection in this ensemble. Then, for any quantity involving $\ket{\psi}$, the gravitational path integral computes the ensemble average. The factorization problem is resolved, as we know that, generally, $\left< \ket{\psi} \ket{\psi} \right> \neq \left<\ket{\psi}\right> \left< \ket{\psi} \right> $.

\paragraph{$n$-replica effects.} In the random state model, we define the $n$-th cumulant of $\ket{\psi}$ as
\begin{equation}
    \left<  \ket{\psi}^{\otimes n} \right>_{\text{conn}} = \partial_J^n \ln \left< \exp (J \ket{\psi}) \right>|_{J=0} ,
\end{equation}
where the right hand side is a formal power series in $J$. For example, we have
\begin{equation}
	\begin{aligned}
		\left<  \ket{\psi}^{\otimes 1} \right>_{\text{conn}} =&
		\left< \ket{\psi} \right> ,\\
		\left<  \ket{\psi}^{\otimes 2} \right>_{\text{conn}} =& \left< \ket{\psi}^{\otimes 2}  \right> - \left< \ket{\psi} \right>^{\otimes 2} , \\
		\left<  \ket{\psi}^{\otimes 3} \right>_{\text{conn}} =& \left< \ket{\psi}^{\otimes 3} \right> - \left(\left< \ket{\psi} \right> \otimes \left< \ket{\psi}^{\otimes 2} \right> + \text{permutations} \right) + 2  \left< \ket{\psi} \right>^{\otimes 3}, \\
		\left<  \ket{\psi}^{\otimes 4} \right>_{\text{conn}} =& \left< \ket{\psi}^{\otimes 4} \right> - \left(\left< \ket{\psi} \right> \otimes \left< \ket{\psi}^{\otimes 3} \right> + \text{permutations} \right) \\
		& - \left(\left< \ket{\psi}^{\otimes 2} \right> \otimes \left< \ket{\psi}^{\otimes 2} \right> + \text{permutations} \right)\\
		&+2  \left(\left< \ket{\psi} \right>^{\otimes 2} \otimes \left< \ket{\psi}^{\otimes 2} \right> + \text{permutations} \right) -6  \left< \ket{\psi} \right>^{\otimes 4}.
	\end{aligned}
\end{equation}
The $n$-th cumulant extracts the connected part of $\left<  \ket{\psi}^{\otimes n} \right>$. It follows that the $n$-th cumulant of $\ket{\psi}$ is equal to the sum of connected $n$-boundary geometries in the gravitational path integral. We call contributions to this cumulant $n$-replica effects.

We introduce the following notation for random state models. Given any tensorial quantity $T$ including $n$ copies of $\ket{\psi}$, we denote the contribution of the $n$-th cumulant of $\ket{\psi}$ to $T$ as $Z_n [T]$. For example,
\begin{equation}
    Z_1 [\ket{\psi}] = \left< \ket{\psi} \right>, \quad Z_2[ \braket{\psi}{\psi}] = \left< \braket{\psi}{\psi} \right> -\left< \bra{\psi} \right> \left< \ket{\psi} \right> .
\end{equation}
Given any tensorial quantity $T$ involving $\ket{\psi}$, we also define the bra-ket disconnected average $\left< T \right>_{\text{disc}}$, where all cumulants connecting bras and kets are dropped. For example,
\begin{equation}
    \left< \braket{\psi}{\psi}^2 \right>_{\text{disc}} = \left< \bra{\psi}^{\otimes 2} \right> \left< \ket{\psi}^{\otimes 2} \right> .
\end{equation}
In comparison to bra-ket disconnected averages, we may sometimes call ordinary averages ``full'', because they contain the full set of replica effects contributing to the average. We can use the same notation in general gravitationally prepared states, where $Z_n[T]$ denotes the contribution of connected $n$-boundary geometries to $T$, and $ \left< T\right>_{\text{disc}}$ denotes the contribution of all geometries that do not connect bras with kets to $T$.

\paragraph{Clarification on entropy bounds.} We now see that, in the framework of random state models of gravitationally prepared states, the entropies computed in \ref{ssec:gpsentropy} are to be understood as ensemble averaged entropies, and thus the following entropy bounds are to be understood as bounds of averaged entropies.

However, we may also consider bounds for entropies computed from individual samples $\ket{\psi}$. To distinguish between bounds of averaged entropies and bounds of sampled entropies, we will use the terms \textit{average entropy bound} and \textit{sample entropy bound}, respectively. We see that entropy bounds computed with the island formula on the gravity side correspond to average entropy bounds on the random state model side.

\section{RMPS models of gravitationally prepared states}

\label{sec:rmps}

\subsection{Definition}

\label{ssec:rmpsdefinition}

We would like to define random state models of two-dimensional gravitationally prepared states.

We first make some simplifying assumptions about $\Sigma$ and $\mathcal{H}_{\text{QFT}, \Sigma}$. To start, we take $\Sigma$ to be a discretized circle of length $L$,
\begin{equation}
    a = 0, ..., L-1, \quad a \approx a+L.
\end{equation}
In other words, $\Sigma$ is a loop of $L$ points. We will study models that take $\Sigma$ to be continuous later.

Each point $a$ in $\Sigma$ has an associated local Hilbert space $\mathcal{H}_a$. We take $\mathcal{H}_a$ to be finite-dimensional,
\begin{equation}
    \dim \mathcal{H}_a = k.
\end{equation}
Then, $\mathcal{H}_a$ has a finite set of basis states, $\ket{1}, ..., \ket{k}$. As a result, we can write the Hilbert space of the matter fields on $\Sigma$ as
\begin{equation}
    \mathcal{H}_{\text{QFT}, \Sigma} = \prod^{L}_{a=1} \mathcal{H}_a = \mathbb{C}^{k^L}.
\end{equation}
In other words, the matter degrees of freedom on $\Sigma$ are given by a lattice of $k$-level systems.

We consider random state models of the following type. Consider $k$ random $N \times N$ Hermitian matrices,
\begin{equation}
    \{A_i,\text{ }i = 1,..., k\}.
\end{equation}
These matrices are selected from a Hermitian ensemble of the form
\begin{equation}
\label{eq:probdist}
    d\mu(\{A_i\}) \propto d \{A_i\}  \exp \left(-N \Tr \left( V(\{A_i\} ) \right) \right) ,
\end{equation}
where $d\{A_i\} $ is the standard flat measure of $\{A_i\}$, and $V(\{A_i\} )$ is a positive function of $\{A_i\}$. In $\mathcal{H}_{\text{QFT}, \Sigma}$, we define the following state:
\begin{equation}
    \ket{\psi} =  \sum^k_{i_{L-1} = 1}... \sum^k_{i_0=1} \Tr \left( A_{i_{L-1}} ... A_{i_0} \right)\ket{i_{L-1} ... i_0}.
\end{equation}
$\ket{\psi}$ is a random matrix product state of bond dimension $N$. Expositions on matrix product states can be found in \cite{Perez-Garcia:2006nqo, verstraete_matrix_2008, Orus:2013kga, Bridgeman:2016dhh}.

We remark on the rationale for the $N$-dependence of the exponent in the probability distribution eq.~\eqref{eq:probdist}. The $N$-dependence given here allows single trace observables of the type $\Tr (f(\{A_i\}))$, where $f$ is some function of $\{A_i\}$, to have a topological expansion in $N$ \cite{hooft_planar_1974, borot_asymptotic_2013}. Furthermore, it ensures that the matrices $\{A_i\}$ have eigenvalues of $\mathcal{O}(1)$ and that these eigenvalues form approximately continuous spectra with gaps between nearest eigenvalues being $\mathcal{O}(N^{-1})$.

In section~\ref{ssec:rmpsmotivation} we give three physical arguments for why such a random matrix product state (RMPS) is a good candidate for a toy model of a gravitationally prepared state. In the remainder of this subsection we will summarize basic parameter matching results that can be obtained from those arguments, and address basic questions about the continuum limit and the unitary condition.

\paragraph{Identification of parameters.} Through the three arguments of section~\ref{ssec:rmpsmotivation} we arrive at the same identification
\begin{equation}
    \label{eq:parid}
    N = e^{\phi_0},
\end{equation}
which we will use as the canonical identification when comparing results in RMPS to 2D gravity. We remark that, in gravity, we take $\phi_0$ to be large. Thus, in RMPS models of gravitationally prepared states, $N$ must be very large.

In the RMPS model, $L$ corresponds to the spatial size of the universe like the parameter of the same name in the JT+CFT model. It does not need to be additionally scaled to match the JT+CFT model.

The factor $\ln k$ corresponds to the number of matter degrees of freedom per unit length in the RMPS model. In this sense, we conjecture that when comparing to JT+CFT gravitationally prepared state calculations, it should be proportional to the CFT central charge $c$. Thus, when we are studying RMPS as a model of the JT+CFT gravitationally prepared state where we do not necessarily scale $c$ with $\phi_0$, it is not necessary to send $k$ to infinity. However, in the context of double holography, as introduced in \cite{almheiri_page_2020}, $c$ is scaled such that $c/\phi_0$ remains constant. In this case, one would scale $k$ such that $\ln k / \ln N$ remains constant.

In this paper, we are considering the dependence of physical quantities of RMPS as we change the parameters $k$ and $N$, in particular considering the large $N$ expansion. Thus we are working in the regime of large but finite $N$. In this paper, we do not take a double-scaling limit of $k$ and $N$, instead only taking $N$ to be large following the parameter identification in eq.~\eqref{eq:parid}. However, taking $k$ to infinity is useful for studying properties of the model analytically, as we shall see in section~\ref{ssec:rmpsok}.

\paragraph{Continuum limit.} RMPS models are defined on discrete space. In section~\ref{sec:crmps}, we construct continuum models of RMPS. However, as we explain in that section, those models do not match the short-distance behavior of relativistic quantum fields. Thus, the question of constructing continuum limits of RMPS models such that short-distance behaviors of relativistic matter fields are reproduced remains an open question. In that regard, RMPS models should be understood as toy models of gravitationally prepared states.

\paragraph{Remark on the unitary condition.} We remark that random matrix product state models were studied outside of the context of gravitationally prepared states in the literature \cite{garnerone_typicality_2010, Garnerone:2010lqy}. However, the models considered were generally those where the matrices $\{A_i\}$ are non-Hermitian and $\sum_i A_i A_i^\dagger = I_{N \times N}$. This condition is called the unitary condition, and ensures that $\braket{\psi}{\psi} \rightarrow 1$ in the large $L$ limit.

However, for gravitationally prepared states, the unitary condition is unphysical. This is because the inner product $\braket{\psi}{\psi}$ must fluctuate even in the large $L$ limit for the RMPS model to be consistent with off-shell wormhole physics. This will be explained in detail in section~\ref{sec:offshell}.

\subsection{Motivation of the RMPS model}

\label{ssec:rmpsmotivation}

Here we give three arguments that suggest that RMPS is a toy model of a two-dimensional gravitationally prepared state.

\paragraph{Topological expansion.} First, as a single-trace observable of the matrices $\{A_i\} $, $\ket{\psi}$ has a topological expansion in terms of $N$. Contributions to $\left< \ket{\psi}^{\otimes n} \right>$ can be grouped in terms of topologies of 2D surfaces connecting the $n$ boundaries. The contribution associated with a connected $n$-boundary surface of genus $g$ is proportional to $N^{2-2g-n}$. The topological expansion of RMPS matches that of 2D gravity if we relate $N$ to the parameter $\phi_0$ by $ N = e^{\phi_0}$.

We explain the topological expansion in further detail. For convenience, we first consider the case where $V(\{A_i\} )$ contains a quadratic term. In this case, the contributions to $\left< \ket{\psi}^{\otimes n} \right>$ can be written as 't Hooft diagrams. Each 't Hooft diagram is understood as a discretized 2D surface, and the topological expansion of $\left< \ket{\psi}^{\otimes n} \right>$ is an expansion in terms of the Euler characteristic of said surface.

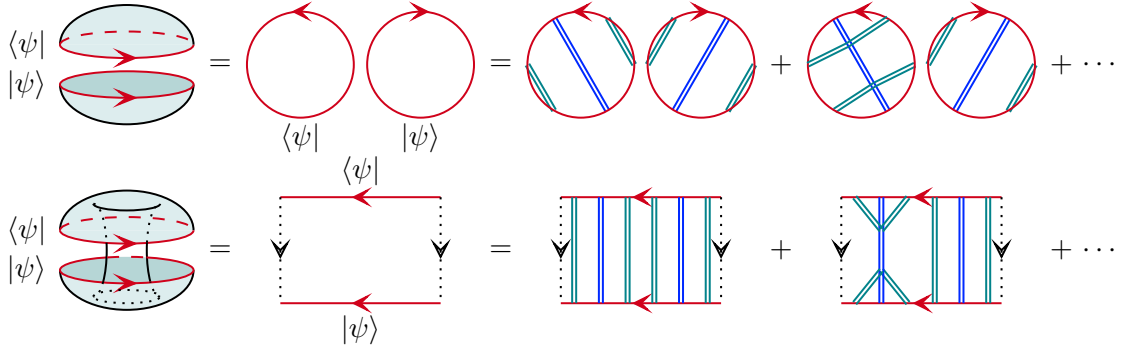
\begin{figure}[t]
    \begin{center}
    \input{figures/tikz/discgeo}
    \end{center}
    \caption{Top: a contribution to the RMPS inner product $\braket{\psi}{\psi}$ in the double disk topology. Bottom: a contribution to the inner product $\braket{\psi}{\psi}$ in the bra-ket wormhole topology. Here $L=6$, $k=2$, and $V(A_1,A_2) = A_1^2 + A_2^2 + A_1^2 A_2^2$. We use 't Hooft's double line notation. The matrix label index $i=1,2$ is indicated by blue and turquoise double lines, respectively.}
    \label{fig:discgeo}
\end{figure}

Furthermore, one can view the index labeling the matrices, $i = 1, ..., k$, as a discrete valued field that lives on the 't Hooft diagram. This is done by denoting the $A_i$ to $A_j$ propagator as a line that has the field value $i$ on its left half and the field value $j$ on its right half. Then, each 't Hooft diagram gives a discretized 2D surface along with a discrete valued field configuration. In computing a specific coefficient $\braket{i_1 ... i_{N_a}}{\psi}$ of $\ket{\psi}$, we are choosing a boundary condition for the discrete valued field on $\Sigma$. The sum over 't Hooft diagrams is a function of such boundary conditions, which gives us the wavefunction of $\ket{\psi}$. See figure~\ref{fig:discgeo}.

Thus, we can interpret $\ket{\psi}$ as a toy model of a gravitationally prepared state, and quantities involving $\ket{\psi}$ as toy models of gravitational path integrals.

Generally, $V(\{A_i\})$ may not contain a quadratic term. In this case, there is still a topological expansion in $\left< \ket{\psi}^{\otimes n} \right>$, but one that cannot be represented in terms of 't Hooft diagrams \cite{borot_asymptotic_2013}. We expect that there is some analogous sum over surfaces and field configurations that is occurring in this scenario.

\paragraph{Entropy bounding.} Second, as a matrix product state with bond dimension $N$, $\ket{\psi}$ has the following entropy bounding property: given the spatial subregion $R$ consisting of $m$ disjoint intervals, i.e. $m$ sequences of one or more successive integers with at least one integer in between two neighboring sequences, we have
\begin{equation}
    S[R] \leq 2m \ln N.
\end{equation}
This can be viewed as an area law, $S[R] \leq |\partial R| \ln N$. The same bounding property holds for the $n$-Renyi entropy, $S_n[R] \leq |\partial R| \ln N$.

We remark that the entropy bounding property described above is a sample entropy bound. It implies an average entropy bound. At large $N$, the entropy bounding property of RMPS matches that of 2D gravity if we relate $N$ to the parameter $\phi_0$ by $ N = e^{\phi_0}$.

We explain the argument for this entropy bounding behavior. Consider the spatial subregion $R$ consisting of $m$ disjoint intervals. We denote the number of points in the $i$-th interval as $d_i$ and the number of points in between the $i$-th and $i+1$-th intervals as $\ell_i$. Without loss of generality, we can assume the first interval starts at $x= 0$.

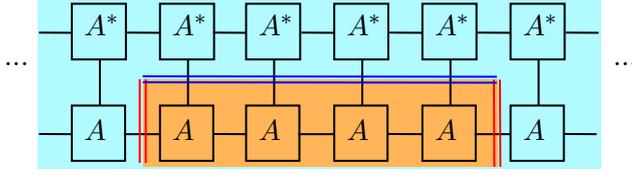
\begin{figure}[t]
    \begin{center}
    \input{figures/tikz/tenentropy}
    \end{center}
    \caption{Tensor diagram for a spatial subregion $R$ containing the points $x= 0,1,2,3$ in the inner product $\braket{\psi}{\psi}$. The matrix $P_R$ is represented in orange and the matrix $Q_R$ is represented in blue. Cutting along the horizontal blue line gives $\rho_R = P_R Q_R$, and cutting along the vertical red lines give $\rho_{R,\text{compressed}} = Q_R P_R$, which is an $N^2 \times N^2$ matrix, implying the entropy bound $S[R] \leq 2 \ln N$.}
    \label{fig:tenentropy}
\end{figure}

The reduced density matrix of $R$ is given by
\begin{equation}
    \begin{aligned}
        \rho_R =&  \left( \sum^k_{i_{L-1} = 1} ... \sum^k_{i_0=1} \right)\left(  \sum^k_{j_{L-\ell_m -1} = 1} ... \sum^k_{j_{L-\ell_m -d_m-1}=1}  \right) ...  \left( \sum^k_{j_{d_1-1} = 1} ... \sum^k_{j_0=1} \right)  \\
        &\times\Tr \left( A^*_{i_{d_1-1}} ... A^*_{i_0} \right)\\
        &\times\Tr \left( A_{i_{L-1 }} ... A_{i_{L-\ell_m}} A_{j_{L-\ell_m -1}}... A_{j_{L-\ell_m -d_m-1}} ... A_{i_{d_1 +\ell_1 -1 }} ... A_{i_{d_1}} A_{j_{d_1-1}} ... A_{j_0} \right)\\
        &\times\left(\ket{j_{L-\ell_m -1}...j_{L-\ell_m -d_m-1}} \otimes  ... \otimes \ket{j_{d_1-1} ... j_0} \right)\\
        &\times\left(\bra{i_{L-\ell_m -1}...i_{L-\ell_m -d_m-1}} \otimes  ... \otimes \bra{i_{d_1-1} ... i_0} \right),
    \end{aligned}
\end{equation}
so that $\rho_R$ is a $k^{d_1 + ... + d_m} \times k^{d_1 + ... + d_m} $ matrix. We note that we can write $\rho_R$ as a product of a $k^{d_1 + ... + d_m} \times N^{2m} $ matrix $P_R$ with a $N^{2m} \times k^{d_1 + ... + d_m}$ matrix $Q_R$, given by
\begin{equation}
    \begin{aligned}
        P_R =& \left( \sum^N_{\mu_m = 1}... \sum^N_{\mu_1 = 1} \right)\left( \sum^N_{\nu_m = 1}... \sum^N_{\nu_1 = 1} \right)\\
        &\times\left(  \sum^k_{j_{L-\ell_m -1} = 1} ... \sum^k_{j_{L-\ell_m -d_m-1}=1}  \right) ...  \left( \sum^k_{j_{d_1-1} = 1} ... \sum^k_{j_0=1} \right) \\
        &\times\bra{\mu_m} A_{j_{L-\ell_m -1}}... A_{j_{L-\ell_m -d_m-1}} \ket{\nu_m}  ... \bra{\mu_1} A_{j_{d_1-1}} ... A_{j_0} \ket{\nu_1} \\
        &\times\left(\ket{j_{L-\ell_m -1}...j_{L-\ell_m -d_m-1}} \otimes  ... \otimes \ket{j_{d_1-1} ... j_0} \right) \left(\bra{\mu_m\nu_m} \otimes  ... \otimes \bra{\mu_1\nu_1} \right),
    \end{aligned}
\end{equation}
\begin{equation}
    \begin{aligned}
        Q_R =&\left( \sum^N_{\mu^\prime_m = 1}... \sum^N_{\mu^\prime_1 = 1} \right)  \left( \sum^N_{\nu^\prime_m = 1}... \sum^N_{\nu^\prime_1 = 1} \right) \left( \sum^k_{i_{L-1} = 1} ... \sum^k_{i_0=1} \right)\\
        &\times\Tr \left( A^*_{i_{d_1-1}} ... A^*_{i_0} \right) \bra{\nu^\prime_1} A_{i_{L-1}}... A_{i_{L-\ell_m}} \ket{\mu^\prime_m}  ... \bra{\nu^\prime_2} A_{i_{d_1 +\ell_1 -1}} ... A_{i_{d_1}} \ket{\mu^\prime_1}\\
        &\times \left(\ket{\mu^\prime_m\nu^\prime_m} \otimes  ... \otimes \ket{\mu^\prime_1\nu^\prime_1} \right) \left(\bra{i_{L-\ell_m -1}...i_{L-\ell_m -d_m-1}} \otimes  ... \otimes \bra{i_{d_1-1} ... i_0} \right).
    \end{aligned}
\end{equation}
By linear algebra, the matrices $P_RQ_R$ and $Q_RP_R$ have the same nonzero eigenvalues. Thus, we can use $Q_RP_R$ in place of $\rho_R$ when computing the entanglement or $n$-Renyi entropy. Simplifying, we get
\begin{equation}
    \begin{aligned}
        &\left(\bra{\mu^\prime_m\nu^\prime_m} \otimes  ... \otimes \bra{\mu^\prime_1\nu^\prime_1} \right)  Q_RP_R\left(\ket{\mu_m\nu_m} \otimes  ... \otimes \ket{\mu_1\nu_1} \right) \\
        &= \begin{aligned}[t]\Tr \bigg( (I_{N\times N} \otimes \ket{\nu_1}  \bra{\nu^\prime_1}) \bigg( \sum^k_{i=1} A_i^* \otimes A_i \bigg)^{\ell_m} (I_{N\times N} \otimes \ket{\mu^\prime_m}  \bra{\mu_m}) \bigg( \sum^k_{i=1} A_i^* \otimes A_i \bigg) ^{d_m}&\\
        \dots (I_{N\times N} \otimes \ket{\nu_2}  \bra{\nu^\prime_2}) \bigg( \sum^k_{i=1} A_i^* \otimes A_i \bigg) ^{\ell_1} (I_{N\times N} \otimes \ket{\mu^\prime_1}  \bra{\mu_1}) \bigg( \sum^k_{i=1} A_i^* \otimes A_i \bigg) ^{d_1}  &\bigg).\end{aligned}
    \end{aligned}
\end{equation}
We call this matrix the compressed density matrix $\rho_{R, \text{compressed}}$ of $R$. We note that $\rho_{R, \text{compressed}}$ is generally not Hermitian but has nonnegative real eigenvalues. The fact that $\rho_{R, \text{compressed}}$ is a $N^{2m} \times N^{2m}$ matrix with real eigenvalues implies the entropy bound. This argument can be represented more clearly as a tensor diagram, given in figure~\ref{fig:tenentropy}.

\paragraph{Holographic reasoning.} We also give a holographic reasoning for modeling two-dimensional gravitationally prepared states with random matrix product states.

Consider the AdS JT+CFT prepared state. We reinterpret the spatial dimension of $\Sigma$ as a Euclidean time dimension. Then, the gravitational path integral defining the gravitationally prepared state becomes like the thermal partition function of an AdS black hole with CFT matter. The inverse temperature is given by $L$.

Note that we have to give some boundary conditions for the CFT matter at $\Sigma$. To keep coherence with the thermal partition function interpretation, we must take the inner product of $\ket{\psi}$ with some translation invariant state. Two reasonable options are the CFT ground state $\ket{0_{\text{CFT}}}$ and $\ket{\psi}$ itself.

In the first case, we see that $\braket{0_{\text{CFT}}}{\psi}$ is equal to the thermal partition function of an AdS black hole exchanging CFT matter with a flat half-space bath. In the second case, we see that $\braket{\psi}{\psi}$ is equal to the thermal partition function of two AdS black holes exchanging CFT matter.

In any case, by the holographic principle \cite{susskind_world_1995, maldacena_large_1999, witten_anti_1998}, we can think of the AdS black holes as being holographically dual to a quantum mechanical system with Hilbert space dimension $\sim e^{\phi_0}$. Thus, we can replace $\ket{\psi}$ with the 0+1-dimensional path integral of a quantum mechanical system with Hilbert space dimension $\sim e^{\phi_0}$.

Note that this quantum mechanical system must still interact with the CFT matter in the bra, so the Hamiltonian of the quantum mechanical system will depend on the value of the conformal fields at each point along $\Sigma$. Discretizing $\Sigma$ and discretizing the value of the conformal fields at each point in $\Sigma$, we get a matrix product state structure.

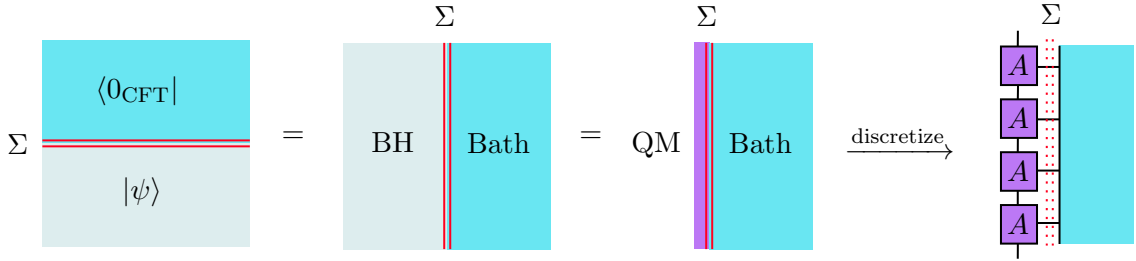
\begin{figure}[t]
    \begin{center}
    \input{figures/tikz/holographic}
    \end{center}
    \caption{Holographic argument for representing the AdS JT+CFT prepared state as a matrix product state. From left to right: we reinterpret $\Sigma$ as a Euclidean time dimension; we apply the central dogma to get a holographic dual quantum mechanics system; we discretize $\Sigma$ to get a matrix product state structure.}
    \label{fig:holographic}
\end{figure}

In this matrix product state, defined with the matrices $\{A_i\}$, the $i = 1, ..., k$ index represents the value of the conformal fields at each point and the matrix $\{A_i\}$ represents a Euclidean transition matrix for the quantum mechanical system. Thus, the bond dimension $N$ of the MPS must be equal to the Hilbert space dimension of the quantum mechanical system, which is $\sim e^{\phi_0}$. See figure~\ref{fig:holographic}.

In black hole physics, we know that black holes are chaotic quantum systems, and that their Hamiltonians can be approximated by random matrices \cite{saad_jt_2019, maxfield_path_2021, jafferis_jt_2023}. Thus, it makes sense for us to take the matrices $\{A_i\}$ to be random, and so we get an RMPS model of a gravitationally prepared state.

We also remark that the holographic argument given above gives an argument for modeling two dimensional gravitationally prepared states but also an argument for the entropy bounding behavior of gravitationally prepared states. Together with the strong subadditivity argument in \cite{chen_bra-ket_2021}, we get an argument for the universality of the bra-ket wormhole phase transition and entropy bounding in gravitationally prepared states.

For a further discussion of the holographic interpretation of RMPS models, see appendix~\ref{sec:holography}.

\subsection{Transfer matrix technology}

\label{ssec:rmpstransfer}

We now discuss methods of computing various observables in RMPS, all involving the concept of the transfer matrix. These ideas are used generally in the study of matrix product states \cite{verstraete_matrix_2008, garnerone_typicality_2010, Garnerone:2010lqy}. The transfer matrix will be essential to understanding various properties of the RMPS in comparison to gravity.

The inner product $\braket{\psi}{\psi}$ can be written as
\begin{equation}
     \braket{\psi}{\psi} = \sum^k_{i_{L-1} = 1} ... \sum^k_{i_0=1}  \Tr \left( A^*_{i_{L-1}} ... A^*_{i_0} \right)\Tr \left( A_{i_{L-1}} ... A_{i_0} \right)= \Tr \left( M^L\right),
\end{equation}
where $M$ is an $N^2 \times N^2$ matrix given by
\begin{equation}
     M = \sum^k_{i=1} A^*_i \otimes A_i.
\end{equation}
We call $M$ the transfer matrix of the RMPS.

The $n$-point correlator of a local operator
\begin{equation}
    O = \sum^k_{i=1} \sum^k_{j=1} O_{ij} \ket{i} \bra{j}
\end{equation}
can be computed as
\begin{equation}
     \begin{aligned}
         &\bra{\psi} O(x_n) ... O(x_1) \ket{\psi} \\
         &\begin{aligned}[t]=& \left(  \sum^k_{i_{L-1}=1} ... \sum^k_{i_1 = 1} \right) \left( \sum^k_{j_{x_n} = 1} ... \sum^k_{j_{x_1}=1} \right) O_{i_{x_n} j_{x_n}} ... O_{i_{x_1} j_{x_1}}\\
         &\times\Tr \left( A^*_{i_{L-1}} ... A^*_{i_0} \right)\Tr \left( A_{i_{L-1}} ... A_{i_{x_n+1}} A_{j_{x_n}} A_{i_{x_n-1}} ... A_{i_{x_1+1}} A_{j_{x_1}} A_{i_{x_1-1}}  ... A_{i_0} \right) \\
         =& \Tr \left( M^{L-x_n-1} M_O M^{x_n - x_{n-1} -1} M_O ... M^{x_2 - x_1 - 1} M_O M^{x_1} \right),\end{aligned}
     \end{aligned}
\end{equation}
where $M_O$ is an $N^2 \times N^2$ matrix given by
\begin{equation}
      M_O = \sum^k_{i=1} \sum^k_{j=1} O_{ij} A^*_i \otimes A_j.
\end{equation}
We call $M_O$ the effective matter operator associated with the local operator $O$.

\begin{figure}[t]
    \begin{center}
    \resizebox{\linewidth}{!}{\input{figures/tikz/tentransfer}}
    \end{center}
    \caption{Left: tensor diagram for the transfer matrix $M$ in the inner product $\braket{\psi}{\psi}$. Right: tensor diagram for the effective matter operator $M_O$ in the quantity $\bra{\psi}O\ket{\psi}$.}
    \label{fig:tentransfer}
\end{figure}
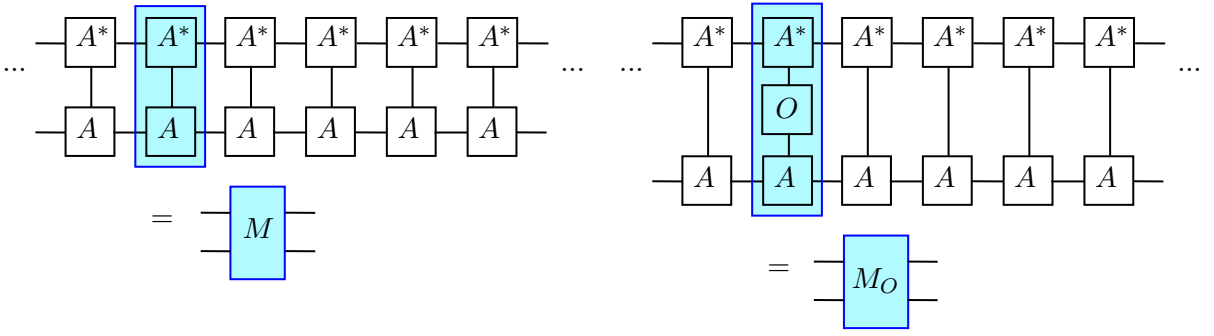

We can understand these results using tensor diagrams. See figure~\ref{fig:tentransfer}. We remark that statistically disconnected quantities such as
\begin{equation}
    Z_1[\bra{\psi}] Z_1[\ket{\psi}]
\end{equation}
can generally be computed selecting the matrices $\{A_i\}$ independently for statistically disconnected copies of $\ket{\psi}$.

In particular, bra-ket disconnected averages can be computed by selecting the matrices $A_i$ independently for bras and kets. Thus, in computing bra-ket disconnected averages of quantities, we simply need to replace $M$ and $M_O$ with
\begin{equation}
    M_{\text{disc}} = \sum^k_{i=1} A^{\prime *}_i \otimes A_i, \quad M_{O,\text{disc}} = \sum^k_{i=1} \sum^k_{j=1} O_{ij} A^{\prime *}_i \otimes A_j
\end{equation}
where $\{A_i\}$ and $\{A^\prime_i\}$ are two independent selections from the matrix ensemble $d\mu(\{A_i\})$. We call $M_{\text{disc}}$ the disconnected transfer matrix and $M_{O,\text{disc}}$ the disconnected effective matter operator. In comparison to the disconnected $M$ and $M_O$ we sometimes call the ordinary $M$ and $M_O$ the full transfer matrix and full effective matter operator. This is because it encodes the full set of replica effects.

We can also use the transfer matrix to compute the entropies of subregions. We consider the spatial subregion $R$ consisting of $m$ disjoint intervals. We denote the number of points in the $i$-th interval as $d_i$ and the number of points in between the $i$-th and $i+1$-th intervals as $\ell_i$. The compressed density matrix $\rho_{R, \text{compressed}}$ of $R$ is given by
\begin{equation}
    \begin{aligned}
        &\left(\bra{\mu^\prime_m\nu^\prime_m} \otimes  ... \otimes \bra{\mu^\prime_1\nu^\prime_1} \right)  Q_RP_R\left(\ket{\mu_m\nu_m} \otimes  ... \otimes \ket{\mu_1\nu_1} \right) \\
        & =\Tr\bigg(
        \begin{aligned}[t]
            &(I_{N\times N} \otimes \ket{\nu_1}  \bra{\nu^\prime_1}) M^{\ell_m} (I_{N\times N} \otimes \ket{\mu^\prime_m}  \bra{\mu_m}) M^{d_m}\\
            &\dots(I_{N\times N} \otimes \ket{\nu_2}  \bra{\nu^\prime_2}) M^{\ell_1} (I_{N\times N} \otimes \ket{\mu^\prime_1}  \bra{\mu_1}) M^{d_1}\bigg).
        \end{aligned}
    \end{aligned}
\end{equation}
Unlike $\rho_R$, the dimensionality of $\rho_{R, \text{compressed}}$ grows with the number of points in $\partial R$, not the number of points in $R$, which makes it more viable to compute the entropy of $R$ when $R$ contains large intervals.

\subsection{Spectral gapping property}

\label{ssec:rmpsspecgap}

So far we've noted that, for RMPS models, the properties of topological expansion and entropy bounding are guaranteed. The bra-ket wormhole phase transition is another key property of the AdS JT+CFT prepared state. As we mentioned in section~\ref{ssec:gpsbkwh}, arguments based on the strong subadditivity of entropy suggest that the existence of a bra-ket wormhole phase transition should be a general property of gravitationally prepared states for which the semiclassical approximation is applicable. Thus, for an RMPS model to be a model of a realistic gravitationally prepared state, it should show a bra-ket wormhole phase transition.

\textbf{Spectral gapping property.} We will give a mathematically well-defined condition for RMPS models, called the \textit{spectral gapping property}, such that RMPS models satisfying this property show characteristic features expected of a bra-ket wormhole transition.

This statement is given in terms of the transfer matrix $M$ and the disconnected transfer matrix $M_{\text{disc}}$ of $\ket{\psi}$, as follows.
\begin{equation}
    \begin{aligned}
        &\text{The highest eigenvalue of $M$ is gapped.}\\
        &\text{The highest eigenvalue of $M_{\text{disc}}$ is ungapped.}
    \end{aligned}
\end{equation}
Note that $M$ or $M_{\text{disc}}$ generally has $N^2$ eigenvalues. Thus, by saying that the spectrum is ungapped at a certain point, we mean that the distance between neighboring eigenvalues is $\mathcal{O}(N^{-2})$ at that point. By saying that the spectrum is gapped at a certain point, we mean that the distance between neighboring eigenvalues is $\mathcal{O}(N^0)$ at that point.

\textbf{Motivation of the spectral gapping property.} We will now motivate the spectral gapping property, explaining why RMPS models satisfying this property show characteristic features expected of a bra-ket wormhole transition.

In the discussion below we assume we have subtracted the averaged one-point function $\left<  \frac{\bra{\psi} O(0)\ket{\psi}}{\braket{\psi}{\psi}}  \right>  $ from the local operator $O(x)$. In other words, we subtract tadpoles to look at the connected two-point functions.

We list characteristic features expected of a bra-ket wormhole transition. These properties are lifted from what we know in the AdS JT+CFT model. First, $\ket{\psi}$ should satisfy
\begin{equation}
\label{eq:bkwhdependence}
    \frac{Z_2[\braket{\psi}{\psi}]}{Z_1[\bra{\psi}]Z_1[ \ket{\psi}]} \propto N^{-2} \exp(\alpha L) + \mathcal{O}(N^{-4})
\end{equation}
at large $L$, for some constant $\alpha>0$. The physical meaning of this property is that, for large $L$, the local bra-ket wormhole geometry is independent of $L$. Thus the effective action of the bra-ket wormhole geometry depends linearly on $L$, and the contribution of the bra-ket wormhole geometry depends exponentially on $L$. Furthermore, we must have $\alpha>0$ due to the negative Casimir energy in the Euclidean periodicity direction of the bra-ket wormhole, and so the bra-ket wormhole geometry dominates over the disk geometry at large $L$.

The critical length $L_{\text{crit}}$ is the value of $L$ for which the above quantity becomes $1$. We see that
\begin{equation}
\label{eq:bkwhcrit}
    L_{\text{crit}} = \frac{2 \ln N}{\alpha} + \mathcal{O}(N^0),
\end{equation}
and so $L_{\text{crit}} \sim \mathcal{O}(\ln N)$.

Second, to highest order in $N$, for $L > L_{\text{crit}}$, the correlator in $\ket{\psi}$ must show exponential decay at large distances, i.e.
\begin{equation}
    \left<\frac{ \bra{\psi}  O(x) O(0)  \ket{\psi} }{\braket{\psi}{\psi}} \right> \propto \exp \left(-\beta |x| \right) + \mathcal{O}(N^{-2}) \quad (|x| \gg 1,\text{ } L > L_{\text{crit}})
\end{equation}
for some constant $\beta$. The physical meaning of this property is that the bra-ket wormhole geometry prepares a thermal state. Thermal CFT states have exponentially decaying correlators, and so, in the bra-ket wormhole dominant phase, the correlator in $\ket{\psi}$ must show exponential decay at large distances.

Third, to highest order in $N$, the bra-ket disconnected correlator must show sub-exponential decay, i.e.
\begin{equation}
    \left<\frac{ \bra{\psi}  O(x) O(0)  \ket{\psi} }{\braket{\psi}{\psi}} \right>_{\text{disc}}  \sim (\text{sub-exponential in } |x|) + \mathcal{O}(N^{-2})  .
\end{equation}
The physical meaning of this property is that the double disk geometry prepares the CFT vacuum state. The CFT vacuum state has subexponentially decaying correlators, and so the bra-ket disconnected correlator in $\ket{\psi}$ must show sub-exponential decay.

To show that the spectral gapping property ensures these three properties, we first note that $\braket{\psi}{\psi}$ and the correlator of $\ket{\psi}$ can be written as
\begin{equation}
\begin{aligned}
    \braket{\psi}{\psi} &= \sum_{\lambda} \lambda^L,\\
    \frac{ \bra{\psi}  O(x) O(0)  \ket{\psi} }{\braket{\psi}{\psi}} &= \frac{\sum_{\lambda_a, \lambda_b}|\bra{\lambda_a}M_O \ket{\lambda_b}|^2  \lambda_a^{L-x-1} \lambda_b^{ x-1} }{\sum_{\lambda} \lambda^L},
\end{aligned}
\end{equation}
where $\lambda, \lambda_a, \lambda_b $, $\ket{\lambda_a}, \ket{\lambda_b}$ are the eigenvalues and eigenkets of either $M$ or $M_{\text{disc}}$, depending on whether we are sampling the full or bra-ket disconnected value on the left hand side. From this formula, we see that the first condition is equivalent to the fact that the highest eigenvalue of $M$ is $\mathcal{O}(N^0)$ higher than that of $M_{\text{disc}}$. The second condition is equivalent to the fact that the highest eigenvalue of $M$ is gapped, and the third condition is equivalent to the fact that the highest eigenvalue of $M_{\text{disc}}$ is ungapped.

We also note that, by the topological expansion of RMPS, the spectrum of $M$ can only differ from the spectrum of $M_{\text{disc}}$ by $\mathcal{O}(N^0)$. By this, we mean that for a given value of $0 \leq m \leq 2$, only $\mathcal{O}(N^m)$ eigenvalues can shift by $\mathcal{O}(N^{-m})$. Thus, the fact that the highest eigenvalue of $M$ is gapped while the highest eigenvalue of $M_{\text{disc}}$ is ungapped implies that $\mathcal{O}(N^0)$ new eigenvalues appear in $M$ that are $\mathcal{O}(N^0)$ higher than the highest eigenvalue in $M_{\text{disc}}$, and thus implies that the highest eigenvalue of $M$ is $\mathcal{O}(N^0)$ higher than that of $M_{\text{disc}}$.

Thus, RMPS models satisfying the spectral gapping property show the three characteristic features of a bra-ket wormhole transition listed above. We will take such models as our objects of interest. Below, we will introduce a family of models that generally satisfy the spectral gapping property.

\subsection{$\mathrm{O}(k)$ models}

\label{ssec:rmpsok}

We consider models with $\mathrm{O}(k)$ symmetry, i.e. models where the potential $V(\{A_i\})$ is invariant under transformations of the type $\{A_i\} \rightarrow \{ R_{ij} A_i\}$, where $R_{ij}$ is an orthogonal matrix. For such models, we can write the potential as
\begin{equation}
    V(\{A_i\}) = k F(|A|^2), \quad |A|^2 = \sum_i A_i^2 .
\end{equation}
We define the large $k$ limit of such models as taking $k$ to infinity while the function $F$ remains constant. The factor of $k$ in $V(A_i)$ allows for the large $k$ limit to be well-behaved. For example by including this factor, we get $M \sim \mathcal{O}(k^0)$ and so the transfer matrix does not diverge or go to zero at $k \rightarrow \infty$. The factor of $k$ is also necessary for the large $k$ analysis we will perform below.

Our claim is that, for sufficiently large $k$, all such $\mathrm{O}(k)$ models satisfy the spectral gapping property. We will prove this claim in three steps. We first show that, for large $k$, the full and disconnected transfer matrices self-average, i.e. higher moments of the transfer matrices go to zero as $k \rightarrow \infty$. This tells us that the spectra of the full and disconnected transfer matrices approach the spectra of their ensemble averages as $k \rightarrow \infty$. We then compute the ensemble averages of the full and disconnected transfer matrices using Haar symmetry. Finally, we argue that, based on the forms of the spectra of the full and disconnected transfer matrices for $k \rightarrow \infty$, the spectral gapping property should be satisfied for sufficiently large $k$.

We note that in our proof we will not assume that $V(\{A_i\})$ contains a quadratic term. Thus, we do not rely on perturbative diagrammatics. Our claim holds just as well when $V(\{A_i\})$ only contains terms quartic and above.

We start with the first step. Given the $k$ $N \times N$ matrices $\{J_i\}$, we define the generating functional
\begin{equation}
    W(\{J_i\}) = \ln \left( \int d \{A_i\} \exp \left[-N k \Tr \left( F(|A|^2) \right)  +  k \sum_i \Tr \left( J_i A_i \right)   \right] \right).
\end{equation}
Then, we have
\begin{equation}
    \left< A_{i_1, a_1b_1} ... A_{i_n, a_nb_n} \right>_{\text{conn}} = \left. k^{-n} \partial_{J_{i_1, b_1 a_1}} ... \partial_{J_{i_n, b_n a_n}} W(\{J_i\}) \right|_{\{J_i\} =0}
\end{equation}
where $\left< \cdot \right>_{\text{conn}} $ denotes the cumulant.

We can understand the sets of $k$ $N \times N$ matrices $\{A_i\}$, $\{J_i\}$ as sets of $N^2$ $k$-dimensional vectors $\{\vec{A}_{ab}\}$, $\{\vec{J}_{ab}\}$. We would first like to show that the norm of $\vec{A}_{ab}$ scales as $\mathcal{O}(k^0)$ in the large $k$ limit. To do this, we estimate the contribution to the matrix integral of $\{A_i\}$ from values of $\vec{A}_{ab}$ with norm scaling as $\mathcal{O}(k^m)$. We see the integration measure $d \{A_i\}$ becomes $\mathcal{O}(k^m)^{\mathcal{O}(k^1)}$, and the exponential $\exp \left(-N k \Tr \left( F(|A|^2) \right) \right)$ becomes $e^{-\mathcal{O}(k^{\max (1, lm+1)})}$, where $l$ is the order of the polynomial $F$. Thus, the combined contribution is $\mathcal{O}(k^m)^{\mathcal{O}(k^1)}$ when $m< 0$ and $e^{-\mathcal{O}(k^{lm+1})}$ when $m>0$. The contribution is maximized when $m=0$, and so the norm of $\vec{A}_{ab}$ must scale as $\mathcal{O}(k^0)$.

We now take the norm of $\vec{J}_{ab}$ to also scale as $\mathcal{O}(k^0)$. It follows that the matrix integral defining $W(\{J_i\})$ is $e^{\mathcal{O}(k^1)}$ and thus $W(\{J_i\})$ is $\mathcal{O}(k^1)$. Therefore, from the above formula for the cumulant, we see that $\left< A_{i_1, a_1b_1} ... A_{i_n, a_nb_n} \right>_{\text{conn}} \sim \mathcal{O}(k^{1-n})$. By $\mathcal{O}(k)$ symmetry, we can further claim that
\begin{equation}
    \begin{aligned}
        &\left< A_{i_1, a_1b_1} ... A_{i_n, a_nb_n} \right>_{\text{conn}}
    &= \begin{cases}
        k^{1-n} (T_n)_{a_1b_1 ... a_n b_n} \sum_{\substack{\text{partitions of $i_1...i_n$}\\\text{into pairs $(i_{c_k}, j_{d_k})$}}} \prod^n_{k=1} \delta_{i_{c_k} j_{d_k}} & (n \text{ even}) \\
        0 &(n \text{ odd})
    \end{cases}
    \end{aligned}
\end{equation}
where $T_n$ is some $\mathcal{O}(k^0)$ tensor. It follows that
\begin{equation}
\begin{aligned}
     &   \left< A_{i_1, a_1b_1} ... A_{i_n, a_nb_n} \right>
    &= \begin{cases}
        k^{-n/2} (S_n)_{a_1b_1 ... a_n b_n} \sum_{\substack{\text{partitions of $i_1...i_n$}\\\text{into pairs $(i_{c_k}, j_{d_k})$}}} \prod^n_{k=1} \delta_{i_{c_k} j_{d_k}} & (n \text{ even}) \\
        0 &(n \text{ odd})
        \end{cases}
    \end{aligned}
\end{equation}
for some tensor $S_n$.

From here it follows from direct calculation that the transfer matrix $M$ satisfies
\begin{equation}
    \left< M_{a_1 b_1 c_1 d_1} ... M_{a_n b_n c_n d_n} \right>_{\text{conn}} \sim \mathcal{O}(k^{1-n}) , \quad a_i, b_i, c_i, d_i = 1,  ..., N
\end{equation}
and that the disconnected transfer matrix $M_{\text{disc}}$ satisfies the same asymptotics as well. The only difference in the calculation for the disconnected transfer matrix $M_{\text{disc}}$ is that $\{A_i\}$ and $\{A^\prime_i\}$ are viewed as independent random variables.

What this means is that, independently of the large $N$ topological expansion, in the large $k$ limit, any function of the transfer matrix $M$ converges to the same quantity computed with $\left< M \right>$, and the same for the disconnected transfer matrix. This implies that the spectrum of $M$, $M_{\text{disc}}$ converges to that of $\left< M \right>$, $\left< M_{\text{disc}} \right>$.

We now move on to the second step, which is to compute $\left< M \right>$, $\left< M_{\text{disc}} \right>$. Note that the random matrix ensemble for $\{A_i\}$ has Haar symmetry, given by
\begin{equation}
    d\mu(\{A_i\}) = d\mu (\{U A_i U^{-1}\}), \quad U \in \mathrm{U}(N).
\end{equation}
Thus we must have
\begin{equation}
    \begin{aligned}
        (U^* \otimes U)\left< M \right> (U^{*-1} \otimes U^{-1}) &= \left< M \right>,\\
        (U^{\prime *} \otimes U)
        \left< M_{\text{disc}} \right> (U^{\prime *-1} \otimes U^{-1}) &= \left< M_{\text{disc}} \right>
    \end{aligned}
\end{equation}
where $U, U^\prime$ are different $N\times N$ unitaries. From this, the $\mathbb{Z}_2$ symmetry $\{A_i\} \rightarrow \{- A_i\}$ of $V(A_i)$, and the fact that the spectra of $\left< M \right>$ and $\left< M_{\text{disc}} \right>$ differ by $\mathcal{O}(N^0)$, it follows that the only possible form of $\left< M \right>$, $\left< M_{\text{disc}} \right>$ is given by
\begin{equation}
    \begin{aligned}
        \left< M \right> &= \lambda_{g;k \rightarrow \infty} \ket{\lambda_{g;k\rightarrow \infty}} \bra{\lambda_{g;k\rightarrow \infty}}, \\
        \left< M_{\text{disc}} \right> &= 0 .
    \end{aligned}
\end{equation}
Here $\ket{\lambda_{g;k\rightarrow \infty}}$ is given by
\begin{equation}
    \ket{\lambda_{g;k\rightarrow \infty}} = \frac{\ket{11} + ... + \ket{NN}}{\sqrt{N}},
\end{equation}
and $\lambda_{g;k\rightarrow \infty}$ is a constant. We see that $\left< \Tr (|A|^2) \right> = \lambda_{g;k\rightarrow \infty}$, so we must have $\lambda_{g;k\rightarrow \infty} > 0$.

It follows that the spectra of $\left< M \right>$ and $\left< M_{\text{disc}} \right>$ are given by
\begin{equation}
    \begin{aligned}
        \left< M \right> &\text{ : eigenvalue $0$ with degeneracy $N^2- 1$, eigenvalue $\lambda_{g;k\rightarrow \infty} >0$ .}  \\
        \left< M_{\text{disc}} \right> &\text{ : eigenvalue $0$ with degeneracy $N^2$.}
    \end{aligned}
\end{equation}
We now proceed to the third step. Compared to the spectra of $M$, $M_{\text{disc}}$ for $k \rightarrow \infty$, we expect the spectra for finite $k$ to be spread out in two ways.

First, in both spectra, the $N^2-1$ degenerate eigenvalues will spread out to form a continuous spectrum with $\mathcal{O}(N^{-2})$ spacing. From the $k$-dependence of the cumulants of the transfer matrices, we see that the continuum spectrum will have width $\mathcal{O}(k^{-1/2})$. We expect the continuum spectra of $M$ and $M_{\text{disc}}$ to differ by $\mathcal{O}(N^0)$. Thus, for large $N$, the largest eigenvalue of the continuum spectra of $M$ and $M_{\text{disc}}$ must be approximately equal.

We denote the largest eigenvalue of $M$ as $\lambda_g$, and the largest eigenvalue of $M_{\text{disc}}$ as $\lambda_{ug}$.

Second, in the spectra of $M$, the gapped highest eigenvalue $\lambda_g$ will statistically fluctuate. By the topological expansion, the variance of $\lambda_g$ will have scale $\mathcal{O}(N^{-2})$. Thus, for large $N$, $\lambda_g$ will be approximately equal to its value at $k \rightarrow \infty$.

Thus we see that the highest eigenvalue of $M_{\text{disc}}$ is ungapped, and the highest eigenvalue of $M$ is gapped if and only if
\begin{equation}
    \lambda_g > \lambda_{ug}.
\end{equation}
Since $\lambda_{ug}$ shrinks as $k$ increases, and $\lambda_g$ stays approximately the same, the spectral gapping property should hold for large enough $k$.

\paragraph{The entropy of a subregion.} The exact average entropy bounds of RMPS models contain subleading terms of $\mathcal{O}(N^0)$ following $|\partial R| \ln N$. For $\mathcal{O}(k)$ models, we can predict that this subleading term is $\mathcal{O}(k^{-1})$.

To show this, we first see that, given a subregion $R$ consisting of $m$ disjoint intervals, in the limit where the lengths $\ell_i$ of the intervals and the distances $d_i$ between them go to infinity, we have
\begin{equation}
    \begin{aligned}
        &\left(\bra{\mu^\prime_m\nu^\prime_m} \otimes  ... \otimes \bra{\mu^\prime_1\nu^\prime_1} \right)  \rho_{R,\text{compressed}}\left(\ket{\mu_m\nu_m} \otimes  ... \otimes \ket{\mu_1\nu_1} \right) \\
        & =\begin{aligned}[t]&\lambda_g^L \bra{\lambda_g} \left( (I_{N\times N} \otimes \ket{\nu_1}  \bra{\nu^\prime_1}) \ket{\lambda_g} \bra{\lambda_g}(I_{N\times N} \otimes \ket{\mu^\prime_m}  \bra{\mu_m}) \ket{\lambda_g} \right.\\
        & \dots \bra{\lambda_g}(I_{N\times N} \otimes \ket{\nu_2}  \bra{\nu^\prime_2}) \ket{\lambda_g} \bra{\lambda_g}(I_{N\times N} \otimes \ket{\mu^\prime_1}  \bra{\mu_1})  \ket{\lambda_g}.\end{aligned}
    \end{aligned}
\end{equation}
Thus, we see that
\begin{equation}
    \text{max}(S_n[R]) = 2m S_n [\rho_N]
\end{equation}
where $S_n[\rho_N]$ is the $n$-Renyi entropy of the matrix $\rho_N$, and $\rho_N$ is the $N\times N$ matrix given by taking the partial trace of $\ket{\lambda_g} \bra{\lambda_g}$:
\begin{equation}
    \bra{\nu} \rho_N \ket{\mu} = \Tr  \left( \ket{\lambda_g} \bra{\lambda_g} (I_{N\times N} \otimes \ket{\mu}  \bra{\nu})  \right).
\end{equation}
Thus, we see that
\begin{equation}
    \text{max}(S_n[R]) = 2m \ln N + \mathcal{O}(N^0)
\end{equation}
where the $\mathcal{O}(N^0)$ term depends on the spectrum of $\rho_N$. But, by our study of the large $k$ expansion, we can see that the spectrum of $\rho_N$ is equal to
\begin{equation}
    (\text{eigenvalues of }\rho_N) \approx \frac{1}{N} \left( 1 + \mathcal{O}(k^{-1}) \right).
\end{equation}
Thus, the $\mathcal{O}(N^0)$ subleading term in the average entropy bound is $\mathcal{O}(k^{-1})$.

\subsection{Numerical results}

\label{ssec:rmpsnumerical}

We numerically verified the spectral gapping property, the bra-ket wormhole phase transition, and entropy bounding for the following $\mathcal{O}(k)$ models:
\begin{equation}
    \begin{aligned}
        &\textbf{Quadratic: }  &&F(|A|^2) = \frac{1}{2} |A|^2, \\
        &\textbf{Quartic: } &&F(|A|^2) = \frac{1}{4} |A|^4, \\
        &\textbf{Quadratic $+$ Quartic: } &&F(|A|^2) = \frac{1}{4} |A|^2 + \frac{1}{8} |A|^4. \\
    \end{aligned}
\end{equation}
These models have been normalized such that $\lambda_g  \approx 1$ at large $N$.

The quadratic model is the easiest to simulate numerically, but in terms of the 't Hooft diagrammatics, it has no vertices, which may lead to issues with the correspondence to the gravity side. The quartic model has no quadratic term, and is indescribable by 't Hooft diagrammatics, which, as we explained previously, is not a problem. The quadratic plus quartic model interpolates between the two.

To sample the random matrices $\{A_i\}$, we used Markov chain Monte Carlo (MCMC) methods using the Metropolis-Hastings algorithm \cite{metropolis_equation_1953, hastings_monte_1970}. The details of our sampling are given in appendix~\ref{sec:mcmc}.

We state the results below. We note that, due to $M$ having negative eigenvalues, the results show a slight discrepancy between the trends of even and odd $L$. This issue will be resolved for the case of continuous RMPS models which we will studied later. For brevity, we present selected figures in this section, focusing on the quadratic plus quartic model. The rest of the figures can be found in appendix~\ref{ssec:supplementary_figures_drmps}.

\paragraph{Spectral gapping property.} We see that all three models satisfy the spectral gapping property with a continuous spectrum for $M_{\text{disc}}$ with width of $\mathcal{O}(k^{-1/2})$ at large $k$, and one extra gapped state for $M$.

We also see that the continuous spectrum of $M_{\text{disc}}$ has a square root edge of the form
\begin{equation}
    \rho(\lambda) \approx \sqrt{\lambda_{ug} - \lambda}, \quad  0 < \lambda_{ug} - \lambda \ll \lambda_{ug},
\end{equation}
where $\lambda_{ug}$ is the maximum eigenvalue of $M_{\text{disc}}$. By the following calculation, we can see a square root edge implies the factor of $L^{-\frac{3}{2}}$:
\begin{equation}
    \begin{aligned}
        Z_1[ \bra{\psi}] Z_1 [\ket{\psi}] &= \sum_{\lambda} \lambda^L\\
        & \approx (\text{const.}) \times  \lambda_{ug}^L \int  d\varepsilon \sqrt{\varepsilon} e^{-\varepsilon L}  \propto L^{-\frac{3}{2}}  \lambda_{ug}^L,
    \end{aligned}
\end{equation}
where $\varepsilon = (\lambda_{ug} - \lambda)/\lambda_{ug}$.
Thus we checked the square root edge by verifying the $L^{-\frac{3}{2}} \exp (- \gamma L)$ dependence of $Z_1[\bra{\psi}]Z_1[ \ket{\psi}] $, where $\gamma$ is a constant.

\begin{figure}[t]
    \centering
    \includegraphics[width=0.45\textwidth]{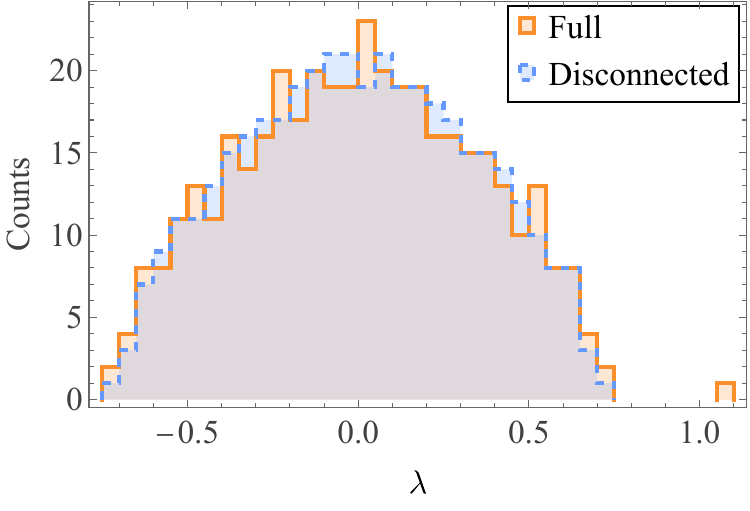}
    \caption{Eigenvalue distribution of a smaple transfer matrix in the quadratic + quartic model with $k=8$, $N=20$. We see a gapped highest eigenvalue in the full transfer matrix.}
    \label{fig:SingleEigQuadQuar}
\end{figure}

In figure~\ref{fig:SingleEigQuadQuar}, we present the eigenvalue distribution of a sample transfer matrix in the quadratic + quartic model. The yellow bars represent the full transfer matrix, and the blue bars represent the disconnected transfer matrix. The green region represents their overlap. We see that the full transfer matrix has an additional gapped highest eigenvalue. Similar results hold for the other two models, which can be found in appendix~\ref{ssec:supplementary_figures_drmps}.

\begin{figure}[t]
    \centering
    \begin{subfigure}[t]{0.45\textwidth}
        \centering
        \includegraphics[width=\textwidth]{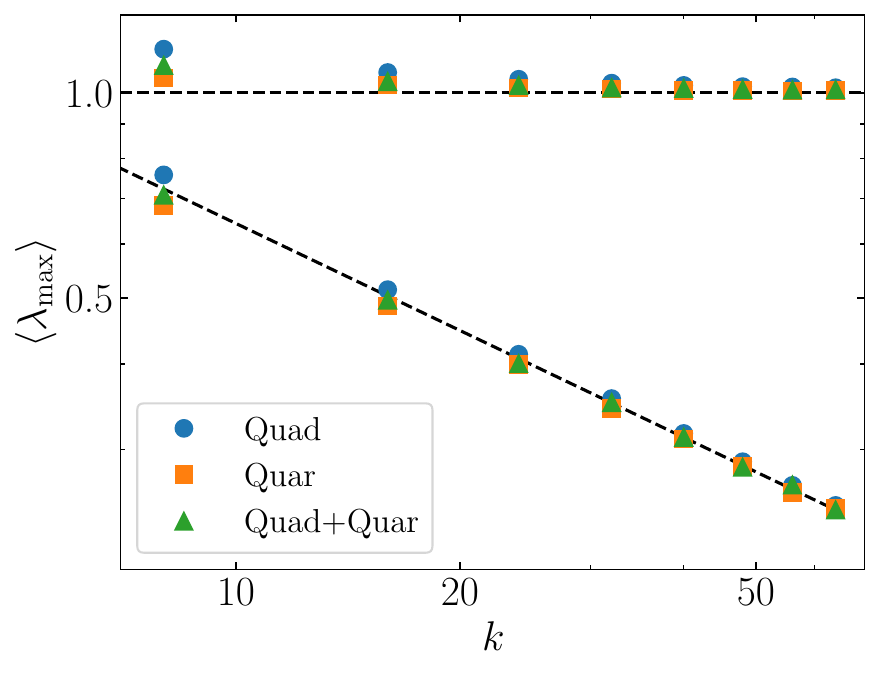}
        \caption{}
    \end{subfigure}
        \hspace{0.05\textwidth}
     \begin{subfigure}[t]{0.45\textwidth}
        \centering
        \includegraphics[width=\textwidth]{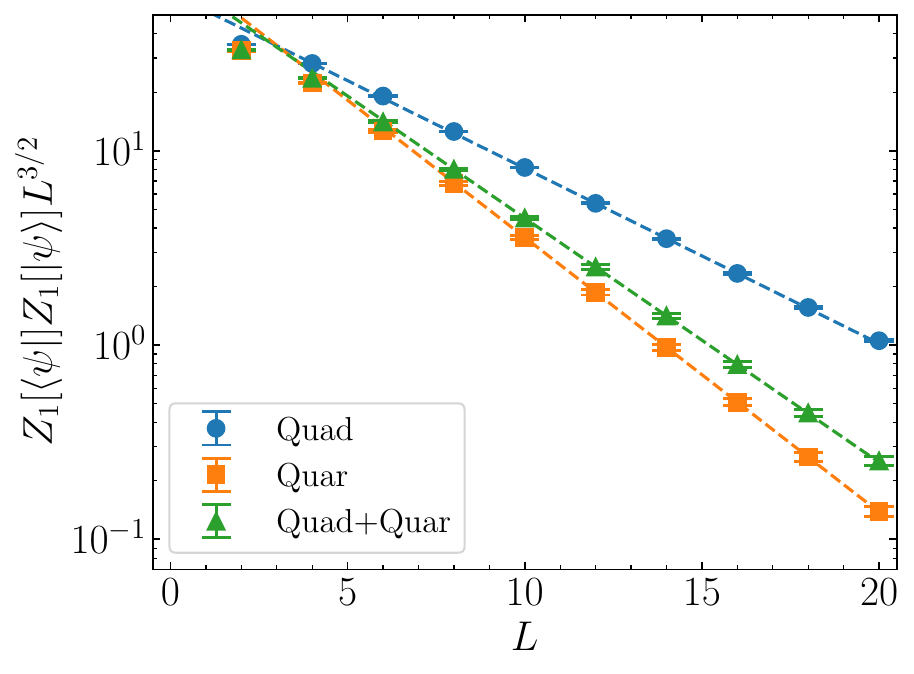}
        \caption{}
    \end{subfigure}
	\caption{Properties of the transfer matrix eigenvalue distributions for the three models. (a) $k^0$-dependence of $\left< \lambda_g \right>$ (top line) and $k^{-\frac{1}{2}}$-dependence of $\left< \lambda_{ug} \right>$ (bottom line). We have $10^3$ samples with $N=10$. (b) $Z_1[\bra{\psi}]Z_1[ \ket{\psi}] L^{\frac{3}{2}}$ as a function of $L$ showing the $L^{-\frac{3}{2}} \exp(-\gamma L)$ dependence of $Z_1[\bra{\psi}]Z_1[ \ket{\psi}]$. We have $10^{4}$ samples with $k=8$ and $N=10$.}
    \label{fig:EigProperties}
\end{figure}

In figure~\ref{fig:EigProperties}, we present properties of the transfer matrix eigenvalue distributions for the three models. In (a), we have the $k^{-\frac{1}{2}}$ dependence of $\left< \lambda_{ug} \right>$ and the $k^0$ dependence of $\left< \lambda_g \right>$ for all three models. In (b), we have the $L^{-\frac{3}{2}} \exp (- \gamma L)$ dependence result verifying the square root edge for all three models.

\paragraph{Bra-ket wormhole phase transition.} We see that all three models satisfy the three conditions of the bra-ket wormhole phase transition. This is expected, as they satisfy the spectral gapping property.

For the $\mathcal{O}(k)$ models, by $\mathcal{O}(k)$ symmetry, any local operator satisfying
\begin{equation}
    O = O^\dagger, \quad \Tr O = 0, \quad \Tr \left(O^2 \right) = 1
\end{equation}
gives the same correlators, and satisfies $\left< \frac{\bra{\psi} O \ket{\psi}}{\braket{\psi}{\psi}}  \right> = 0$. In our calculations we used such an operator.

We note that for the correlators and bra-ket disconnected correlators, there is actually an $\mathcal{O}(N^{-2})$ offset which is explained by higher genus effects beyond the bra-ket wormhole and double disk topologies. To study the decaying part of the correlator we subtracted out this offset. We will study this offset, which is an interesting physical effect, in much more detail in the next section.

We also note that, for large $L$, higher genus effects become important in the computation of $\frac{Z_2[\braket{\psi}{\psi}]}{Z_1[\bra{\psi}]Z_1[ \ket{\psi}]} $ and cause deviations from exponential behavior. We study examples of such higher genus effects in much more detail in the next section. For the purposes of this section, we studied the quantity $\frac{Z_2[\braket{\psi}{\psi}]}{Z_1[\bra{\psi}]Z_1[ \ket{\psi}]} $ in the regime of $L$ where the said higher genus effects are small.

\begin{figure}[t]
    \centering
    \begin{subfigure}[t]{0.45\textwidth}
        \centering
        \includegraphics[width=\textwidth]{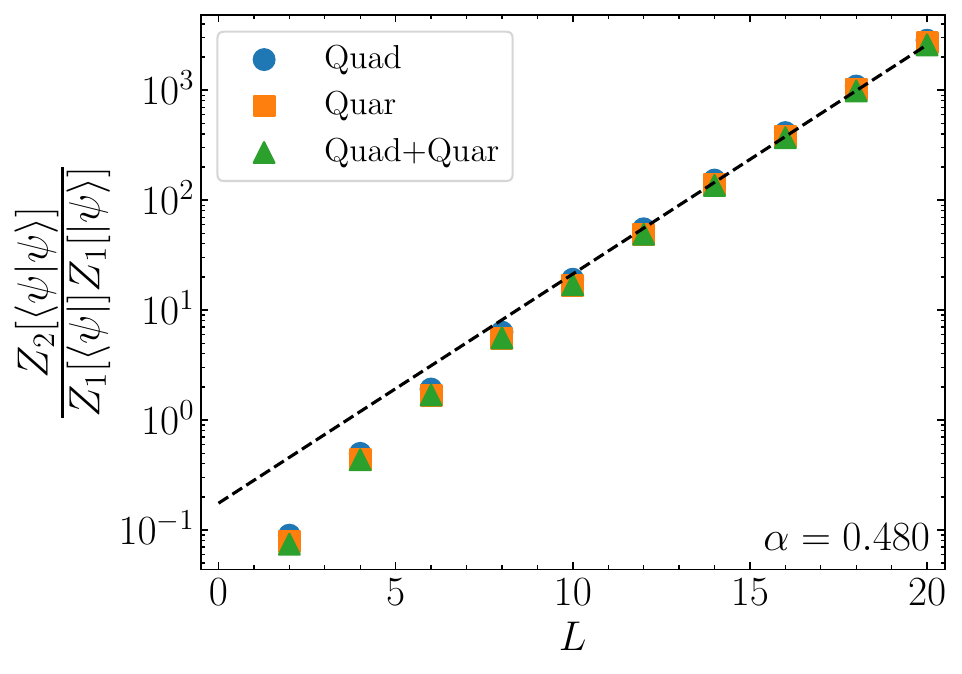}
        \caption{}
    \end{subfigure}
    \hspace{0.05\textwidth}
    \begin{subfigure}[t]{0.45\textwidth}
        \centering
        \includegraphics[width=\textwidth]{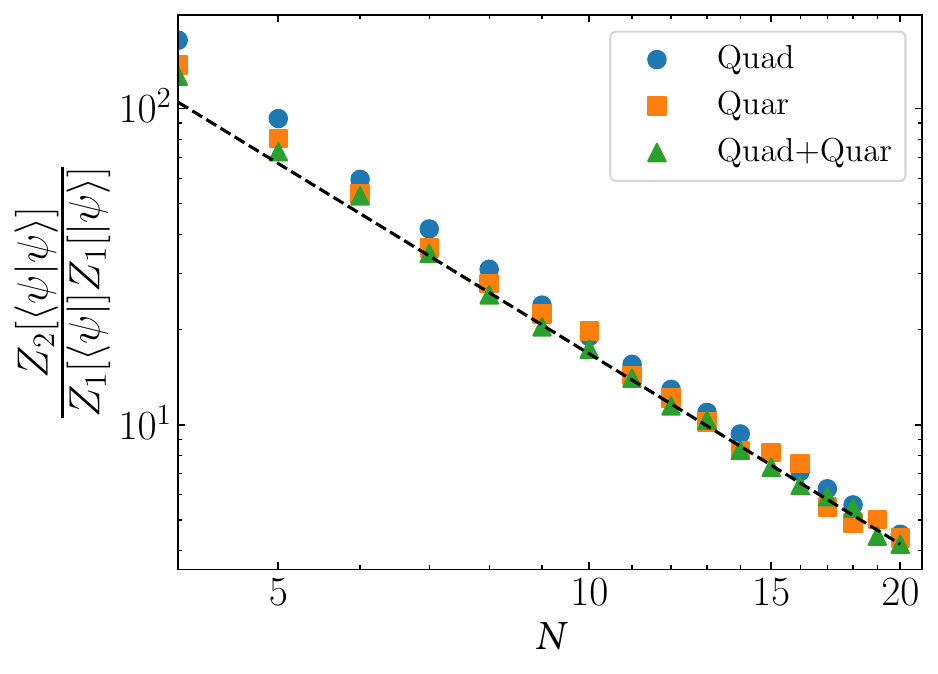}
        \caption{}
    \end{subfigure}
	\caption{Behavior of the quantity $Z_2[\braket{\psi}{\psi}]/(Z_1[\bra{\psi}]Z_1[\ket{\psi}])$ for the three models. (a) $N^{-2}$ dependence. We have $10^{5}$ samples with $k=8$ and $L=10$. (b) $\exp(\alpha L)$ dependence at large $L$. We have $10^{4}$ samples with $k=8$ and $N=10$.}
    \label{fig:DBKWHProperties}
\end{figure}

\begin{figure}[t]
    \centering
    \begin{subfigure}[t]{0.45\textwidth}
        \centering
        \includegraphics[width=\textwidth]{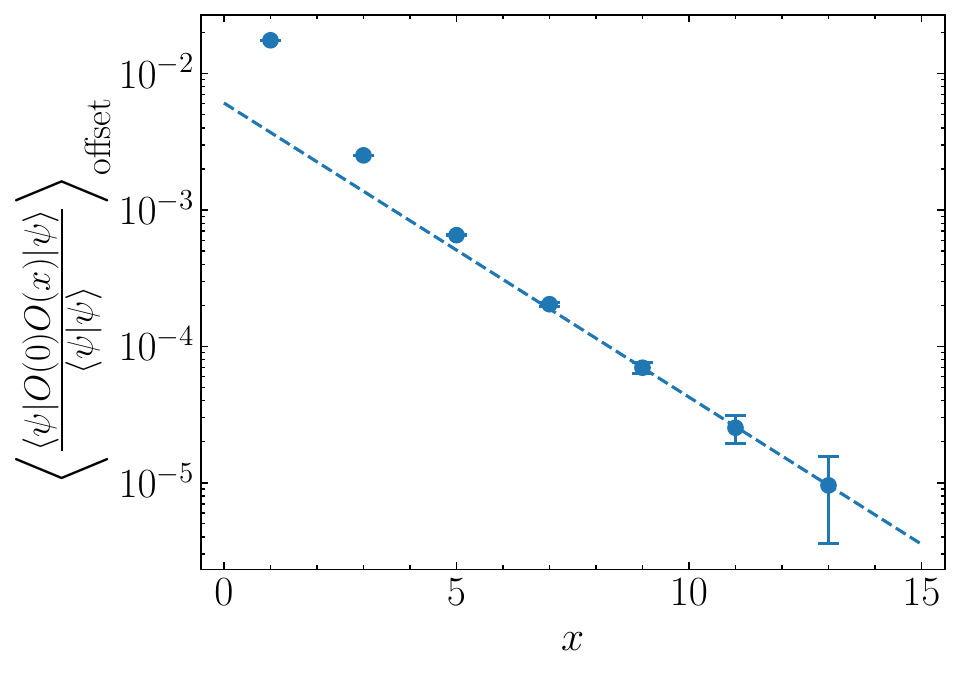}
        \caption{}
    \end{subfigure}
    \hspace{0.05\textwidth}
    \begin{subfigure}[t]{0.45\textwidth}
        \centering
        \includegraphics[width=\textwidth]{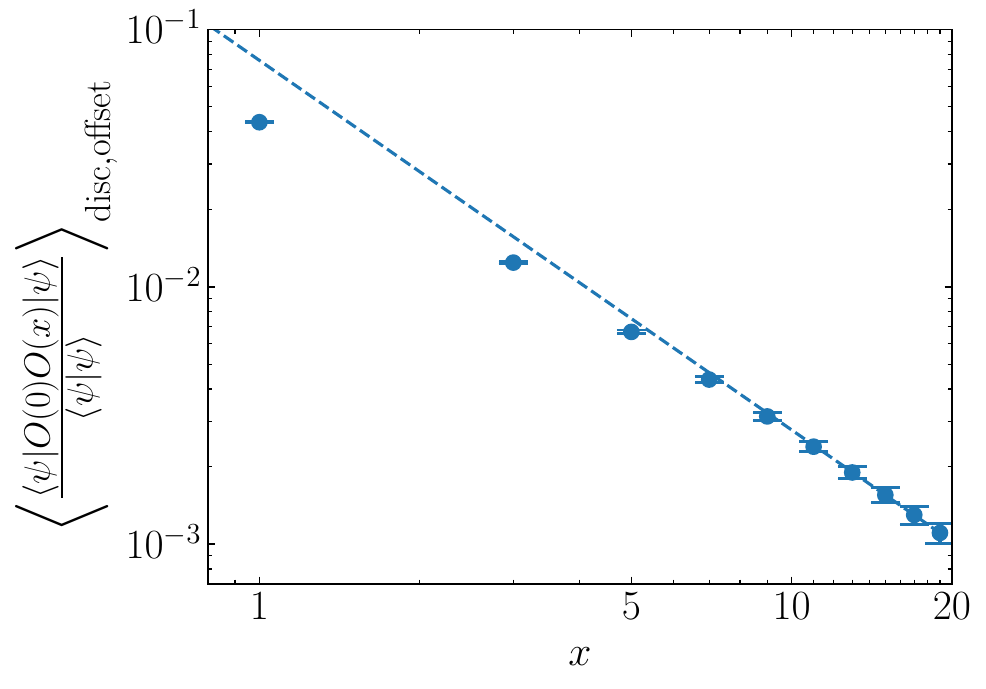}
        \caption{}
    \end{subfigure}
	\caption{Two-point correlators of a local operator in the quadratic + quartic model with the $x \rightarrow \infty$ value subtracted. Here $k=8$, $N=10$, and $L=400$. (a) Full correlators showing exponential decay in $x$. Here we used $10^6$ samples. (b) Bra-ket disconnected correlators showing power law decay in $x$. Here we used $10^5$ samples.}
    \label{fig:DCorrProperties}
\end{figure}

In figure~\ref{fig:DBKWHProperties}, we present the $N^{-2} \exp(\alpha L)$ dependence of the quantity $\frac{Z_2[\braket{\psi}{\psi}]}{Z_1[\bra{\psi}]Z_1[ \ket{\psi}]} $ for all three models, at $L$ large but not too large such that higher genus effects become important.

In figure~\ref{fig:DCorrProperties}, we present the exponential decay of the correlator and the power law decay of the bra-ket disconnected correlator in the quadratic + quartic model, after subtracting the offset at $x \rightarrow \infty$. Similar results hold for the other two models, which can be found in appendix~\ref{ssec:supplementary_figures_drmps}.

\paragraph{Entropy bounding.} We see the expected entropy bounding behavior for a single interval in all three models, with average entropy bound $S_{\text{max},n}$ equal to $2\ln N$ plus an $\mathcal{O}(N^0)$ subleading term. We see the $\mathcal{O}(N^0)$ subleading term has $k^{-1}$ dependence at large $k$.

\begin{figure}[t]
    \centering
    \begin{subfigure}[t]{0.45\textwidth}
        \centering
        \includegraphics[width=\textwidth]{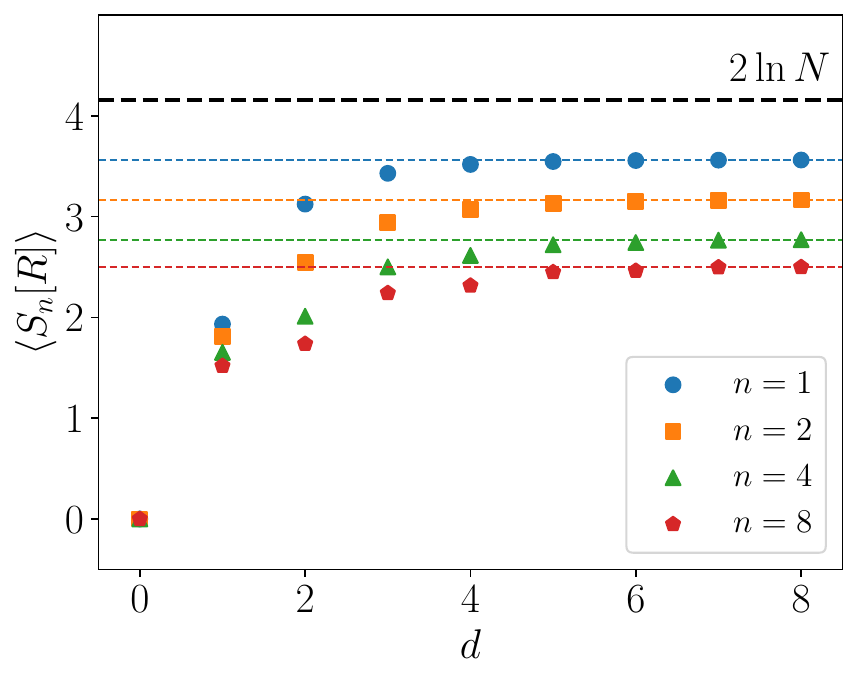}
        \caption{}
    \end{subfigure}
        \hspace{0.05\textwidth}
    \begin{subfigure}[t]{0.45\textwidth}
        \centering
        \includegraphics[width=\textwidth]{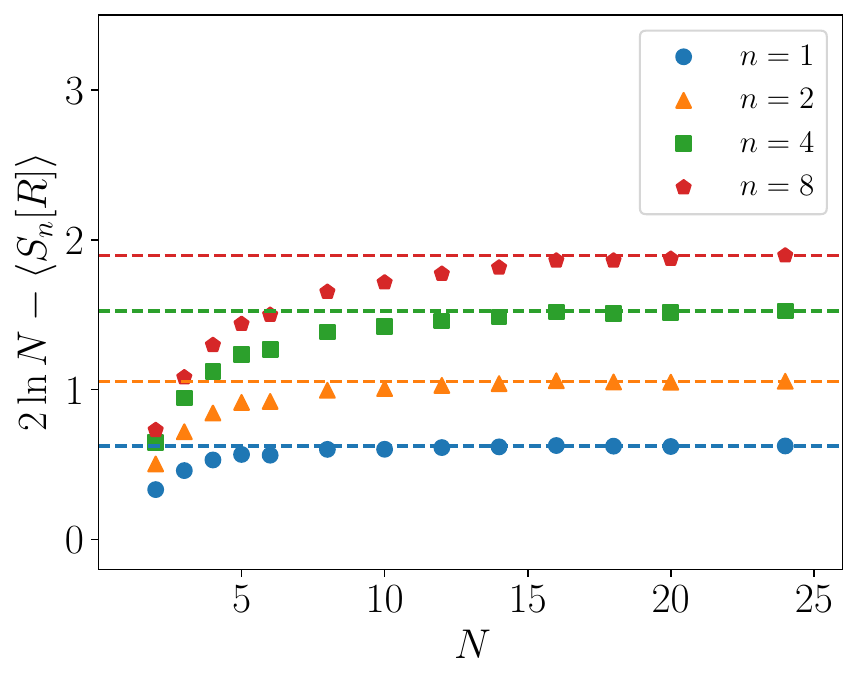}
        \caption{}
    \end{subfigure}
	\caption{
    The bounding behavior of the averaged entropies of an interval in the quadratic + quartic model.
    (a) $S_n[R]$ showing bounding behavior. Here $N=8$ with $10^4$ samples. (b) Convergence of $2\ln N - S_{\text{max}, n}$ at large $N$. We used $k=128$ and $10^3$ samples.
    }
    \label{fig:DEntropyProperties}
\end{figure}

\begin{figure}[!t]
    \centering
    \includegraphics[width=0.48\textwidth]{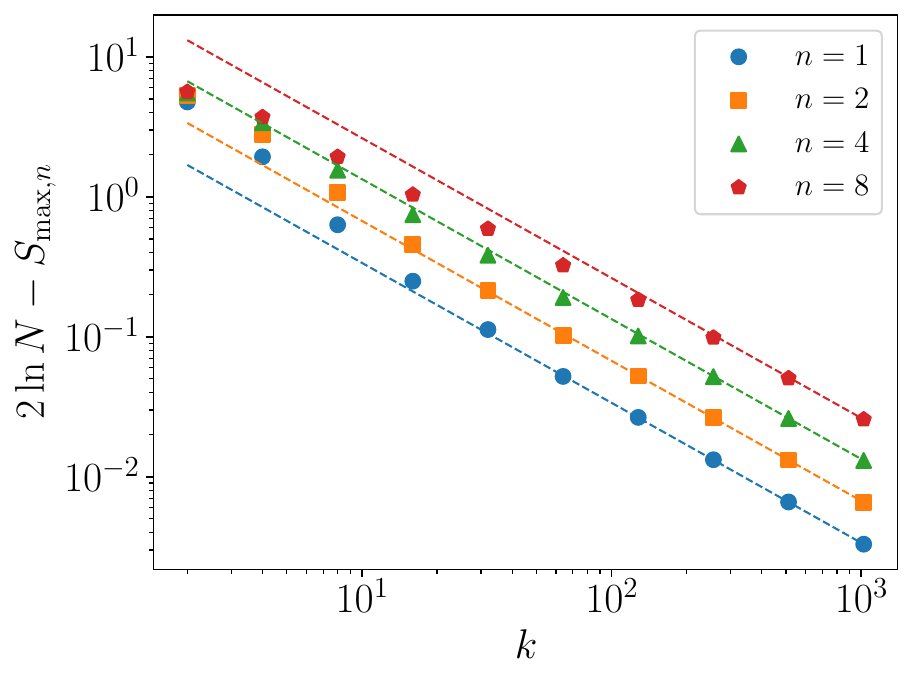}
    \caption{$k^{-1}$-dependence of $2\ln N - S_{\text{max}, n}$ in the quadratic + quartic model. Here $N=24$. We used $10^3$ samples.}
    \label{fig:DEntropykdep}
\end{figure}

In figure~\ref{fig:DEntropyProperties}(a), we present the bounding behavior of the averaged entanglement and Renyi entropies ($n=1,2,4,8$) of an interval in the quadratic + quartic model. In (b) we show that the average entropy bound is equal to $2\ln N$ plus an $\mathcal{O}(N^0)$ subleading term. Similar results hold for the other two models, which can be found in appendix~\ref{ssec:supplementary_figures_drmps}.

In figure~\ref{fig:DEntropykdep} we present the $k^{-1}$ dependence of $2\ln N - S_{\text{max}, n}$ ($n=1,2,4,8$) in the quadratic + quartic model. Similar results hold for the other two models, which can be found in appendix~\ref{ssec:supplementary_figures_drmps}.

\section{Off-shell wormholes}

\label{sec:offshell}

\subsection{Off-shell wormholes}

\label{ssec:offshelldefinition}

We consider the AdS JT+CFT model. The bra-ket wormhole, as well as $n$-Renyi replica wormholes, are examples of higher topology geometries that are also solutions of the AdS JT+CFT field equations. We call such geometries on-shell wormholes. On the other hand, we can have higher topology geometries that are not solutions of the AdS JT+CFT field equations. We call such geometries off-shell wormholes. In general, given a specific boundary configuration, for a given topology $(n, g)$, there may not be a solution of the AdS JT+CFT field equations. In that case, the topology $(n,g)$ can only contribute through off-shell wormholes.

We note that while we may generally denote a higher topology geometry in its entirety as a wormhole, it is sometimes helpful to view a higher topology geometry as the result of adding a wormhole to a lower topology geometry, in the sense of either increasing $n$ or $g$. We will sometimes denote this added wormhole as a wormhole, as opposed to the whole geometry.

We remark that in the AdS JT+CFT model, we can only compute on-shell wormhole effects with the saddlepoint approximation. Due to the Casimir divergence, which we will review below, there is no known procedure for computing off-shell wormhole effects. In RMPS, by contrast, all higher-topology corrections are well defined within the random-state model. This is a novel advantage of using RMPS models of gravitationally prepared states.

However, because the off-shell wormhole effects we are computing have no concretely computable gravitational path integral counterparts, the gravitational interpretations of the computations done in this section should be understood as conjectural interpretations based on basic qualitative behaviors of the gravitational path integral such as the genus expansion and large $L$ dependence. It remains an open question to regulate the Casimir divergence and rigorously calculate off-shell wormhole effects from the gravitational path integral.

\paragraph{The Casimir divergence.} The sum of off-shell wormhole contributions are not well defined in the AdS JT+CFT model, due to an instability in the effective action occurring for wormholes of infinitely small width.

To show this, we consider the simple case of the ket-ket wormhole, which is a geometry connecting two kets, and contributes to
\begin{equation}
    Z_1\left[\bra{\psi}\right]  Z_1\left[\bra{\psi}\right]  Z_2 \left[\ket{\psi} \ket{\psi}\right]  .
\end{equation}
We consider the contribution where two disk geometries in $Z_1\left[\bra{\psi}\right]$ are connected to the ket-ket wormhole geometry in $Z_2 \left[\ket{\psi} \ket{\psi}\right] $. The ket-ket wormhole geometry is given by
\begin{equation}
	ds^2 = \frac{dz^2 + dx^2}{\left( \frac{L}{b} \sin \frac{b  z}{L} \right)^2} ,\quad 0 < z <  \frac{\pi L}{b}, \quad x \approx x+ L.
\end{equation}
$b$ is equal to the width of the ket-ket wormhole. Here we have integrated out the dilaton field $\phi$, as well as the boundary wiggles, leaving the modulus parameter $b$, which is integrated over in the gravitational path integral.

The effective action of the ket-ket wormhole geometry as a function of $b$ is given by
\begin{equation}
    S_{\text{eff}} [b] =  \frac{\phi_r b^2}{2\pi L} + \frac{cb}{24} - \frac{\pi^2 c}{6b} .
\end{equation}
As $b \rightarrow 0$, we have $S_{\text{eff}}  \rightarrow - \infty$. This is due to the term $\frac{\pi^2 c}{6b} $, which is due to the Casimir energy on the wormhole waist. As $b$ goes to zero, the Casimir energy goes to minus infinity, which causes the instability.

We expect that this phenomenon happens generally for off-shell geometries of any topology, as taking the width of an off-shell wormhole to zero always causes an instability due to the diverging Casimir energy. We call this the Casimir divergence.

\subsection{RMPS and the Casimir divergence} RMPS gives us well-defined models of two-dimensional gravitationally prepared states without divergences. It seems that the Casimir divergence is somehow regulated in RMPS. The question is, how?

\paragraph{Gravitational path integral picture.} There are two cases. One is that RMPS regulates the widths of off-shell wormholes. This would amount to having a minimum value of $b$ in the ket-ket wormhole. Another is that RMPS regulates the lengths of off-shell wormholes. The length of a geodesic running across the ket-ket wormhole geometry at constant $x$ is given by $2 \ln \frac{L}{b} + \text{const}$. Thus a wormhole length regulator would amount to having a minimum value of $\frac{b}{L}$ in the ket-ket wormhole. The two cases are related, but are different in that having a wormhole length regulator allows wormholes to have smaller minimum widths when they are closer to the boundary.

We now compute the expected $L$-dependence of
\begin{equation}
    \frac{Z_1\left[\bra{\psi}\right]  Z_1\left[\bra{\psi}\right]  Z_2 \left[\ket{\psi} \ket{\psi}\right] }{\left( Z_1[\bra{\psi}]  Z_1[\ket{\psi}] \right)^2},
\end{equation}
at highest order in $e^{\phi_0}$, for both cases. In either case, the ket-ket wormhole contribution is dominated by the minimum possible value of $b$, where the action is minimum.

For the wormhole width regulated case, we have
\begin{equation}
    S_{\text{eff,min}} \approx \frac{cb_{\text{min}}}{24} - \frac{\pi^2 c}{6b_{\text{min}}}  = \text{const}.
\end{equation}
Furthermore, we get an additional $L^2$ factor from the integrals over boundary wiggles \cite{stanford_fermionic_2017, saad_jt_2019}. In conclusion, we get
\begin{equation}
     \frac{Z_1\left[\bra{\psi}\right]  Z_1\left[\bra{\psi}\right]  Z_2 \left[\ket{\psi} \ket{\psi}\right] }{\left( Z_1[\bra{\psi}]  Z_1[\ket{\psi}] \right)^2} \propto e^{-2\phi_0} L^2   + \mathcal{O}(e^{-4 \phi_0}),
\end{equation}
i.e. $L^2$ dependence.

For the wormhole length regulated case, we have
\begin{equation}
    S_{\text{eff,min}} [b] \approx  \frac{\phi_r L}{2\pi} \left( \frac{b}{L}\right)^2_{\text{min}} + \frac{cL}{24} \left( \frac{b}{L}\right)_{\text{min}} = \sigma L,
\end{equation}
where $\sigma$ is a constant. In conclusion, we get
\begin{equation}
     \frac{Z_1\left[\bra{\psi}\right]  Z_1\left[\bra{\psi}\right]  Z_2 \left[\ket{\psi} \ket{\psi}\right] }{\left( Z_1[\bra{\psi}]  Z_1[\ket{\psi}] \right)^2} \approx  e^{-2\phi_0} L^2 \exp(- \sigma L)+ \mathcal{O}(e^{-4 \phi_0}),
\end{equation}
i.e. exponential decay at large $L$.

\begin{figure}[t]
    \centering
    \begin{subfigure}[t]{0.45\textwidth}
        \centering
        \includegraphics[width=\textwidth]{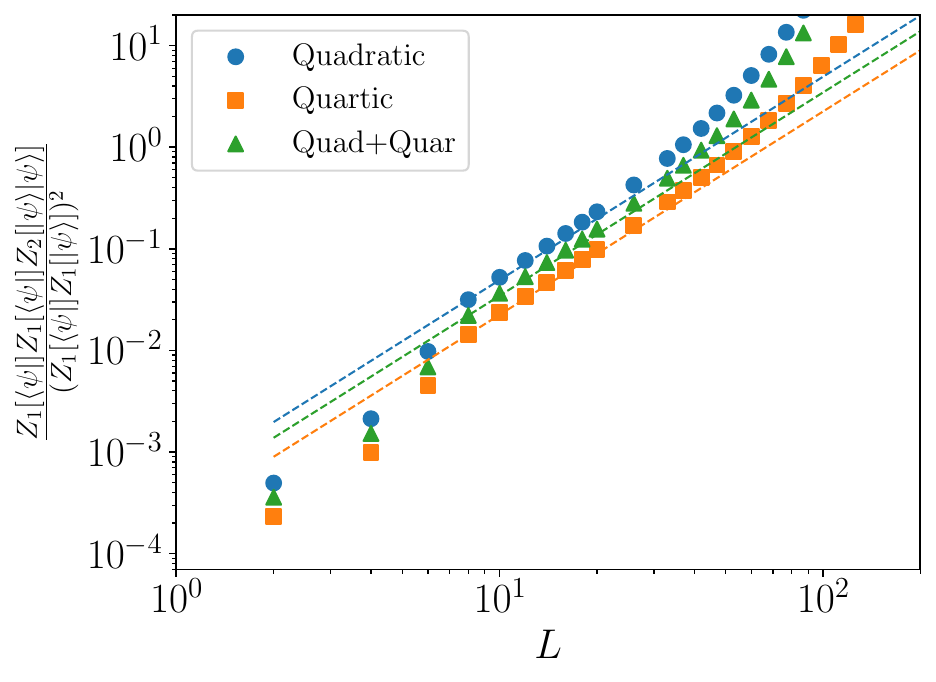}
        \caption{}
    \end{subfigure}
    \hspace{0.05\textwidth}
    \begin{subfigure}[t]{0.45\textwidth}
        \centering
        \includegraphics[width=\textwidth]{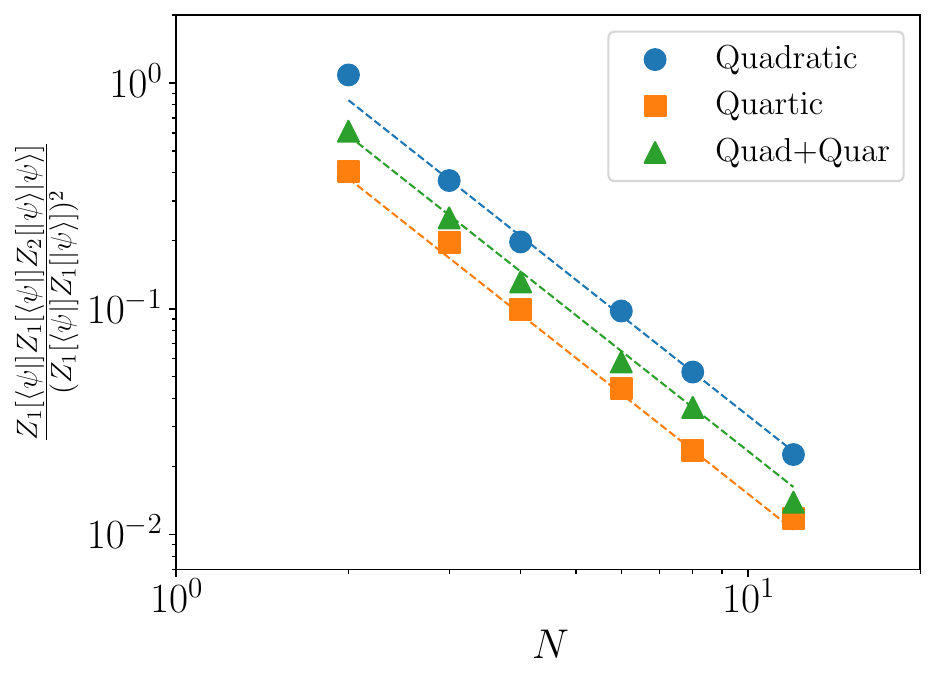}
        \caption{}
    \end{subfigure}

	\caption{$L$ and $N$ dependence of $\frac{Z_1\left[\bra{\psi}\right]  Z_1\left[\bra{\psi}\right]  Z_2 \left[\ket{\psi} \ket{\psi}\right] }{\left( Z_1[\bra{\psi}]  Z_1[\ket{\psi}] \right)^2}$ for quadratic, quartic, and quadratic + quartic models. We used $10^6$ samples with $k=64$. (a) We see $L^{2}$ dependence in the intermediate $L$ region. Here $N=8$. (b) We see $N^{-2}$ dependence. Here $L=10$.}

        \label{fig:Dketket}
\end{figure}

\paragraph{Numerical RMPS results.} In figure~\ref{fig:Dketket}, we present the $L$ and $N$ dependence of the quantity $\frac{Z_1\left[\bra{\psi}\right]  Z_1\left[\bra{\psi}\right]  Z_2 \left[\ket{\psi} \ket{\psi}\right] }{\left( Z_1[\bra{\psi}]  Z_1[\ket{\psi}] \right)^2}$ for the quadratic, quartic, and quadratic + quartic models. In the intermediate $L$ region, we have $N^{-2}$ dependence and $L^2$ dependence.

We remark that at very large $L$, higher genus effects become important in the quantity $\frac{Z_1\left[\bra{\psi}\right]  Z_1\left[\bra{\psi}\right]  Z_2 \left[\ket{\psi} \ket{\psi}\right] }{\left( Z_1[\bra{\psi}]  Z_1[\ket{\psi}] \right)^2}$. We checked the $L$-dependence of this quantity below this $L$ range.

Thus, we see that the numerical results on the RMPS side agree with the wormhole width regulated case. Under the conjectural gravitational interpretation described above, the observed RMPS scaling is consistent with regulating the Casimir divergence through a lower cutoff on wormhole width.

On-shell wormholes such as the bra-ket wormhole and $n$-Renyi replica wormholes give physically meaningful effects. We will see that off-shell wormholes can also give physically meaningful effects. In particular, in gravitationally prepared states, off-shell wormholes give nontrivial contributions to correlators, which become dominant at large distances. We will study this in detail below.

\subsection{Wormhole contributions to the inner product}

\label{ssec:offshellinnerprod}

Here we discuss the story of how fluctuations in the highest eigenvalue of the transfer matrix $\lambda_g$ connect to off-shell wormholes on the bra-ket wormhole. This example illustrates how off-shell wormhole contributions are encoded in RMPS structure.

\textbf{Gravitational path integral picture.} We consider computing $\left< \braket{\psi}{\psi} \right>$, where $L \rightarrow \infty$. This is equal to computing $\left< \lambda_g^L \right>$. Denote the probability distribution of $\lambda_g$ as $P(\lambda_g)  d \lambda_g$. Then, we see that
\begin{equation}
    \left< \braket{\psi}{\psi} \right> = \int d \lambda_g P(\lambda_g) \lambda_g^L.
\end{equation}
Note that we can understand the right hand side as a generating functional for cumulants of $\ln \lambda_g$, i.e.
\begin{equation}
     \ln \left< \braket{\psi}{\psi} \right> \supset \sum^{\infty}_{n=1} \left<(\ln \lambda_g)^n \right>_{\text{conn}} L^n
\end{equation}
as $L \rightarrow \infty$. Here we used $\supset$ instead of $=$ in case the latter is an asymptotic series.

We now compute the right-hand side from the point of view of the gravitational path integral. Here, $\braket{\psi}{\psi}$ is dominated by a bra-ket wormhole geometry, which gives the contribution
\begin{equation}
    \exp( \alpha L) .
\end{equation}
But we can also have contributions of this bra-ket wormhole geometry along with some off-shell wormholes attached to it. We wish to compute the $L$ dependence of such contributions. Each off-shell wormhole has finite size, which, compared to $L \rightarrow \infty$, takes up negligible space along the $x$ axis (i.e. the width direction) of the bra-ket wormhole. Thus we can characterize each off-shell wormhole as connecting to $n$ distinct points along the $x$ axis of the bra-ket wormhole. We call these off-shell wormholes $n$-point wormholes.

\begin{figure}[t]
\begin{center}
\resizebox{\linewidth}{!}{\input{figures/tikz/cnpoint}}
    \end{center}
    \caption{Left: a contribution to $\braket{\psi}{\psi}$ for large $L$ with a 2-point wormhole and a 4-point wormhole. Right: $C_{4-\text{point}}$ as a sum of 4-point wormhole geometries.}
    \label{fig:cnpoint}
\end{figure}
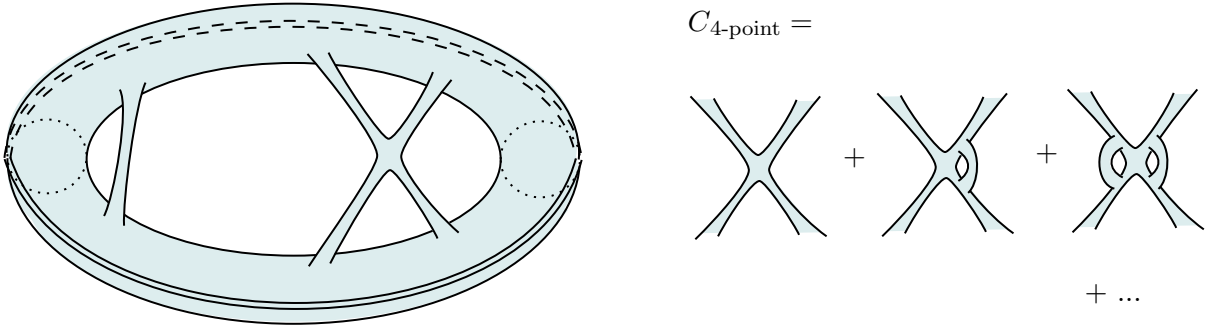

An $n$-point wormhole has an additional topological suppression cost of $N^{2(1-n)}$, or higher, if the $n$-point wormhole has additional nontrivial genus. An $n$-point wormhole also has an associated moduli factor of $L^n$, given by the number of ways to place its $n$ endpoints along the $x$-axis of the bra-ket wormhole. Thus, the total contribution of geometries with one $n$-point wormhole on the bra-ket wormhole is given by
\begin{equation}
    \exp(\alpha  L)  C_{n\text{-point}} L^n,
\end{equation}
where $C_{n\text{-point}}  \sim \mathcal{O}(N^{2(1-n)})$ is the sum over all possible shapes of $n$-point wormholes, including topological suppression costs. An illustration is given in figure~\ref{fig:cnpoint}.

It follows that the total contribution of geometries with $m_n$ $n$-point wormholes for each $n$ is given by
\begin{equation}
    \exp( \alpha  L)  \prod^\infty_{n=1} \left(  \frac{1}{m_n!} C^m_{n\text{-point}} L^{mn} \right).
\end{equation}
Summing all this up, we get
\begin{equation}
    \ln \left< \braket{\psi}{\psi} \right> \supset \sum^{\infty}_{n=1} C_{n\text{-point}} L^n,
\end{equation}
where we have denoted $\alpha = C_{1\text{-point}}$.

Thus, we see that
\begin{equation}
     \left<(\ln \lambda_g)^n \right>_{\text{conn}}  = C_{n\text{-point}}  \sim \mathcal{O}(N^{2(1-n)}).
\end{equation}
This tells us that the $n$-th cumulant of $\ln \lambda_g$ has $N^{2(1-n)}$ dependence at large $N$. Furthermore, the $n$-th cumulant can be understood as the sum over all possible shapes of $n$-point wormholes,  including topological suppression costs.

\paragraph{Numerical RMPS results.} We numerically computed $C_{n-\text{point}}$ for $n=2,3,4$ for the quadratic, quartic, and quadratic + quartic models. For the $n=2,4$ cases of the quadratic model, and in all cases of the quartic and quadratic + quartic models, we verified that $C_{n-\text{point}}$ has $N^{2(1-n)}$ dependence at large $N$. For the $n=3$ case of the quadratic model, $C_{n-\text{point}}$ decayed faster than $N^{-4}$, implying that the leading order $3$-point wormhole topology is absent in this model. We remark that this may be due to the lack of 't Hooft vertices in the quadratic model.\\

\begin{figure}[t]
    \centering
    \includegraphics[width=0.45\textwidth]{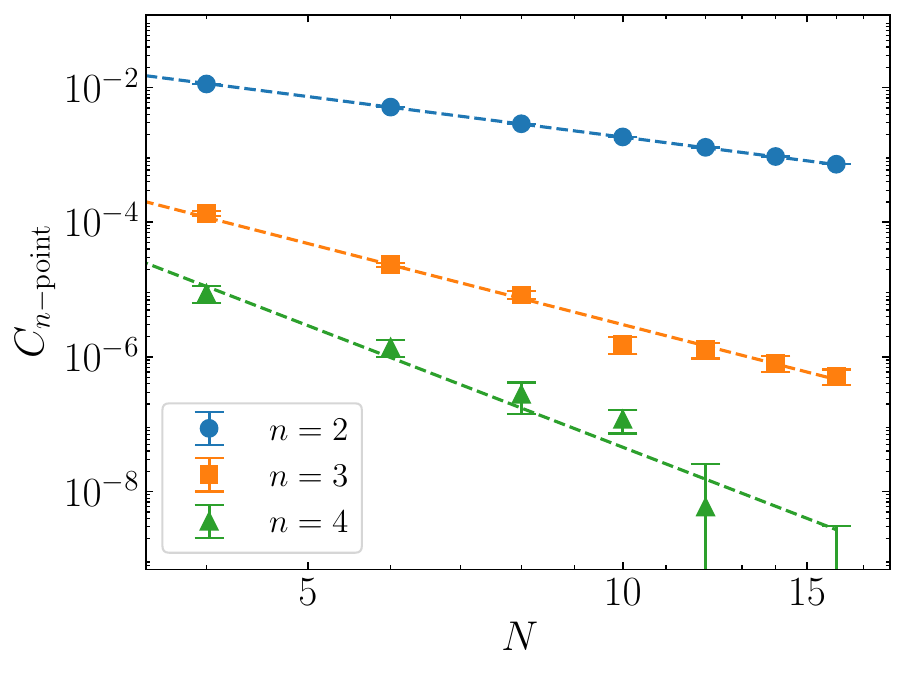}
	\caption{$N$ dependence of $\ln\lambda_g$ cumulants in the quadratic + quartic model. For large $N$, we have $N^{2(1-n)}$ dependence. We used $10^{5}$ samples with $k=8$.}
    \label{fig:EigLogCumu}
\end{figure}

In figure~\ref{fig:EigLogCumu}, we present the $N$ dependence of $n$th $\ln\lambda_g$ cumulants in the quadratic + quartic model. For large $N$, we have $N^{2(1-n)}$ dependence. Results for the other models can be found in appendix~\ref{ssec:supplementary_figures_drmps}.

\paragraph{A remark on self-averaging.} We say that a random quantity is self-averaging if its variance is significantly smaller than its mean squared. Here, we see that, due to off-shell wormhole effects, the quantity $\braket{\psi}{\psi}$ will not be self-averaging at very large $L$, at the order of $L \sim \mathcal{O}(N)$ or higher.

This, however, does not mean that physics breaks down at large $L$. This is because physical observables involve normalized kets, i.e. $\frac{\ket{\psi}}{\sqrt{\braket{\psi}{\psi}}}$. For example, the quantity $\frac{\bra{\psi}O\ket{\psi}}{\braket{\psi}}$, where $O$ is some observable, continues to self-average at large $L$ if $N$ is large. The same is true for the entropy of a subregion. Thus, there is nothing physically alarming about the self-averaging breaking of the inner product at large $L$.

\subsection{Nonzero long-distance correlators}

\label{ssec:offshellcorr}

\textbf{Gravitational path integral picture.} We consider the $n$-point correlator
\begin{equation}
    \left< \frac{\bra{\psi} O(x_n) ... O(x_1) \ket{\psi}}{\braket{\psi}{\psi}} \right>
\end{equation}
in the bra-ket wormhole dominant phase. Here we assume we have subtracted the averaged one-point function $\left<  \frac{\bra{\psi} O(0)\ket{\psi}}{\braket{\psi}{\psi}}  \right>  $ from $O(x)$.

Consider the case where $|x_m - x_l| \gg 1$ for all $m>l$. On the level of the bra-ket wormhole geometry, we predict that the correlator decays exponentially to zero as the distances $|x_m - x_l|$ go to infinity. However, we will also have contributions of $n$-point wormholes connecting two or more points $x_m$, creating shortcuts and thus yielding a nonzero value of the correlator even at infinite distances. When the distances $|x_m - x_l|$ are sufficiently large, these contributions can become the dominant contribution to the correlator. Thus, nonzero long-distance correlators are an example of an off-shell wormhole effect that is physically relevant.

We note that the wormhole topologies studied here are the same as that of \cite{Stanford:2020wkf}, although in a purely Euclidean context. Also, we do not rule out the possibility that for certain types of operator insertions $O(x)$, there may be stable wormholes contributing to the nonzero long-distance correlators. Our results do not depend on whether this happens.

Since the contribution of $n$-point wormholes only depends on the local physics at each point $x_m$, when the distances between operator insertions $|x_m - x_l|$ and $L$ become sufficiently large, the nonzero long-distance correlators become independent of $|x_m - x_l|$ and $L$.

\begin{figure}[t]
    \begin{center}

    \resizebox{\linewidth}{!}{\input{figures/tikz/conpoint}}
    \end{center}
    \caption{Left: a contribution to the nonzero long-distance value of the 6-point function $\left< \frac{\bra{\psi} O(x_6) ... O(x_1) \ket{\psi}}{\braket{\psi}{\psi}} \right>$ with a 2-point wormhole and a 4-point wormhole. Right: $C_{O;4-\text{point}}$ as a sum of 4-point wormhole geometries with operator insertions.}
    \label{fig:conpoint}
\end{figure}
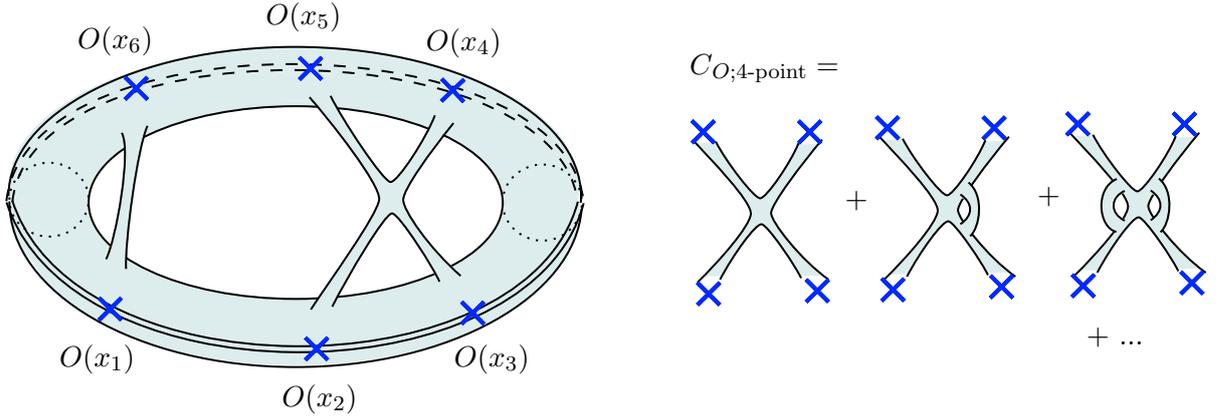

To compute the nonzero long-distance correlators, it helps to take $|x_m -x_l| \rightarrow \infty$, along with $L \rightarrow \infty$. In particular, we would be interested in the contribution of $n$-point wormholes to the nonzero long-distance correlator at $|x_m -x_l| \rightarrow \infty$, $L \rightarrow \infty$. We call this the $n$-point correlator offset, and denote it as $C_{O;n-\text{point}}$. An illustration is given in figure~\ref{fig:conpoint}.

From the gravitational path integral side, it follows from the topological suppression of $n$-point wormholes that
\begin{equation}
\label{eq:conpointndep}
    C_{O;n-\text{point}} \sim \mathcal{O}(N^{2(1-n)}) .
\end{equation}
Thus, the $n$-point correlator offset scales as $N^{2(1-n)}$ for large $N$.

We remark that, for large $N$, in computing the $n$-point function, contributions from products of lower point correlator offsets dominate over that of higher point correlator offsets. The most important effect would be the two-point correlator offset $C_{O;2-\text{point}}$, which is of $\mathcal{O}(N^{-2})$ and gives us the asymptotic value of the two-point function
\begin{equation}
    \left< \frac{\bra{\psi} O(x) O(0) \ket{\psi}}{\braket{\psi}{\psi}} \right> .
\end{equation}
We can imagine, in a two-dimensional expanding universe, measuring such two-point functions of the cosmic microwave background radiation and indeed obtaining an $\mathcal{O}(N^{-2})$ quantity which would be the two-point correlator offset.

\paragraph{Numerical results.} Taking $|x_m -x_l| \rightarrow \infty$ and $L \rightarrow \infty$, we see that
\begin{equation}
    \left< \frac{\bra{\psi} O(x_n) ... O(x_1) \ket{\psi}}{\braket{\psi}{\psi}} \right> = \left< \left(  \frac{\bra{\psi} O(0)\ket{\psi}}{\braket{\psi}{\psi}} \right)^n \right> = \left< \left(\lambda_g^{-1}  \bra{\lambda_g}  M_O \ket{\lambda_g}\right)^n \right>,
\end{equation}
where $\ket{\lambda_g}$ is the highest eigenvalue eigenstate of the transfer matrix $M$. In particular, we see that the $n$-point correlator offset is equal to the $n$-th cumulant of $\lambda_g^{-1}  \bra{\lambda_g}  M_O \ket{\lambda_g}$:
\begin{equation}
    C_{O;n-\text{point}} = \left< \left(\lambda_g^{-1}  \bra{\lambda_g}  M_O \ket{\lambda_g}\right)^n \right>_{\text{conn}}.
\end{equation}
Thus we study these cumulants on the RMPS side.

We numerically computed $C_{O;n-\text{point;cRMPS}}$ for $n=2,3,4$ for the quadratic, quartic, and quadratic+quartic models. In all cases, we verified that $C_{O;n-\text{point}}$ has $N^{2(1-n)}$ dependence at large $N$.

\begin{figure}[t]
    \centering
    \includegraphics[width=0.45\textwidth]{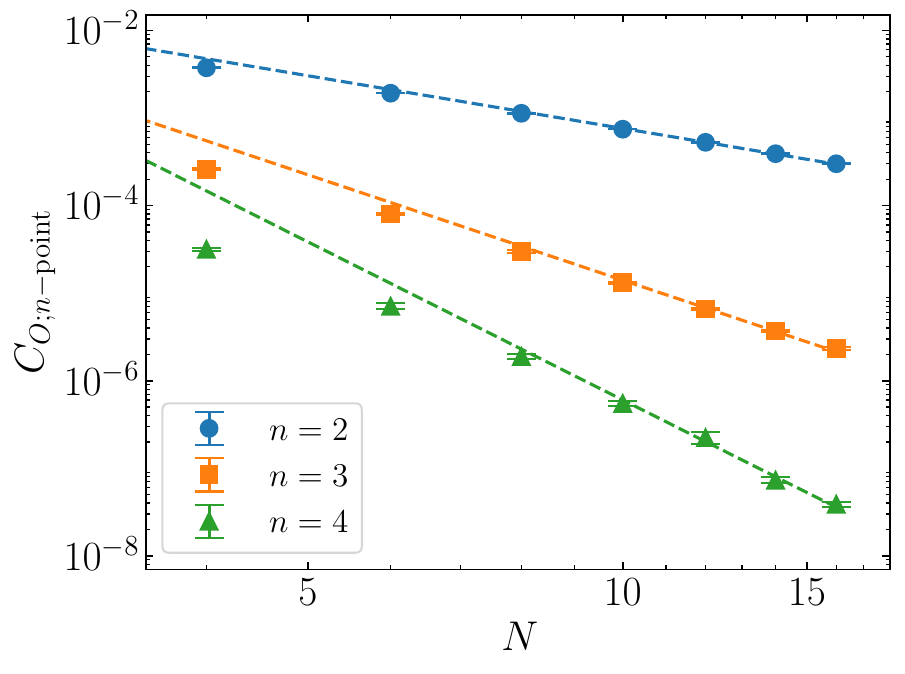}
	\caption{$N$ dependence of $\lambda_g^{-1}\bra{\lambda_g}M_O\ket{\lambda_g}$ cumulants in the quadratic + quartic model. For large $N$, we have $N^{2(1-n)}$ dependence. We used $k=8$ with $10^{5}$ samples.}
    \label{fig:DCorrCumu}
\end{figure}

In figure~\ref{fig:DCorrCumu}, we present the $N$ dependence of $n$th $\lambda_g^{-1}\bra{\lambda_g}M_O\ket{\lambda_g}$ cumulants in the quadratic + quartic model. For large $N$, we have $N^{2(1-n)}$ dependence. Similar results hold for the other models, which can be found in appendix~\ref{ssec:supplementary_figures_drmps}.

\section{Continuous RMPS models}

\label{sec:crmps}

\subsection{Motivation and definition}

\label{ssec:crmpsdefinition}

So far, we have studied models that take $\Sigma$ to be discrete. It is worth asking whether we can have random state models that take $\Sigma$ to be continuous. The answer is yes. We can do this using continuous matrix product states (cMPS) \cite{ganahl_continuous_2017}.

We take $\Sigma$ to be a circle of length $L$, $x= x+L$. We define the local bosonic creation and annihilation operators $c_i(x)$, $c_i^\dagger (x)$ with $i = 1, ..., k$ satisfying
\begin{equation}
    \commutator{c_i(x)}{c_j^\dagger(x^\prime)} = \delta_{ij} \delta(x - x^\prime) .
\end{equation}
We define the Hilbert space $\mathcal{H}_{\text{QFT,}\Sigma}$ of the matter fields on $\Sigma$ as the Hilbert space generated by the states
\begin{equation}
    c^\dagger_{i_m}(x_m) ... c^\dagger_{i_1} (x_1) \ket{0} , \quad x_1, ..., x_m \in \Sigma,
\end{equation}
for all nonnegative integers $m$, where $\ket{0}$ is the zero state satisfying $c_i(x)\ket{0} = 0$ for all $x \in \Sigma$.

Given $k$ random $N \times N$ Hermitian matrices $\{A_i\}$ selected from a unitary ensemble, we define the following state:
\begin{equation}
    \ket{\psi} = \Tr  \left[ \mathcal{T}_x \left\{ \exp \left( \int^L_0 dx  \left( \sum^k_{i=1} A_i c_i^\dagger(x) \right) \right)  \right\} \right] \ket{0}.
\end{equation}
Here $\Tr$ denotes the trace over $\mathbb{C}^N$. $ \mathcal{T}_x $ denotes ordering the $ \sum^k_{i=1} A_i c_i^\dagger(x) $ in increasing order of $x$ from right to left. $\ket{\psi}$ is a cRMPS of bond dimension $N$.

The cRMPS model is a continuous version of the RMPS model. We may also call the RMPS model the discrete RMPS model (dRMPS) in comparison. The cRMPS model is a good candidate for a model of a gravitationally prepared state for the following reasons which are inherited from the discrete model.

First, as a single-trace observable of the matrices $\{A_i\}$, $\ket{\psi}$ has a topological expansion in terms of $N$. Contributions to $\left< \ket{\psi}^{\otimes n} \right>$ can be grouped in terms of topologies of 2D surfaces connecting the $n$ boundaries. The contribution associated with a connected $n$-boundary surface of genus $g$ is proportional to $N^{2-2g-n}$. The topological expansion of cRMPS matches that of 2D gravity if we relate $N$ to the parameter $\phi_0$ by $ N = e^{\phi_0}$.

Second, $\ket{\psi}$ has the following entropy bounding property: given the spatial subregion $R$ consisting of $m$ disjoint intervals, we have
\begin{equation}
    S[R] \leq 2m \ln N.
\end{equation}
We will show this property using the transfer Hamiltonian below. This is a sample entropy bound, which implies an average entropy bound. At large $N$, the entropy bounding property of cRMPS matches that of 2D gravity if we relate $N$ to the parameter $\phi_0$ by $ N = e^{\phi_0}$.

In both of these reasons we have the same identification
\begin{equation}
    N = e^{\phi_0},
\end{equation}
which we will use as the canonical identification when comparing results in cRMPS to 2D gravity. We remark that, in gravity, we take $\phi_0$ to be large. Thus, in cRMPS models of gravitationally prepared states, $N$ must be very large.

We remark on how subregion entropies are defined in cRMPS. We define the Hilbert space $\mathcal{H}_{\text{QFT}, R}$ associated with the spatial subregion $R$ as the Hilbert space generated by the states
\begin{equation}
    c^\dagger_{i_m}(x_m) ... c^\dagger_{i_1} (x_1) \ket{0_R} , \quad x_1, ..., x_m \in R,
\end{equation}
for all nonnegative integers $m$, where $\ket{0_R}$ is the zero state satisfying $c_i(x)\ket{0_R} = 0$ for all $x \in R$. Given two disjoint regions $R_1, R_2$, we have
\begin{equation}
    \mathcal{H}_{\text{QFT}, R_1 \cup R_2} = \mathcal{H}_{\text{QFT}, R_1} \otimes \mathcal{H}_{\text{QFT}, R_2}, \quad \ket{0_{R_1 \cup R_2}} = \ket{0_{R_1}} \otimes \ket{0_{R_2}}.
\end{equation}
For example, $\mathcal{H}_{\text{QFT}, \Sigma} = \mathcal{H}_{\text{QFT}, R} \otimes \mathcal{H}_{\text{QFT}, R^c}$ and $\ket{0} = \ket{0_R} \otimes \ket{0_{R^c}}$. We can now define the entropy of a spatial subregion as the Von Neumann entropy of the reduced density matrix in the Hilbert space of that spatial subregion.

We note that, in continuum quantum field theory, there is generally no well-defined Hilbert space associated with a spatial subregion $R$. Indeed, the local behavior of the cRMPS model described here differs from that of a relativistic quantum field. We remark that models of continuous relativistic matrix product states have been suggested \cite{tilloy_relativistic_2021}. However, these models do not a priori satisfy the desired entropy bounding properties, intuitively because $\ket{0}$ does not factorize for a relativistic quantum field theory. Thus we will not dwell on such constructions here.

\paragraph{Transfer Hamiltonian technology.} We compute the inner product $\braket{\psi}{\psi}$. For the transfer Hamiltonian
\begin{equation}
H = - \sum^k_{i=1} A^*_i \otimes A_i,
\end{equation}
we see that
\begin{equation}
    \begin{aligned}
        &\braket{\psi}{\psi}\\
        &= \bra{0} \Tr  \left[ \mathcal{T}_x \left\{ \exp \left( \int^L_0 dx  \left( \sum^k_{i=1} ((A^*_i \otimes I_{N \times N} ) c_i(x) + (I_{N \times N} \otimes A_i) c_i^\dagger(x) \right) \right)  \right\} \right] \ket{0} \\
        &=  \Tr  \left(e^{-L H} \right).
    \end{aligned}
\end{equation}
Furthermore, we see that
\begin{equation}
    \begin{aligned}
        &\bra{\psi} \Tr  \left[\exp\left(  \int^L_0 dx \left(\sum^k_{i=1} \left(J^*_i(x)  c_i (x) + J_i(x) c_i^\dagger(x) \right)\right)\right) \right] \ket{\psi}\\
        &= \Tr  \left[ \mathcal{T}_x \left\{ \exp \left( \int^L_0 dx  \left( \sum^k_{i=1} ((A^*_i +J^*_i (x)) \otimes (A_i +J _i (x))  )   \right) \right)  \right\} \right]
    \end{aligned}
\end{equation}
and so, for $O = c, c^\dagger$,
\begin{equation}
    \begin{aligned}
        \bra{\psi} O_{i_n} (x_n) ... O_{i_1} (x_1) \ket{\psi} &= \Tr  \left(e^{-(L-x_n) H} \Phi_{O; i_n} ...  e^{-(x_2 - x_1)H } \Phi_{O;i_1} e^{-L x_1} \right),\\
        \Phi_{c;i} &= A^*_i \otimes I_{N\times N},\\
        \Phi_{c^\dagger; i}  &=  I_{N\times N} \otimes A_i.
    \end{aligned}
\end{equation}
We call $\Phi_{c; i}$ and $\Phi_{c^\dagger;i}$ the effective matter operators associated with the local operators $c$ and $c^\dagger$, respectively.

We define the Hermitian field operator $\varphi_i = c_i + c_i^\dagger$. We see that
\begin{equation}
    \begin{aligned}
        \bra{\psi} \varphi_{i_n} (x_n) ... \varphi_{i_1} (x_1) \ket{\psi} &= \Tr  \left(e^{-(L-x_n) H} \Phi_{i_n} ...  e^{-(x_2 - x_1)H} \Phi_{i_1} e^{-L x_1} \right),\\
        \Phi_i &= A^*_i \otimes I_{N\times N}+  I_{N\times N} \otimes A_i.
    \end{aligned}
\end{equation}
We call $\Phi_i $ the real effective matter operator.

We also have the disconnected transfer Hamiltonian and disconnected real effective matter operator
\begin{equation}
    H_{\text{disc}} = - \sum^k_{i=1} A^{\prime *}_i \otimes A_i, \quad \Phi_{\text{disc},i} = A^{\prime *}_i \otimes I_{N\times N} + I_{N\times N} \otimes A_i,
\end{equation}
which can be used to compute bra-ket disconnected quantities. Here $\{A_i\}$ and $\{A^\prime_i\}$ are two independent selections from the matrix ensemble $d\mu(\{A_i\})$.

We can now use the transfer Hamiltonian to compute the entropies of spatial subregions. We consider the spatial subregion $R$ consisting of $m$ disjoint intervals. We denote the length of the $i$-th interval as $d_i$ and the spacing between the $i$-th and $i+1$-th intervals as $\ell_i$. Without loss of generality, we can assume the first interval starts at $x= 0$. The reduced density matrix of $R$ is given by
\begin{equation}
        \rho_R =    \bra{\psi}  I_{\mathcal{H}_{\text{QFT,} R}} \ket{\psi},
\end{equation}
where $I_{\mathcal{H}_{\text{QFT,} R}}$ is the identity operator on $\mathcal{H}_{\text{QFT,} R}$. Here the non-contracted tensor indices in $\mathcal{H}_{\text{QFT,} R^c}$ on the right-hand side constitute the free tensor indices on the left-hand side.

$\rho_R$ is a $ \infty \times \infty $ matrix. We note that we can write $\rho_R$ as a product of an $ \infty \times N^{2m} $ matrix $P_R$ with a $N^{2m} \times \infty$ matrix $Q_R$, given by
\begin{equation}
    \begin{aligned}
        P_R =& \left( \sum^N_{\mu_m = 1}... \sum^N_{\mu_1 = 1} \right)    \left( \sum^N_{\nu_m = 1}... \sum^N_{\nu_1 = 1} \right)  \\
        &\times\bra{\mu_m}  \mathcal{T}_x \left\{ \exp \left( \int^{L-\ell_m }_{L-\ell_m -d_m} dx  \left( \sum^k_{i=1} A_i c^\dagger_i(x) \right) \right)  \right\} \ket{\nu_m} \ket{0_{(L-\ell_m -d_m, L - \ell_m)}}\\
        &... \bra{\mu_1} \mathcal{T}_x \left\{ \exp \left( \int^{d_1}_{0} dx  \left( \sum^k_{i=1} A_i c^\dagger_i(x) \right) \right)  \right\}  \ket{\nu_1} \ket{0_{(0,d_1)}} \left(\bra{\mu_m\nu_m}  ...\otimes  \bra{\mu_1\nu_1} \right),
    \end{aligned}
\end{equation}
\begin{equation}
    \begin{aligned}
        Q_R =&\left( \sum^N_{\mu^\prime_m = 1}... \sum^N_{\mu^\prime_1 = 1} \right)  \left( \sum^N_{\nu^\prime_m = 1}... \sum^N_{\nu^\prime_1 = 1} \right) \left(\ket{\mu^\prime_m\nu^\prime_m} \otimes  ... \otimes \ket{\mu^\prime_1\nu^\prime_1} \right) \\
        &\times\bra{\psi} \bra{\nu^\prime_1} \mathcal{T}_x \left\{ \exp \left( \int^{L }_{L-\ell_m } dx  \left( \sum^k_{i=1} A_i c^\dagger_i(x) \right) \right)  \right\}  \ket{\mu^\prime_m}  \\
        &... \bra{\nu^\prime_2} \mathcal{T}_x \left\{ \exp \left( \int^{d_1 + \ell_1}_{d_1} dx  \left( \sum^k_{i=1} A_i c^\dagger_i(x) \right) \right)  \right\}   \ket{\mu^\prime_1}.
    \end{aligned}
\end{equation}
By linear algebra, the matrices $P_RQ_R$ and $Q_RP_R$ have the same nonzero eigenvalues. Thus, we can use $Q_RP_R$ in place of $\rho_R$ when computing the entanglement or $n$-Renyi entropy. Simplifying, we get
\begin{equation}
    \begin{aligned}
        &\left(\bra{\mu^\prime_m\nu^\prime_m} \otimes  ... \otimes \bra{\mu^\prime_1\nu^\prime_1} \right)  Q_RP_R\left(\ket{\mu_m\nu_m} \otimes  ... \otimes \ket{\mu_1\nu_1} \right)  \\
        & =\begin{aligned}[t]\Tr \Big( (I_{N\times N} \otimes \ket{\nu_1}  \bra{\nu^\prime_1}) e^{-\ell_m H} (I_{N\times N} \otimes \ket{\mu^\prime_m}  \bra{\mu_m}) e^{-d_m H} &\\
        \dots(I_{N\times N} \otimes \ket{\nu_2}  \bra{\nu^\prime_2}) e^{-\ell_1 H}(I_{N\times N} \otimes \ket{\mu^\prime_1}  \bra{\mu_1}) e^{-d_1 H} & \Big) .\end{aligned}
    \end{aligned}
\end{equation}
We call this matrix the compressed density matrix $\rho_{R, \text{compressed}}$ of $R$. We note that $\rho_{R, \text{compressed}}$ is generally not Hermitian but has nonnegative real eigenvalues. The fact that $\rho_{R, \text{compressed}}$ is a $N^{2m} \times N^{2m}$ matrix with real eigenvalues implies the entropy bound,
\begin{equation}
    S[R], \text{ } S_n[R] \leq 2m \ln N.
\end{equation}
Furthermore, unlike $\rho_R$, the dimensionality of $\rho_{R, \text{compressed}}$ is finite, which makes it possible to compute the entropy of $R$ numerically.

We remark that, for the $\mathcal{O}(k)$ models, like the RMPS models studied previously, we expect an $\mathcal{O}(N^0)$ subleading term in the average entropy bound of size $\mathcal{O}(k^{-1})$.

\paragraph{Spectral gapping property.} Following the logic of ~\ref{ssec:rmpsspecgap}, in cRMPS, we have the following spectral gapping property that ensures characteristic behaviors of the bra-ket wormhole phase transition:
\begin{equation}
    \begin{aligned}
        &\text{The lowest eigenvalue of $H$ is gapped.}\\
        &\text{The lowest eigenvalue of $H_{\text{disc}}$ is ungapped.}
    \end{aligned}
\end{equation}
We are interested in cRMPS models satisfying the spectral gapping property.

But since $H$ in cRMPS is equal to $-M$ in dRMPS, if we find an ensemble for $\{A_i\}$ satisfying the spectral gapping property for dRMPS, it will satisfy the spectral gapping property for cRMPS, and vice versa. Thus, the $\mathrm{O}(k)$ cRMPS models with $V(\{A_i\}) = k F(|A|^2)$ satisfy the spectral gapping property at large enough $k$, and we will primarily study said models.

\paragraph{Off-shell wormholes.} The content here mirrors that of section~\ref{sec:offshell}.

Like dRMPS, we can compute $\frac{Z_1\left[\bra{\psi}\right]  Z_1\left[\bra{\psi}\right]  Z_2 \left[\ket{\psi} \ket{\psi}\right] }{\left( Z_1[\bra{\psi}]  Z_1[\ket{\psi}] \right)^2} $ for cRMPS at large $L$ and check the $L$-dependence to determine which regularization for off-shell wormholes is coherent with the numerical results. We see that the results show $L^2$ dependence and thus agree with a cutoff on the wormhole width. We give the detailed numerical results below.

We denote the lowest eigenvalue of $H$ as $ - \lambda_g$ and the corresponding eigenstate as $\ket{- \lambda_g}$. We see that, for $L \rightarrow \infty$,
\begin{equation}
    \ln \left<  \braket{\psi}{\psi} \right> \supset \sum^\infty_{n=1} \left< \lambda_g^n \right>_{\text{conn}} L^n.
\end{equation}
Thus, from the gravitational path integral, we can argue that
\begin{equation}
    \left< \lambda_g^n \right>_{\text{conn}} = C_{n\text{-point;cRMPS}} \sim \mathcal{O}(N^{2(1-n)}),
\end{equation}
where $C_{n\text{-point;cRMPS}} $ is the sum over all possible shapes of $n$-point wormholes in cRMPS. We will check this behavior numerically below.

We can also consider nonzero long-distance correlators, which are given by
\begin{equation}
    C_{\Phi_i;n\text{-point;cRMPS}} = \left< \bra{-\lambda_g}\Phi_i \ket{-\lambda_g}^n \right>_{\text{conn}}.
\end{equation}
We remark that by $\mathbb{Z}_2$ symmetry of the matrix ensemble under $\{A_i\} \rightarrow \{- A_i\}$, $C_{\Phi_i;n\text{-point;cRMPS}}$ is zero for odd $n$. For even $n$, from the gravitational path integral, we can argue that
\begin{equation}
    C_{\Phi_i;n\text{-point;cRMPS}} \sim \mathcal{O}(N^{2(1-n)}).
\end{equation}
We will check this behavior numerically below.

\subsection{Numerical results}

\label{ssec:crmpsnumerical}

For brevity, we present selected figures in this section. The rest of the figures can be found in appendix~\ref{ssec:supplementary_figures_crmps}.

Since the transfer Hamiltonian $H$ of the cRMPS models is equal to the transfer matrix $M$ of the RMPS models, we do not need to recalculate the spectra of the cRMPS models.

\paragraph{Bra-ket wormhole phase transition.} Here we remark that $\left< \frac{\bra{\psi} \Phi_i \ket{\psi}}{\braket{\psi}{\psi}}  \right> = 0$ by the $\mathbb{Z}_2$ symmetry of $V(\{A_i\})$.

We see that all three models satisfy the three conditions of the bra-ket wormhole phase transition. This is expected, as they satisfy the spectral gapping property.

\begin{figure}[t]
    \centering
    \begin{subfigure}[t]{0.45\textwidth}
        \centering
        \includegraphics[width=\textwidth]{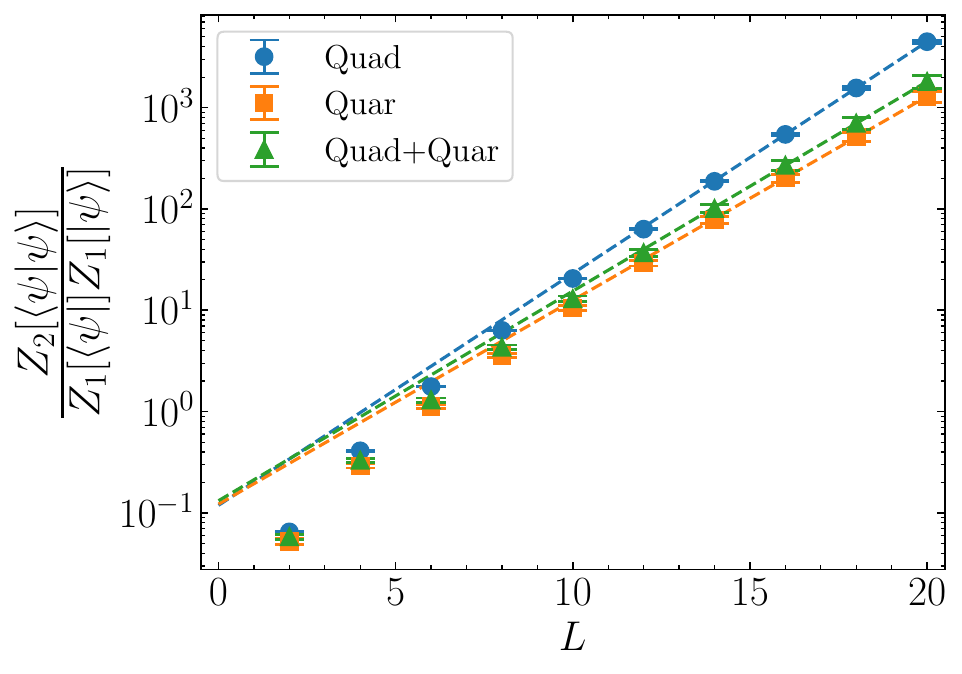}
        \caption{}
    \end{subfigure}
    \hspace{0.05\textwidth}
    \begin{subfigure}[t]{0.45\textwidth}
        \centering
        \includegraphics[width=\textwidth]{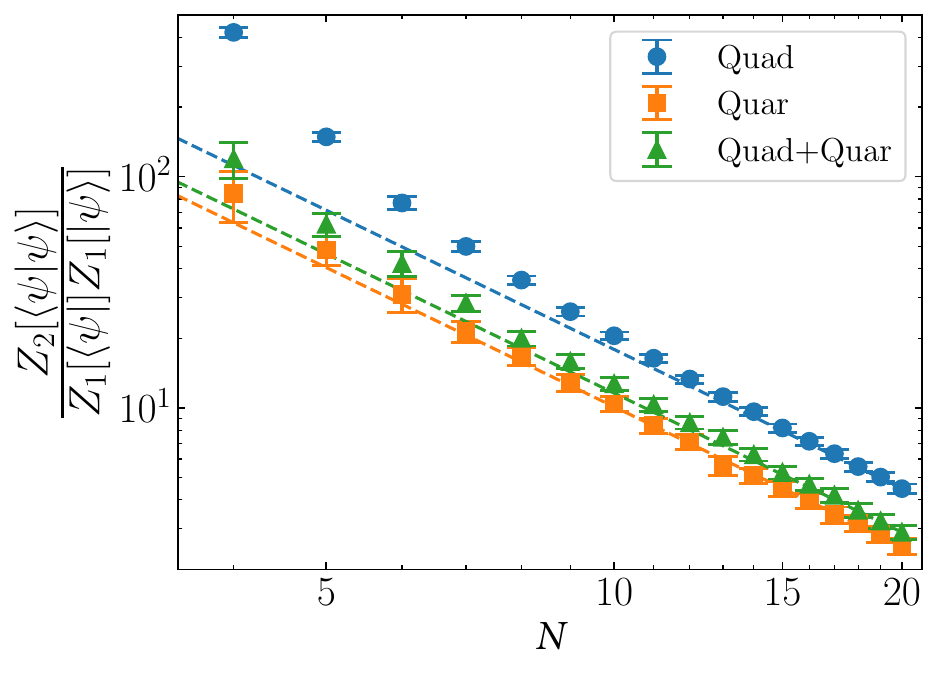}
        \caption{}
    \end{subfigure}
	\caption{Behavior of the quantity $Z_2[\braket{\psi}{\psi}]/(Z_1[\bra{\psi}]Z_1[\ket{\psi}])$ for the three continuous models. (a) $N^{-2}$ dependence. We have $10^{4}$ samples with $k=8$ and $L=10$. (b) $\exp(\alpha L)$ dependence at large $L$. We have $10^{4}$ samples with $k=8$ and $N=10$.}
    \label{fig:CBKWHProperties}
\end{figure}

In figure~\ref{fig:CBKWHProperties}, we present the $N^{-2} \exp(\alpha L)$ dependence of the quantity $\frac{Z_2[\braket{\psi}{\psi}]}{Z_1[\bra{\psi}]Z_1[ \ket{\psi}]} $ for all three models, at $L$ large but not too large such that higher genus effects become important.

\begin{figure}[t]
    \centering
    \begin{subfigure}[t]{0.45\textwidth}
        \centering
        \includegraphics[width=\textwidth]{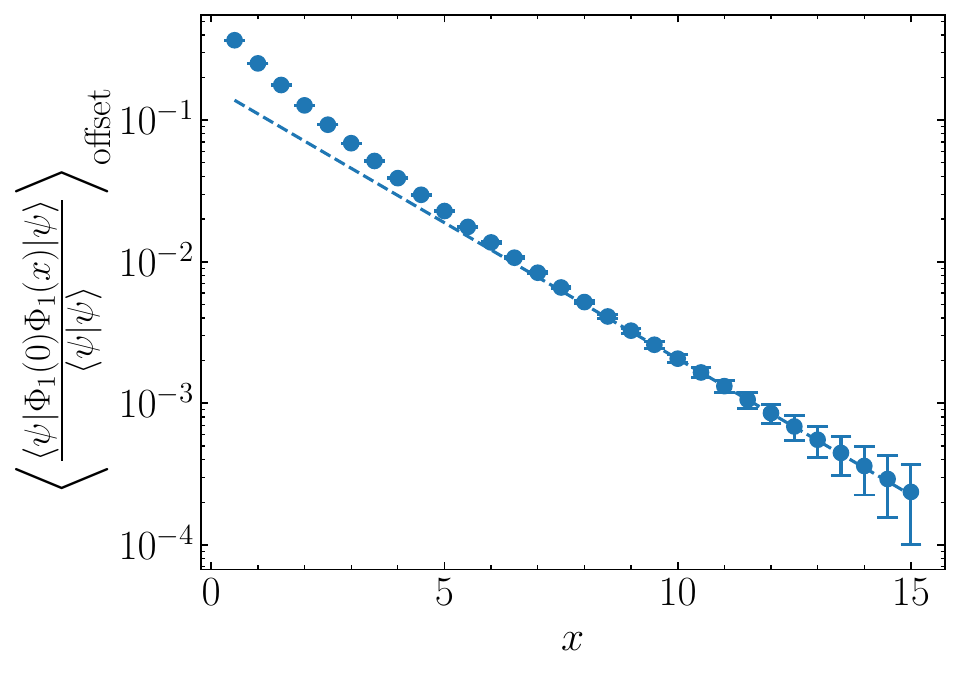}
        \caption{}
    \end{subfigure}
    \hspace{0.05\textwidth}
    \begin{subfigure}[t]{0.45\textwidth}
        \centering
        \includegraphics[width=\textwidth]{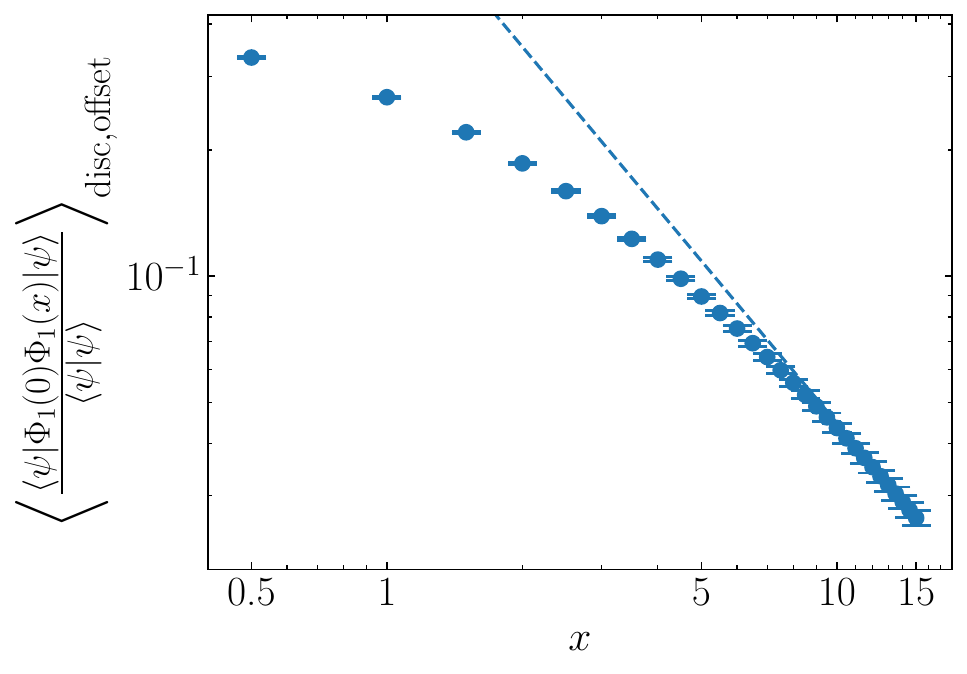}
        \caption{}
    \end{subfigure}
	\caption{Two-point correlations of a local operator in the continuous quadratic + quartic model with the $x \rightarrow \infty$ value subtracted. Here $k=8$, $N=10$, and $L=400$. (a) Full correlators showing exponential decay in $x$. Here we used $10^6$ samples. (b) Bra-ket disconnected correlators showing power law decay in $x$. Here we used $10^5$ samples.}
    \label{fig:CCorrProperties}
\end{figure}

In figure~\ref{fig:CCorrProperties}, we present the exponential decay of the correlator and the power law decay of the bra-ket disconnected correlator in the quadratic + quartic model, after subtracting the offset at $x \rightarrow \infty$. Similar results hold for the other models, which can be found in appendix~\ref{ssec:supplementary_figures_crmps}.

\paragraph{Entropy bounding.} We see the expected entropy bounding behavior for a single interval in all three continuous models, with average entropy bound $S_{\text{max},n}$ equal to $2\ln N$ plus an $\mathcal{O}(N^0)$ subleading term. We see the $\mathcal{O}(N^0)$ subleading term in the average entropy bound has $k^{-1}$ dependence at large $k$.

\begin{figure}[t]
    \centering
    \begin{subfigure}[t]{0.45\textwidth}
        \centering
        \includegraphics[width=\textwidth]{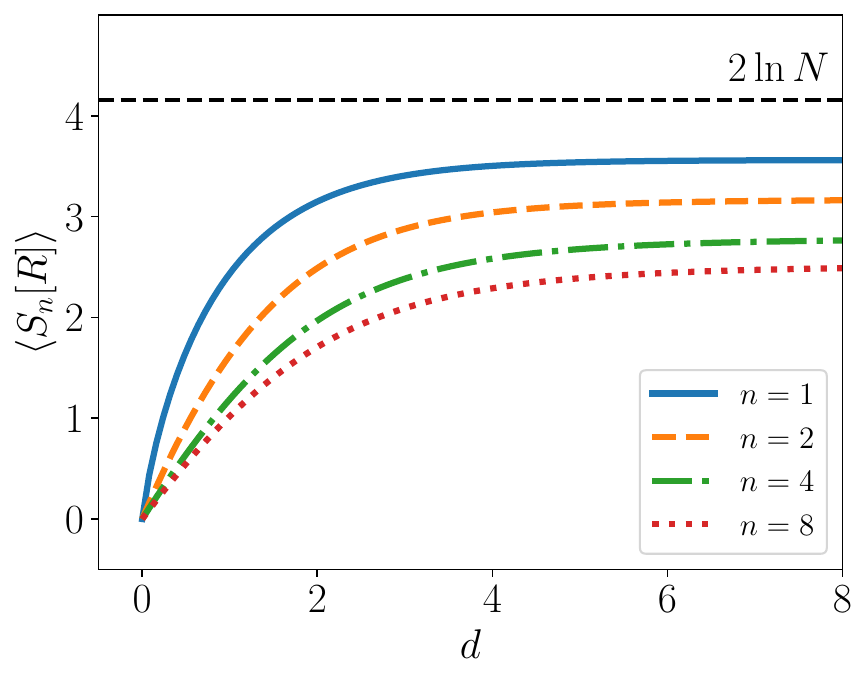}
        \caption{}
    \end{subfigure}
        \hspace{0.05\textwidth}
    \begin{subfigure}[t]{0.45\textwidth}
        \centering
        \includegraphics[width=\textwidth]{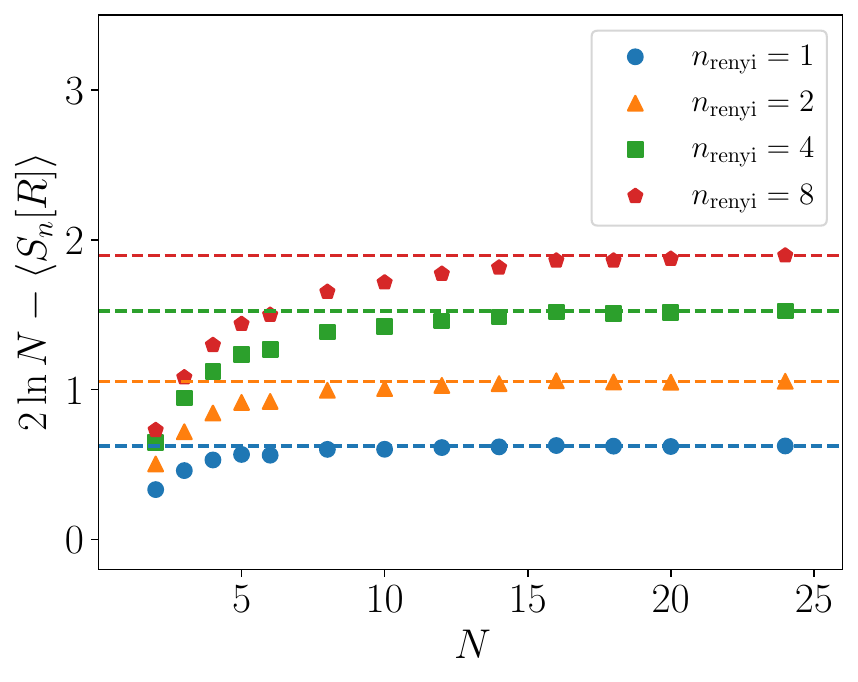}
        \caption{}
    \end{subfigure}
	\caption{The bounding behavior of the averaged entanglement and Renyi entropies of an interval in the continuous quadratic + quartic model. (a) $S_n[R]$ showing bounding behavior. Here $N=8$ with $10^4$ samples. (b) Convergence of $2\ln N - S_{\text{max}, n}$ at large $N$. We used $k=128$ and $10^3$ samples.}
    \label{fig:CEntropyProperties}
\end{figure}

In figure~\ref{fig:CEntropyProperties}(a), we present the bounding behavior of the averaged entanglement and Renyi entropies of an interval in the continuous quadratic + quartic model. In (b) we show that the entropy bound is equal to $2\ln N$ plus an $\mathcal{O}(N^0)$ subleading term. Similar results hold for the other two models, which can be found in appendix~\ref{ssec:supplementary_figures_crmps}.

\begin{figure}[t]
    \centering
    \begin{subfigure}[t]{0.45\textwidth}
        \centering
        \includegraphics[width=\textwidth]{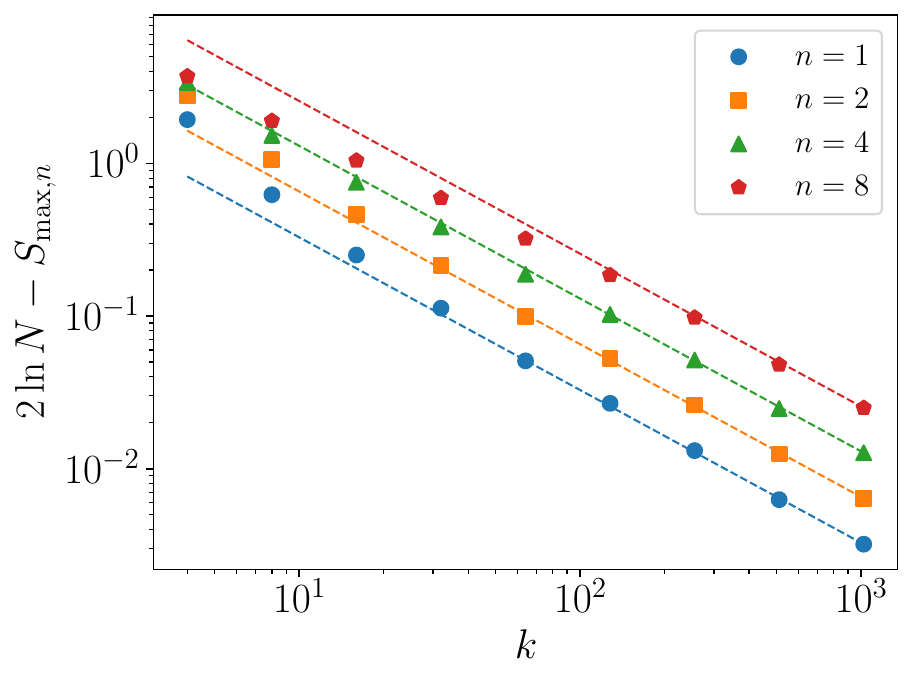}
    \end{subfigure}
	\caption{$k^{-1}$-dependence of $2\ln N - S_{\text{max}, n}$ in the continuous quadratic + quartic model. Here $N=24$. We used $10^3$ samples.
    }
    \label{fig:CEntropykdep}
\end{figure}

In figure~\ref{fig:CEntropykdep} we present the $k^{-1}$ dependence of $2\ln N - S_{\text{max}, n}$ ($n=1,2,4,8$) in the continuous quadratic + quartic model. Similar results hold for the other two models, which can be found in appendix~\ref{ssec:supplementary_figures_crmps}.

\paragraph{Off-shell wormholes.} We numerically computed $\frac{Z_1\left[\bra{\psi}\right]  Z_1\left[\bra{\psi}\right]  Z_2 \left[\ket{\psi} \ket{\psi}\right] }{\left( Z_1[\bra{\psi}]  Z_1[\ket{\psi}] \right)^2}$ for the three continuous models. In all cases, in the intermediate $L$ region, we have $N^{-2}$ dependence and $L^2$ dependence.

We remark that at very large $L$, higher genus effects become important in the quantity $\frac{Z_1\left[\bra{\psi}\right]  Z_1\left[\bra{\psi}\right]  Z_2 \left[\ket{\psi} \ket{\psi}\right] }{\left( Z_1[\bra{\psi}]  Z_1[\ket{\psi}] \right)^2}$. We checked the $L$-dependence of this quantity below this $L$ range.

We numerically computed $C_{n-\text{point;cRMPS}}$ for $n=2,3,4$ for the three continuous models. In all cases, we verified that $C_{n-\text{point}}$ has $N^{2(1-n)}$ dependence at large $N$.

We also numerically computed $C_{\Phi_1;n-\text{point;cRMPS}}$ for $n=2, 4$ for the three continuous models. In all cases, we verified that $C_{\Phi_1; n-\text{point}}$ has $N^{2(1-n)}$ dependence at large $N$.

\begin{figure}[t]
    \centering
    \begin{subfigure}[t]{0.45\textwidth}
        \centering
        \includegraphics[width=\textwidth]{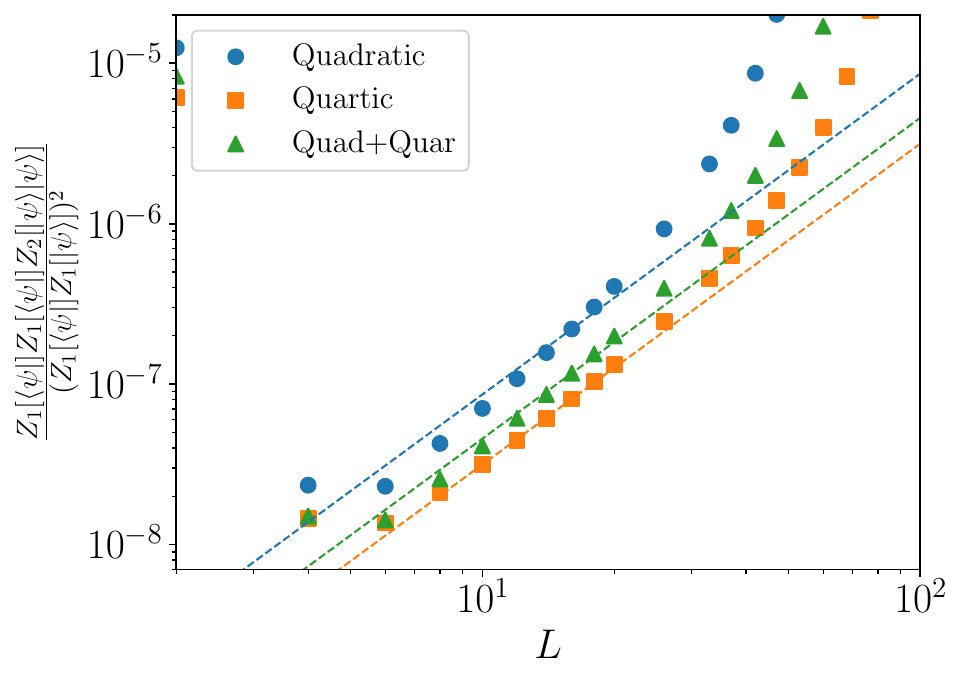}
        \caption{}
    \end{subfigure}
        \hspace{0.05\textwidth}
    \begin{subfigure}[t]{0.45\textwidth}
        \centering
        \includegraphics[width=\textwidth]{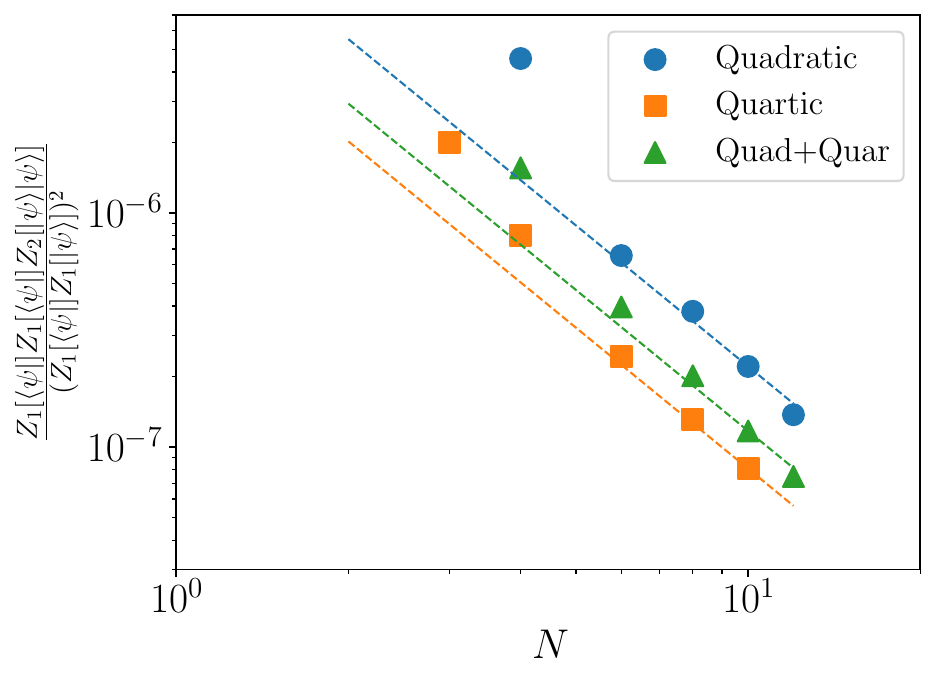}
        \caption{}
    \end{subfigure}

	\caption{$L$ and $N$ dependence of $\frac{Z_1\left[\bra{\psi}\right]  Z_1\left[\bra{\psi}\right]  Z_2 \left[\ket{\psi} \ket{\psi}\right] }{\left( Z_1[\bra{\psi}]  Z_1[\ket{\psi}] \right)^2}$ for quadratic, quartic, and quadratic + quartic cRMPS models. We used $10^6$ samples with $k=256$. (a) We see $L^{2}$ dependence in the intermediate $L$ region. Here $N=8$. (b) We see $N^{-2}$ dependence. Here $L=16$.}
    \label{fig:CKetket}
\end{figure}

In figure~\ref{fig:CKetket}, we present the $L$ and $N$ dependence of $\frac{Z_1\left[\bra{\psi}\right]  Z_1\left[\bra{\psi}\right]  Z_2 \left[\ket{\psi} \ket{\psi}\right] }{\left( Z_1[\bra{\psi}]  Z_1[\ket{\psi}] \right)^2}$ for the quadratic, quartic, and quadratic + quartic cRMPS models. In the intermediate $L$ region, we have $N^{-2}$ dependence and $L^2$ dependence.

\begin{figure}[t]
    \centering
    \begin{subfigure}[t]{0.45\textwidth}
        \centering
        \includegraphics[width=\textwidth]{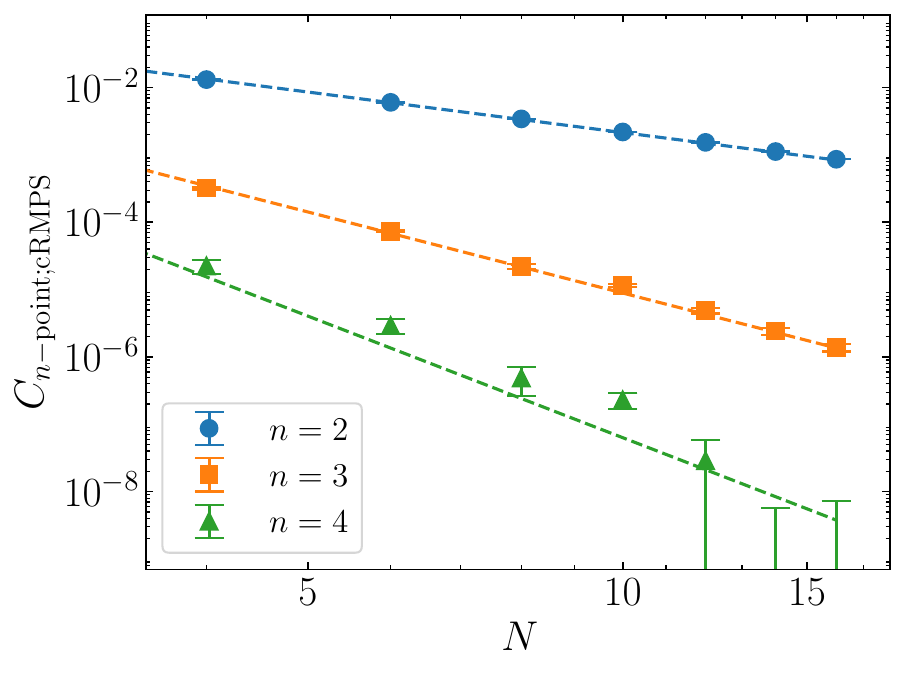}
        \caption{}
    \end{subfigure}
        \hspace{0.05\textwidth}
    \begin{subfigure}[t]{0.45\textwidth}
        \centering
        \includegraphics[width=\textwidth]{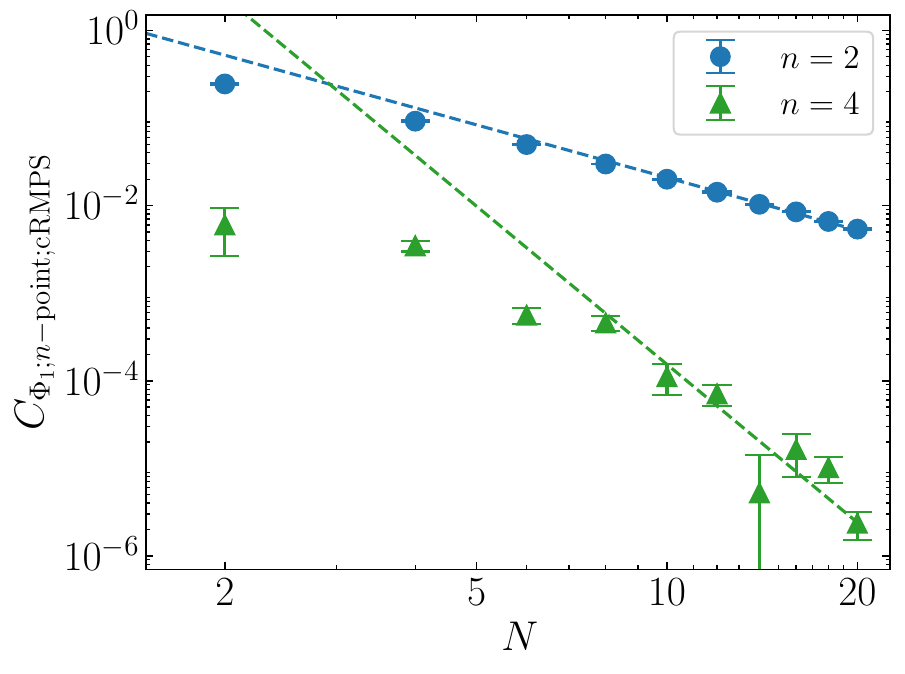}
        \caption{}
    \end{subfigure}
	\caption{(a) $N$ dependence of $\lambda_g$ cumulants in the continuous quadratic + quartic model. We used $k=8$ with $10^{5}$ samples. (b) $N$ dependence of $\langle -\lambda_g|\Phi_i|-\lambda_g\rangle$ cumulants in the same model. We used $k=8$ with $10^{4}$ samples. For large $N$, we have $N^{2(1-n)}$ dependence for both.}
    \label{fig:EigCumu}
\end{figure}

In figure~\ref{fig:EigCumu}(a), we present the $N$ dependence of $n$th $\lambda_g$ cumulants in the quadratic + quartic model. For large $N$, we have $N^{2(1-n)}$ dependence. Similar results hold for the other models, which can be found in appendix~\ref{ssec:supplementary_figures_crmps}.

In figure~\ref{fig:EigCumu}(b), we present the $N$ dependence of $n$th $\lambda_g^{-1}\bra{\lambda_g}M_O\ket{\lambda_g}$ cumulants in the quadratic + quartic model. For large $N$, we have $N^{2(1-n)}$ dependence. Similar results hold for the other models, which can be found in appendix~\ref{ssec:supplementary_figures_crmps}.

\section{dS gravitationally prepared states and RMPS}

\label{sec:dsrmps}

\subsection{dS gravitationally prepared states}

\label{ssec:dsanalytical}

Considering the cosmological motivation for studying gravitationally prepared states, it would be interesting to study gravitationally prepared states in de Sitter space. We now study gravitationally prepared states in de Sitter space and their implications for RMPS.

\paragraph{The dS JT+CFT model.} The dS JT+CFT model \cite{chen_bra-ket_2021} is a model of a gravitationally prepared state in $d=2$ dimensions. We consider $\text{dS}_2$ JT gravity \cite{maldacena_two_2021, cotler_emergent_2025, cotler_isometric_2023, cotler_non-perturbative_2025} (i.e. JT gravity with a positive cosmological constant $\Lambda = 1$) coupled to conformal matter fields $\varphi_i$. In other words, the action is given by
\begin{equation}
    S_{\text{JT+CFT}}\left[g, \phi, \varphi_i\right] = - \phi_0 \chi(g) - \frac{1}{4\pi} \int d^2x \sqrt{g}   \phi (R - 2) - \frac{1}{2\pi} \int_{\partial g} dx \sqrt{h} \phi K + S_{\text{CFT}} [\varphi_i, g] .
\end{equation}
Here $\chi(g)$ is the Euler characteristic of $g$. $h$ is the induced metric on $\partial g$, and $K$ is the extrinsic curvature of $\partial g$.

We note that two dimensional Euclidean geometries with constant positive scalar curvature do not have asymptotic boundaries. This becomes a problem in constructing the dS JT+CFT model as a gravitationally prepared state.

A solution to the problem is to work with negative de Sitter geometries, i.e. de Sitter geometries with purely negative metric signature \cite{cotler_non-perturbative_2025}. Such geometries are equal to Anti de Sitter geometries with the sign of the metric flipped, and thus have asymptotic boundaries.

One thing to note for negative geometries is that we should define $\sqrt{g} = - \sqrt{|g|}$, which is the choice consistent with analytic continuation from Euclidean geometries.

We remark that negative geometries do not satisfy the KSW condition \cite{kontsevich_wick_2021, witten_note_2022} for non-Euclidean metrics, and thus general quantum field theories are not well defined on negative geometries. However, we note that CFTs are perfectly well-defined on negative metrics, as a negative metric can be obtained by Weyl rescaling a Euclidean metric by $-1$.

We thus define the dS JT+CFT model as a path integral over negative geometries. In the dS JT+CFT model, the gravitational degrees of freedom are $g$ and $\phi$. Thus we need to fix boundary conditions for $g$ and $\phi$. This is given by
\begin{equation}
    ds^2 = - \frac{dz^2+ dx^2 }{z^2}, \quad  \phi = \frac{\phi_r}{z} \quad \text{as }z \rightarrow 0
\end{equation}
along with periodicity conditions for $x$:
\begin{equation}
    x \approx x + L.
\end{equation}
The boundary is placed at $z = \epsilon$, where $\epsilon$ is infinitesimal.

We can now define the dS JT+CFT gravitationally prepared state
\begin{equation}
    \ket{\psi} = \int Dg  D\phi  |_{(L, \phi_r)} \prod_i D\varphi_i  \exp \left( -S_{\text{JT+CFT}}\left[g, \phi, \varphi_i\right] \right)  \ket{\varphi_{i, \Sigma} },
\end{equation}
where $(L, \phi_r)$ indicate the corresponding boundary conditions for $g, \phi$.

It turns out that the dS JT+CFT gravitationally prepared state as defined above is mathematically equivalent to the AdS JT+CFT gravitationally prepared state, as multiplying $-1$ to the metric in the dS JT+CFT action gives us the AdS JT+CFT action, and vice versa.

However, there is a physical difference between the two models. As we will see below, in the dS JT+CFT model, it is more physical for $\phi_r$ to be purely imaginary with positive imaginary part than to be real.

\paragraph{Analytical continuation of the disk geometry.} In the dS JT+CFT model, we have the saddlepoint disk geometry
\begin{equation}
	ds^2 = - \frac{dz^2 + dx^2}{\frac{L^2}{4\pi^2} \sinh^2 \frac{2\pi z}{L}}, \quad \phi = \frac{c}{12} + \frac{\phi_r}{\frac{L}{2\pi} \tanh \frac{2\pi z}{L}}, \quad x = x + L.
\end{equation}
Here $c$ is the central charge of the matter CFT.

\begin{figure}[t]
    \begin{center}

    \input{figures/tikz/contours}
    \end{center}
    \caption{Two contours in the $z$ plane and their corresponding saddlepoint geometries in the dS JT+CFT model. The top contour gives us a hemisphere connected to Lorentzian de Sitter space. The bottom contour gives us $-1$ times the Euclidean AdS disk geometry.}
    \label{fig:contours}
\end{figure}
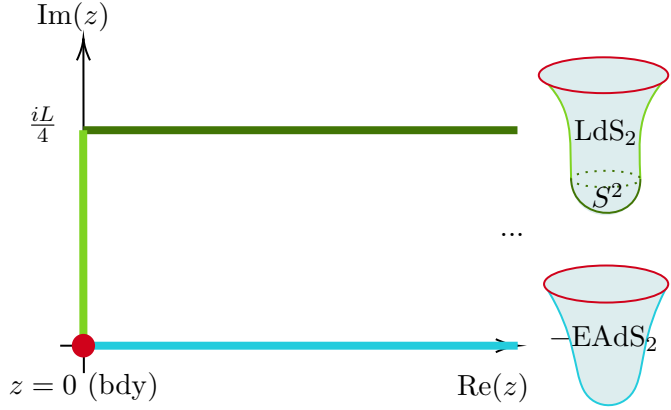

We note that this geometry can be analytically continued to a hybrid Euclidean-Lorentzian manifold \cite{hertog_holographic_2012, maldacena_two_2021, cotler_non-perturbative_2025}. This is done by taking the following contour of $z$:
\begin{equation}
	z = \begin{cases}
		i \sigma & 0  < \sigma < \frac{L}{4}\\
		\frac{(i-1)L}{4} + \sigma & \frac{L}{4} < \sigma .
	\end{cases}
\end{equation}
As a result, we get the geometry
\begin{equation}
	\begin{aligned}
	    ds^2 &= \begin{cases}
		\frac{-d\sigma^2 + dx^2}{\frac{L^2}{4\pi^2} \sin^2 \frac{2\pi \sigma}{L} } &  0  < \sigma < \frac{L}{4}\\
		\frac{d \sigma^2 + dx^2}{\frac{L^2}{4\pi^2}  \cosh^2(\frac{2\pi \sigma}{L}  - \frac{\pi}{2})} & \frac{L}{4} < \sigma ,
	\end{cases}\\
    \phi &=  \begin{cases}
		\frac{c}{12} - i \frac{\phi_r}{\frac{L}{2\pi} \tan \frac{2\pi \sigma}{L} } & 0  < \sigma < \frac{L}{4} \\
		\frac{c}{12} + \frac{2\pi \phi_r}{L}  \tanh (\frac{2\pi \sigma}{L}  - \frac{\pi}{2})& \frac{L}{4} < \sigma ,
	\end{cases}
	\end{aligned}
    \label{eq:dsgeo}
\end{equation}
which consists of a hemisphere ($\frac{L}{4} < \sigma $) connected to Lorentzian de Sitter space ($0  < \sigma < \frac{L}{4} $). See figure~\ref{fig:contours}.

The analytically continued geometry describes the Bunch-Davies vacuum evolving into Lorentzian de Sitter space. Thus, we can interpret the gravitationally prepared state as the quantum state of a closed inflationary universe.

We note that the real-time regularized value of the dilaton field near the boundary is given by
\begin{equation}
    \phi_{\text{real-time}} =-i \phi_r.
\end{equation}
Thus, for $\phi_{\text{real-time}}$ to be real and positive, $\phi_r$ must be purely imaginary with a positive imaginary part.

\subsection{Bra-ket wormhole divergence problem}

We have just seen that, in the dS JT+CFT model, it makes sense for $\phi_r$ must be purely imaginary with a positive imaginary part on physical grounds.

However, it turns out that the bra-ket wormhole solution becomes singular in the case where $\phi_r$ is purely imaginary \cite{chen_bra-ket_2021}. We can see this result as follows. Assume $\phi_r$ is complex with a positive real part. Then, we have the saddlepoint bra-ket wormhole geometry given by
\begin{equation}
	ds^2 = \frac{dz^2 + dx^2}{\left( \frac{\beta}{\pi} \sin \frac{\pi  z}{\beta} \right)^2}, \quad
	\phi = \frac{c}{3} + \frac{c}{4} \frac{\frac{\pi}{2} - \frac{\pi  z}{\beta} }{\tan  \frac{\pi  z}{\beta} }  + \frac{i \Im \phi_r}{\frac{\beta}{\pi} \tan  \frac{\pi  z}{\beta} } ,\quad 0 < z <  \beta, \quad x \approx x+ L,
\end{equation}
where the periodicity in the $z$ direction is given by
\begin{equation}
	\beta = \frac{8 \Re \phi_r}{c } .
\end{equation}
Thus, when $\Re \phi_r \rightarrow 0$, $\beta \rightarrow 0$, and the temperature of the bra-ket wormhole solution diverges. We call this the \textit{bra-ket wormhole divergence problem} of the dS JT+CFT prepared state.

At the moment some methods of resolving the bra-ket wormhole divergence problem have been suggested \cite{fumagalli_sitter_2025, dsgps_new}. In any case, if we have a well-defined gravitationally prepared state in 2D de Sitter space, we expect it to be described by a random matrix product state. We can expect, though, that analogously to the property of $\phi_r$ being purely imaginary in the dS JT+CFT model, the defining matrices $\{A_i\}$ may not be Hermitian.

\paragraph{Bra-ket disconnected physics.} Although the bra-ket wormhole divergence problem leaves the dS JT+CFT model, without modifications, unable to make predictions about GPS quantities in general, it does not affect bra-ket disconnected quantities. Thus, despite the bra-ket wormhole divergence problem, we can still discuss bra-ket disconnected physics. In appendix~\ref{sec:discphys}, we have computed the bra-ket disconnected entropy in RMPS models.

\subsection{Cosmological times in RMPS models}

\paragraph{Two times for RMPS.} In \cite{dsgps_new}, gravitationally prepared states were studied in the context of dS universes. Here $\Sigma$ is understood as the reheating surface where inflation ends, and the value of $L$ is fixed to be $2\pi$.

In this work, two different cosmological times were defined for gravitationally prepared states. The first is the length of inflation, which is parametrized by a UV cutoff $k_{\text{max}}$ for the momenta of field excitations on $\Sigma$:
\begin{equation}
    t_{\text{inf}} \sim \ln k_{\text{max}}.
\end{equation}
$k_{\text{max}}$ can be understood as the maximum momentum that was redshifted to sub-Planckian energies and thus made observable during inflation. This cutoff is employed when evaluating the gravitationally prepared state, as modes above $k_{\text{max}}$ are physically unmeasurable.

The second is the time elapsed since reheating, which is parametrized by a IR cutoff $k_{\text{min}}$ for the momenta of field excitations on $\Sigma$:
\begin{equation}
    t_{\text{obs}} \sim  \frac{1}{k_{\text{min}}}.
\end{equation}
$k_{\text{min}}$ can be understood as the minimum momentum that reentered the cosmological horizon and thus became observable after inflation. This cutoff is employed when calculating expectation values of observables in the gravitationally prepared state, as modes below $k_{\text{min}}$ have yet to be observed.

We will now define these two times in RMPS models, interpreting them as models of gravitationally prepared states in inflating universes. In the RMPS model, we note that we already have an effective UV cutoff. For dRMPS, this is due to the discrete lattice spacing. For cRMPS, this is due to the regulation of UV divergences intrinsic to the definition of the model, which allows entropies to be well-defined for spatial subregions.

However, this UV cutoff is fixed. Instead of scaling the UV cutoff, we are allowed to relatively scale the value of the parameter $L$ in the model. It is natural to interpret $L$ as the size of the universe which grows exponentially during inflation. Thus, we suggest that the length of inflation can be parametrized in RMPS models by the parameter $L$:
\begin{equation}
    t_{\text{inf}} \sim  \ln L.
\end{equation}
Then, the UV cutoff relative to the total size of $\Sigma$ grows exponentially in $t_{\text{inf}}$, as expected.

The time elapsed since reheating can be understood simply as the length of the worldline of an observer in the flat space region to the future of $\Sigma$. Such an observer can only measure the quantum fields in a subregion of size $2 t_{\text{inf}}$ in $\Sigma$. Thus, we can relate the time elapsed since reheating to the characteristic length scale $d$ of observables measured in the RMPS:
\begin{equation}
    t_{\text{obs}} \sim  d.
\end{equation}
For example, $d$ can be the distance probed in a correlator measurement, or the length of an interval in an entropy calculation.

\paragraph{Scrambling time.} In \cite{dsgps_new}, it was shown that the critical value of $t_{\text{inf}}$ for which bra-ket wormhole dominance occurs is given by
\begin{equation}
    t_{\text{inf, crit}} \sim \ln S_{\text{dS}} .
\end{equation}
This corresponds to the scrambling time in quantum chaos, and thus suggests that the bra-ket wormhole phase transition is the result of quantum chaos.

By eqs.~\eqref{eq:bkwhdependence},~\eqref{eq:bkwhcrit}, we see that for RMPS, we have the same result: $L_{\text{crit}} \propto 2 \ln N$ and so $t_{\text{inf, crit}} \sim \ln S_{\text{dS}}$, where $S_{\text{dS}} = 2\ln N$. Thus, the bra-ket wormhole phase transition for RMPS may be the result of quantum chaos in some de Sitter bulk.

We also remark on the critical times elapsed since reheating for which subregion entropy bounding and the dominance of correlators by off-shell wormhole effects can be measured. In both cases, the critical time elapsed is proportional to the de Sitter entropy, or, in other words, exponential in the scrambling time:
\begin{equation}
    t_{\text{obs, entropy bounding}} \propto S_{\text{dS}}, \quad t_{\text{obs, off-shell WH}} \propto S_{\text{dS}}.
\end{equation}
This follows from eqs.~\eqref{eq:entropydcrit},~\eqref{eq:conpointndep},~\eqref{eq:corrdepgps}. Thus, such effects become visible in the very far future after reheating.

\section{Conclusion and discussion}

\label{sec:discussion}
We recap on our work. We have presented a program of modeling gravitationally prepared states in two dimensions with random matrix product states.

In section~\ref{sec:gps}, we reviewed two important features of gravitationally prepared states --- the bra-ket wormhole phase transition and entropy bounding. From our holographic argument in section~\ref{ssec:rmpsdefinition} and the strong subadditivity argument in \cite{chen_bra-ket_2021}, we expect these features to be universal.

In section~\ref{sec:rmps}, we motivated the program of modeling 2D gravitationally prepared states with random matrix product states based on three arguments --- the topological expansion behavior of random matrices, the entropy bounding behavior of matrix product states, and the holographic argument. We then defined the \textit{spectral gapping condition} for an RMPS model which ensures that it shows characteristic features of a bra-ket wormhole transition. We then presented a broad class of models --- the $\mathrm{O}(k)$ models --- for which the spectral gapping condition holds. Furthermore, the parameter $k$ gave us additional control in computing ensemble averages both analytically and numerically, with the large $k$ limit playing a helpful role. We analytically and numerically verified the bra-ket wormhole phase transition and entropy bounding properties of $\mathrm{O}(k)$ models, focusing on three specific models (quadratic, quartic, quadratic + quartic) on the numerical side.

In section~\ref{sec:offshell}, we utilized the novelty of the RMPS models by computing quantities that can be interpreted on the gravity side as effects of off-shell wormholes. Most higher topologies in a gravitationally prepared state contribute purely off-shell, and the effects of off-shell wormholes cannot be computed using the semiclassical limit. RMPS models provide a novel setup where all higher topology contributions can be computed. We first showed that RMPS models agree with the interpretation of the Casimir divergence of thin wormholes being regulated by regulating the wormhole width. We then computed the contributions of off-shell wormholes to the inner product and to correlators of local operators. In particular, we see a nonzero contribution to correlators at long distances in the RMPS state which can be interpreted gravitationally as wormholes creating shortcuts between local operator insertions. Such nonzero long distance correlations may have operational meaning as nonperturbative quantum gravitational effects that can be measured in the background radiation of gravitating universes.

In section~\ref{sec:crmps}, we discussed an RMPS model with continuous space. In this model, there is a transfer Hamiltonian instead of a transfer matrix. Furthermore, the transfer Hamiltonian of the continuous RMPS model relates to the transfer matrix of the discrete RMPS model. The continuous model satisfies all the properties of gravitationally prepared states that we checked for the discrete model, with the added benefit that it gives physical results for all real-valued distances. The continuous RMPS model, however, does not exactly have the local structure of a relativistic QFT. For example, entropies of spatial subregions are well-defined without momentum cutoffs. Finding an RMPS model of a gravitationally prepared state with the local structure of a relativistic QFT would be an interesting task.

In section~\ref{sec:dsrmps}, we discussed gravitationally prepared states in two dimensional de Sitter space. Such states can be defined as a sum over negative definite metrics. These negative definite metrics can be analytically continued to Euclidean-Lorentzian hydrid metrics. It follows that the normalized boundary dilaton field value, which was real for the AdS JT+CFT model, must be purely imaginary in the dS JT+CFT model. However, this leads to the bra-ket wormhole geometry becoming singular. Following \cite{dsgps_new}, we defined the duration of inflation and the time elapsed since reheating for RMPS, and showed that the critical timescale for the bra-ket wormhole phase transition during inflation is roughly the scrambling time for de Sitter space, and that the critical timescale for observing subregion entropy bounding or off-shell wormhole dominance of correlators is exponential in the scrambling time for de Sitter space.

A future question of interest is studying gravitationally prepared states in higher dimensions. As explained above, the bra-ket wormhole phase transition and entropy bounding must generally also hold for higher dimensional gravitationally prepared states. A possible candidate for models of gravitationally prepared states in higher dimensions is that of random cluster states, which generalize random matrix product states to higher dimensions and satisfy an area-law bound for the entropies of spatial subregions. Such models may teach us about the physics of gravitationally prepared states in higher dimensions.

Another question of interest is finding matrix ensembles $V(A_i)$ that give us RMPS models that are more precisely dual to gravitationally prepared states of known gravity theories. It is possible that the ensembles considered in \cite{saad_jt_2019, jafferis_jt_2023} will be relevant to this question.

Another question is understanding de Sitter gravitationally prepared states and its relation to physics in the static patch of de Sitter space. Recently, de Sitter quantum gravity has been studied in relation to the double-scaled SYK model \cite{susskind_entanglement_2021, chandrasekaran_algebra_2023}, through algebraic methods \cite{chandrasekaran_algebra_2023, kolchmeyer_chaos_2024}, and by involving observers \cite{Witten:2023qsv, chandrasekaran_algebra_2023, harlow_quantum_2025}. Such approaches usually focus on the static patch of de Sitter space, as opposed to the future boundary, which the gravitationally prepared state approach concerns itself with. It would be interesting to see how the two approaches, that of the static patch and that of the boundary, inform and complement each other.

We can also wonder if we can use RMPS methods to learn about baby universes \cite{Giddings:1988wv, Hawking:1988ae, hawking_quantum_1987}. As noted in \cite{chen_bra-ket_2021}, the bra-ket wormhole geometry can be thought of as preparing a semiclassical state for the CFT on the spatial slice $\Sigma$ as well as an entangled baby universe which has spatial geometry equal to the width of the bra-ket wormhole. It is interesting to ask if RMPS models can model different types of baby universes in other scenarios.

We note that RMPS models give pure states that approximately look like mixed states. This is an example of spoofing entanglement \cite{Engelhardt:2024lnd}, which occurs in the context of Hawking radiation. Thus, a natural question to ask is whether RMPS can be used to model the quantum state of Hawking radiation, and examples of spoofing entanglement in general.

Finally, it would be interesting to further understand the nonperturbative behavior of gravitationally prepared states from the perspective of quantum chaos. Quantum chaos has a close relation to nonperturbative effects in quantum black holes \cite{Sekino_2008, shenker_black_2014, saad_semiclassical_2019, chen_comments_2023}. Recently, de Sitter quantum gravity has been studied from the perspective of quantum chaos theory \cite{susskind_entanglement_2021, kolchmeyer_chaos_2024}. It is reasonable to ask if the characteristic non-perturbative physics of gravitationally prepared states, such as the entropy bounding behavior as well as the nonzero large distance correlators, have a chaos theoretic interpretation.

\acknowledgments
We thank Seok Kim, Junghwan Lee, Hong Liu, Takil Yannet Son, and Washington Taylor for useful comments and discussions. We are supported by Grant Korea NRF2019R1C1C1010050 and RS-2024-00342093.

\appendix

\section{CFT stress-energy and entanglement entropy}

\label{sec:cft}

Here are some useful formulae for computing the stress-energy tensor and entanglement entropy for a 2D CFT with central charge $c$.

For an exposition of computing Weyl contributions to the stress tensor from the conformal anomaly, see \cite{birrell_quantum_1982}. For more detailed references on 2D CFT entropy, see \cite{calabrese_entanglement_2004, casini_entanglement_2005-1, caraglio_entanglement_2008, casini_entanglement_2009}.

\paragraph{Casimir energy.} We consider the flat cylinder
\begin{equation}
	ds^2  = d \rho^2 + d\theta^2, \quad \theta \approx \theta + 2\pi.
\end{equation}
The stress-energy tensor on the cylinder is given by
\begin{equation}
	\begin{aligned}
		&T^{\text{Casimir}}_{\rho\rho} = - T^{\text{Casimir}}_{\theta\theta}  = \frac{c}{24\pi},\\
		&T^{\text{Casimir}}_{\rho\theta} = T^{\text{Casimir}}_{\theta\rho} =0 .
	\end{aligned}
\end{equation}
In particular we have
\begin{equation}
	E = - \int d\theta T^{\text{Casimir}}_{\rho\rho} = - \frac{c}{12},
\end{equation}
which is the Casimir energy.

\paragraph{Effective action in curved space.} We consider a metric $g$ and a conformally related metric $\Omega^2 g$ on the manifold $M$. Then, we can relate the CFT effective action on the $\Omega^2 g$ metric to that on the $g$ metric by
\begin{equation}
	\begin{aligned}
		&S_{\text{eff}} [\Omega^2 g] -S_{\text{eff}} [g] \\
		& = - \frac{c}{24 \pi} \int_M ( (\nabla\ln \Omega)^2 + R \ln \Omega )  -  \frac{c}{12 \pi} \int_{\partial M} K  \ln \Omega  \quad  \text{(computed on the metric $g$)}.
	\end{aligned}
\end{equation}
This is called the Weyl anomaly action. The integral is evaluated on the $g$ metric.

\paragraph{Weyl contributions to the stress tensor.} We consider curved Euclidean space
\begin{equation}
	ds^2 = \Omega^2 (dx^2 + dy^2).
\end{equation}
We see that the CFT stress tensor relates to that of flat space by
\begin{equation}
	\begin{aligned}
		&T_{ab} = T^{\text{flat}}_{ab} + T^{\text{Weyl}}_{ab},\\
		&T^{\text{Weyl}}_{00} = \frac{c}{24\pi} \left(\frac{3 (\partial_1 \Omega)^2 - (\partial_0 \Omega)^2}{\Omega^2} - \frac{2 \partial^2_1 \Omega}{\Omega} \right),\\
		&T^{\text{Weyl}}_{11} = \frac{c}{24\pi} \left(\frac{3 (\partial_0 \Omega)^2 - (\partial_1 \Omega)^2}{\Omega^2} - \frac{2 \partial^2_0 \Omega}{\Omega} \right),\\
		&T^{\text{Weyl}}_{01} = T^{\text{Weyl}}_{10} = \frac{c}{12\pi} \left(- \frac{2 (\partial_0 \Omega)(\partial_1 \Omega)}{\Omega^2} + \frac{\partial_0 \partial_1 \Omega}{\Omega}  \right).
	\end{aligned}
\end{equation}
This follows from the Weyl anomaly action. In particular the trace of $T^\text{Weyl}_{ab}$ is the Weyl anomaly, given by
\begin{equation}
	(T^a_a)^\text{Weyl} = \frac{c}{12\pi} \left(\frac{ (\partial_1 \Omega)^2 + (\partial_0 \Omega)^2}{\Omega^4} - \frac{ (\partial^2_0 +\partial^2_1) \Omega}{\Omega^3} \right) = \frac{c}{24\pi} R
\end{equation}
where we have used the formula
\begin{equation}
	R = -2 \Omega^{-2} (\partial^2_0 +\partial^2_1) (\ln \Omega)
\end{equation}
for the Ricci scalar in conformal coordinates.

\paragraph{Entanglement entropy.} The $n$-Renyi entropy of an interval in flat space is given by
\begin{equation}
	\frac{(n+1)c}{6n} \ln |a-b|
\end{equation}
where $a$, $b$ are the endpoints of the interval.

For a massless Dirac field, given multiple intervals with positively oriented endpoints $a_i$ and negatively oriented endpoints $b_i$, we have the entropy of the intervals
\begin{equation}
	S_n = \frac{(n+1)c}{6n} \left( \sum_{i \neq j} \ln \frac{|a_i - b_j|}{\varepsilon} -  \sum_{i < j} \ln \frac{|a_i - a_j|}{\varepsilon} - \sum_{i < j} \ln \frac{|b_i - b_j|}{\varepsilon} \right).
\end{equation}
In the limit where $a_i$ is very close to $b_{s(i)}$ for some permutation $s(i)$, we have
\begin{equation}
	S = \frac{c}{3} \sum_{i} \ln \frac{|a_i - b_{s(i)}|}{\varepsilon}
\end{equation}
for all 2D CFTs.

In curved Euclidean space
\begin{equation}
	ds^2 = \Omega^2 (x,y) (dx^2 + dy^2),
\end{equation}
the entropy of multiple intervals is given by
\begin{equation}
	S_n = \frac{(n+1)c}{12n} \sum_i \ln (\Omega(a_i) \Omega(b_i)) + S_{\text{flat}}
\end{equation}
where we have the endpoints $a_i$, $b_i$.

Applying these results, the entropy of an interval on a cylinder
\begin{equation}
	ds^2  = d \rho^2 + d\theta^2, \quad \theta \approx \theta + 2\pi
\end{equation}
with endpoints $a$, $b$ is given by
\begin{equation}
	S_n =  \frac{(n+1)c}{12n} \ln f(a,b), \quad f(a,b) =  \frac{4}{\varepsilon^2}\left( \sin^2 \frac{|\theta_b - \theta_a|}{2} + \sinh^2\frac{|\rho_b - \rho_a|}{2}\right).
\end{equation}
This can be found by conformally mapping the cylinder to flat space
\begin{equation}
	x+iy = e^{\rho + i \theta}
\end{equation}
and using the curved space formula.

\section{Markov-Chain-Monte-Carlo methods}

\label{sec:mcmc}

\subsection{MCMC setups}

\label{ssec:mcmcconcepts}

In calculating ensemble-averaged RMPS quantities of the form $\langle f(\{A_i\})\rangle$, we perform Monte-Carlo integrations:
	\begin{equation}
		\langle f(\{A_i\})\rangle=\int d\mu(\{A_i\})\,f(\{A_i\})\simeq\frac{1}{n}\sum_{s=1}^{n}f(\{A_i\}_s).
	\end{equation}
Here we assume that the samples $\{A_i\}_s$ follow the probability distribution $P(\{A_i\})$ induced from $d\mu(\{A_i\})$. It is desired that each sample is independent from each other. However, for models containing non-quadratic terms, such sampling is not generally feasible.

A family of alternative methods is given by Markov-Chain-Monte-Carlo(MCMC) methods, where we can sample arbitrary distributions while sacrificing the independence of each sample. In this work, we used the Metropolis-Hastings algorithm, and here is a brief review of our procedure.

Given the $s$th sample $x_{s}$, we sample a candidate $x'_{s+1}$ from a conditional probability distribution $g(x'_{s+1}|x_{s})$, where sampling from the distribution $g$ is generally easy. Then, we compute the acceptance $\alpha$, which is defined as
	\begin{equation}
		\alpha:=\frac{P(x'_{s+1})}{P(x_{s})}\frac{g(x_{s}|x'_{s+1})}{g(x'_{s+1}|x_{s})}.
	\end{equation}
If $\alpha\geq1$, we accept the candidate such that $x_{s+1}=x'_{s+1}$. If not, we accept the candidate with probability $\alpha$, and reject it otherwise, in which case $x_{s+1}=x_{s}$. It has been shown that the resulting sequence $\{x_{s}\}$ follows the target distribution $P(x)$ as the number of samples goes to infinity.

The efficiency of the Metropolis-Hastings algorithm depends significantly on the choice of the proposal function $g$. For example, if $g$ is too narrow, then it takes a lot of samples to explore the whole parameter space. Otherwise, if $g$ is too broad, then the acceptance $\alpha$ will generally be very low, so we cannot advance to a different sample.

\paragraph{Single MCMC.} In order to sample general $\mathrm{O}(k)$ models, we have built a heuristic sampler, which depends on two parameters $\delta_{1}$ and $\delta_{2}$. The sampler picks two kinds of random variables. First, it picks a number $X$ which follows the normal distribution $X\sim N(0,1)$. Second, it picks a set of $N\times N$ Hermitian matrices $\{B_{i}\}_{i=1,\dots,N}$ where each matrix $B_{i}$ is independently sampled from the probability density proportional to $\exp(-\mathrm{Tr}[B_{i}^{2}]/2)$. Then the candidate sample $\{A'_{i}\}$ is determined from the previous sample $\{A_{i}\}$ as follows:
	\begin{equation}
		A'_{i}=\frac{(\delta_{1}/\sqrt{kN})X+\sqrt{\sum_{j=1}^{k}\mathrm{Tr}[A_{j}^{2}]}}{\sqrt{\sum_{j=1}^{k}\mathrm{Tr}[M_{j}^{2}]}}M_{i},\quad M_{i}:=\frac{1}{\sqrt{\sum_{j=1}^{k}\mathrm{Tr}[A_{j}^{2}]}}A_{i}+\delta_{2}B_{i}.
	\end{equation}
Since $X$ and $\{B_{i}\}$ are random, we have a conditional probability distribution $g(\{A'_{i}\}|\{A_{i}\})$.
It can be shown that
	\begin{equation}
		\frac{g(\{A_{i}\}|\{A'_{i}\})}{g(\{A'_{i}\}|\{A_{i}\})}=\left(\frac{\sqrt{\sum_{j=1}^{k}\mathrm{Tr}[A'^{2}_{j}]}}{\sqrt{\sum_{j=1}^{k}\mathrm{Tr}[A_{j}^{2}]}}\right)^{kN^{2}}=\left(\frac{(\delta_{1}/\sqrt{kN})X+\sqrt{\sum_{j=1}^{k}\mathrm{Tr}[A_{j}^{2}]}}{\sqrt{\sum_{j=1}^{k}\mathrm{Tr}[A_{j}^{2}]}}\right)^{kN^{2}}.
	\end{equation}
The parameters $\delta_{1}$ and $\delta_{2}$ can be determined as `broadnesses' of the sampler, so they should be tuned for each target distribution. We found that $\delta_{1}=1/2$ and $\delta_{2}=1/kN$ works pretty well for the Quartic and Quadratic + Quartic models. We call this setup \textit{single MCMC}.

To be precise, we note that the above formula for $g(\{A_{i}\}|\{A'_{i}\})/g(\{A'_{i}\}|\{A_{i}\})$ is not correct if $(\delta_{1}/\sqrt{kN})X+\sqrt{\sum_{j=1}^{k}\mathrm{Tr}[A_{j}^{2}]}$ can be less than zero. Assuming that $k$ and $N$ are reasonably large, the quantity $\sqrt{\sum_{j=1}^{k}\mathrm{Tr}[A_{j}^{2}]}$ is typically very large while $\delta_{1}/\sqrt{kN}$ is very small, so we do not have to worry about this.

\paragraph{Double MCMC.} In computing bra-ket disconnected quantities, we have to select the matrices $A_{i}$ independently for bras and kets.

In this case, we regard a collection of two sets of matrices $(\{A^{1}_{i}\},\{A^{2}_{i}\})$ as one sample. Then we draw a candidate $(\{A'^{1}_{i}\},\{A'^{2}_{i}\})$ from the proposal functions $g(\{A'^{1}_{i}\}|\{A^{1}_{i}\})$ and $g(\{A'^{2}_{i}\}|\{A^{2}_{i}\})$. The acceptance $\alpha$ is given by
	\begin{equation}
		\alpha=\frac{P(\{A'^{1}_{i}\})P(\{A'^{2}_{i}\})}{P(\{A^{1}_{i}\})P(\{A^{2}_{i}\})}\frac{g(\{A'^{1}_{i}\}|\{A^{1}_{i}\})g(\{A'^{2}_{i}\}|\{A^{2}_{i}\})}{g(\{A^{1}_{i}\}|\{A'^{1}_{i}\})g(\{A^{2}_{i}\}|\{A'^{2}_{i}\})}.
	\end{equation}
We call this setup \textit{double MCMC}.

\subsection{Autocorrelation and effective sample sizes}

\label{ssec:mcmcautocorr}

Because of autocorrelation in MCMC, the effective sample size of the MCMC chain is given by the total sample size divided by the factor $\nu$, where
\begin{equation}
    \nu:=1+2\sum_{n=1}^{\infty}\rho_{n}.
\end{equation}
Here $\rho_{n}$ is the autocorrelation of the chain at lag $n$, given by
\begin{equation}
    \rho_{n}:=\mathrm{Cor}(x_{0},x_{n}).
\end{equation}
This means that if we obtain $S$ samples using the MCMC method, the sample mean roughly follows the distribution
\begin{equation}
    \bar{x}_{\text{sample}}\sim N(\bar{x},\nu\sigma^{2}/S),
\end{equation}
i.e. a normal distribution with variance $\nu\sigma^{2}/S$.

\begin{figure}[t]
    \centering
    \begin{subfigure}[t]{0.45\textwidth}
        \centering
        \includegraphics[width=\textwidth]{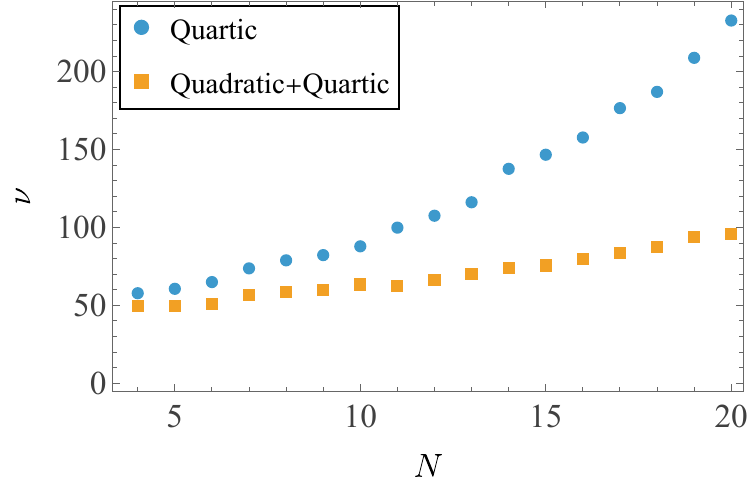}
        \caption{}
    \end{subfigure}
    \hspace{0.05\textwidth}
    \begin{subfigure}[t]{0.45\textwidth}
        \centering
        \includegraphics[width=\textwidth]{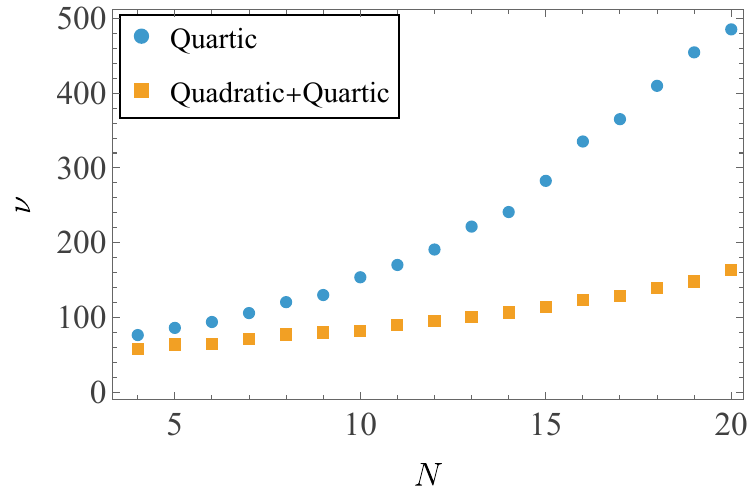}
        \caption{}
    \end{subfigure}
    \caption{
    Autocorrelation factors of MCMC on RMPS models with $k=8$. (a) is for single MCMC, and (b) is for double MCMC.
    }
    \label{fig:autocorr}
\end{figure}

In many cases, the autocorrelation $\rho_{n}$ is difficult to compute analytically. Instead, we may estimate $\rho_{n}$ using many sample chains.
From our MCMC settings in RMPS models with non-quadratic terms, we obtained estimated values of $\nu$, which are shown in figure~\ref{fig:autocorr}.

Sometimes it is beneficial to apply a post-processing called \textit{thinning}, where we take every $p$th sample and discard others for a fixed number $p$.
This method effectively reduces autocorrelation, and is efficient when calculation times of quantities in interest are much longer than MPS sampling times.
For the simulations presented in this study, we chose the thinning parameter $p$ heuristically, with values ranging from 5 to 2000.

\section{Bra-ket disconnected subregion entropy}

\label{sec:discphys}

\subsection{Bra-ket disconnected subregion entropy}

\label{ssec:discentropy}

We compute bra-ket disconnected subregion entropies $\left<S_n[R]\right>_{\text{disc}}$, i.e. the average entropy of a spatial subregion $R$ in the pseudo density matrix
\begin{equation}
    \ket{\psi} \bra{\psi^\prime}
\end{equation}
where $\ket{\psi}$ and $\ket{\psi^\prime}$ are independently selected states in the RMPS ensemble. We write the subregion entropy of a given sample $(\ket{\psi}, \ket{\psi^\prime})$ as $S_{\ket{\psi} \bra{\psi^\prime};n}[R]$.

Note that the bra-ket disconnected subregion entropy is a pseudo entropy \cite{nakata_holographic_2021, mollabashi_pseudo_2021, mollabashi_aspects_2021}, and, for a given selection of $\ket{\psi}$ and $\ket{\psi^\prime}$, is generally complex-valued. In fact, for a given selection of $\ket{\psi}$ and $\ket{\psi^\prime}$, the pseudo density matrix of $R$ given by
\begin{equation}
    \rho_{\ket{\psi} \bra{\psi^\prime};R} = \Tr_{R^c} (\ket{\psi} \bra{\psi^\prime})
\end{equation}
generally has negative complex eigenvalues, and thus we do not a priori expect from the RMPS structure that $|S_{\ket{\psi} \bra{\psi^\prime};n}[R]|$ to be bounded by $|\partial R| \ln N$.

However, in the limit where the lengths of the intervals in $R$ and the spacings between them go to infinity, we have
\begin{equation}
    S_{\ket{\psi} \bra{\psi^\prime};n}[R] =  |\partial R| S_n[\rho_{N; \text{disc}}]
\end{equation}
where $S_n[\rho_{N; \text{disc}}]$ is the entropy of the matrix $\rho_{N; \text{disc}}$ and $\rho_{N; \text{disc}}$ is an $N\times N$ matrix given by the partial trace of $M_{\text{disc}}^L$ for dRMPS and $e^{- L H_{\text{disc}}}$ for cRMPS:
\begin{equation}
    \bra{\nu} \rho_{N;\text{disc}} \ket{\mu} =
    \begin{cases}
        \Tr  \left( M_{\text{disc}}^L (I_{N\times N} \otimes \ket{\mu}  \bra{\nu})  \right) & (\text{dRMPS})\\
        \Tr  \left(e^{ -L H_{\text{disc}}} (I_{N\times N} \otimes \ket{\mu}  \bra{\nu})  \right) & (\text{cRMPS}).
    \end{cases}
\end{equation}
In this case $\rho_{N; \text{disc}}$ is positive semi-definite and thus we do expect $S_{\ket{\psi} \bra{\psi^\prime};n}[R]$ to be real and bounded by $2 \ln N$. We call this the asymptotic value of $S_{\ket{\psi} \bra{\psi^\prime};n}[R]$.

We remark that for the single interval case with $d$ even in dRMPS, or for cRMPS, $S_{\ket{\psi} \bra{\psi^\prime};n}[R]$ will generally be real for $n=2,3,4,...$. This is because we can write
\begin{equation}
    \Tr (\rho_{\ket{\psi} \bra{\psi^\prime};R}^n  ) = \begin{cases}
        \Tr(M_{\text{disc}}^{L-d} O^\dagger_n M_{\text{disc}}^d O_n ) & (\text{dRMPS})\\
        \Tr(e^{ -(L-d) H_{\text{disc}}}O^\dagger_n e^{ -d H_{\text{disc}}} O_n ) & (\text{cRMPS}),
    \end{cases}
\end{equation}
where
\begin{equation}
    O_n = \sum^N_{\mu_1 , ..., \mu_n  = 1}(I \otimes \ket{\mu_1} \bra{\mu_2}) \otimes (I \otimes \ket{\mu_2} \bra{\mu_3}) ... \otimes (I \otimes \ket{\mu_n} \bra{\mu_1})
\end{equation}
is an RMPS twist operator. The above asserted reality result follows from the fact that, given positive operators $A, B$,
\begin{equation}
    \Tr (AO^\dagger BO) = \Tr(X^\dagger X), \quad X = A^{\frac{1}{2}} O B^{\frac{1}{2}}.
\end{equation}
However, due to the negative eigenvalues of $\rho_{\ket{\psi} \bra{\psi^\prime};R} $, $S_{\ket{\psi} \bra{\psi^\prime}}[R]$ may still be complex even in those cases.

\paragraph{Bra-ket disconnected entropy bounding in JT+CFT.} In \cite{chen_bra-ket_2021, teresi_islands_2022} a calculation was done that we can now interpret as the bounding of asymptotic bra-ket disconnected entropy in the AdS JT+CFT and dS JT+CFT models. Here we will review their results.

We compute the asymptotic value of $\left<S_n[R]\right>_{\text{disc}}$ in the AdS JT+CFT model. We will use a version of the island formula. As a result, we get bounding behavior with the bound given by $|\partial R| \ln N + O(N^0)$.

We must note beforehand that the island formula is derived from the gravitational path integral assuming the replica trick, and that the replica trick generally may not hold for pseudo density matrices with complex eigenvalues. However, by the argument above, we can assume the island formula holds when computing the asymptotic entropy.

The island formula for the bra-ket disconnected entropy is given by
\begin{equation}
	\left<S[R]\right>_{\text{disc}} = \min  \Eext \left(  S_{\text{gen}} [I \cup R]  \right)
\end{equation}
where the island $I$ exists in the dominant geometry of the bra-ket disconnected inner product, i.e. the double disk geometry. In the general case where $S_{\text{gen}}[I \cup R]$ is complex, we take the extremal value with minimum real part.

We consider the general case of $R$ consisting of $m$ disjoint intervals. We denote the intervals as $x_{2i-1} < x < x_{2i} $ with $i = 1 ... m$. Then, we find an extremal choice of $I$ given by the $n$ line segments
\begin{equation}
    z = \frac{6 \phi_r}{c} , \quad x_{2i-1} <x< x_{2i} .
    \label{eq:discisland}
\end{equation}
Each line segment can either be on the ket disc or the bra disc; taking all cases into account simply adds a constant factor $m \ln 2$ to the entropy. We will ignore this factor.

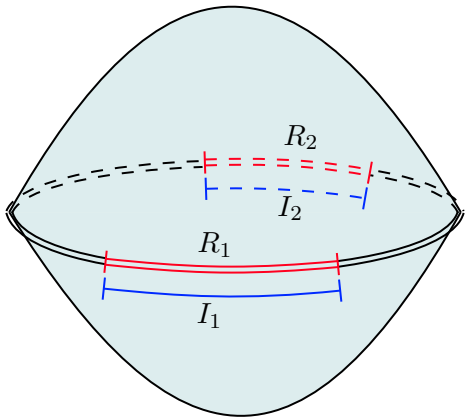
\begin{figure}[t]
    \begin{center}

    \input{figures/tikz/discisland}
    \end{center}
    \caption{Picture of a spatial subregion $R$ given by the union of two intervals $R_1$, $R_2$ and a dominant choice of island $I$ for the bra-ket disconnected entropy given by the union of two line segments $I_1$, $I_2$ atop the double disc of the AdS JT+CFT prepared state. We assume the lengths of $R_1$, $R_2$ and the distances between them are very large.}
    \label{fig:discisland}
\end{figure}

The orientation of $I$ relative to $R$ is chosen so that $I$ purifies the CFT entropy of $R$. In computing the asymptotic entropy, this island configuration is the dominant configuration. We see that
\begin{equation}
	S_{\text{gen}}[I \cup R] = m S_{\text{disc;max}},
\end{equation}
where
\begin{equation}
    S_{\text{disc;max}} = 2 \phi_0 +  \frac{c}{2} + \frac{c}{3} \ln \frac{6\phi_r}{c}.
\end{equation}
Thus, we see that the asymptotic entropy is $m S_{\text{disc;max}}$. See figure~\ref{fig:discisland}.

We remark that by using cosmic branes \cite{dong_gravity_2016, almheiri_replica_2020}, we could also derive a bounding behavior $\left<S[R]\right>_{\text{disc}} < m S_{n;\text{disc;max}}$ for the average $n$-Renyi bra-ket disconnected entropy. In this case the results would be generally valid even in the non-asymptotic case. However, the calculation would be much more involved than that of the entanglement entropy.

By the relation of the AdS JT+CFT and dS JT+CFT models, we get the same results for dS JT+CFT, just with $\phi_r$ replaced with $ i \phi_{\text{real-time}}$.

\paragraph{Numerical results.} We computed $|\left<S_n[R]\right>_{\text{disc}}|$ where $R$ is a single interval. The results show an average entropy bound $S_{n;\text{disc;max}}$ at large $d$, with bound $2\ln N + \mathcal{O}(N^0)$. We check that, other than the case of odd $d$ for dRMPS, the $n$-Renyi entropies for $n>1$ (but not the entanglement entropy) were purely real.

\begin{figure}[t]
    \centering
    \begin{subfigure}[t]{0.45\textwidth}
        \centering
        \includegraphics[width=\textwidth]{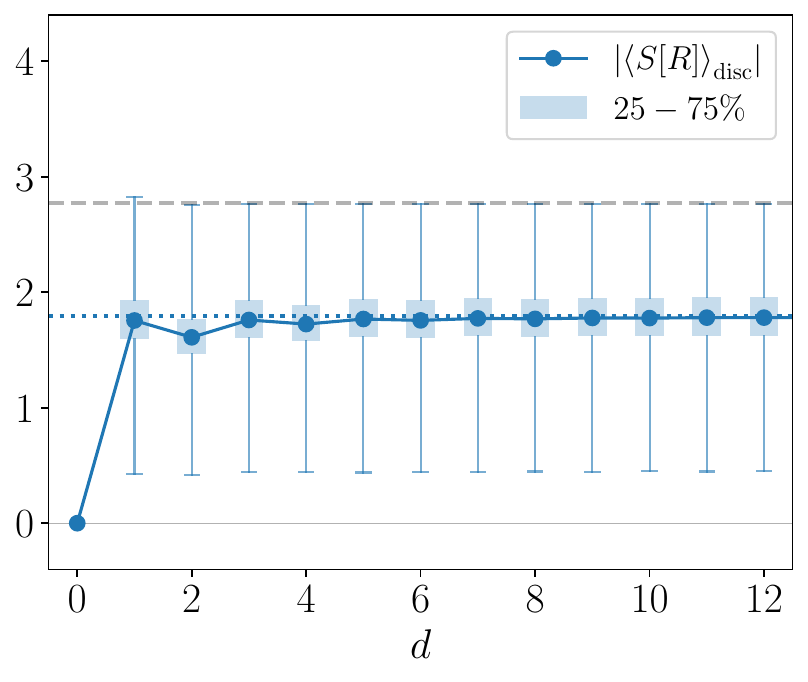}
        \caption{}
    \end{subfigure}
    \hspace{0.05\textwidth}
    \begin{subfigure}[t]{0.45\textwidth}
        \centering
        \includegraphics[width=\textwidth]{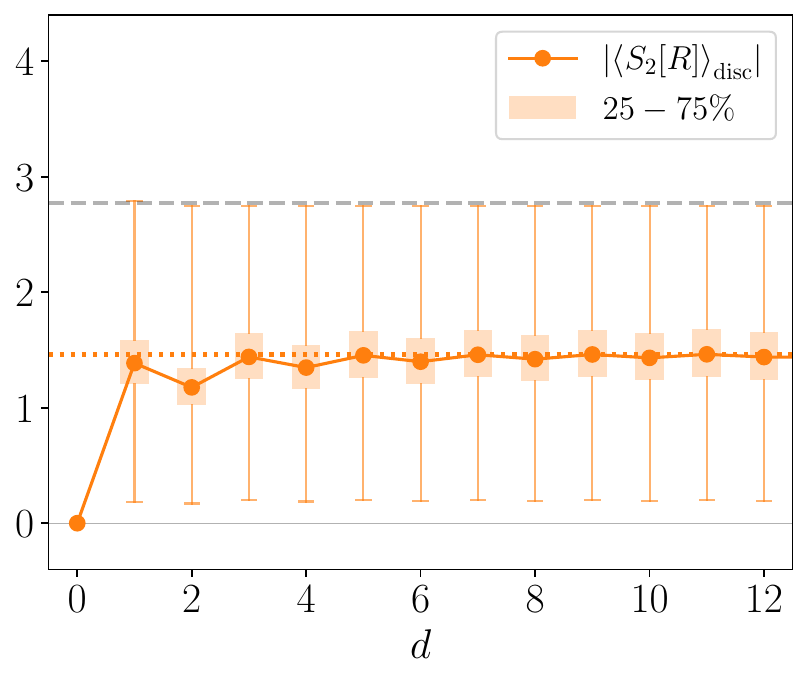}
        \caption{}
    \end{subfigure}
    \begin{subfigure}[t]{0.45\textwidth}
        \centering
        \includegraphics[width=\textwidth]{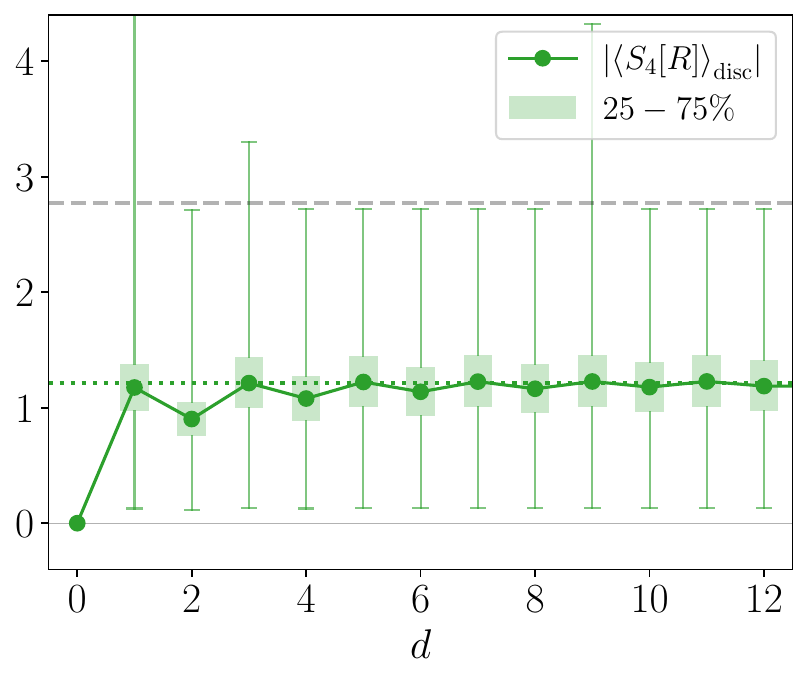}
        \caption{}
    \end{subfigure}
    \hspace{0.05\textwidth}
    \begin{subfigure}[t]{0.45\textwidth}
        \centering
        \includegraphics[width=\textwidth]{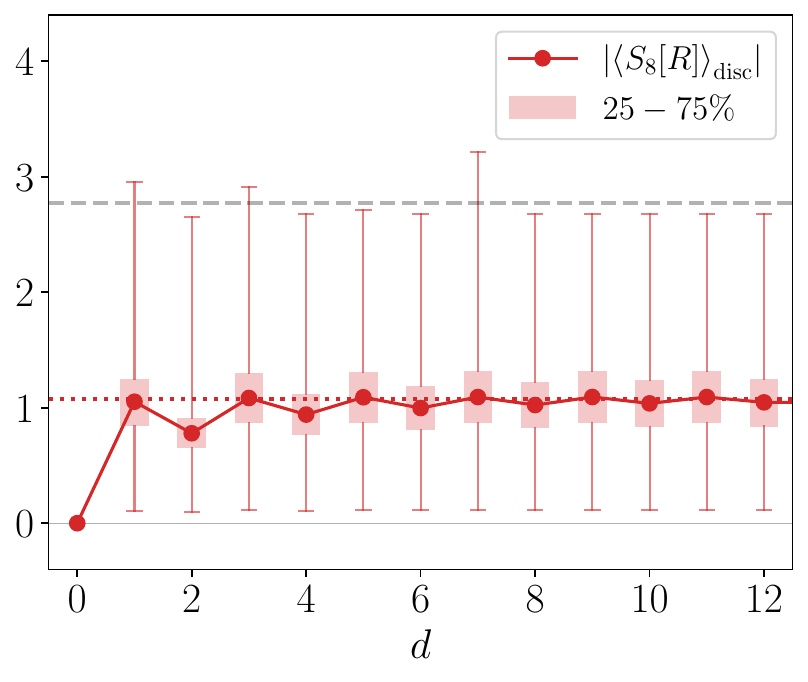}
        \caption{}
    \end{subfigure}
    \caption{
    Disconnected entropy $\langle S_n[R]\rangle_{\rm disc}$ for dRMPS in the quadratic+quartic model. We have $N=4$, $k=64$, and $L=256$.
    Each point was obtained from $10^3$ samples.
    }
    \label{fig:disc_entropy_dRMPS}
\end{figure}

In figure~\ref{fig:disc_entropy_dRMPS}, we present the behavior of the averaged entanglement and Renyi bra-ket disconnected entropies of an interval in the quadratic + quartic model. The dashed lines indicate $2\ln N$, and the dotted lines indicate $S_{n;\text{disc;max}}$. The vertical lines indicate the maximum and minimum values from the simulations, and the boxes represent the 25–75\% quartiles for each $d$.
In the discrete RMPS model, we noticed that when the length $d$ of $R$ was odd, we had $|S_{\ket{\psi} \bra{\psi^\prime};n}[R]| > 2\ln N$ for some samples. Thus, from the numerics, although we see an average entropy bound, we generally do not see a sample entropy bound. Similar results hold for the other two models, which can be found in appendix~\ref{ssec:supplementary_figures_drmps}.

\begin{figure}[t]
    \centering
    \begin{subfigure}[t]{0.45\textwidth}
        \centering
        \includegraphics[width=\textwidth]{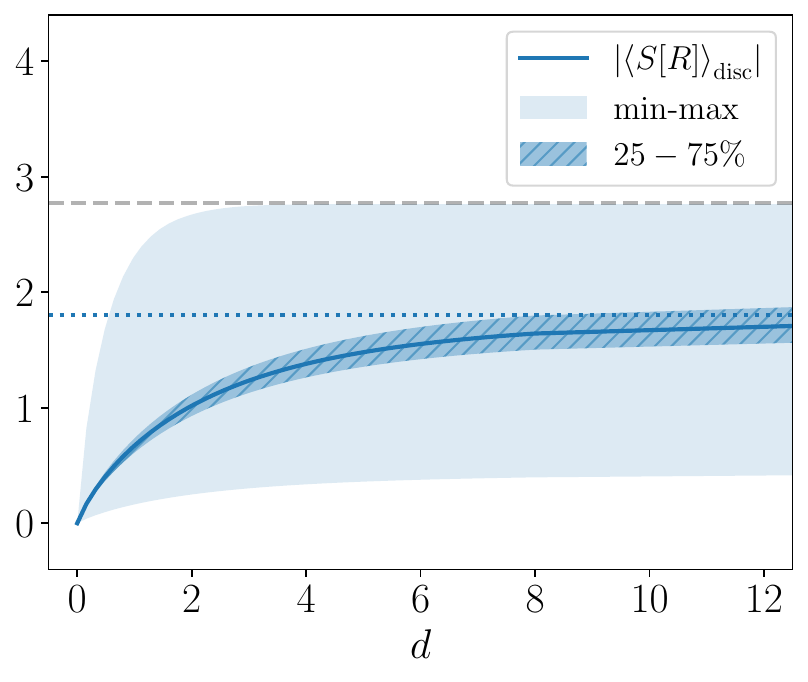}
        \caption{}
    \end{subfigure}
    \hspace{0.05\textwidth}
    \begin{subfigure}[t]{0.45\textwidth}
        \centering
        \includegraphics[width=\textwidth]{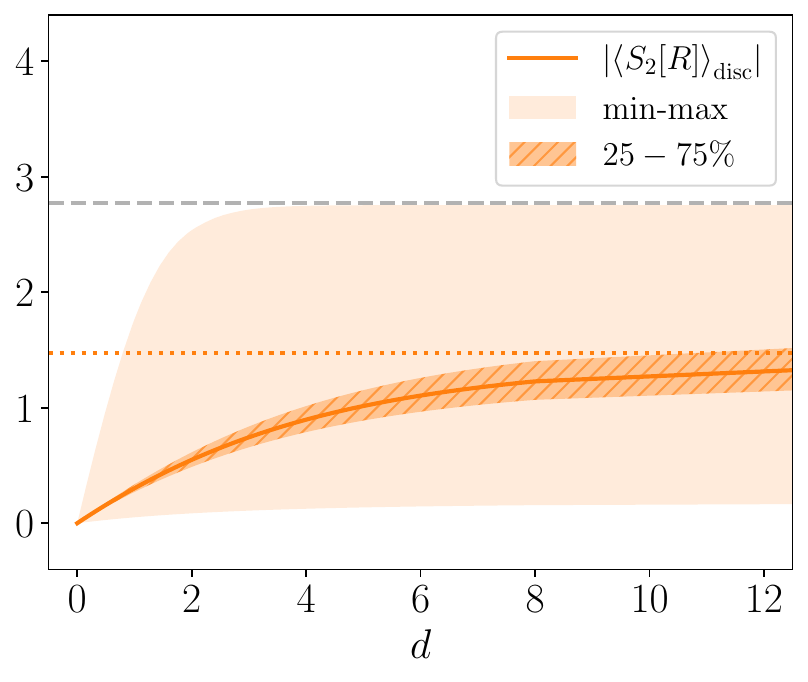}
        \caption{}
    \end{subfigure}
    \begin{subfigure}[t]{0.45\textwidth}
        \centering
        \includegraphics[width=\textwidth]{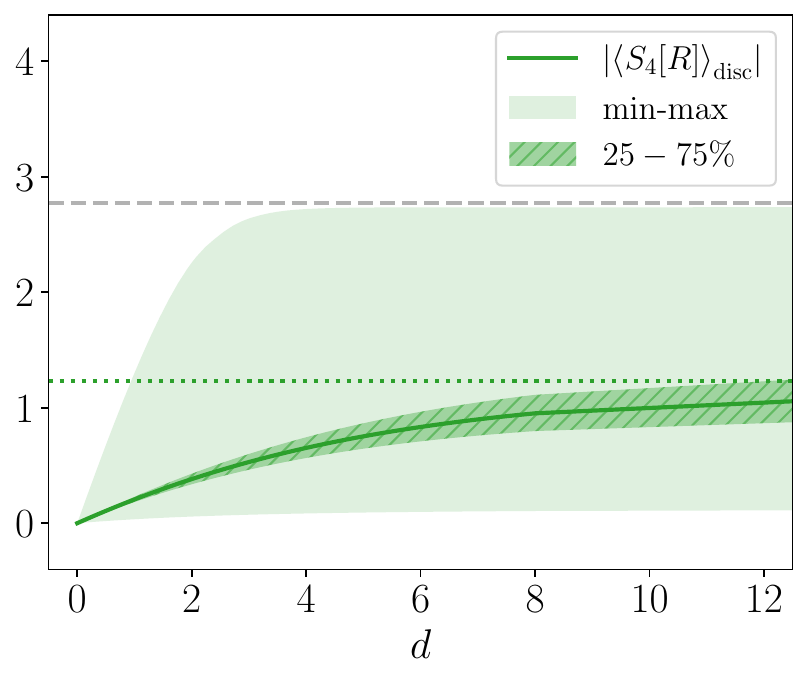}
        \caption{}
    \end{subfigure}
    \hspace{0.05\textwidth}
    \begin{subfigure}[t]{0.45\textwidth}
        \centering
        \includegraphics[width=\textwidth]{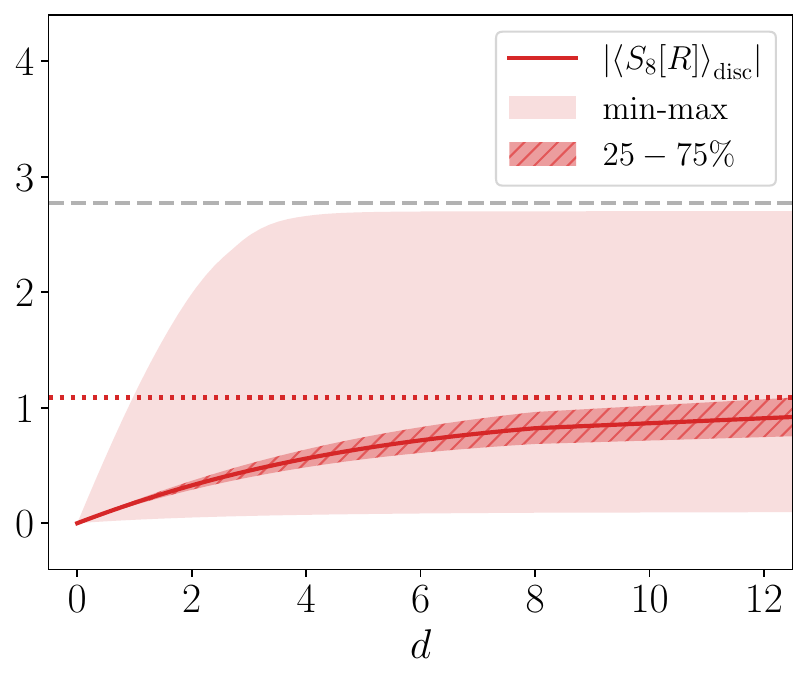}
        \caption{}
    \end{subfigure}
    \caption{
    Disconnected entropy $\langle S_n[R]\rangle_{\rm disc}$ for cRMPS in the quadratic+quartic model. We have $N=4$, $k=64$, and $L=256$.
    Each point was obtained from $10^3$ samples.
    }
    \label{fig:disc_entropy_cRMPS}
\end{figure}

In figure~\ref{fig:disc_entropy_cRMPS}, we present the behavior of the averaged entanglement and Renyi bra-ket disconnected entropies of an interval in the continuous model. The dashed lines indicate $2\ln N$, and the dotted lines indicate $S_{n;\text{disc;max}}$. The solid-colored regions represent the maximum and minimum values from the simulations, and the hatched regions correspond to the 25--75\% quartiles. From the numerics, we see both an average entropy bound, and a sample entropy bound. Similar results hold for the other two models, which can be found in appendix~\ref{ssec:supplementary_figures_crmps}.

\begin{figure}[t]
    \centering
    \begin{subfigure}[t]{0.45\textwidth}
        \centering
        \includegraphics[width=\textwidth]{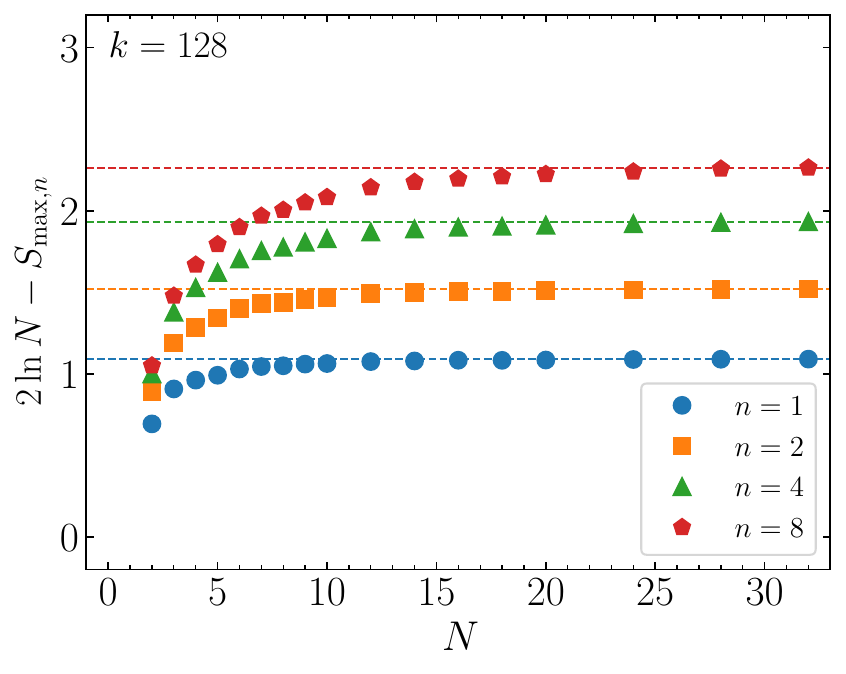}
        \caption{}
    \end{subfigure}
    \hspace{0.05\textwidth}
    \begin{subfigure}[t]{0.45\textwidth}
        \centering
        \includegraphics[width=\textwidth]{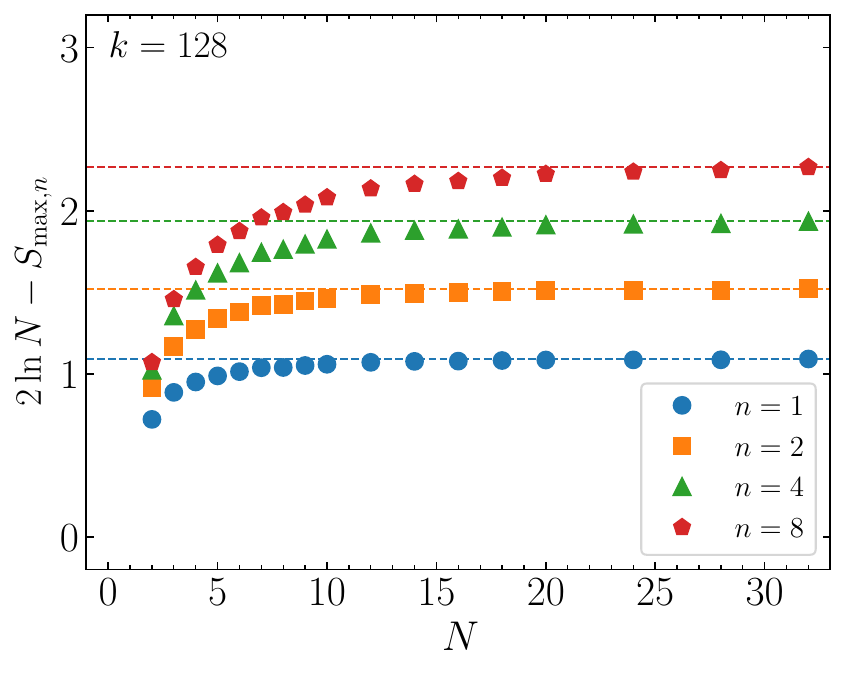}
        \caption{}
    \end{subfigure}
    \caption{
       Convergence of $2 \ln N - S_{{\rm disc;max},n}$ at large $N$ for the (a) dRMPS (b) cRMPS quadratic + quartic models. We used $k=128$ and $10^3$ samples for each point.
    }
    \label{fig:disc_N_entropy}
\end{figure}

In figure~\ref{fig:disc_N_entropy} we show that the disconnected average entropy bound is equal to $2\ln N$ plus an $\mathcal{O}(N^0)$ subleading term for (a) dRMPS, (b) cRMPS in the quadratic + quartic model. Similar results hold for the other two models, which can be found in Appendices~\ref{ssec:supplementary_figures_drmps},~\ref{ssec:supplementary_figures_crmps}.

\FloatBarrier

\subsection{Complex island condition}

It is worth noting that, in the dS JT+CFT model, the bra-ket disconnected entropy island lies in real time, that is, at
\begin{equation}
    \sigma = \frac{6 \phi_{\text{real-time}}}{c}
\end{equation}
in the geometry \eqref{eq:dsgeo}. This is a \textit{complex island} in the sense that it lies on the complexification of the negative-definite disk saddlepoint geometry.

It is interesting to ask which complex islands are allowed. We give a general idea for computing such criteria based on the KSW condition \cite{kontsevich_wick_2021, witten_note_2022}. Not all complex islands can be allowed due to issues defining the matter path integral on complex replica geometries. We suggest that complex islands should be included in the island formula as long as they satisfy a certain condition that we call the good island condition.

We consider the AdS JT+CFT model with $\phi_r$ complex and take $L = 2\pi$. We note that $M$ is a hyperboloid in $\mathbb{R}^3$:
\begin{equation}
	\begin{aligned}
		& M = \left\{\left(x^i\right) \in \mathbb{R}^3  \text{ : }\eta_{i j} w^i w^j= - 1  \right\}, \quad  d s^2=\eta_{i j} dw^i d w^j,\\
		& w^0= \coth z, \quad w^1=\csch z \cos x, \quad w^2=\csch z\sin x
	\end{aligned}.
\end{equation}
We can define the complexification $M_{\mathbb{C}}$ of the disk geometry $M$ as a complex manifold with 4 real dimensions containing $M$ as a submanifold such that the metric on $M_{\mathbb{C}}$ is a holomorphic extension of the metric on $M$. Thus we can find $M_{\mathbb{C}}$ by complexifying $\mathbb{R}^3$ to $\mathbb{C}^3$:
\begin{equation}
	\begin{aligned}
		& M_{\mathbb{C}} = \left\{\left(X^i\right) \in \mathbb{C}^3  \text{ : }\eta_{i j} W^i XW^j= - 1  \right\}, \quad  d s^2=\eta_{i j} d W^i d W^j,\\
		& W^0= \coth Z, \quad W^1=\csch Z \cos X, \quad W^2=\csch Z\sin X.
	\end{aligned}
\end{equation}
In $(X,Z)$ coordinates the metric can be written as
\begin{equation}
	\begin{aligned}
		&ds^2 = \frac{dZ^2 + dX^2}{\sinh^2 Z}, \quad \phi = \frac{c}{12} + \frac{\phi_r}{\tanh Z}, \quad Z, X \in \mathbb{C},\\
		&X \approx X + 2\pi, \quad Z \approx Z + 2\pi i , \quad (X, Z) \approx (X+\pi, Z + \pi i).
	\end{aligned}
\end{equation}
We call points in $M_{\mathbb{C}}$ complex spacetime points.

In discussing complex islands it is helpful to consider 2-dimensional submanifolds in $M_{\mathbb{C}}$ obtained by a deformation of $M$. We call such submanifolds contours in $M_{\mathbb{C}}$ or analytic deformations of $M$.

To define the matter path integral on such a contour, we must define the spacetime action on it, and thus we must know the volume form on it. Given the orientation $dz \wedge dx$ on $M$, we have the volume form on $M$,
\begin{equation}
	\dvol_{g} =\begin{cases}
		&\frac{dz \wedge dx}{\sinh^2 z} \quad (\text{compact universe}) \\
		&\frac{dz \wedge dx}{z^2} \quad (\text{noncompact universe}) .
	\end{cases}
\end{equation}
The volume form on $M_{\mathbb{C}}$ is the holomorphic continuation of $\dvol_{g}$:
\begin{equation}
	\dvol_{g_{\mathbb{C}}} = \begin{cases}
		&\frac{dZ \wedge dX}{\sinh^2 Z} \quad (\text{compact universe}) \\
		&\frac{dZ \wedge dX}{Z^2} \quad (\text{noncompact universe}) .
	\end{cases}
\end{equation}
The volume form on a contour $M^\prime$ is the pullback of $\dvol_{g_{\mathbb{C}}}$ onto $M^\prime$.

Generally the metric and volume form on $M^\prime$ will be complex. However, for such general complex geometries, the CFT path integral may not be well defined. In general, the QFT path integral is only well defined on complex geometries satisfying specific conditions.

The KSW condition gives a condition for the QFT path integral on a complex metric $g_{ab}$ to be defined for all p-form QFTs. To check the condition we can first diagonalize the metric at each point to find the square roots $\lambda_i$ of the complex eigenvalues of the metric satisfying
\begin{equation}
	\dvol_g = \prod^{D-1}_{i=0} \lambda_i dx^0 \wedge ... \wedge dx^{D-1},
\end{equation}
where $x^i$ are the coordinates diagonalizing the metric at this point and $ dx^1 \wedge ... \wedge dx^{D_1}$ is the orientation of the manifold. Then, at each point, we require that
\begin{equation}
	\sum_i | \arg \lambda_i | < \frac{\pi}{2}
\end{equation}
is satisfied. We say that such metrics are KSW-good.

For 2D CFTs we should be able to relax this condition because we can multiply an arbitrary complex number to the metric at each point, by analytical continuation of the Weyl rescaling. Thus for 2D CFTs we get
\begin{equation}
	| \arg \lambda_1 - \arg \lambda_2 | < \frac{\pi}{2}
\end{equation}
as a condition for good metrics.

In general we say that a metric is good if for the given matter content of the model we could define the path integral on it. We say that a contour in $M_{\mathbb{C}}$ is good if it has a good metric with respect to the matter CFT.

In computing the pseudo $n$-Renyi entropy for general complex $\phi_r$, we can find $n$-replica saddles by gluing together complex geometries with cosmic branes. In the $n\rightarrow 1$ limit this becomes equivalent to finding a complex island on a contour in $M_{\mathbb{C}}$. Since the matter path integral on the $n$-replica saddle must be well defined, the complex geometries with cosmic branes must have good metrics, and in the $n\rightarrow 1$ limit the contour containing the complex island must have a good metric; i.e. it must be a good contour. Thus for a complex island to be allowed, it must lie on a good contour in $M_{\mathbb{C}}$.
\begin{equation}
	\text{complex island $I$ allowed} \Leftrightarrow  \exists \text{ good contour } M^\prime \in  M_{\mathbb{C}} \text{ s.t. } I \subset M^\prime.
\end{equation}
We call such complex islands good islands. We remark that even though an island isn't good, it might be homologous to a good island, which is allowed and gives the same entropy. Thus in general we will check if an island is homologous to a good island.

We say that a set of points is good if there is a good contour $M^\prime$ containing these points. Assuming that the space of good contours is path connected, we see that we can find a good island with the endpoints
\begin{equation}
	\{ p_i \}
\end{equation}
if and only if $ \{ p_i \}$ is good. Note that it is a necessary condition for any subset of $ \{ p_i \}$ to be good for $\{ p_i \}$ to be good.

We suggest that a set of points is good if and only if for a certain ordering of the points, the following quantum state is well defined for any choice of local operators $O_i$:
\begin{equation}
	\prod_i O_i(p_i) \ket{0_{\text{CFT}}}.
\end{equation}
This is because if there is a good contour $M^\prime$ passing through these points, we should be able to compute the above state  as a CFT path integral on $M^\prime$ with operator insertions at $p_i$:
\begin{equation}
	\prod_i O_i(p_i) \ket{0_{\text{CFT}}} = \int_{M^\prime } \exp (-S_{\text{CFT}}) \prod_i O_i(p_i).
\end{equation}
Thus we can use this method generally to see if a given set of points lies on a good contour.

For example, we can use this method to compute the set
\begin{equation}
	M_{\mathbb{C} ; \text { good }}
\end{equation}
of points reachable by a good contour in $M_{\mathbb{C}}$. We have for the primary operator $O$ with conformal dimension $\Delta$,
\begin{equation}
	\begin{aligned}
		O(Z, W) \ket{0_{\text{CFT}}} &= \Omega^{-\Delta} (Z)\text{ }e^{-Z H - i X P} O(0, 0) \ket{0_{\text{CFT}}}\\
		&= \Omega^{-\Delta}  (Z)\sum_{E,p} e^{-Z E - i X p} \ket{E, p} \bra{E,p} O(0, 0) \ket{0_{\text{CFT}}}.
	\end{aligned}
\end{equation}
Here $\Omega(Z)$ is just the conformal scale factor of the metric and $\ket{E,p}$ are the energy and momentum eigenstates of the CFT. Assuming the CFT has a density of states depending sub-exponentially on the energy-momentum, it follows that for this expression to converge for a general choice of the operator $O$ we require
\begin{equation}
	-Z E - i X p \rightarrow - \infty \quad \text{when} \quad  E, |p| \rightarrow \infty .
\end{equation}
We know $E \geq p$ as a standard property of local quantum field theory, and thus the condition that $(Z,X) \in M_{\mathbb{C}}$ is good is given by
\begin{equation}
	\Re Z > |\Im X|.
\end{equation}
For the compact universe case, this is equivalent to
\begin{equation}
	\Re W^{0} > \sqrt{(\Re W^{1})^2 + (\Re W^{2})^2} ,
\end{equation}
and thus we can check that this region is invariant under the $SO(1,2)$ symmetry of $M_{\mathbb{C}}$.

Using this criterion, for the island $I$ (\ref{eq:discisland}) we found in computing the bra-ket disconnected entropy of an interval, we see that it is a good island for $\Re \phi_r > 0$. The case where $\Re \phi_r = 0$ (as in the dS JT+CFT model) is marginal.

\section{Holographic interpretation of RMPS models}

\label{sec:holography}

Following the reinterpretation of the spatial dimension of $\Sigma$ as a Euclidean time dimension in section~\ref{ssec:rmpsdefinition}, we see that RMPS models of gravitationally prepared states provide holographic duals of 2D gravitating systems that can exchange matter particles with a nongravitating bath.

\begin{figure}[t]
	\begin{center}
    \input{figures/tikz/holointer}
	\end{center}
	\caption{Left: the path integral of the partition function of a 2D gravity region coupled to a nongravitating bath. The Euclidean timeline boundary $\Sigma$ is denoted in red. Right: the Penrose diagram of the Lorentzian geometry of the corresponding thermofield double state. The Lorentzian timelike boundaries of the black hole region are denoted in red.}
    \label{fig:holointer}
\end{figure}
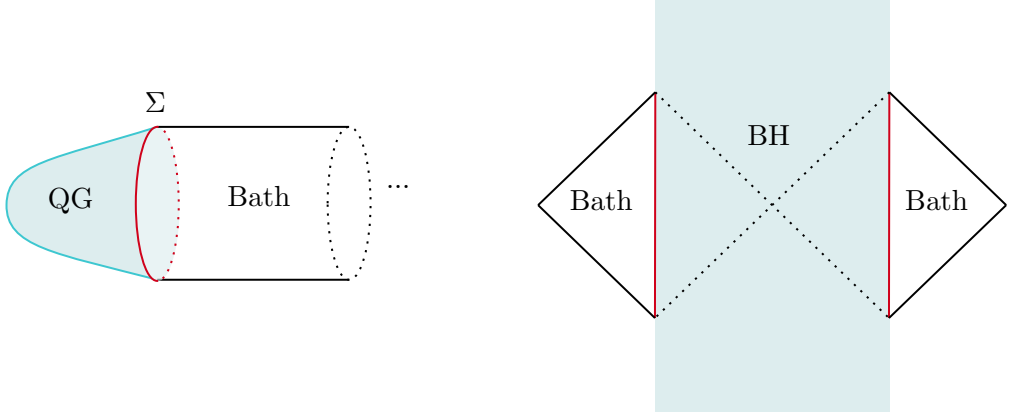

We can make this statement precise in the following way. The partition function $Z_{\text{QG+bath}}$ of a 2D gravity region coupled to a nongravitating bath at a Euclidean timelike boundary $\Sigma$ of length $L$ is given by
\begin{equation}
    Z_{\text{QG+bath}} = \bra{\psi}\ket{\psi_{\text{bath}}}
\end{equation}
where $\bra{\psi}$ is a RMPS state with boundary length $L$, and $\ket{\psi_{\text{bath}}}$ is the matter state prepared on $\Sigma$ by the Euclidean path integral of the bath region. This partition function can be understood as the inner product of a thermofield double state of a double-sided eternal black hole. See figure~\ref{fig:holointer}.

We can match the parameter $\ln N$ of the RMPS model with the approximate black hole entropy $\phi_0$ on the gravity side. We conjecture that the parameter $\ln k$ is proportional to the matter CFT central charge $c$ on the gravity side.

As an example of a concrete observable that can be computed in this framework, we consider the entropy of a subregion in the non-gravitating region, which we call the \textit{bath entropy}. This was a key observable studied in early papers on the island formula and the Page curve in the black hole information paradox, for example in \cite{almheiri_replica_2020}. Using the RMPS model, we could achieve an exact calculation of this entropy in certain circumstances.

In particular, we see that the bra-ket disconnected entropy that was computed in appendix~\ref{sec:discphys} is physically similar to the bath entropy, with $\ket{\psi_{\text{bath}}}$ replaced with an independent copy of $\ket{\psi}$. We see that the island bounding the bra-ket disconnected entropy is topologically identical to the island found in \cite{almheiri_replica_2020}. From the island formula, we expect that the bra-ket disconnected entropy in the case where $d = \frac{L}{2}$ is approximately equal to the entropy of the entire bath region in the thermofield double state of a double-sided eternal black hole.

\section{Supplementary figures}
\label{sec:supplementary_figures}

\subsection{Supplementary figures for dRMPS}

\label{ssec:supplementary_figures_drmps}

\begin{figure}[t]
    \centering
    \begin{subfigure}[t]{0.45\textwidth}
        \centering
        \includegraphics[width=\textwidth]{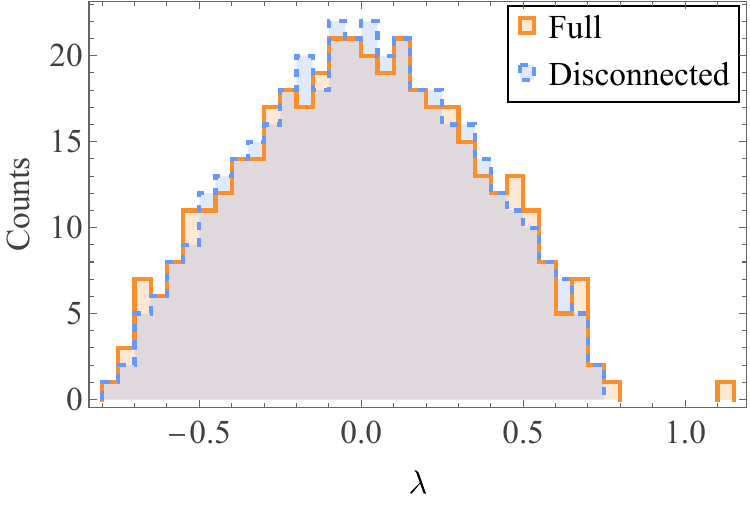}
        \caption{}
    \end{subfigure}
    \hspace{0.05\textwidth}
    \begin{subfigure}[t]{0.45\textwidth}
        \centering
        \includegraphics[width=\textwidth]{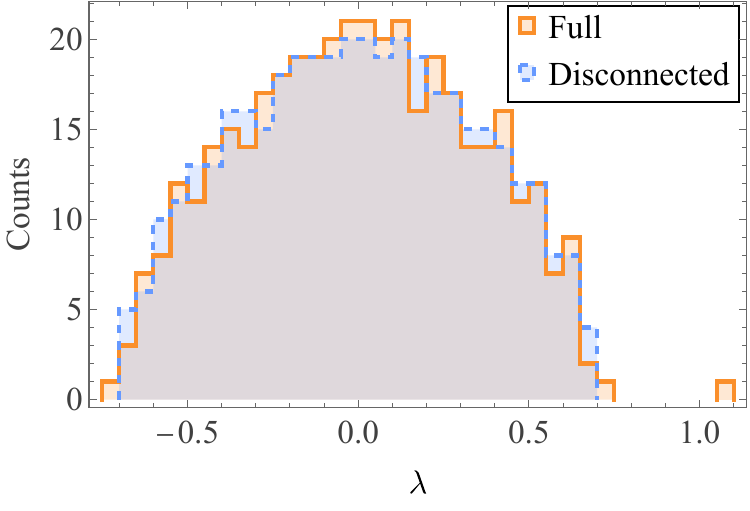}
        \caption{}
    \end{subfigure}
	\caption{Eigenvalue distribution of a sample transfer matrix in the discrete (a) quadratic (b) quartic models with $k=8$, $N=20$. We see a gapped highest eigenvalue in the full transfer matrix.}
    \label{fig:singleeig}
\end{figure}

\paragraph{Spectral gapping property.} In figure~\ref{fig:singleeig} (a), (b), we present the eigenvalue distribution of a typical transfer matrix in the quadratic and quartic models. The yellow bars represent the full transfer matrix, and the blue bars represent the disconnected transfer matrix. The green region represents their overlap. We see that the full transfer matrix has an additional gapped highest eigenvalue.

\begin{figure}[t]
    \centering
    \begin{subfigure}[t]{0.45\textwidth}
        \centering
        \includegraphics[width=\textwidth]{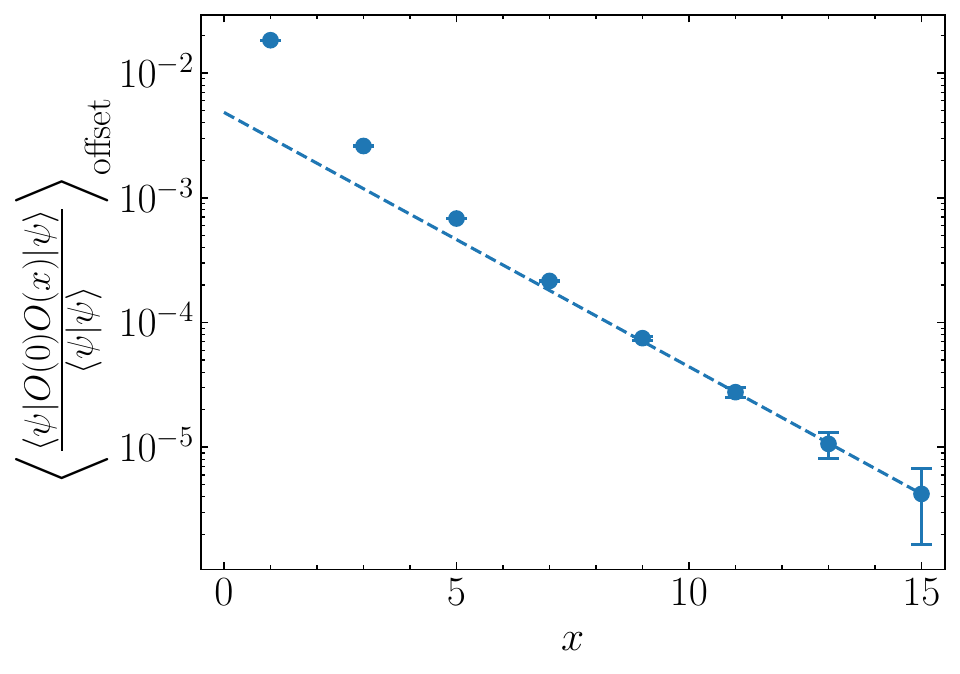}
        \caption{}
    \end{subfigure}
    \hspace{0.05\textwidth}
    \begin{subfigure}[t]{0.45\textwidth}
        \centering
        \includegraphics[width=\textwidth]{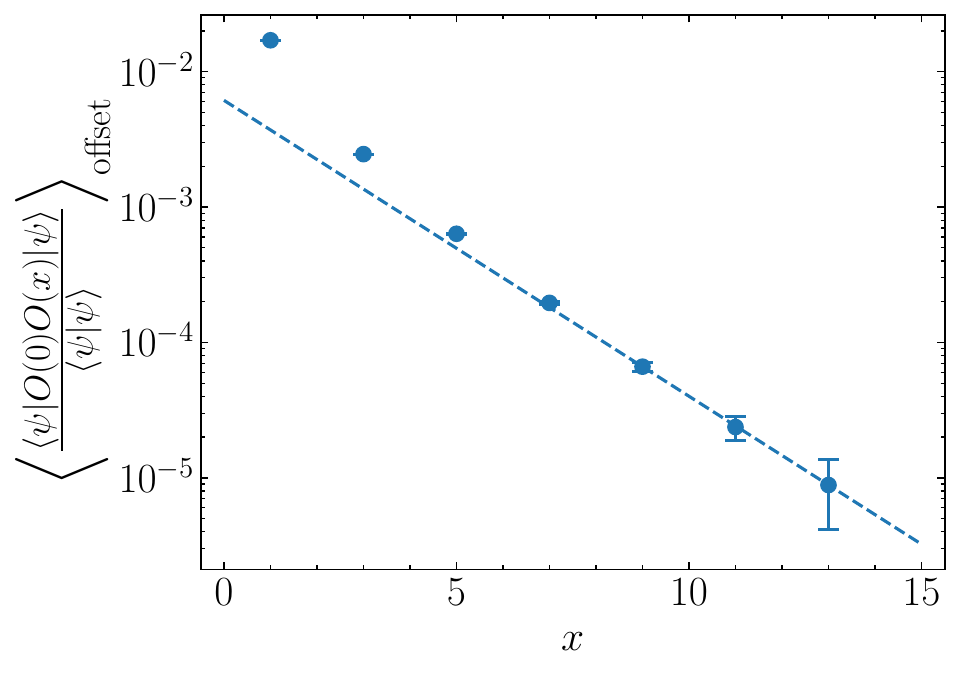}
        \caption{}
    \end{subfigure}
    \begin{subfigure}[t]{0.45\textwidth}
        \centering
        \includegraphics[width=\textwidth]{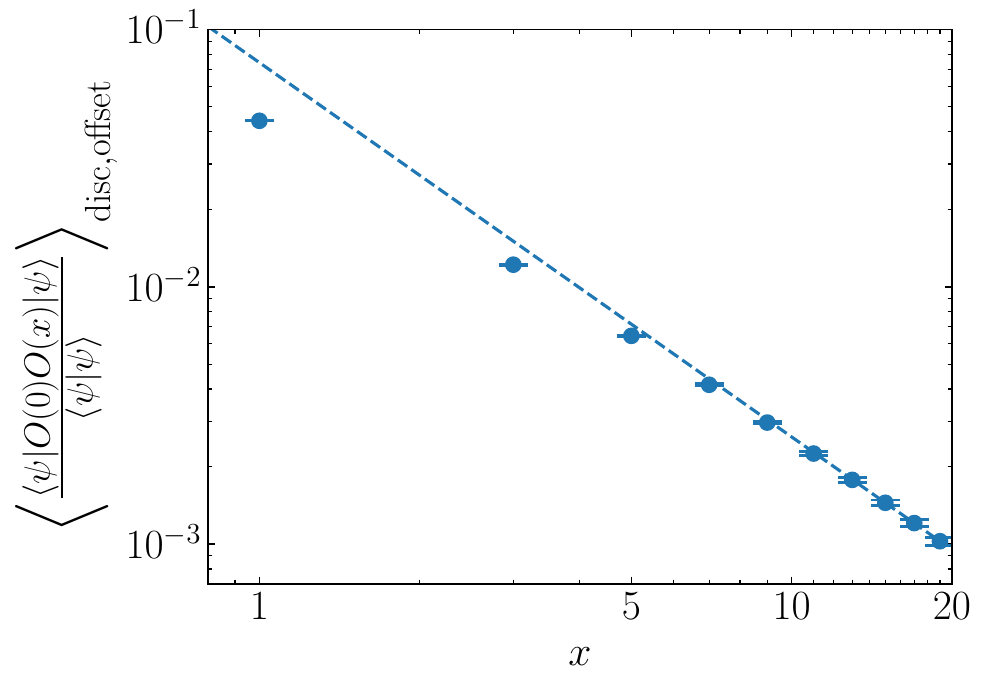}
        \caption{}
    \end{subfigure}
    \hspace{0.05\textwidth}
    \begin{subfigure}[t]{0.45\textwidth}
        \centering
        \includegraphics[width=\textwidth]{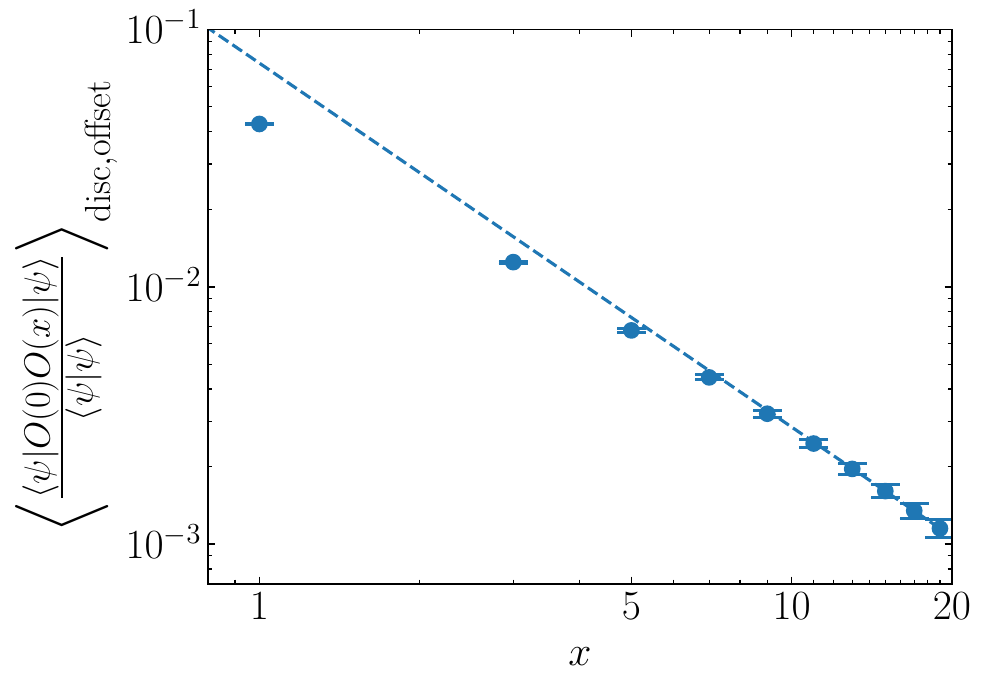}
        \caption{}
    \end{subfigure}
	\caption{Two-point correlations of a local operator in the discrete quadratic and quartic models with the $x \rightarrow \infty$ value subtracted. Here $k=8$, $N=10$, and $L=400$. We have full correlators showing exponential decay in $x$ for the discrete (a) quadratic (b) quartic models. Here we used $10^5$ samples. We have bra-ket disconnected correlators showing power law decay in $x$ for the discrete (c) quadratic (d) quartic models. Here we used $10^4$ samples.}
    \label{fig:dtwomodelCorrProperties}
\end{figure}

\paragraph{Bra-ket wormhole phase transition.} In figure~\ref{fig:dtwomodelCorrProperties} (a), (b), we present the exponential decay of the correlator in the quadratic and quartic models after subtracting the offset at $x \rightarrow \infty$. In (c), (d), we present the power law decay of the bra-ket disconnected correlator in the quartic and quadratic + quartic models after subtracting the offset at $x \rightarrow \infty$.

\begin{figure}[t]
    \centering
    \begin{subfigure}[t]{0.45\textwidth}
        \centering
        \includegraphics[width=\textwidth]{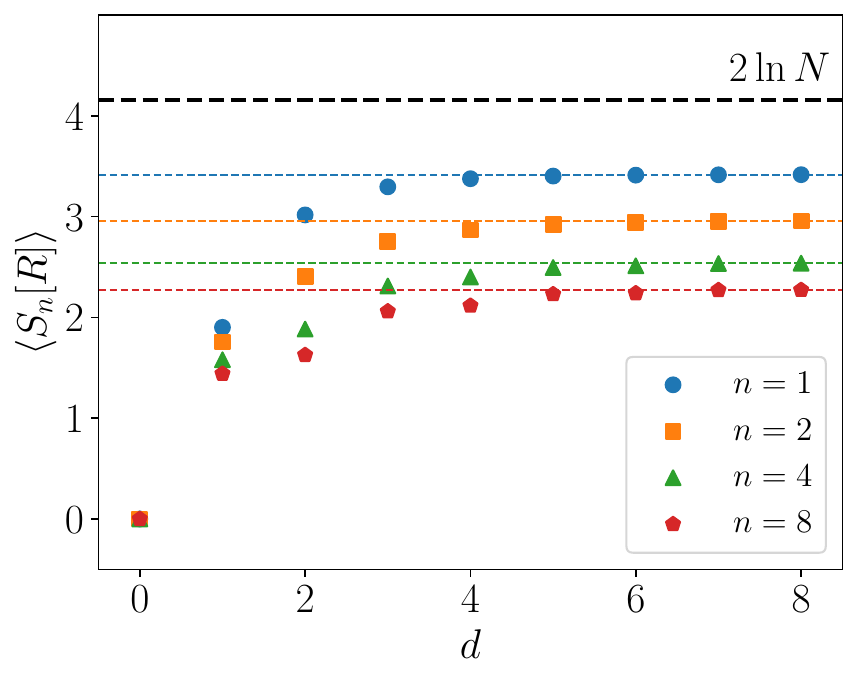}
        \caption{}
    \end{subfigure}
        \hspace{0.05\textwidth}
    \begin{subfigure}[t]{0.45\textwidth}
        \centering
        \includegraphics[width=\textwidth]{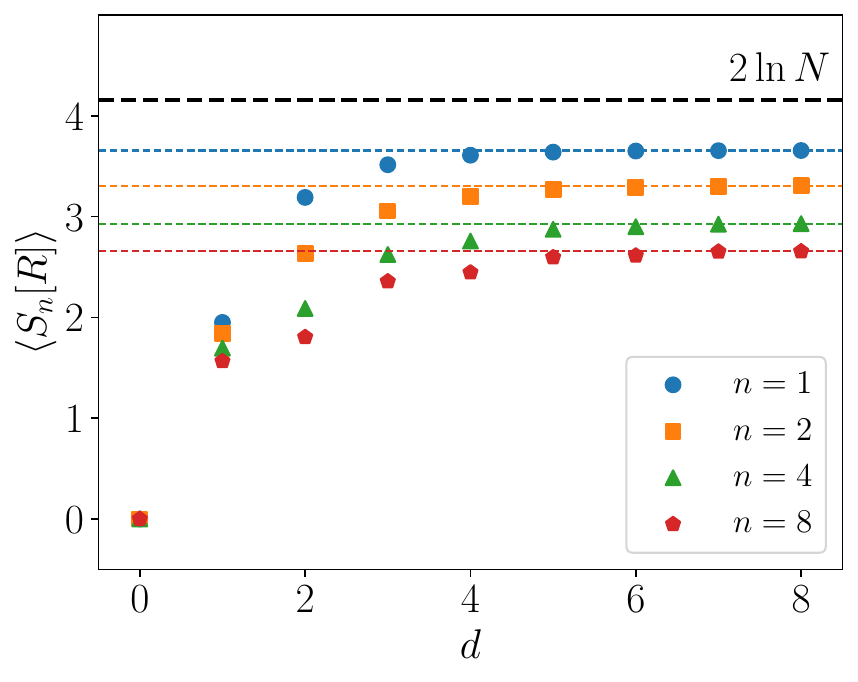}
        \caption{}
    \end{subfigure}
    \begin{subfigure}[t]{0.45\textwidth}
        \centering
        \includegraphics[width=\textwidth]{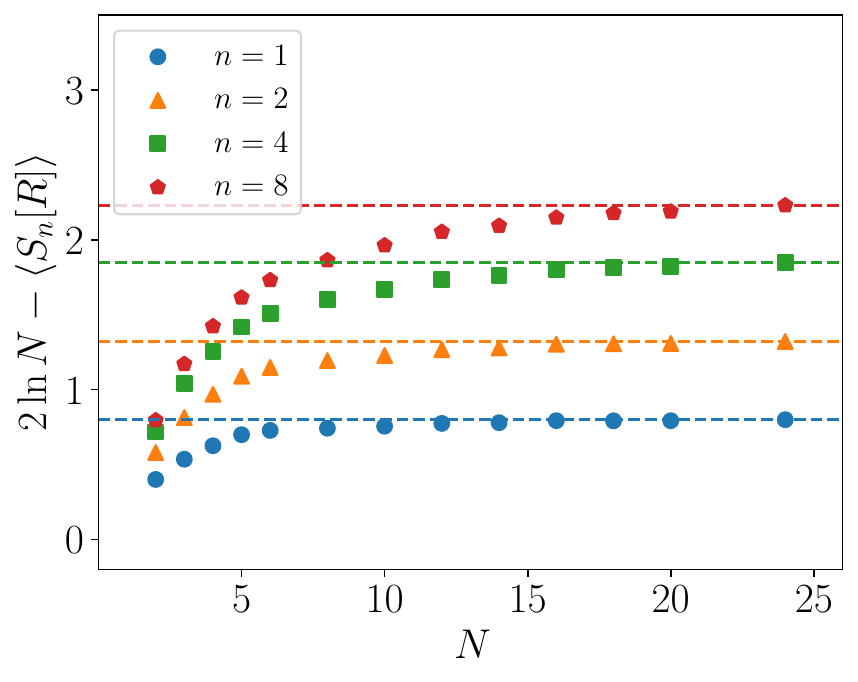}
        \caption{}
    \end{subfigure}
        \hspace{0.05\textwidth}
    \begin{subfigure}[t]{0.45\textwidth}
        \centering
        \includegraphics[width=\textwidth]{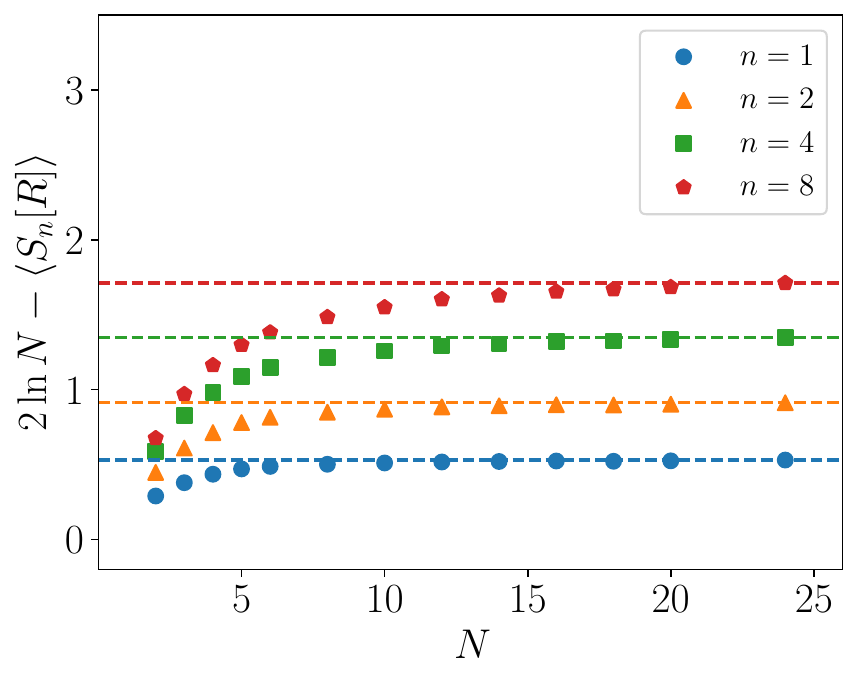}
        \caption{}
    \end{subfigure}
	\caption{
    The bounding behavior of the averaged entropies of an interval in the discrete quadratic and quartic models.
    $S_n[R]$ showing bounding behavior in the (a) quadratic (b) quartic model. Here $N=8$ with $10^4$ samples. Convergence of $2\ln N - S_{\text{max}, n}$ at large $N$ in the (c) quadratic (d) quartic model.  We used $k=128$ and $10^3$ samples.
    }
    \label{fig:drmps_entropy}
\end{figure}

\paragraph{Entropy bounding.} In ~\ref{fig:drmps_entropy} (a), (b), we present the bounding behavior of the averaged entanglement and Renyi entropies ($n=1,2,4,8$) of an interval in the discrete quadratic and quartic models. In (c), (d) we show that the average entropy bound is equal to $2\ln N$ plus an $\mathcal{O}(N^0)$ subleading term in the discrete quadratic and quartic models.

\begin{figure}[t]
    \centering
    \begin{subfigure}[t]{0.45\textwidth}
        \centering
        \includegraphics[width=\textwidth]{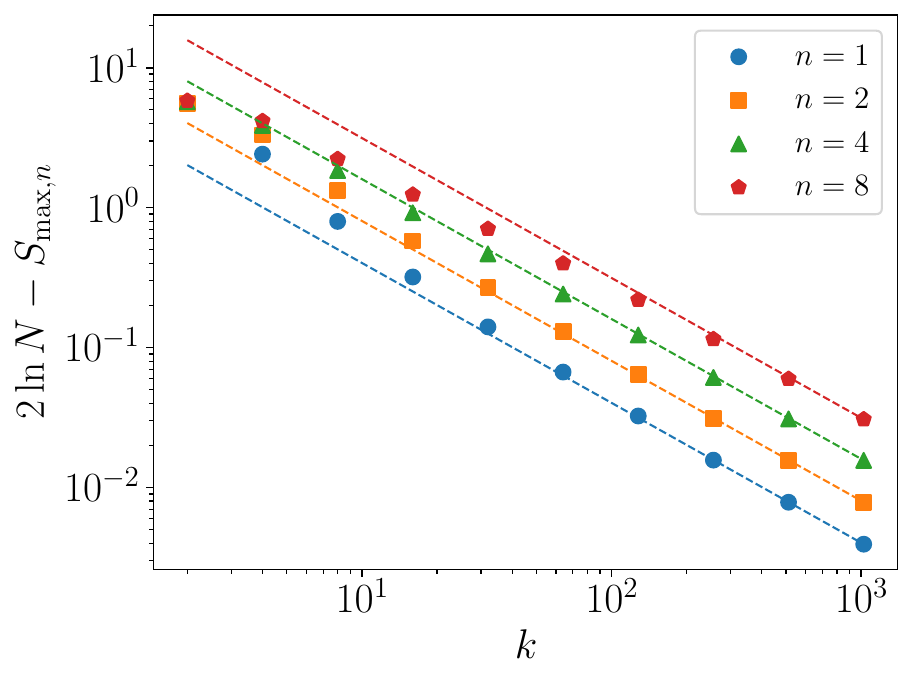}
        \caption{}
    \end{subfigure}
    \hspace{0.05\textwidth}
    \begin{subfigure}[t]{0.45\textwidth}
        \centering
        \includegraphics[width=\textwidth]{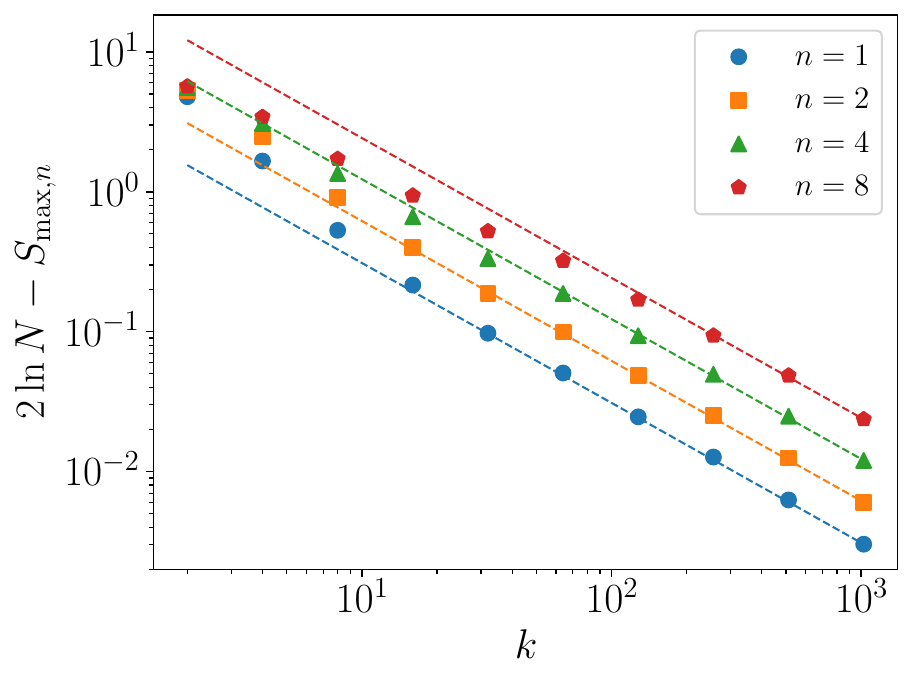}
        \caption{}
    \end{subfigure}
	\caption{$k^{-1}$-dependence of $2\ln N - S_{\text{max}, n}$ in the discrete (a) quadratic (b) quartic model. Here $N=24$. We used $10^3$ samples.
    }
    \label{fig:drmps_entropy_kdep}
\end{figure}

In figure~\ref{fig:drmps_entropy_kdep} (a), (b), we present the $k^{-\frac{1}{2}}$ dependence of $2\ln N - S_{\text{max}, n}$ in the discrete quadratic and quartic models.

\begin{figure}[t]
    \centering
    \begin{subfigure}[t]{0.45\textwidth}
        \centering
        \includegraphics[width=\textwidth]{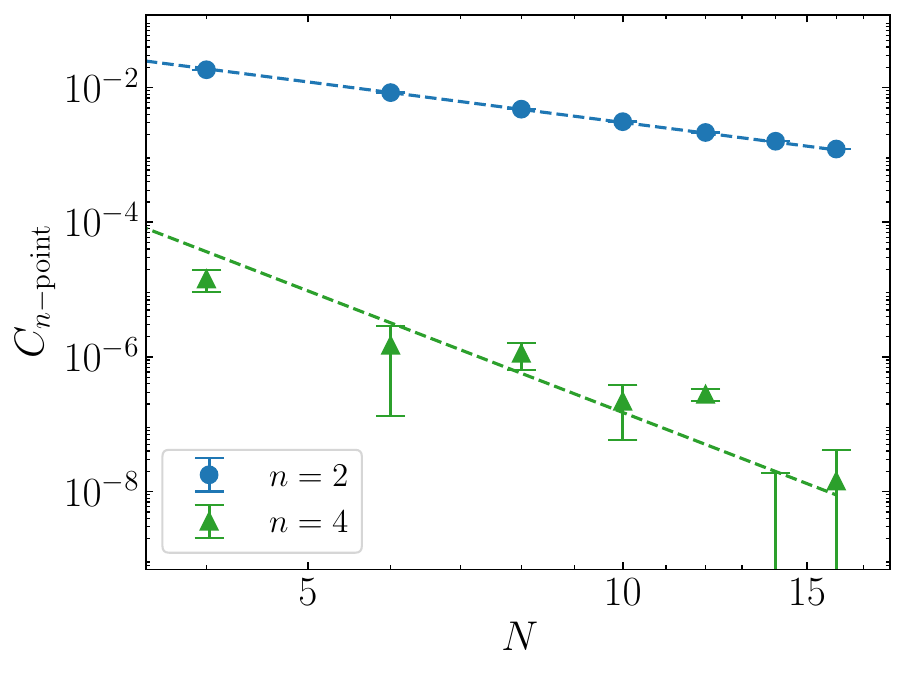}
        \caption{}
    \end{subfigure}
    \hspace{0.05\textwidth}
    \begin{subfigure}[t]{0.45\textwidth}
        \centering
        \includegraphics[width=\textwidth]{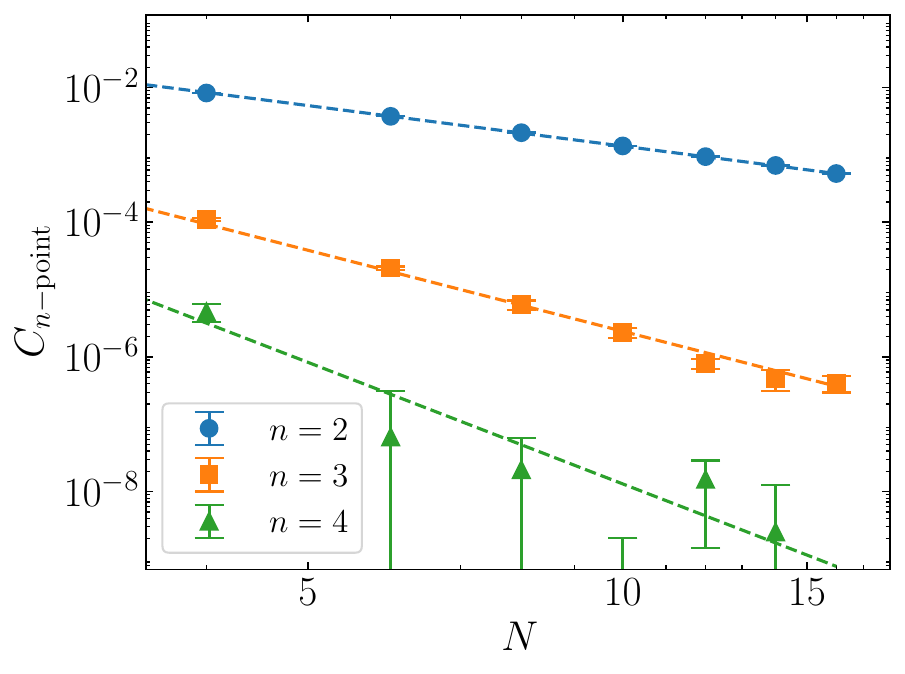}
        \caption{}
    \end{subfigure}
    \begin{subfigure}[t]{0.45\textwidth}
        \centering
        \includegraphics[width=\textwidth]{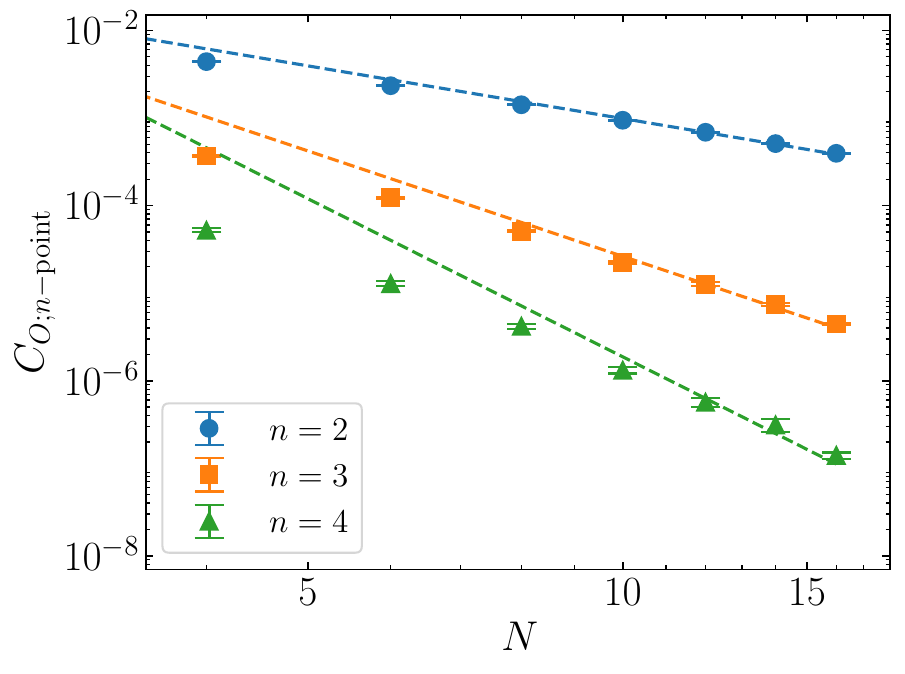}
        \caption{}
    \end{subfigure}
    \hspace{0.05\textwidth}
    \begin{subfigure}[t]{0.45\textwidth}
        \centering
        \includegraphics[width=\textwidth]{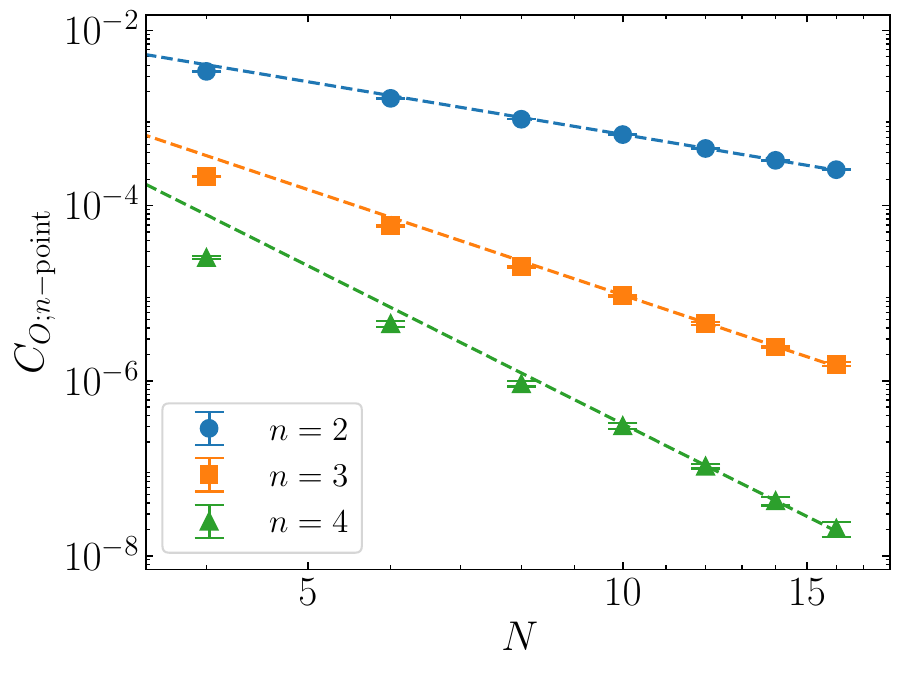}
        \caption{}
    \end{subfigure}
	\caption{$N$ dependence of $\ln\lambda_g$ cumulants for (a) quadratic, (b) quartic models. $N$ dependence of $\lambda_g^{-1}\bra{\lambda_g}M_O\ket{\lambda_g}$ cumulants for (c) quadratic, (d) quartic models. Trendlines represent the expected $N$ dependence of cumulants. In every setting, we used $k=8$ with $10^{5}$ samples.}
    \label{fig:drmps_corr}
\end{figure}

\paragraph{Off-shell wormholes.} In figure~\ref{fig:drmps_corr} (a), (b), we present the $N$ dependence of $n$th $\ln \lambda_g$ cumulants in the quadratic and quartic models. In (c), (d), we present the $N$ dependence of $n$th $\lambda_g^{-1}\bra{\lambda_g}M_O\ket{\lambda_g}$ cumulants in the quadratic and quartic models. For large $N$, we have $N^{2(1-n)}$ dependence for both, excluding $n=3$ for the quadratic model as explained in section~\ref{ssec:rmpsnumerical}.

\begin{figure}[t]
    \centering
    \begin{subfigure}[t]{0.45\textwidth}
        \centering
        \includegraphics[width=\textwidth]{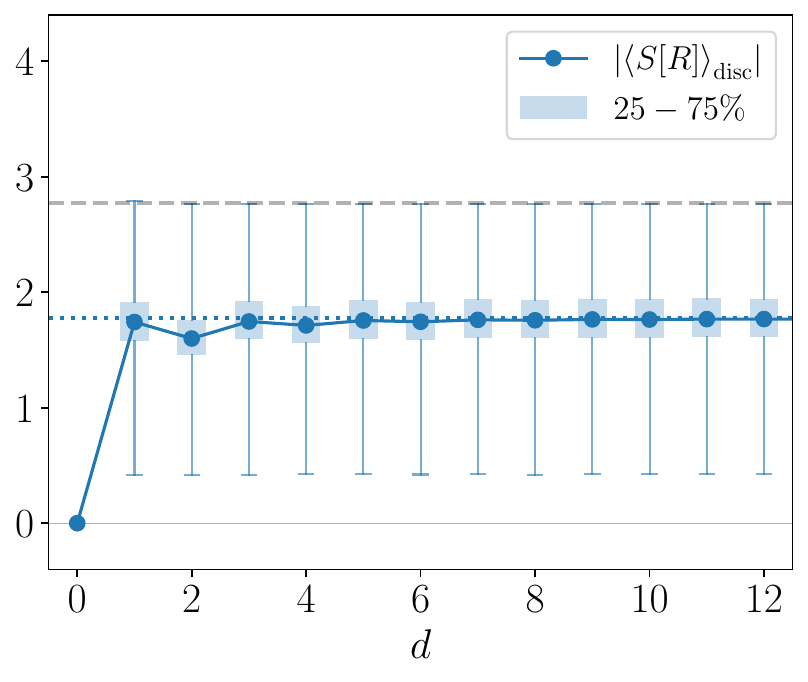}
        \caption{}
    \end{subfigure}
    \hspace{0.05\textwidth}
    \begin{subfigure}[t]{0.45\textwidth}
        \centering
        \includegraphics[width=\textwidth]{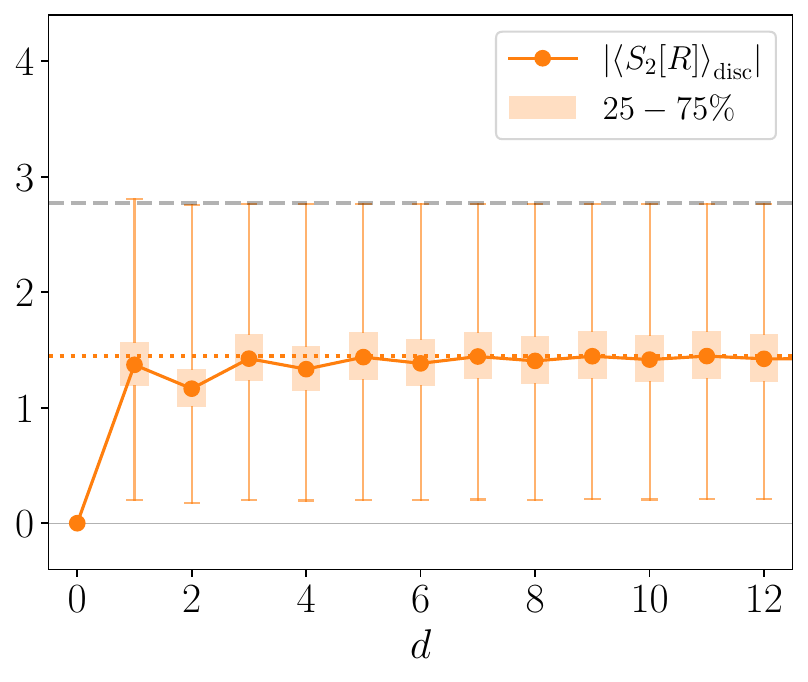}
        \caption{}
    \end{subfigure}
    \begin{subfigure}[t]{0.45\textwidth}
        \centering
        \includegraphics[width=\textwidth]{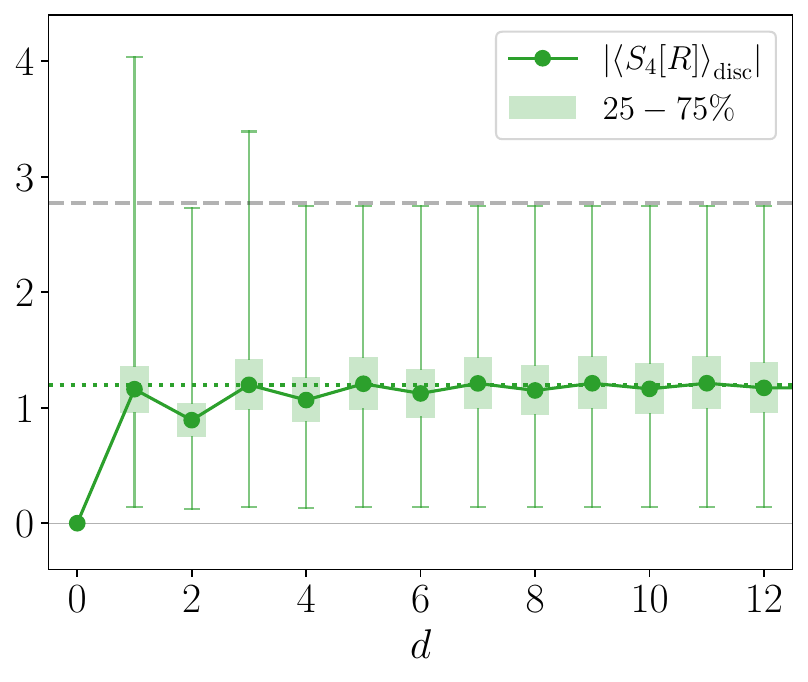}
        \caption{}
    \end{subfigure}
    \hspace{0.05\textwidth}
    \begin{subfigure}[t]{0.45\textwidth}
        \centering
        \includegraphics[width=\textwidth]{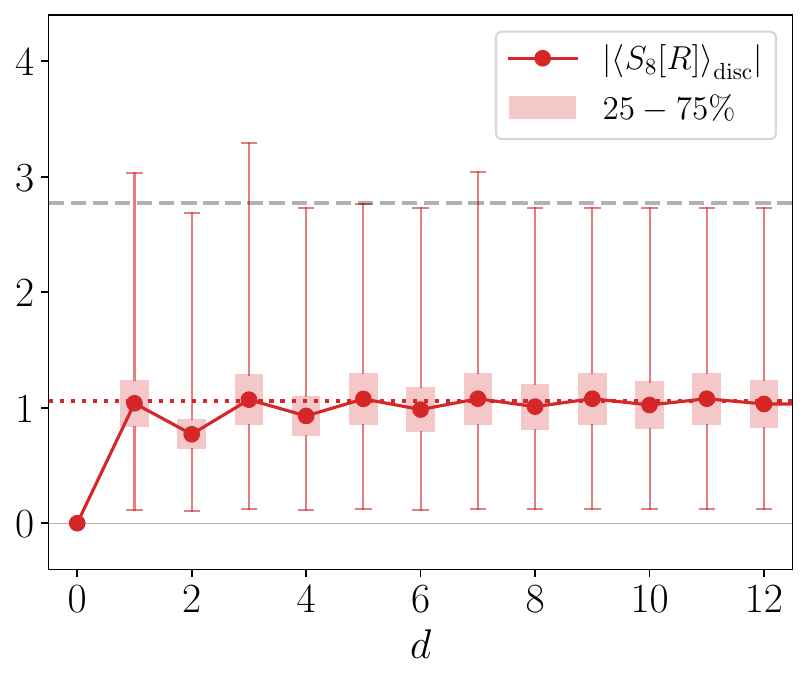}
        \caption{}
    \end{subfigure}
    \caption{
    Disconnected entropy $\langle S_n[R]\rangle_{\rm disc}$ for dRMPS in the quadratic model. We have $N=4$, $k=64$, and $L=256$.
    Each point was obtained from $10^3$ samples.
    }
    \label{fig:disc_entropy_dRMPS_quad}
\end{figure}

\begin{figure}[t]
    \centering
    \begin{subfigure}[t]{0.45\textwidth}
        \centering
        \includegraphics[width=\textwidth]{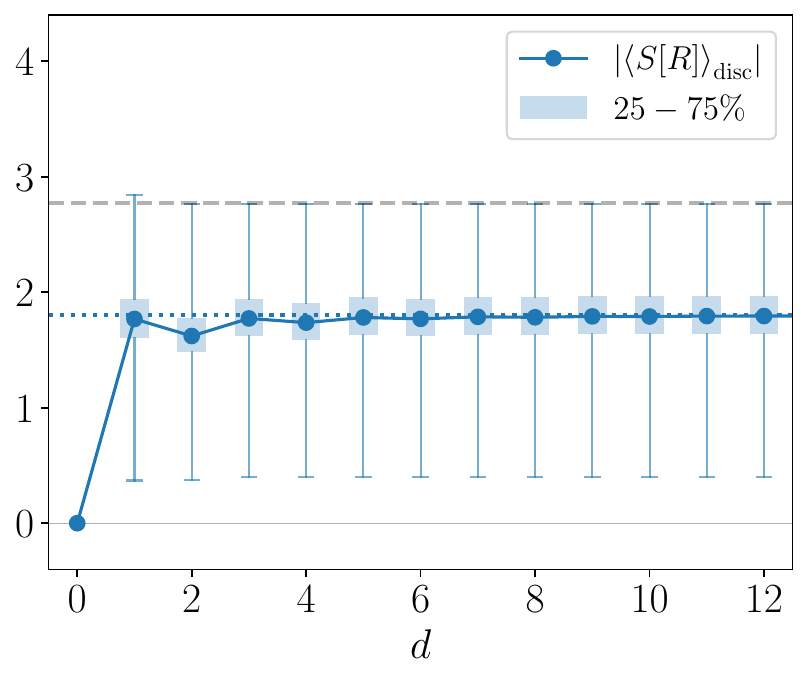}
        \caption{}
    \end{subfigure}
    \hspace{0.05\textwidth}
    \begin{subfigure}[t]{0.45\textwidth}
        \centering
        \includegraphics[width=\textwidth]{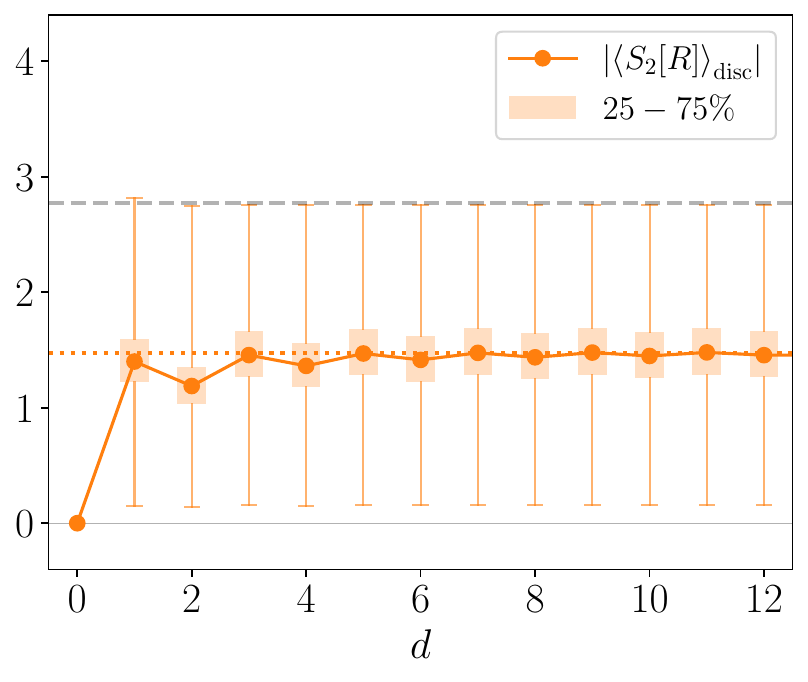}
        \caption{}
    \end{subfigure}
    \begin{subfigure}[t]{0.45\textwidth}
        \centering
        \includegraphics[width=\textwidth]{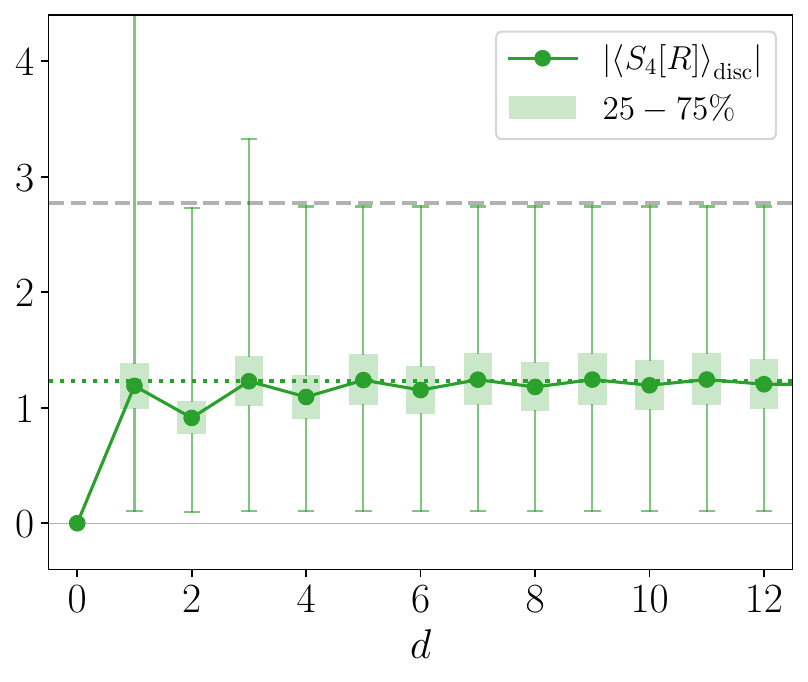}
        \caption{}
    \end{subfigure}
    \hspace{0.05\textwidth}
    \begin{subfigure}[t]{0.45\textwidth}
        \centering
        \includegraphics[width=\textwidth]{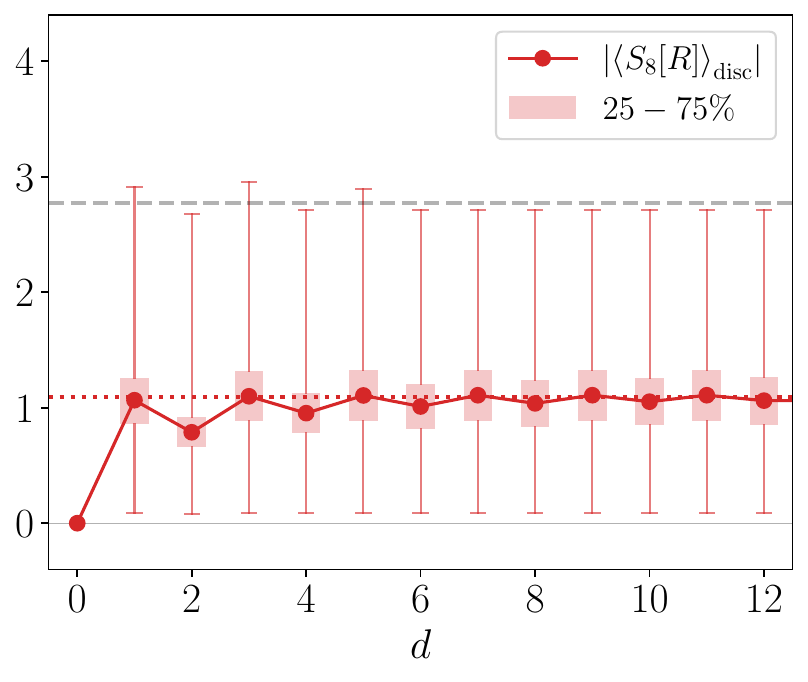}
        \caption{}
    \end{subfigure}
    \caption{
    Disconnected entropy $\langle S_n[R]\rangle_{\rm disc}$ for dRMPS in the quartic model. We have $N=4$, $k=64$, and $L=256$.
    Each point was obtained from $10^3$ samples.
    }
    \label{fig:disc_entropy_dRMPS_quar}
\end{figure}

\paragraph{Bra-ket disconnected entropy.} In figures~\ref{fig:disc_entropy_dRMPS_quad},~\ref{fig:disc_entropy_dRMPS_quar}, we present the behavior of the averaged entanglement and Renyi bra-ket disconnected entropies of an interval in the quadratic and quartic models. The dashed lines indicate $2\ln N$, and the dotted lines indicate $S_{n;\text{disc;max}}$. The vertical lines indicate the maximum and minimum values from the simulations, and the boxes represent the 25–75\% quartiles for each $d$. We noticed that when the length $d$ of $R$ was odd, we had $|S_{\ket{\psi} \bra{\psi^\prime};n}[R]| > 2\ln N$ for some samples. Thus, from the numerics, although we see an average entropy bound, we generally do not see a sample entropy bound for both models.

\begin{figure}[t]
    \centering
    \begin{subfigure}[t]{0.45\textwidth}
        \centering
        \includegraphics[width=\textwidth]{figures/dRMPS/dRMPS_disc_N_S_quad_.pdf}
        \caption{}
    \end{subfigure}
    \hspace{0.05\textwidth}
    \begin{subfigure}[t]{0.45\textwidth}
        \centering
        \includegraphics[width=\textwidth]{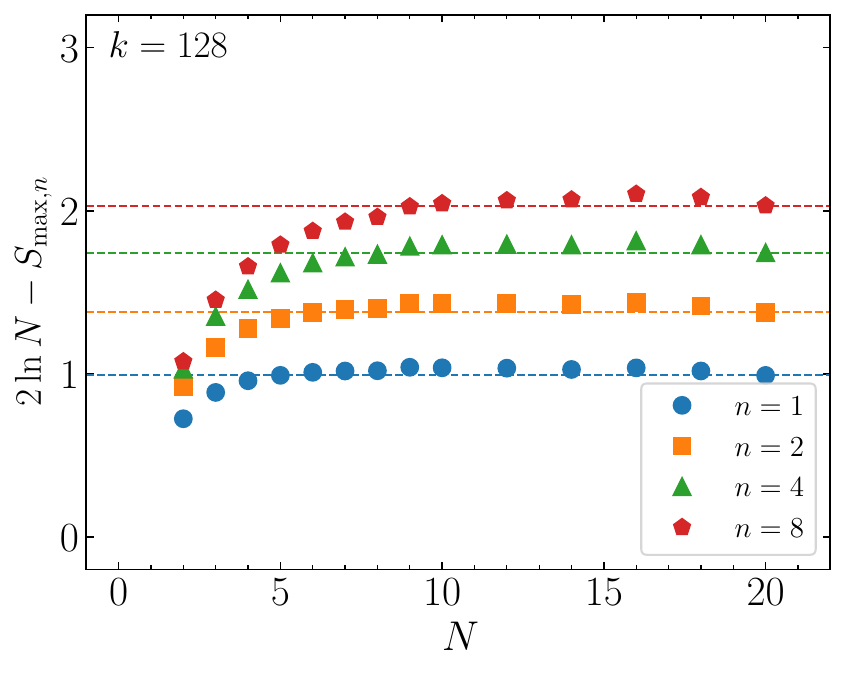}
        \caption{}
    \end{subfigure}
    \caption{
        Convergence of $2 \ln N - S_{{\rm max;disc},n}$ at large $N$ in the discrete (a) quadratic (b) quartic models. We used $k=128$ and $10^3$ samples for each point.
    }
    \label{fig:disc_N_entropy_d}
\end{figure}

In figure~\ref{fig:disc_N_entropy_d} we show that the disconnected average entropy bound is equal to $2\ln N$ plus an $\mathcal{O}(N^0)$ subleading term for dRMPS in the (a) quadratic (b) quartic models.

\FloatBarrier

\subsection{Supplementary figures for cRMPS}

\label{ssec:supplementary_figures_crmps}

\begin{figure}[t]
    \centering
    \begin{subfigure}[t]{0.45\textwidth}
        \centering
        \includegraphics[width=\textwidth]{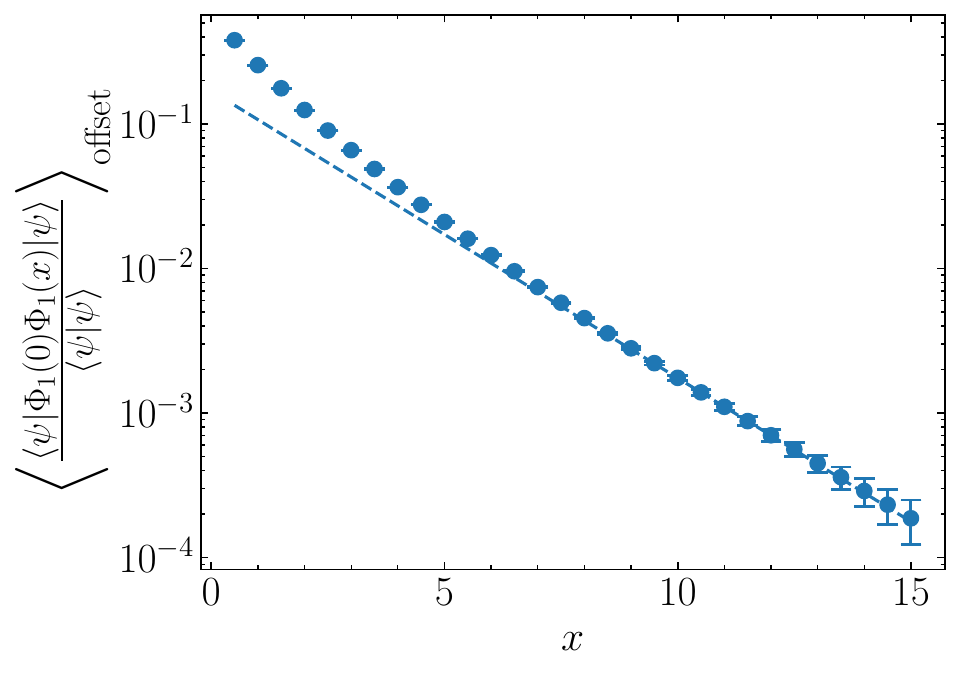}
        \caption{}
    \end{subfigure}
    \hspace{0.05\textwidth}
    \begin{subfigure}[t]{0.45\textwidth}
        \centering
        \includegraphics[width=\textwidth]{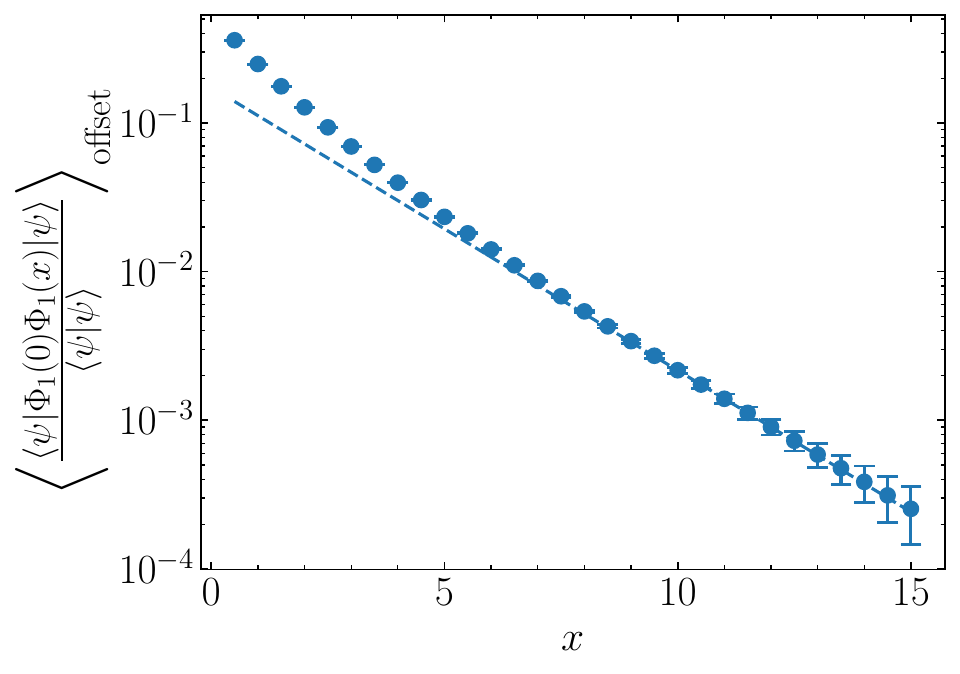}
        \caption{}
    \end{subfigure}
    \begin{subfigure}[t]{0.45\textwidth}
        \centering
        \includegraphics[width=\textwidth]{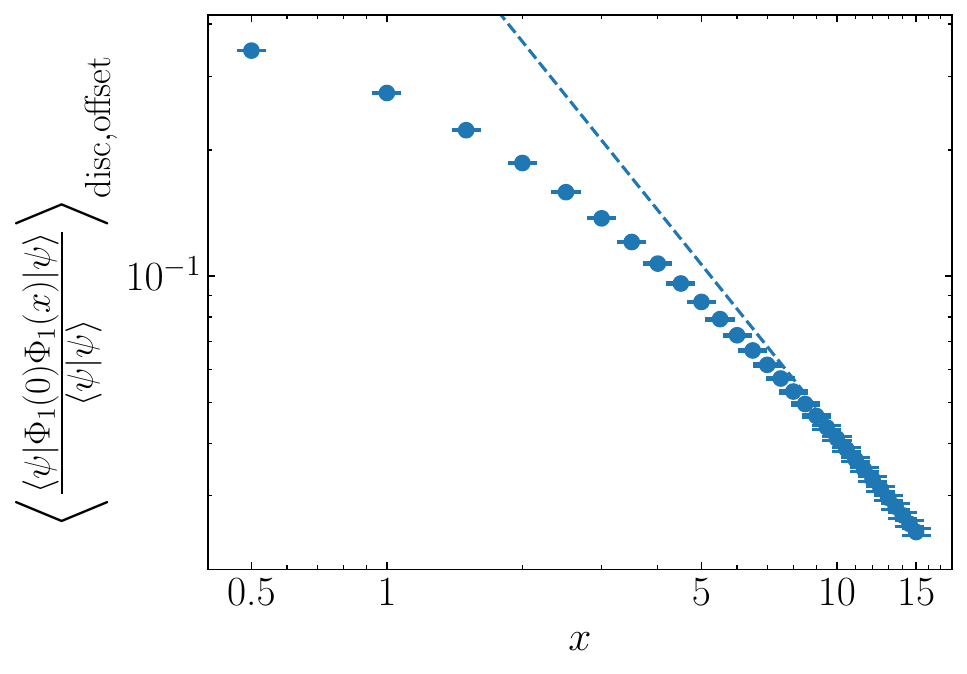}
        \caption{}
    \end{subfigure}
    \hspace{0.05\textwidth}
    \begin{subfigure}[t]{0.45\textwidth}
        \centering
        \includegraphics[width=\textwidth]{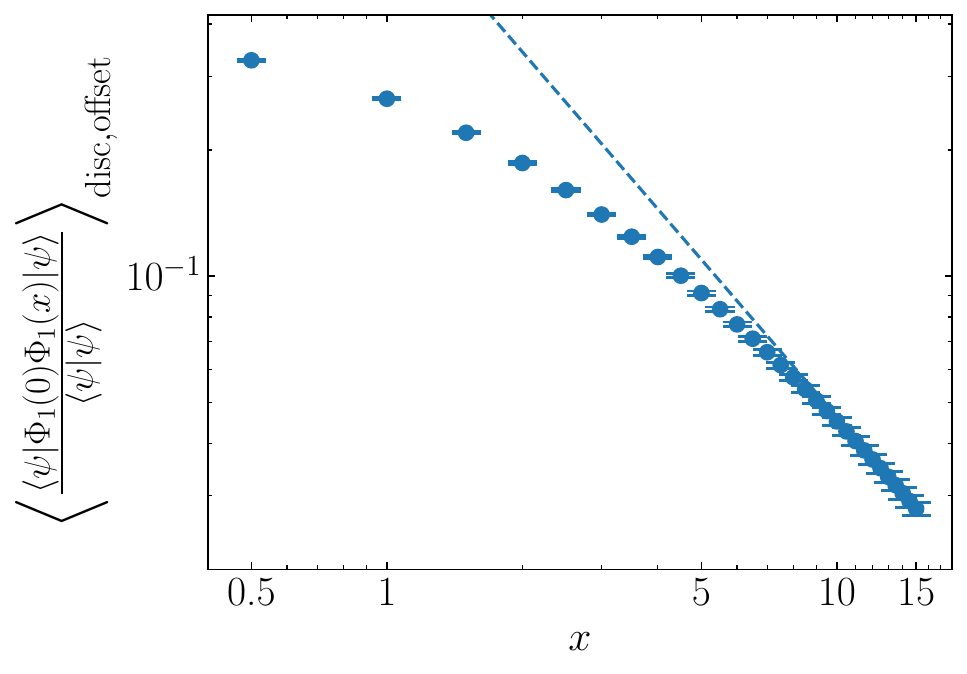}
        \caption{}
    \end{subfigure}
	\caption{Two-point correlations of a local operator in the continuous quadratic and quartic models with the $x \rightarrow \infty$ value subtracted. Here $k=8$, $N=10$, and $L=400$. We have full correlators showing exponential decay in $x$ for the continuous (a) quadratic (b) quartic models. Here we used $10^5$ samples. We have bra-ket disconnected correlators showing power law decay in $x$ for the continuous (c) quadratic (d) quartic models. Here we used $10^4$ samples.}
    \label{fig:ctwomodelCorrProperties}
\end{figure}

\paragraph{Bra-ket wormhole phase transition.} In figure~\ref{fig:ctwomodelCorrProperties} (a), (b), we present the exponential decay of the correlator in the quadratic and quartic models after subtracting the offset at $x \rightarrow \infty$. In (c), (d), we present the power law decay of the bra-ket disconnected correlator in the quadratic and quartic models after subtracting the offset at $x \rightarrow \infty$.

\begin{figure}[t]
    \centering
    \begin{subfigure}[t]{0.45\textwidth}
        \centering
        \includegraphics[width=\textwidth]{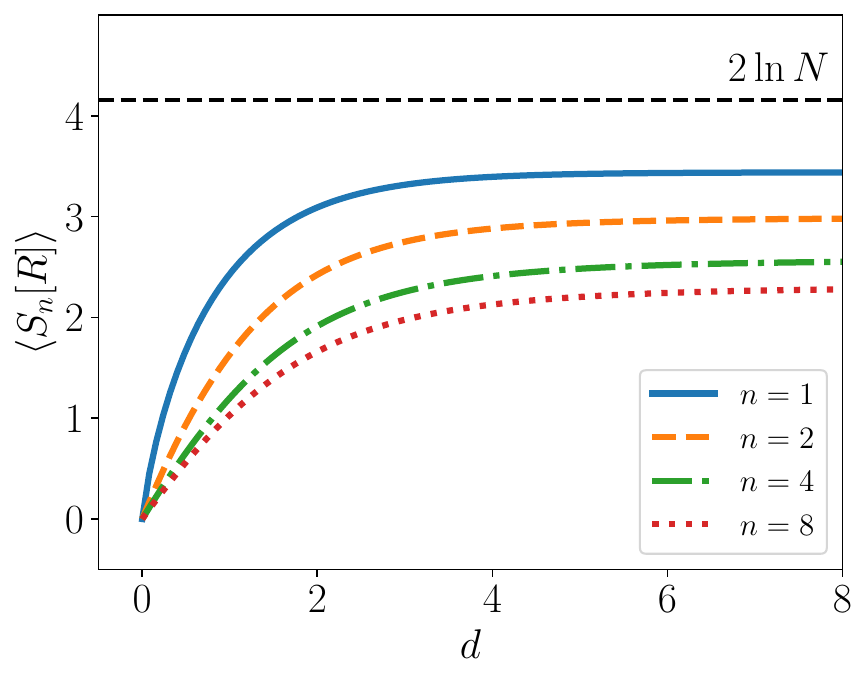}
        \caption{}
    \end{subfigure}
        \hspace{0.05\textwidth}
    \begin{subfigure}[t]{0.45\textwidth}
        \centering
        \includegraphics[width=\textwidth]{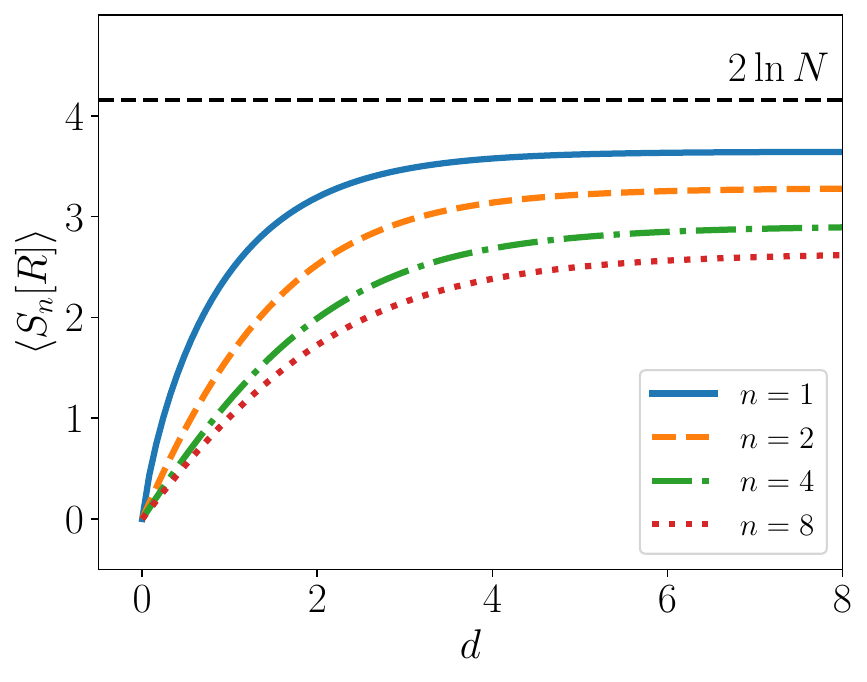}
        \caption{}
    \end{subfigure}
    \begin{subfigure}[t]{0.45\textwidth}
        \centering
        \includegraphics[width=\textwidth]{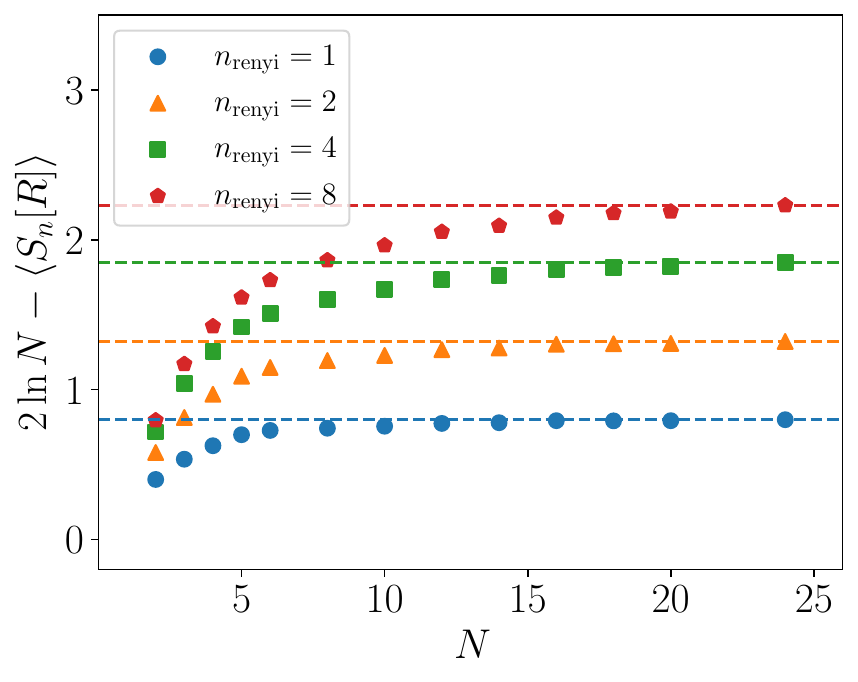}
        \caption{}
    \end{subfigure}
        \hspace{0.05\textwidth}
    \begin{subfigure}[t]{0.45\textwidth}
        \centering
        \includegraphics[width=\textwidth]{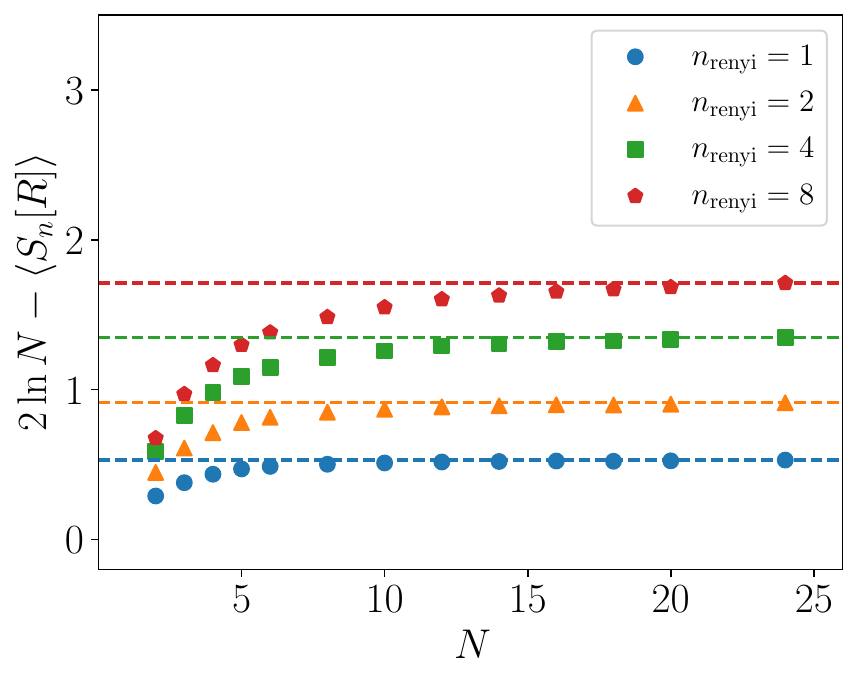}
        \caption{}
    \end{subfigure}
	\caption{
    The bounding behavior of the averaged entropies of an interval in the continous quadratic, quartic models.
    $S_n[R]$ showing bounding behavior in the (a) quadratic (b) quartic model. Here $N=8$ with $10^4$ samples. Convergence of $2\ln N - S_{\text{max}, n}$ at large $N$ in the (c) quadratic (d) quartic model.  We used $k=128$ and $10^3$ samples.
    }
    \label{fig:crmps_entropy}
\end{figure}

\paragraph{Entropy bounding.} In ~\ref{fig:crmps_entropy} (a), (b), we present the bounding behavior of the averaged entanglement and Renyi entropies ($n=1,2,4,8$) of an interval in the continuous quadratic and quartic models. In (c), (d) we show that the average entropy bound is equal to $2\ln N$ plus an $\mathcal{O}(N^0)$ subleading term in the continuous quadratic and quartic models.

\begin{figure}[t]
    \centering
    \begin{subfigure}[t]{0.45\textwidth}
        \centering
        \includegraphics[width=\textwidth]{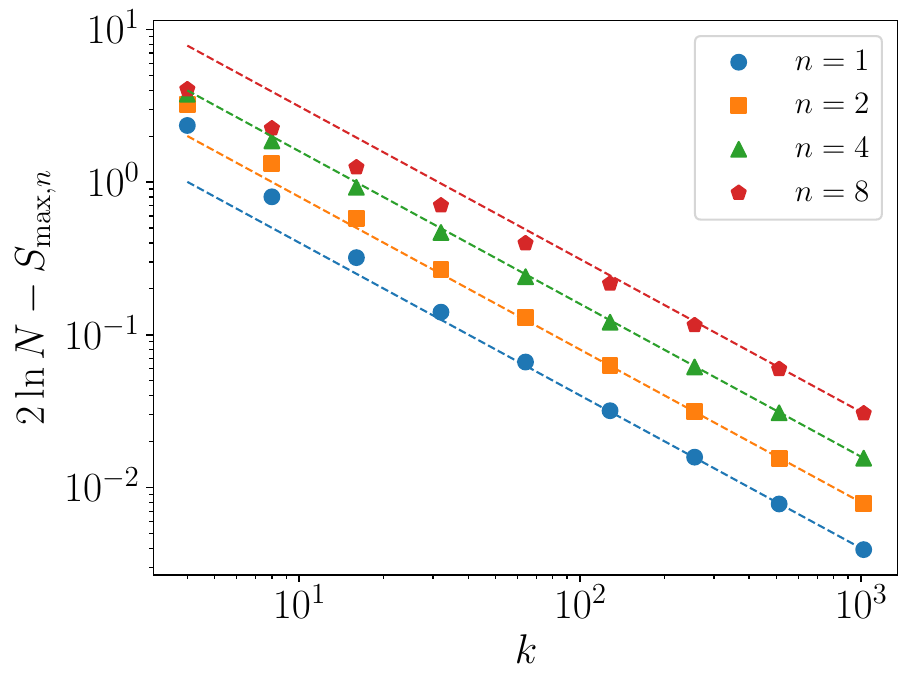}
        \caption{}
    \end{subfigure}
    \hspace{0.05\textwidth}
    \begin{subfigure}[t]{0.45\textwidth}
        \centering
        \includegraphics[width=\textwidth]{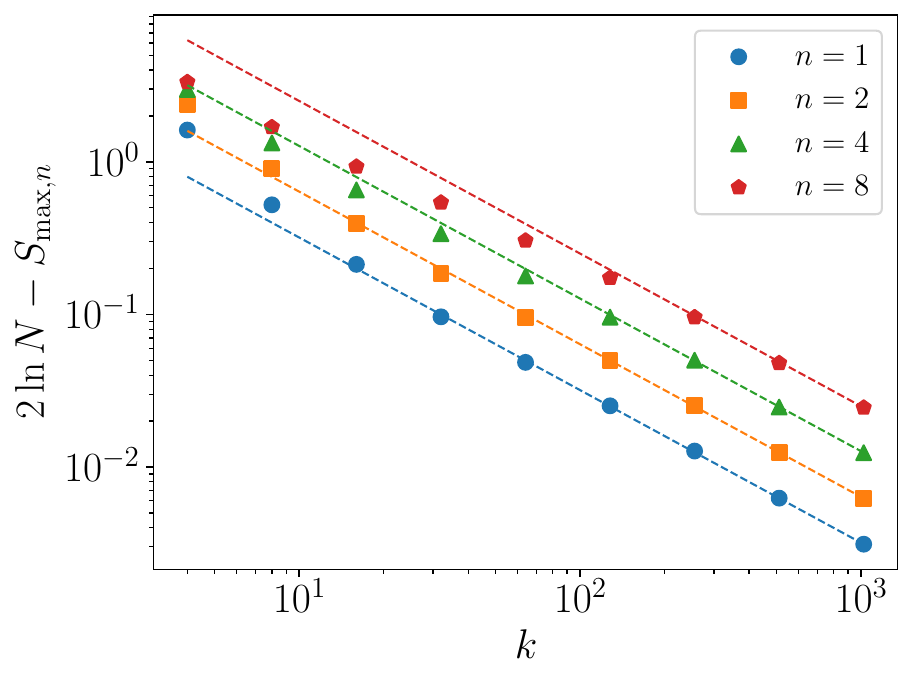}
        \caption{}
    \end{subfigure}
	\caption{$k^{-1}$-dependence of $2\ln N - S_{\text{max}, n}$ in the continous (a) quadratic (b) quartic model. Here $N=24$. We used $10^3$ samples.
    }
    \label{fig:crmps_entropy_kdep}
\end{figure}

In figure~\ref{fig:crmps_entropy_kdep} (a), (b), we present the $k^{-\frac{1}{2}}$ dependence of $2\ln N - S_{\text{max}, n}$ in the continuous quadratic and quartic models.

\begin{figure}[t]
    \centering
    \begin{subfigure}[t]{0.45\textwidth}
        \centering
        \includegraphics[width=\textwidth]{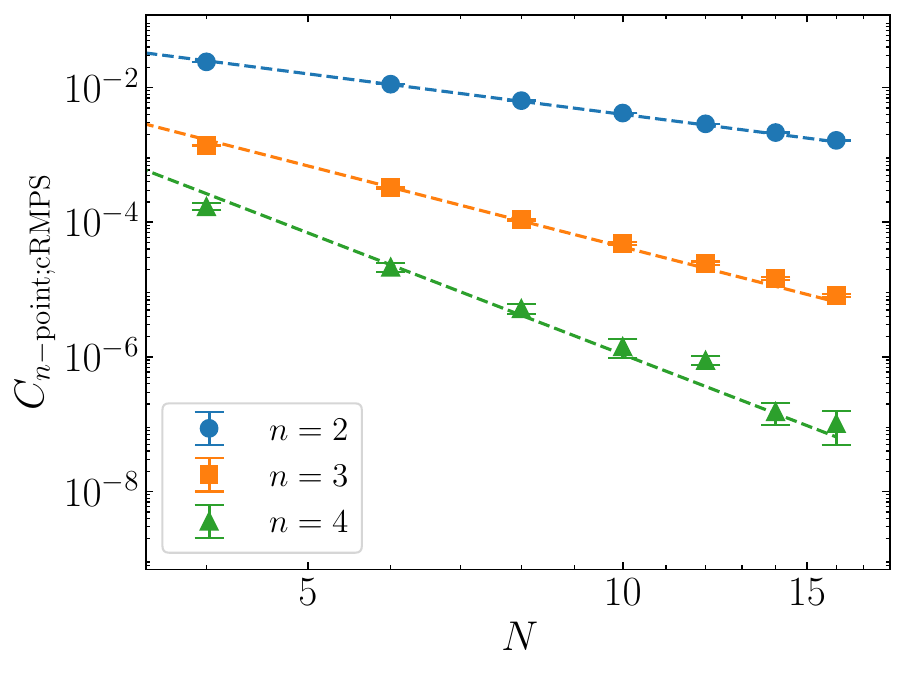}
        \caption{}
    \end{subfigure}
    \hspace{0.05\textwidth}
    \begin{subfigure}[t]{0.45\textwidth}
        \centering
        \includegraphics[width=\textwidth]{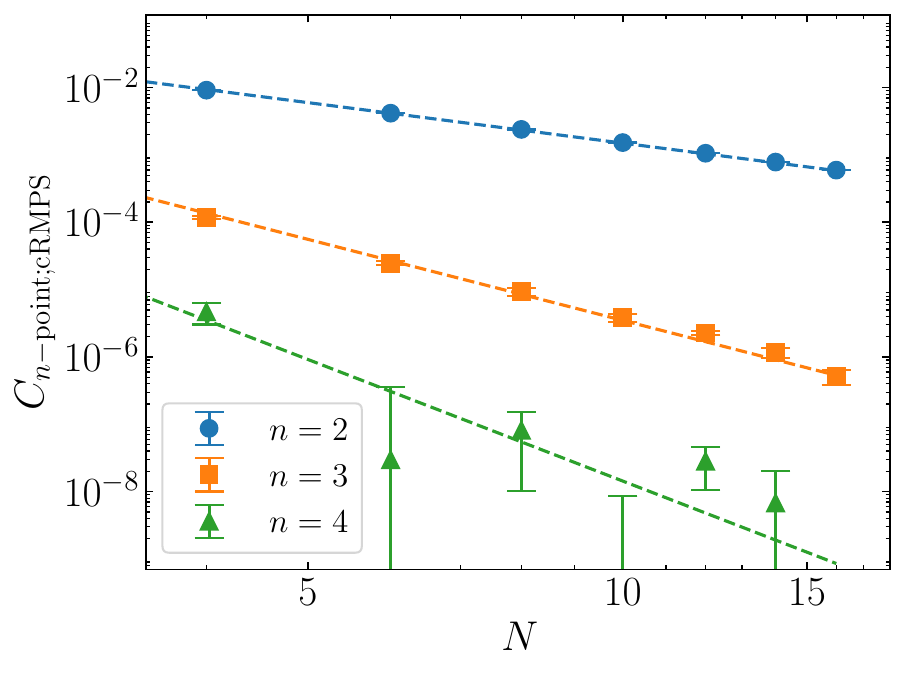}
        \caption{}
    \end{subfigure}
    \begin{subfigure}[t]{0.45\textwidth}
        \centering
        \includegraphics[width=\textwidth]{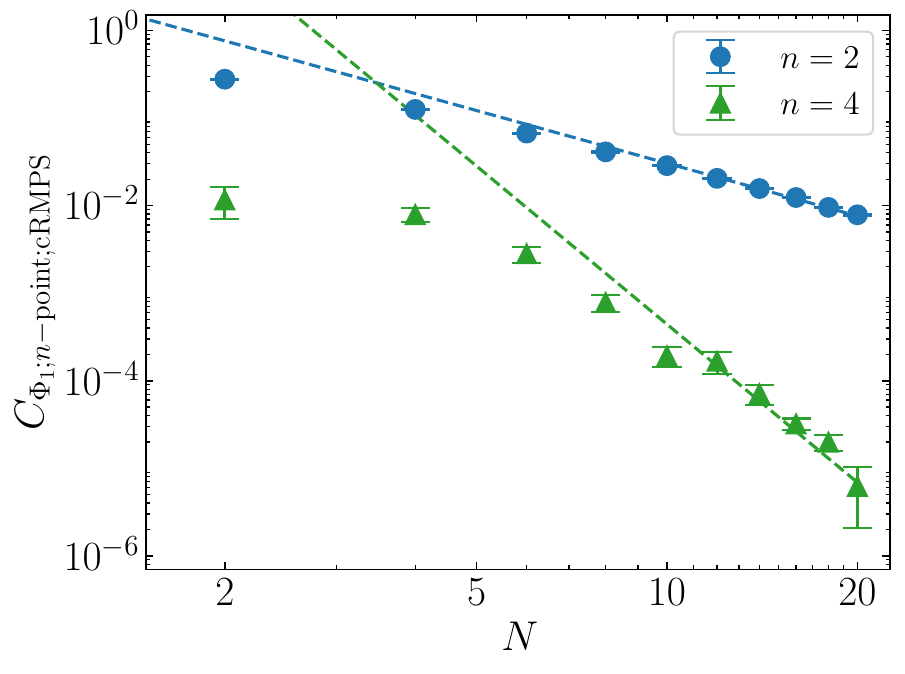}
        \caption{}
    \end{subfigure}
    \hspace{0.05\textwidth}
    \begin{subfigure}[t]{0.45\textwidth}
        \centering
        \includegraphics[width=\textwidth]{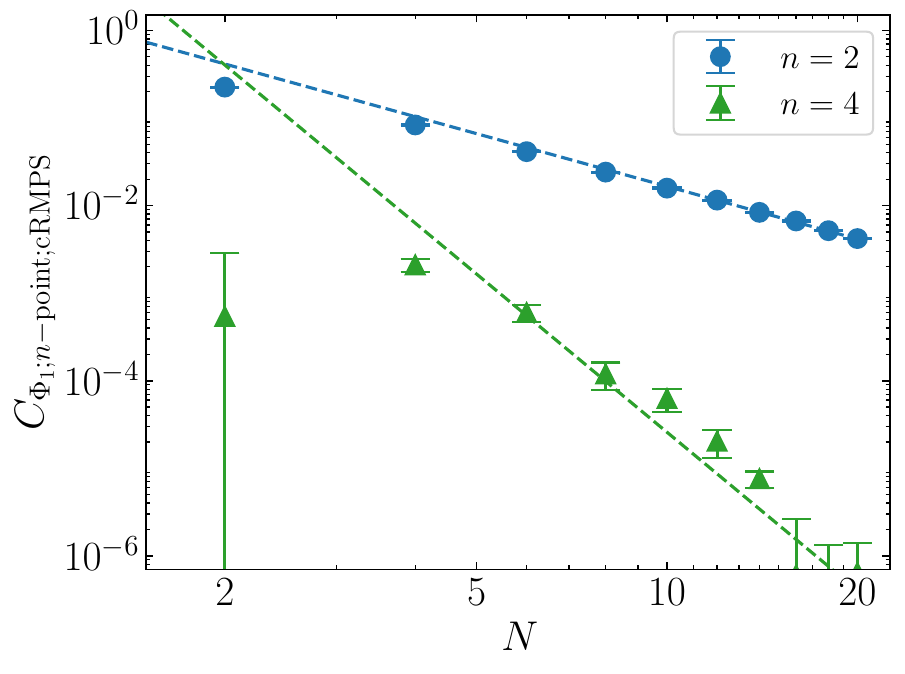}
        \caption{}
    \end{subfigure}
	\caption{$N$ dependence of $\lambda_g$ cumulants for continuous (a) quadratic, (b) quartic models. $N$ dependence of long-distance correlators $(\langle -\lambda_g|\Phi_i|-\lambda_g\rangle^n)_{\rm conn}$ for continuous (c) quadratic, (d) quartic models. Trendlines represent the expected $N$ dependence of cumulants ($N^{2(1-n)}$). We used $k=8$ with $10^5$ samples for (a) and (b), and $10^4$ samples for (c) and (d).}
    \label{fig:crmps_corr}
\end{figure}

\paragraph{Off-shell wormholes.} In figure~\ref{fig:crmps_corr} (a), (b), we present the $N$ dependence of $n$th $\lambda_g$ cumulants in the quadratic and quartic models. In (c), (d), we present the $N$ dependence of $n$th $\lambda_g^{-1}\bra{\lambda_g}M_O\ket{\lambda_g}$ cumulants in the quadratic and quartic models. For large $N$, we have $N^{2(1-n)}$ dependence for both.

\begin{figure}[t]
    \centering
    \begin{subfigure}[t]{0.45\textwidth}
        \centering
        \includegraphics[width=\textwidth]{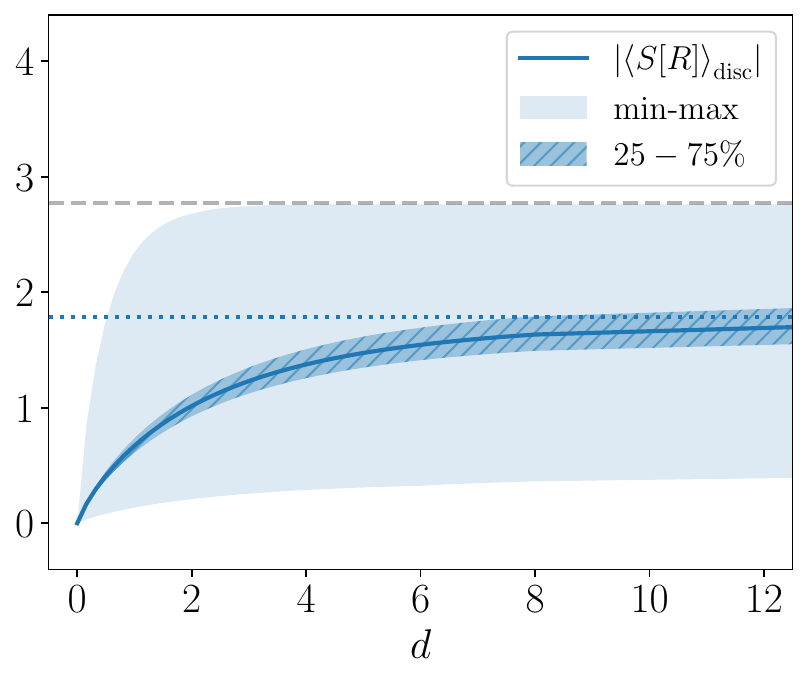}
        \caption{}
    \end{subfigure}
    \hspace{0.05\textwidth}
    \begin{subfigure}[t]{0.45\textwidth}
        \centering
        \includegraphics[width=\textwidth]{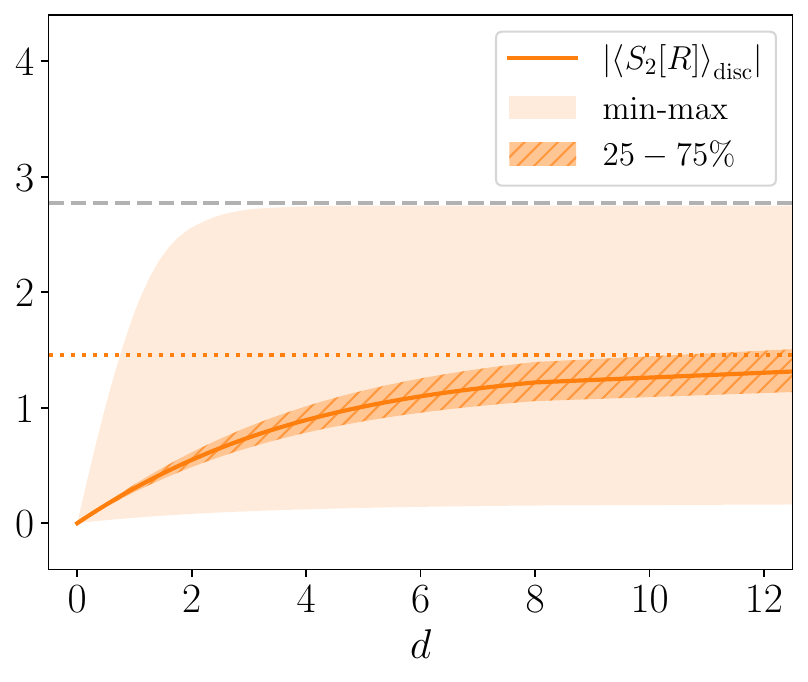}
        \caption{}
    \end{subfigure}
    \begin{subfigure}[t]{0.45\textwidth}
        \centering
        \includegraphics[width=\textwidth]{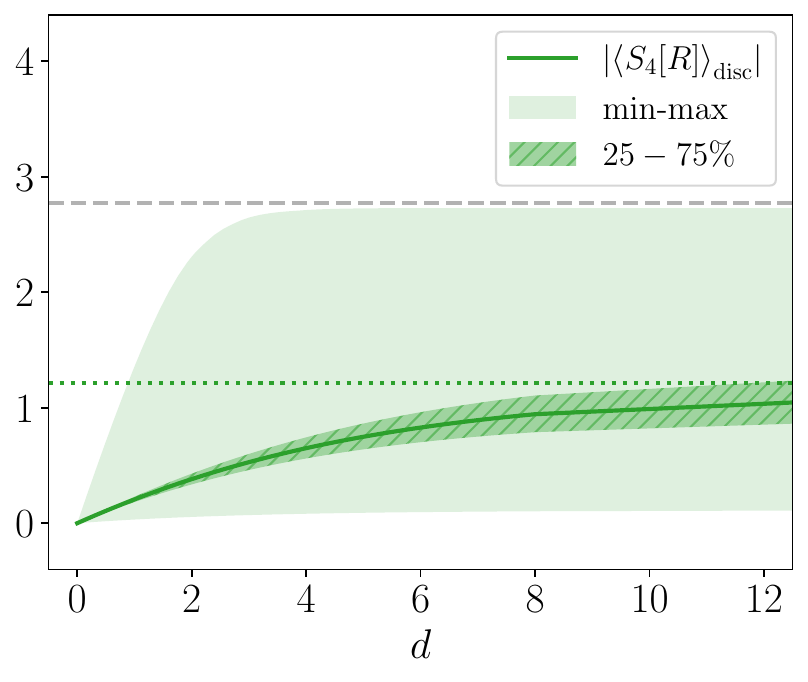}
        \caption{}
    \end{subfigure}
    \hspace{0.05\textwidth}
    \begin{subfigure}[t]{0.45\textwidth}
        \centering
        \includegraphics[width=\textwidth]{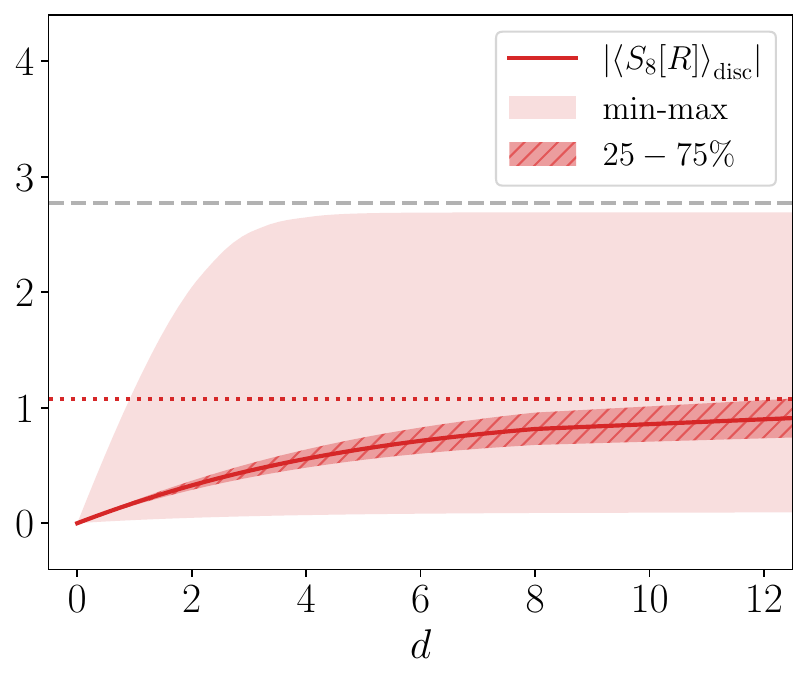}
        \caption{}
    \end{subfigure}
    \caption{
   Disconnected entropy $\langle S_n[R]\rangle_{\rm disc}$ for cRMPS in the quadratic model. We have $N=4$, $k=64$, and $L=256$. Each point was obtained from $10^3$ samples.
    }
    \label{fig:disc_entropy_cRMPS_quad}
\end{figure}

\begin{figure}[t]
    \centering
    \begin{subfigure}[t]{0.45\textwidth}
        \centering
        \includegraphics[width=\textwidth]{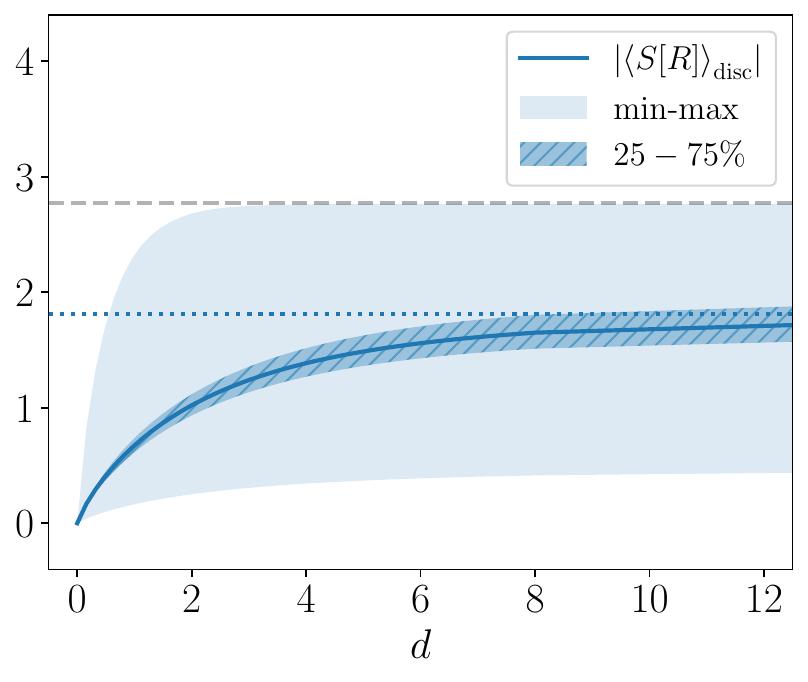}
        \caption{}
    \end{subfigure}
    \hspace{0.05\textwidth}
    \begin{subfigure}[t]{0.45\textwidth}
        \centering
        \includegraphics[width=\textwidth]{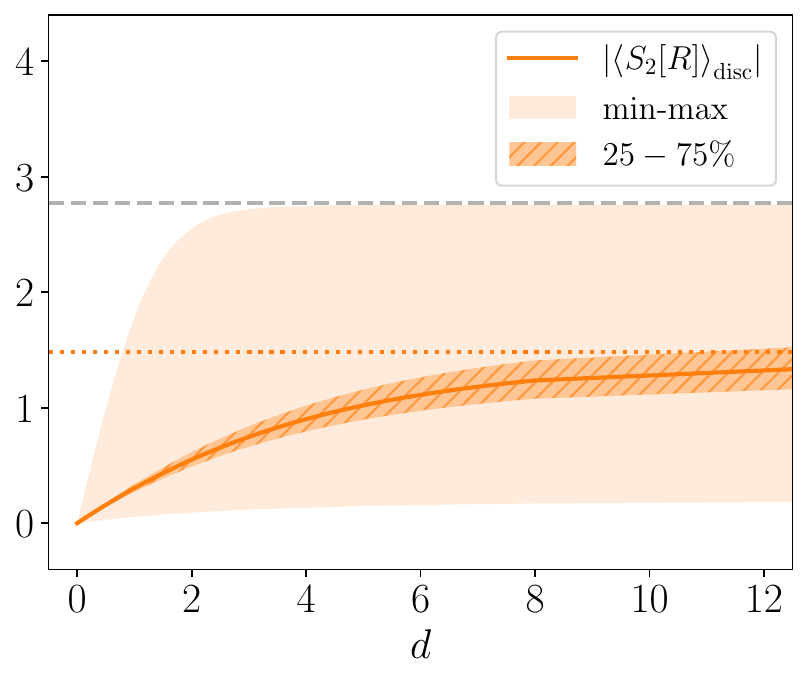}
        \caption{}
    \end{subfigure}
    \begin{subfigure}[t]{0.45\textwidth}
        \centering
        \includegraphics[width=\textwidth]{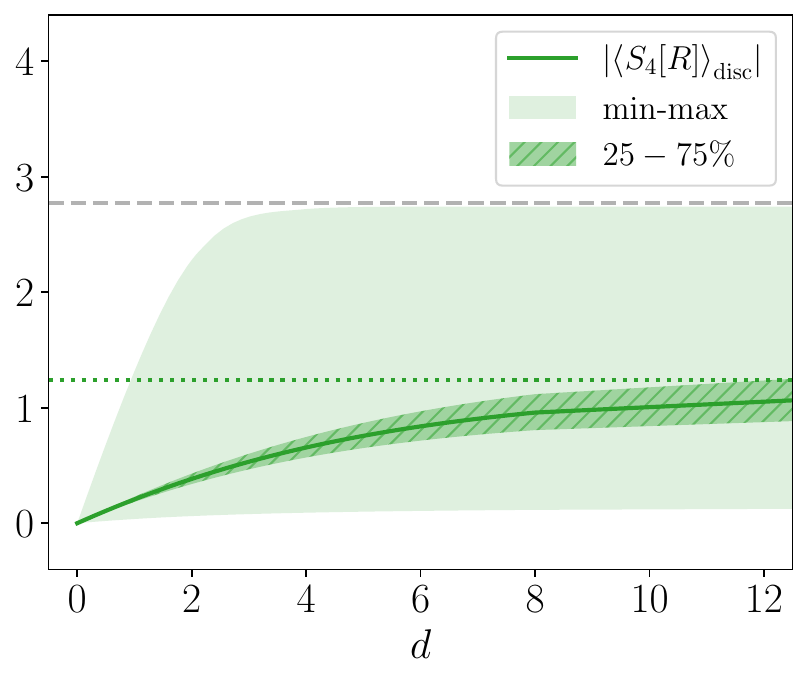}
        \caption{}
    \end{subfigure}
    \hspace{0.05\textwidth}
    \begin{subfigure}[t]{0.45\textwidth}
        \centering
        \includegraphics[width=\textwidth]{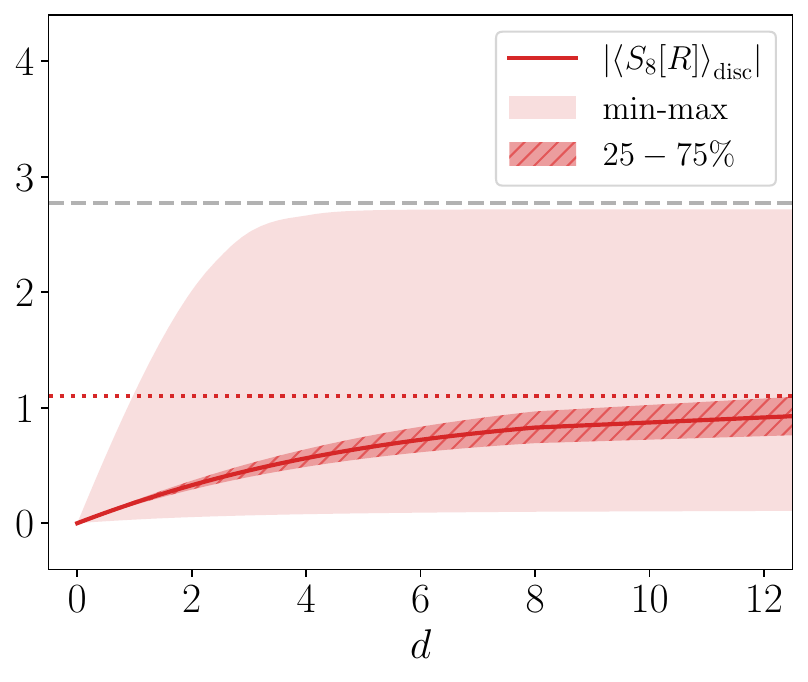}
        \caption{}
    \end{subfigure}
    \caption{
    Disconnected entropy $\langle S_n[R]\rangle_{\rm disc}$ for cRMPS in the quartic model. We have $N=4$, $k=64$, and $L=256$. Each point was obtained from $10^3$ samples.
    }
    \label{fig:disc_entropy_cRMPS_quar}
\end{figure}

\paragraph{Bra-ket disconnected entropy.} In figures~\ref{fig:disc_entropy_cRMPS_quad},~\ref{fig:disc_entropy_cRMPS_quar}, we present the behavior of the averaged entanglement and Renyi bra-ket disconnected entropies of an interval in the continuous quadratic and quartic models. The dashed lines indicate $2\ln N$, and the dotted lines indicate $S_{n;\text{disc;max}}$. The solid-colored regions represent the maximum and minimum values from the simulations, and the hatched regions correspond to the 25--75\% quartiles. From the numerics, we see both an average entropy bound and a sample entropy bound for both models.

\begin{figure}[t]
    \centering
    \begin{subfigure}[t]{0.45\textwidth}
        \centering
        \includegraphics[width=\textwidth]{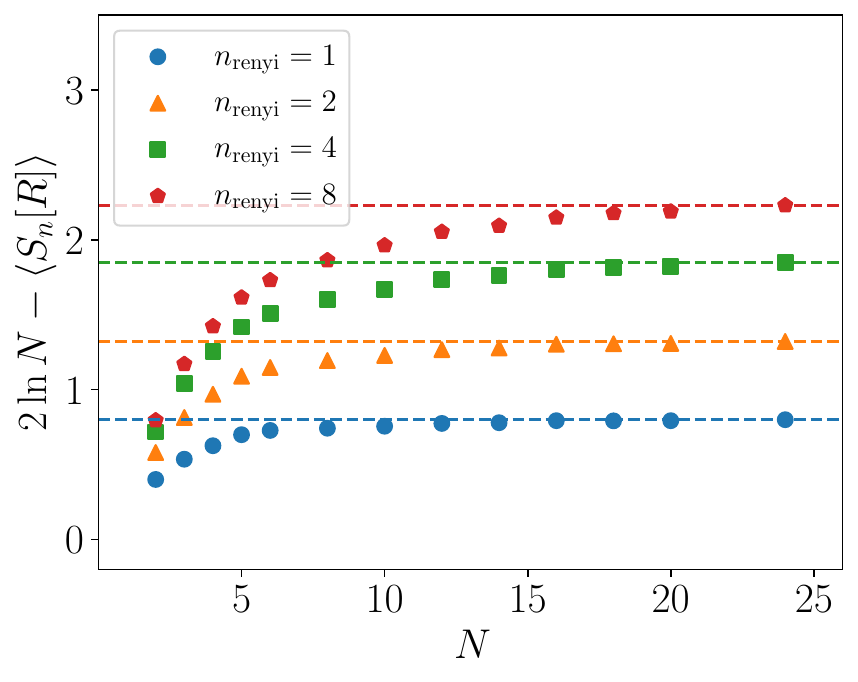}
        \caption{}
    \end{subfigure}
    \hspace{0.05\textwidth}
    \begin{subfigure}[t]{0.45\textwidth}
        \centering
        \includegraphics[width=\textwidth]{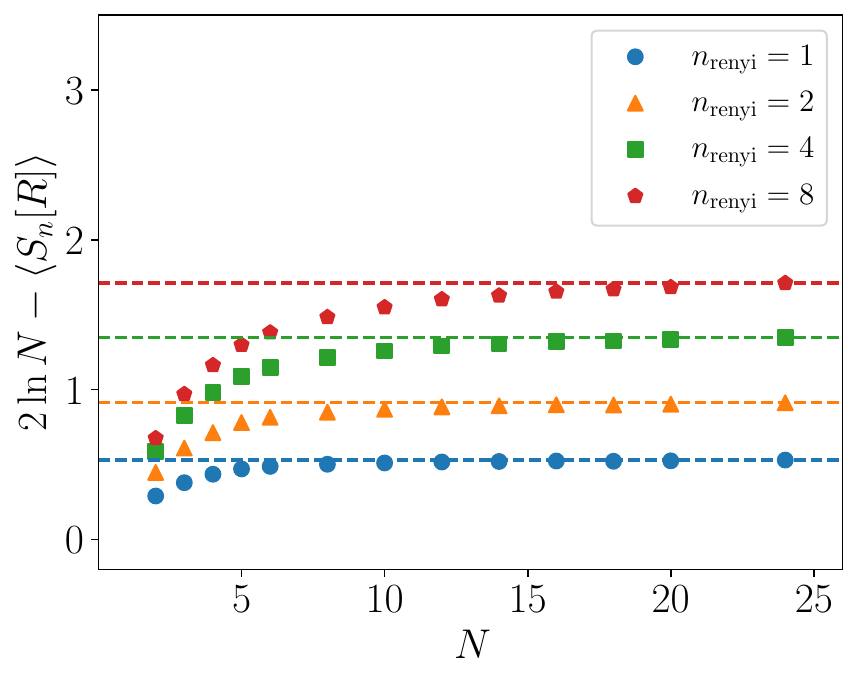}
        \caption{}
    \end{subfigure}
    \caption{
        Convergence of $2 \ln N - S_{{\rm max;disc},n}$ at large $N$ in the continuous (a) quadratic (b) quartic models. We used $k=128$ and $10^3$ samples for each point.
    }
    \label{fig:disc_N_entropy_c}
\end{figure}

In figure~\ref{fig:disc_N_entropy_c} we show that the disconnected average entropy bound is equal to $2\ln N$ plus an $\mathcal{O}(N^0)$ subleading term for cRMPS in the (a) quadratic (b) quartic models.

\clearpage
\paragraph{Data Availability Statement.} This article has no associated data or the data will not be deposited.

\paragraph{Code Availability Statement.} This article has no associated code or the code will not be deposited.

\bibliographystyle{RMPS}
\bibliography{biblio}

\end{document}

%% file: figures/tikz/diskbkwh.tex
\tikzset{every picture/.style={line width=0.75pt}} %

\begin{tikzpicture}[x=0.75pt,y=0.75pt,yscale=-1,xscale=1]

\draw  [draw opacity=0][fill={rgb, 255:red, 221; green, 237; blue, 238 }  ,fill opacity=1 ] (306.35,136.36) .. controls (306.35,112.83) and (345.28,93.76) .. (393.31,93.76) .. controls (441.33,93.76) and (480.26,112.83) .. (480.26,136.36) .. controls (480.26,159.89) and (441.33,178.97) .. (393.31,178.97) .. controls (345.28,178.97) and (306.35,159.89) .. (306.35,136.36) -- cycle ;
\draw  [draw opacity=0][fill={rgb, 255:red, 221; green, 237; blue, 238 }  ,fill opacity=1 ] (301.19,131.63) .. controls (301.19,110.71) and (342.43,93.76) .. (393.31,93.76) .. controls (444.18,93.76) and (485.42,110.71) .. (485.42,131.63) .. controls (485.42,152.54) and (444.18,169.49) .. (393.31,169.49) .. controls (342.43,169.49) and (301.19,152.54) .. (301.19,131.63) -- cycle ;
\draw  [draw opacity=0][fill={rgb, 255:red, 221; green, 237; blue, 238 }  ,fill opacity=1 ] (300.9,122.92) .. controls (300.9,98.71) and (342.14,79.08) .. (393.02,79.08) .. controls (443.89,79.08) and (485.13,98.71) .. (485.13,122.92) .. controls (485.13,147.13) and (443.89,166.75) .. (393.02,166.75) .. controls (342.14,166.75) and (300.9,147.13) .. (300.9,122.92) -- cycle ;
\draw  [color={rgb, 255:red, 0; green, 0; blue, 0 }  ,draw opacity=1 ][fill={rgb, 255:red, 255; green, 255; blue, 255 }  ,fill opacity=1 ] (343.06,130.75) .. controls (343.06,119.99) and (365.51,111.28) .. (393.2,111.28) .. controls (420.89,111.28) and (443.34,119.99) .. (443.34,130.75) .. controls (443.34,141.5) and (420.89,150.22) .. (393.2,150.22) .. controls (365.51,150.22) and (343.06,141.5) .. (343.06,130.75) -- cycle ;
\draw  [dash pattern={on 0.84pt off 2.51pt}] (301.19,127.62) .. controls (301.19,117.9) and (310.69,110.01) .. (322.41,110.01) .. controls (334.14,110.01) and (343.64,117.9) .. (343.64,127.62) .. controls (343.64,137.35) and (334.14,145.24) .. (322.41,145.24) .. controls (310.69,145.24) and (301.19,137.35) .. (301.19,127.62) -- cycle ;
\draw  [draw opacity=0] (301.19,127.62) .. controls (301.19,156.35) and (342.56,179.64) .. (393.6,179.64) .. controls (443.22,179.64) and (483.71,157.62) .. (485.91,129.99) -- (393.6,127.62) -- cycle ; \draw   (301.19,127.62) .. controls (301.19,156.35) and (342.56,179.64) .. (393.6,179.64) .. controls (443.22,179.64) and (483.71,157.62) .. (485.91,129.99) ;
\draw  [draw opacity=0] (300.9,125.92) .. controls (300.9,100.05) and (342.27,79.08) .. (393.31,79.08) .. controls (444.34,79.08) and (485.71,100.05) .. (485.71,125.92) .. controls (485.71,126.65) and (485.68,127.38) .. (485.61,128.11) -- (393.31,125.92) -- cycle ; \draw   (300.9,125.92) .. controls (300.9,100.05) and (342.27,79.08) .. (393.31,79.08) .. controls (444.34,79.08) and (485.71,100.05) .. (485.71,125.92) .. controls (485.71,126.65) and (485.68,127.38) .. (485.61,128.11) ;
\draw  [dash pattern={on 0.84pt off 2.51pt}] (443.18,127.93) .. controls (443.18,118.2) and (452.68,110.32) .. (464.4,110.32) .. controls (476.12,110.32) and (485.63,118.2) .. (485.63,127.93) .. controls (485.63,137.66) and (476.12,145.55) .. (464.4,145.55) .. controls (452.68,145.55) and (443.18,137.66) .. (443.18,127.93) -- cycle ;
\draw  [dash pattern={on 4.5pt off 4.5pt}]  (299.79,127.09) .. controls (306.9,108.51) and (331.33,96.56) .. (360.72,91.7) .. controls (370.89,90.02) and (381.62,89.18) .. (392.37,89.18) .. controls (392.81,89.18) and (393.25,89.18) .. (393.69,89.19) .. controls (433.07,89.44) and (472.2,101.07) .. (485.13,123.59) .. controls (485.83,124.81) and (486.46,126.06) .. (487.01,127.35)(302.59,128.16) .. controls (309.42,110.31) and (333.21,99.29) .. (361.21,94.66) .. controls (371.22,93.01) and (381.78,92.18) .. (392.37,92.18) .. controls (392.81,92.18) and (393.24,92.18) .. (393.67,92.19) .. controls (431.7,92.43) and (469.97,103.2) .. (482.53,125.08) .. controls (483.17,126.2) and (483.75,127.35) .. (484.25,128.52) ;
\draw [fill={rgb, 255:red, 221; green, 237; blue, 238 }  ,fill opacity=1 ]   (112,120.92) .. controls (159.02,223.32) and (194.26,224.06) .. (237.09,120.92) ;
\draw [fill={rgb, 255:red, 221; green, 237; blue, 238 }  ,fill opacity=1 ]   (237.09,120.92) .. controls (191.76,16.83) and (154.83,17.77) .. (112,120.92) ;
\draw    (302.69,127.61) .. controls (302.79,136.28) and (308.99,143.41) .. (318.71,149.07) .. controls (336.08,159.18) and (364.78,164.6) .. (393.46,164.9) .. controls (394.28,164.91) and (395.1,164.91) .. (395.92,164.91) .. controls (432.79,164.91) and (469.42,156.53) .. (480.77,138.03) .. controls (482.68,134.92) and (483.84,131.52) .. (484.13,127.82)(299.69,127.64) .. controls (299.8,137.22) and (306.32,145.33) .. (317.2,151.66) .. controls (334.92,161.98) and (364.16,167.6) .. (393.43,167.9) .. controls (394.26,167.91) and (395.09,167.91) .. (395.92,167.91) .. controls (434.19,167.91) and (471.63,158.66) .. (483.32,139.6) .. controls (485.48,136.08) and (486.8,132.23) .. (487.12,128.05) ;
\draw  [dash pattern={on 4.5pt off 4.5pt}]  (110.56,120.51) .. controls (112.36,114.11) and (121.07,108.95) .. (133.75,105.63) .. controls (145.27,102.62) and (159.9,101.02) .. (174.54,100.92) .. controls (175.02,100.92) and (175.5,100.91) .. (175.98,100.91) .. controls (203.85,100.91) and (230.93,106.6) .. (237.33,117.63) .. controls (237.87,118.56) and (238.28,119.53) .. (238.54,120.54)(113.44,121.32) .. controls (115.07,115.55) and (123.43,111.44) .. (134.51,108.53) .. controls (145.82,105.57) and (160.19,104.02) .. (174.56,103.92) .. controls (175.03,103.92) and (175.5,103.91) .. (175.98,103.91) .. controls (202.25,103.91) and (228.61,108.57) .. (234.74,119.14) .. controls (235.14,119.83) and (235.45,120.55) .. (235.64,121.3) ;
\draw    (113.47,120.64) .. controls (114.78,127.51) and (121.9,132.45) .. (131.68,135.89) .. controls (142.89,139.83) and (157.76,141.73) .. (172.76,141.73) .. controls (173.35,141.73) and (173.94,141.73) .. (174.53,141.72) .. controls (200.9,141.46) and (227.64,135.52) .. (234.37,123.86) .. controls (234.97,122.83) and (235.39,121.75) .. (235.62,120.62)(110.53,121.2) .. controls (111.98,128.87) and (119.52,134.8) .. (130.69,138.72) .. controls (142.17,142.76) and (157.39,144.73) .. (172.76,144.73) .. controls (173.36,144.73) and (173.96,144.73) .. (174.56,144.72) .. controls (202.43,144.44) and (229.95,137.52) .. (236.97,125.36) .. controls (237.73,124.04) and (238.27,122.66) .. (238.56,121.22) ;

\end{tikzpicture}

%% file: figures/tikz/island.tex
\tikzset{every picture/.style={line width=0.75pt}} %

\begin{tikzpicture}[x=0.75pt,y=0.75pt,yscale=-1,xscale=1]

\draw  [draw opacity=0][fill={rgb, 255:red, 221; green, 237; blue, 238 }  ,fill opacity=1 ] (470.17,210.88) -- (492.22,184.75) -- (502.57,193.48) -- (480.51,219.61) -- cycle ;
\draw  [draw opacity=0][fill={rgb, 255:red, 221; green, 237; blue, 238 }  ,fill opacity=1 ] (121.8,175.83) .. controls (121.8,121.92) and (206.14,78.21) .. (310.19,78.21) .. controls (414.23,78.21) and (498.57,121.92) .. (498.57,175.83) .. controls (498.57,229.74) and (414.23,273.45) .. (310.19,273.45) .. controls (206.14,273.45) and (121.8,229.74) .. (121.8,175.83) -- cycle ;
\draw  [draw opacity=0][fill={rgb, 255:red, 221; green, 237; blue, 238 }  ,fill opacity=1 ] (110.63,164.98) .. controls (110.63,117.06) and (199.97,78.21) .. (310.19,78.21) .. controls (420.4,78.21) and (509.74,117.06) .. (509.74,164.98) .. controls (509.74,212.9) and (420.4,251.74) .. (310.19,251.74) .. controls (199.97,251.74) and (110.63,212.9) .. (110.63,164.98) -- cycle ;
\draw  [draw opacity=0][fill={rgb, 255:red, 221; green, 237; blue, 238 }  ,fill opacity=1 ] (110,145.02) .. controls (110,89.55) and (199.35,44.57) .. (309.56,44.57) .. controls (419.77,44.57) and (509.12,89.55) .. (509.12,145.02) .. controls (509.12,200.49) and (419.77,245.47) .. (309.56,245.47) .. controls (199.35,245.47) and (110,200.49) .. (110,145.02) -- cycle ;
\draw  [color={rgb, 255:red, 0; green, 0; blue, 0 }  ,draw opacity=1 ][fill={rgb, 255:red, 255; green, 255; blue, 255 }  ,fill opacity=1 ] (166.02,156.61) .. controls (166.02,118.31) and (230.85,87.26) .. (310.81,87.26) .. controls (390.78,87.26) and (455.61,118.31) .. (455.61,156.61) .. controls (455.61,194.91) and (390.78,225.96) .. (310.81,225.96) .. controls (230.85,225.96) and (166.02,194.91) .. (166.02,156.61) -- cycle ;
\draw  [dash pattern={on 0.84pt off 2.51pt}] (110.63,155.8) .. controls (110.63,141.12) and (123.02,129.22) .. (138.31,129.22) .. controls (153.6,129.22) and (166,141.12) .. (166,155.8) .. controls (166,170.48) and (153.6,182.38) .. (138.31,182.38) .. controls (123.02,182.38) and (110.63,170.48) .. (110.63,155.8) -- cycle ;
\draw  [draw opacity=0] (110.63,155.8) .. controls (110.63,155.8) and (110.63,155.8) .. (110.63,155.8) .. controls (110.63,221.63) and (200.25,275) .. (310.81,275) .. controls (418.49,275) and (506.3,224.38) .. (510.82,160.93) -- (310.81,155.8) -- cycle ; \draw   (110.63,155.8) .. controls (110.63,155.8) and (110.63,155.8) .. (110.63,155.8) .. controls (110.63,221.63) and (200.25,275) .. (310.81,275) .. controls (418.49,275) and (506.3,224.38) .. (510.82,160.93) ;
\draw  [draw opacity=0] (110,151.89) .. controls (110,151.89) and (110,151.89) .. (110,151.89) .. controls (110,92.62) and (199.63,44.57) .. (310.19,44.57) .. controls (420.75,44.57) and (510.37,92.62) .. (510.37,151.89) .. controls (510.37,153.48) and (510.31,155.06) .. (510.18,156.64) -- (310.19,151.89) -- cycle ; \draw   (110,151.89) .. controls (110,151.89) and (110,151.89) .. (110,151.89) .. controls (110,92.62) and (199.63,44.57) .. (310.19,44.57) .. controls (420.75,44.57) and (510.37,92.62) .. (510.37,151.89) .. controls (510.37,153.48) and (510.31,155.06) .. (510.18,156.64) ;
\draw  [dash pattern={on 4.5pt off 4.5pt}]  (109.14,155.65) .. controls (112.43,123.39) and (137.1,99.06) .. (173.01,82.73) .. controls (209.94,65.93) and (258.74,57.6) .. (307.75,57.6) .. controls (308.65,57.6) and (309.54,57.6) .. (310.44,57.6) .. controls (393,58.12) and (475.32,82.37) .. (502.57,129.09) .. controls (507.46,137.46) and (510.59,146.54) .. (511.68,156.35)(112.12,155.95) .. controls (115.32,124.64) and (139.51,101.26) .. (174.25,85.46) .. controls (210.85,68.82) and (259.19,60.6) .. (307.75,60.6) .. controls (308.64,60.6) and (309.53,60.6) .. (310.42,60.6) .. controls (391.64,61.11) and (473.09,84.51) .. (499.98,130.6) .. controls (504.65,138.61) and (507.66,147.3) .. (508.7,156.68) ;
\draw    (112.06,155.35) .. controls (128.42,207) and (180.67,239.33) .. (242.06,253.05) .. controls (263.74,257.9) and (286.58,260.42) .. (309.41,260.65) .. controls (310.53,260.66) and (311.64,260.67) .. (312.76,260.67) .. controls (390.5,260.67) and (467.73,234.08) .. (498.36,181.32) .. controls (502.9,173.5) and (506.41,165.1) .. (508.74,156.14)(109.2,156.26) .. controls (125.85,208.81) and (178.82,241.99) .. (241.41,255.98) .. controls (263.29,260.87) and (286.33,263.42) .. (309.38,263.65) .. controls (310.51,263.66) and (311.64,263.67) .. (312.76,263.67) .. controls (391.74,263.67) and (469.9,236.32) .. (500.95,182.83) .. controls (505.63,174.77) and (509.25,166.13) .. (511.64,156.89) ;
\draw  [draw opacity=0][fill={rgb, 255:red, 221; green, 237; blue, 238 }  ,fill opacity=1 ] (244,237) -- (368,237) -- (368,264.5) -- (244,264.5) -- cycle ;
\draw  [draw opacity=0][fill={rgb, 255:red, 221; green, 237; blue, 238 }  ,fill opacity=1 ] (211.16,225.74) -- (250.77,234.76) -- (244,264.5) -- (204.39,255.48) -- cycle ;
\draw [color={rgb, 255:red, 255; green, 0; blue, 31 }  ,draw opacity=1 ]   (207.47,242.58) .. controls (237.94,252.73) and (270.55,257.87) .. (303.17,258.84) .. controls (307.06,258.96) and (310.95,259.02) .. (314.84,259.02) .. controls (332.69,259.02) and (350.46,257.8) .. (367.8,255.51)(206.53,245.42) .. controls (237.26,255.67) and (270.17,260.85) .. (303.07,261.84) .. controls (307,261.96) and (310.92,262.02) .. (314.84,262.02) .. controls (332.82,262.02) and (350.73,260.79) .. (368.2,258.49) ;
\draw [shift={(368,257)}, rotate = 172.48] [color={rgb, 255:red, 255; green, 0; blue, 31 }  ,draw opacity=1 ][line width=0.75]    (0,5.59) -- (0,-5.59)   ;
\draw [shift={(207,244)}, rotate = 198.43] [color={rgb, 255:red, 255; green, 0; blue, 31 }  ,draw opacity=1 ][line width=0.75]    (0,5.59) -- (0,-5.59)   ;
\draw [color={rgb, 255:red, 0; green, 42; blue, 255 }  ,draw opacity=1 ][fill={rgb, 255:red, 255; green, 255; blue, 255 }  ,fill opacity=1 ]   (220,211) .. controls (259,225) and (315,233) .. (360,222) ;
\draw [shift={(360,222)}, rotate = 166.26] [color={rgb, 255:red, 0; green, 42; blue, 255 }  ,draw opacity=1 ][line width=0.75]    (0,5.59) -- (0,-5.59)   ;
\draw [shift={(220,211)}, rotate = 199.75] [color={rgb, 255:red, 0; green, 42; blue, 255 }  ,draw opacity=1 ][line width=0.75]    (0,5.59) -- (0,-5.59)   ;
\draw  [dash pattern={on 0.84pt off 2.51pt}] (455.61,156.61) .. controls (455.61,141.48) and (468,129.22) .. (483.29,129.22) .. controls (498.58,129.22) and (510.98,141.48) .. (510.98,156.61) .. controls (510.98,171.74) and (498.58,184) .. (483.29,184) .. controls (468,184) and (455.61,171.74) .. (455.61,156.61) -- cycle ;
\draw [color={rgb, 255:red, 0; green, 42; blue, 255 }  ,draw opacity=1 ][fill={rgb, 255:red, 255; green, 255; blue, 255 }  ,fill opacity=1 ]   (287,88) .. controls (321,85) and (367,89.5) .. (400,102) ;
\draw [shift={(400,102)}, rotate = 200.75] [color={rgb, 255:red, 0; green, 42; blue, 255 }  ,draw opacity=1 ][line width=0.75]    (0,5.59) -- (0,-5.59)   ;
\draw [shift={(287,88)}, rotate = 174.96] [color={rgb, 255:red, 0; green, 42; blue, 255 }  ,draw opacity=1 ][line width=0.75]    (0,5.59) -- (0,-5.59)   ;
\draw  [draw opacity=0][fill={rgb, 255:red, 221; green, 237; blue, 238 }  ,fill opacity=1 ] (281,53) -- (377,53) -- (377,71) -- (281,71) -- cycle ;
\draw  [draw opacity=0][fill={rgb, 255:red, 221; green, 237; blue, 238 }  ,fill opacity=1 ] (316,62) -- (412,62) -- (412,80) -- (316,80) -- cycle ;
\draw [color={rgb, 255:red, 255; green, 0; blue, 31 }  ,draw opacity=1 ] [dash pattern={on 4.5pt off 4.5pt}]  (280.93,58.5) .. controls (291.93,57.97) and (302.04,57.69) .. (311.57,57.69) .. controls (328.64,57.69) and (343.88,58.6) .. (359.12,60.62) .. controls (375.75,62.84) and (392.38,66.39) .. (411.39,71.55)(281.07,61.5) .. controls (292.02,60.97) and (302.08,60.69) .. (311.57,60.69) .. controls (328.5,60.69) and (343.61,61.59) .. (358.73,63.6) .. controls (375.23,65.8) and (391.74,69.32) .. (410.61,74.45) ;
\draw [shift={(411,73)}, rotate = 195.2] [color={rgb, 255:red, 255; green, 0; blue, 31 }  ,draw opacity=1 ][line width=0.75]    (0,5.59) -- (0,-5.59)   ;
\draw [shift={(281,60)}, rotate = 177.25] [color={rgb, 255:red, 255; green, 0; blue, 31 }  ,draw opacity=1 ][line width=0.75]    (0,5.59) -- (0,-5.59)   ;

\draw (289,201.4) node [anchor=north west][inner sep=0.75pt]    {$I_{1}$};
\draw (327,90.4) node [anchor=north west][inner sep=0.75pt]    {$I_{2}$};
\draw (283,235.4) node [anchor=north west][inner sep=0.75pt]    {$R_{1}$};
\draw (331,65.4) node [anchor=north west][inner sep=0.75pt]    {$R_{2}$};

\end{tikzpicture}

%% file: figures/tikz/nrenyiwh.tex
\tikzset{every picture/.style={line width=0.75pt}} %

\begin{tikzpicture}[x=0.75pt,y=0.75pt,yscale=-1,xscale=1]

\draw  [draw opacity=0][fill={rgb, 255:red, 221; green, 237; blue, 238 }  ,fill opacity=1 ] (422.3,186.97) -- (427.85,182.45) -- (442.11,199.32) -- (436.56,203.84) -- cycle ;
\draw  [draw opacity=0][fill={rgb, 255:red, 221; green, 237; blue, 238 }  ,fill opacity=1 ] (229,145) .. controls (249,135) and (261,134) .. (241,154) .. controls (221,174) and (242,193) .. (246,196) .. controls (250,199) and (241,228) .. (221,198) .. controls (201,168) and (209,155) .. (229,145) -- cycle ;
\draw  [draw opacity=0][fill={rgb, 255:red, 221; green, 237; blue, 238 }  ,fill opacity=1 ] (450.04,176.35) .. controls (451.86,203.48) and (417.59,129.48) .. (323.8,130.24) .. controls (230,131) and (195.55,200.99) .. (197.4,172.3) .. controls (199.24,143.62) and (215,132) .. (232,122) .. controls (249,112) and (305.12,102.58) .. (321,102) .. controls (336.88,101.42) and (382.08,107.84) .. (407,120) .. controls (431.92,132.16) and (448.21,149.22) .. (450.04,176.35) -- cycle ;
\draw  [draw opacity=0][fill={rgb, 255:red, 221; green, 237; blue, 238 }  ,fill opacity=1 ] (197,169.88) .. controls (195,142.77) and (231.07,215.36) .. (323.54,215.18) .. controls (416,215) and (451.32,143.6) .. (449.66,172.3) .. controls (448,201) and (432,208) .. (414,224) .. controls (396,240) and (342.62,244.95) .. (326.74,245.64) .. controls (310.87,246.32) and (268,241) .. (243,229) .. controls (218,217) and (199,197) .. (197,169.88) -- cycle ;
\draw  [draw opacity=0][fill={rgb, 255:red, 221; green, 237; blue, 238 }  ,fill opacity=1 ] (415.32,199.99) .. controls (395.29,209.93) and (395.7,207.91) .. (409.35,186.95) .. controls (423,166) and (402.46,151.95) .. (398.47,148.94) .. controls (394.48,145.93) and (403.57,116.95) .. (423.48,147.01) .. controls (443.39,177.07) and (435.35,190.05) .. (415.32,199.99) -- cycle ;
\draw  [dash pattern={on 0.84pt off 2.51pt}] (197.4,172.3) .. controls (197.4,163.23) and (205.21,155.87) .. (214.84,155.87) .. controls (224.48,155.87) and (232.29,163.23) .. (232.29,172.3) .. controls (232.29,181.38) and (224.48,188.74) .. (214.84,188.74) .. controls (205.21,188.74) and (197.4,181.38) .. (197.4,172.3) -- cycle ;
\draw  [draw opacity=0] (197.4,172.3) .. controls (197.4,172.3) and (197.4,172.3) .. (197.4,172.3) .. controls (197.4,213) and (253.87,246) .. (323.54,246) .. controls (391.35,246) and (446.66,214.74) .. (449.56,175.54) -- (323.54,172.3) -- cycle ; \draw   (197.4,172.3) .. controls (197.4,172.3) and (197.4,172.3) .. (197.4,172.3) .. controls (197.4,213) and (253.87,246) .. (323.54,246) .. controls (391.35,246) and (446.66,214.74) .. (449.56,175.54) ;
\draw  [draw opacity=0] (197,169.88) .. controls (197,169.88) and (197,169.88) .. (197,169.88) .. controls (197,133.24) and (253.47,103.53) .. (323.14,103.53) .. controls (392.8,103.53) and (449.28,133.24) .. (449.28,169.88) .. controls (449.28,170.89) and (449.24,171.88) .. (449.15,172.87) -- (323.14,169.88) -- cycle ; \draw   (197,169.88) .. controls (197,169.88) and (197,169.88) .. (197,169.88) .. controls (197,133.24) and (253.47,103.53) .. (323.14,103.53) .. controls (392.8,103.53) and (449.28,133.24) .. (449.28,169.88) .. controls (449.28,170.89) and (449.24,171.88) .. (449.15,172.87) ;
\draw  [dash pattern={on 4.5pt off 4.5pt}]  (195.9,172.15) .. controls (197.96,152.43) and (213,137.41) .. (235.11,127.22) .. controls (258.6,116.39) and (290.03,111.01) .. (321.6,111.01) .. controls (322.17,111.01) and (322.74,111.01) .. (323.3,111.02) .. controls (375.57,111.33) and (427.6,126.49) .. (444.84,155.49) .. controls (447.96,160.73) and (449.96,166.43) .. (450.66,172.57)(198.89,172.46) .. controls (200.84,153.67) and (215.43,139.6) .. (236.37,129.94) .. controls (259.52,119.27) and (290.49,114.01) .. (321.6,114.01) .. controls (322.16,114.01) and (322.73,114.01) .. (323.29,114.02) .. controls (374.21,114.33) and (425.38,128.64) .. (442.26,157.02) .. controls (445.16,161.9) and (447.03,167.19) .. (447.67,172.91) ;
\draw    (198.82,171.84) .. controls (209.08,203.61) and (241.87,223.43) .. (280.33,231.87) .. controls (293.96,234.86) and (308.31,236.42) .. (322.66,236.55) .. controls (323.36,236.56) and (324.06,236.56) .. (324.76,236.56) .. controls (373.52,236.56) and (422.02,220.27) .. (441.24,187.78) .. controls (444.07,182.99) and (446.26,177.85) .. (447.71,172.36)(195.97,172.76) .. controls (206.51,205.42) and (240.02,226.1) .. (279.69,234.8) .. controls (293.51,237.83) and (308.07,239.41) .. (322.63,239.55) .. controls (323.34,239.56) and (324.05,239.56) .. (324.76,239.56) .. controls (374.76,239.56) and (424.17,222.52) .. (443.82,189.31) .. controls (446.79,184.28) and (449.09,178.89) .. (450.61,173.12) ;
\draw  [draw opacity=0][fill={rgb, 255:red, 221; green, 237; blue, 238 }  ,fill opacity=1 ] (279,220) -- (359.57,220) -- (359.57,240.5) -- (279,240.5) -- cycle ;
\draw  [draw opacity=0][fill={rgb, 255:red, 221; green, 237; blue, 238 }  ,fill opacity=1 ] (260.74,215.54) -- (285.7,221.12) -- (281.44,239.51) -- (256.47,233.93) -- cycle ;
\draw [color={rgb, 255:red, 255; green, 0; blue, 31 }  ,draw opacity=1 ]   (258.59,225.41) .. controls (277.73,231.67) and (298.23,234.83) .. (318.73,235.44) .. controls (321.18,235.51) and (323.63,235.55) .. (326.07,235.55) .. controls (337.3,235.55) and (348.47,234.8) .. (359.38,233.38)(257.65,228.26) .. controls (277.07,234.61) and (297.86,237.83) .. (318.64,238.44) .. controls (321.12,238.51) and (323.6,238.55) .. (326.07,238.55) .. controls (337.43,238.55) and (348.73,237.79) .. (359.76,236.36) ;
\draw [shift={(359.57,234.87)}, rotate = 172.62] [color={rgb, 255:red, 255; green, 0; blue, 31 }  ,draw opacity=1 ][line width=0.75]    (0,5.59) -- (0,-5.59)   ;
\draw [shift={(258.12,226.83)}, rotate = 198.11] [color={rgb, 255:red, 255; green, 0; blue, 31 }  ,draw opacity=1 ][line width=0.75]    (0,5.59) -- (0,-5.59)   ;
\draw  [dash pattern={on 0.84pt off 2.51pt}] (414.77,172.3) .. controls (414.77,163.23) and (422.58,155.87) .. (432.22,155.87) .. controls (441.85,155.87) and (449.66,163.23) .. (449.66,172.3) .. controls (449.66,181.38) and (441.85,188.74) .. (432.22,188.74) .. controls (422.58,188.74) and (414.77,181.38) .. (414.77,172.3) -- cycle ;
\draw  [color={rgb, 255:red, 0; green, 0; blue, 0 }  ,draw opacity=1 ] (232.3,172.3) .. controls (232.3,148.62) and (273.15,129.42) .. (323.54,129.42) .. controls (373.92,129.42) and (414.77,148.62) .. (414.77,172.3) .. controls (414.77,195.98) and (373.92,215.18) .. (323.54,215.18) .. controls (273.15,215.18) and (232.3,195.98) .. (232.3,172.3) -- cycle ;

\draw  [draw opacity=0][fill={rgb, 255:red, 221; green, 237; blue, 238 }  ,fill opacity=1 ] (258.12,177.83) .. controls (254.12,160.83) and (365.14,173.24) .. (359.57,185.87) .. controls (354,198.5) and (353,211.5) .. (359.57,225.63) .. controls (366.14,239.76) and (252.74,239.54) .. (260.74,215.54) .. controls (262.21,211.14) and (263.18,207.66) .. (263.77,204.8) .. controls (266.4,192.07) and (261.39,191.72) .. (258.12,177.83) -- cycle ;
\draw    (258.12,177.83) .. controls (266.12,188.83) and (266.21,208.36) .. (257.95,217.9) ;
\draw    (359.57,185.87) .. controls (351,199) and (355,215.63) .. (359.57,225.63) ;
\draw  [draw opacity=0][fill={rgb, 255:red, 221; green, 237; blue, 238 }  ,fill opacity=1 ] (422.3,137.97) -- (427.85,133.45) -- (442.11,150.32) -- (436.56,154.84) -- cycle ;
\draw  [draw opacity=0][fill={rgb, 255:red, 221; green, 237; blue, 238 }  ,fill opacity=1 ] (229,96) .. controls (249,86) and (261,85) .. (241,105) .. controls (221,125) and (242,144) .. (246,147) .. controls (250,150) and (241,179) .. (221,149) .. controls (201,119) and (209,106) .. (229,96) -- cycle ;
\draw  [draw opacity=0][fill={rgb, 255:red, 221; green, 237; blue, 238 }  ,fill opacity=1 ] (450.04,127.35) .. controls (451.86,154.48) and (417.59,80.48) .. (323.8,81.24) .. controls (230,82) and (195.55,151.99) .. (197.4,123.3) .. controls (199.24,94.62) and (215,83) .. (232,73) .. controls (249,63) and (305.12,53.58) .. (321,53) .. controls (336.88,52.42) and (382.08,58.84) .. (407,71) .. controls (431.92,83.16) and (448.21,100.22) .. (450.04,127.35) -- cycle ;
\draw  [draw opacity=0][fill={rgb, 255:red, 221; green, 237; blue, 238 }  ,fill opacity=1 ] (197,120.88) .. controls (195,93.77) and (231.07,166.36) .. (323.54,166.18) .. controls (416,166) and (451.32,94.6) .. (449.66,123.3) .. controls (448,152) and (432,159) .. (414,175) .. controls (396,191) and (342.62,195.95) .. (326.74,196.64) .. controls (310.87,197.32) and (268,192) .. (243,180) .. controls (218,168) and (199,148) .. (197,120.88) -- cycle ;
\draw  [draw opacity=0][fill={rgb, 255:red, 221; green, 237; blue, 238 }  ,fill opacity=1 ] (415.32,150.99) .. controls (395.29,160.93) and (395.7,158.91) .. (409.35,137.95) .. controls (423,117) and (402.46,102.95) .. (398.47,99.94) .. controls (394.48,96.93) and (403.57,67.95) .. (423.48,98.01) .. controls (443.39,128.07) and (435.35,141.05) .. (415.32,150.99) -- cycle ;
\draw  [dash pattern={on 0.84pt off 2.51pt}] (197.4,123.3) .. controls (197.4,114.23) and (205.21,106.87) .. (214.84,106.87) .. controls (224.48,106.87) and (232.29,114.23) .. (232.29,123.3) .. controls (232.29,132.38) and (224.48,139.74) .. (214.84,139.74) .. controls (205.21,139.74) and (197.4,132.38) .. (197.4,123.3) -- cycle ;
\draw  [draw opacity=0] (197.4,123.3) .. controls (197.4,123.3) and (197.4,123.3) .. (197.4,123.3) .. controls (197.4,164) and (253.87,197) .. (323.54,197) .. controls (391.35,197) and (446.66,165.74) .. (449.56,126.54) -- (323.54,123.3) -- cycle ; \draw   (197.4,123.3) .. controls (197.4,123.3) and (197.4,123.3) .. (197.4,123.3) .. controls (197.4,164) and (253.87,197) .. (323.54,197) .. controls (391.35,197) and (446.66,165.74) .. (449.56,126.54) ;
\draw  [draw opacity=0] (197,120.88) .. controls (197,84.24) and (253.47,54.53) .. (323.14,54.53) .. controls (392.8,54.53) and (449.28,84.24) .. (449.28,120.88) .. controls (449.28,121.89) and (449.24,122.88) .. (449.15,123.87) -- (323.14,120.88) -- cycle ; \draw   (197,120.88) .. controls (197,84.24) and (253.47,54.53) .. (323.14,54.53) .. controls (392.8,54.53) and (449.28,84.24) .. (449.28,120.88) .. controls (449.28,121.89) and (449.24,122.88) .. (449.15,123.87) ;
\draw  [dash pattern={on 4.5pt off 4.5pt}]  (195.9,123.15) .. controls (197.96,103.43) and (213,88.41) .. (235.11,78.22) .. controls (258.6,67.39) and (290.03,62.01) .. (321.6,62.01) .. controls (322.17,62.01) and (322.74,62.01) .. (323.3,62.02) .. controls (375.57,62.33) and (427.6,77.49) .. (444.84,106.49) .. controls (447.96,111.73) and (449.96,117.43) .. (450.66,123.57)(198.89,123.46) .. controls (200.84,104.67) and (215.43,90.6) .. (236.37,80.94) .. controls (259.52,70.27) and (290.49,65.01) .. (321.6,65.01) .. controls (322.16,65.01) and (322.73,65.01) .. (323.29,65.02) .. controls (374.21,65.33) and (425.38,79.64) .. (442.26,108.02) .. controls (445.16,112.9) and (447.03,118.19) .. (447.67,123.91) ;
\draw    (198.82,122.84) .. controls (209.08,154.61) and (241.87,174.43) .. (280.33,182.87) .. controls (293.96,185.86) and (308.31,187.42) .. (322.66,187.55) .. controls (323.36,187.56) and (324.06,187.56) .. (324.76,187.56) .. controls (373.52,187.56) and (422.02,171.27) .. (441.24,138.78) .. controls (444.07,133.99) and (446.26,128.85) .. (447.71,123.36)(195.97,123.76) .. controls (206.51,156.42) and (240.02,177.1) .. (279.69,185.8) .. controls (293.51,188.83) and (308.07,190.41) .. (322.63,190.55) .. controls (323.34,190.56) and (324.05,190.56) .. (324.76,190.56) .. controls (374.76,190.56) and (424.17,173.52) .. (443.82,140.31) .. controls (446.79,135.28) and (449.09,129.89) .. (450.61,124.12) ;
\draw  [draw opacity=0][fill={rgb, 255:red, 221; green, 237; blue, 238 }  ,fill opacity=1 ] (279,171) -- (359.57,171) -- (359.57,191.5) -- (279,191.5) -- cycle ;
\draw  [draw opacity=0][fill={rgb, 255:red, 221; green, 237; blue, 238 }  ,fill opacity=1 ] (260.74,166.54) -- (285.7,172.12) -- (281.44,190.51) -- (256.47,184.93) -- cycle ;
\draw [color={rgb, 255:red, 255; green, 0; blue, 31 }  ,draw opacity=1 ]   (258.59,176.41) .. controls (277.73,182.67) and (298.23,185.83) .. (318.73,186.44) .. controls (321.18,186.51) and (323.63,186.55) .. (326.07,186.55) .. controls (337.3,186.55) and (348.47,185.8) .. (359.38,184.38)(257.65,179.26) .. controls (277.07,185.61) and (297.86,188.83) .. (318.64,189.44) .. controls (321.12,189.51) and (323.6,189.55) .. (326.07,189.55) .. controls (337.43,189.55) and (348.73,188.79) .. (359.76,187.36) ;
\draw [shift={(359.57,185.87)}, rotate = 172.62] [color={rgb, 255:red, 255; green, 0; blue, 31 }  ,draw opacity=1 ][line width=0.75]    (0,5.59) -- (0,-5.59)   ;
\draw [shift={(258.12,177.83)}, rotate = 198.11] [color={rgb, 255:red, 255; green, 0; blue, 31 }  ,draw opacity=1 ][line width=0.75]    (0,5.59) -- (0,-5.59)   ;
\draw  [dash pattern={on 0.84pt off 2.51pt}] (414.77,123.3) .. controls (414.77,114.23) and (422.58,106.87) .. (432.22,106.87) .. controls (441.85,106.87) and (449.66,114.23) .. (449.66,123.3) .. controls (449.66,132.38) and (441.85,139.74) .. (432.22,139.74) .. controls (422.58,139.74) and (414.77,132.38) .. (414.77,123.3) -- cycle ;
\draw  [color={rgb, 255:red, 0; green, 0; blue, 0 }  ,draw opacity=1 ] (232.3,123.3) .. controls (232.3,99.62) and (273.15,80.42) .. (323.54,80.42) .. controls (373.92,80.42) and (414.77,99.62) .. (414.77,123.3) .. controls (414.77,146.98) and (373.92,166.18) .. (323.54,166.18) .. controls (273.15,166.18) and (232.3,146.98) .. (232.3,123.3) -- cycle ;

\draw  [draw opacity=0][fill={rgb, 255:red, 221; green, 237; blue, 238 }  ,fill opacity=1 ] (277.55,71.2) .. controls (272.1,59.91) and (386.57,62.37) .. (381,75) .. controls (375.43,87.63) and (375,105.5) .. (379,119) .. controls (383,132.5) and (275,129.5) .. (279,119.5) .. controls (283,109.5) and (285.2,105.67) .. (285,97) .. controls (284.8,88.33) and (283,82.5) .. (277.55,71.2) -- cycle ;
\draw [color={rgb, 255:red, 255; green, 0; blue, 31 }  ,draw opacity=1 ] [dash pattern={on 4.5pt off 4.5pt}]  (282.8,113.51) .. controls (295.36,111.78) and (308.94,110.68) .. (322.95,110.68) .. controls (333.15,110.68) and (343.58,111.26) .. (354,112.62) .. controls (362.5,113.72) and (371.01,115.33) .. (379.38,117.55)(283.2,116.49) .. controls (295.64,114.77) and (309.08,113.68) .. (322.95,113.68) .. controls (333.02,113.68) and (343.32,114.26) .. (353.62,115.59) .. controls (361.99,116.68) and (370.36,118.26) .. (378.62,120.45) ;
\draw [shift={(379,119)}, rotate = 194.84] [color={rgb, 255:red, 255; green, 0; blue, 31 }  ,draw opacity=1 ][line width=0.75]    (0,5.59) -- (0,-5.59)   ;
\draw [shift={(283,115)}, rotate = 172.15] [color={rgb, 255:red, 255; green, 0; blue, 31 }  ,draw opacity=1 ][line width=0.75]    (0,5.59) -- (0,-5.59)   ;
\draw  [draw opacity=0][fill={rgb, 255:red, 221; green, 237; blue, 238 }  ,fill opacity=1 ] (281,60) -- (355,60) -- (355,80) -- (281,80) -- cycle ;
\draw  [draw opacity=0][fill={rgb, 255:red, 221; green, 237; blue, 238 }  ,fill opacity=1 ] (303,63) -- (377,63) -- (377,83) -- (303,83) -- cycle ;
\draw [color={rgb, 255:red, 255; green, 0; blue, 31 }  ,draw opacity=1 ] [dash pattern={on 4.5pt off 4.5pt}]  (280.8,64.51) .. controls (293.36,62.78) and (306.94,61.68) .. (320.95,61.68) .. controls (331.15,61.68) and (341.58,62.26) .. (352,63.62) .. controls (360.5,64.72) and (369.01,66.33) .. (377.38,68.55)(281.2,67.49) .. controls (293.64,65.77) and (307.08,64.68) .. (320.95,64.68) .. controls (331.02,64.68) and (341.32,65.26) .. (351.62,66.59) .. controls (359.99,67.68) and (368.36,69.26) .. (376.62,71.45) ;
\draw [shift={(377,70)}, rotate = 194.84] [color={rgb, 255:red, 255; green, 0; blue, 31 }  ,draw opacity=1 ][line width=0.75]    (0,5.59) -- (0,-5.59)   ;
\draw [shift={(281,66)}, rotate = 172.15] [color={rgb, 255:red, 255; green, 0; blue, 31 }  ,draw opacity=1 ][line width=0.75]    (0,5.59) -- (0,-5.59)   ;
\draw    (381,75) .. controls (372.43,88.13) and (375.43,120) .. (380,130) ;
\draw    (277.55,71.2) .. controls (287.55,85.7) and (286,110) .. (279,125) ;

\draw (302.13,218.23) node [anchor=north west][inner sep=0.75pt]    {$R_{1}$};
\draw (302.13,169.23) node [anchor=north west][inner sep=0.75pt]    {$R_{1}$};
\draw (322.13,35.23) node [anchor=north west][inner sep=0.75pt]    {$R_{2}$};
\draw (321.13,90.23) node [anchor=north west][inner sep=0.75pt]    {$R_{2}$};

\end{tikzpicture}

%% file: figures/tikz/discgeo.tex
\tikzset{every picture/.style={line width=0.75pt}} %

\begin{tikzpicture}[x=0.5pt,y=0.5pt,yscale=-1,xscale=1]

\draw  [color={rgb, 255:red, 208; green, 2; blue, 27 }  ,draw opacity=1 ][line width=0.75]  (180,55) .. controls (180,32.91) and (197.91,15) .. (220,15) .. controls (242.09,15) and (260,32.91) .. (260,55) .. controls (260,77.09) and (242.09,95) .. (220,95) .. controls (197.91,95) and (180,77.09) .. (180,55) -- cycle ;
\draw  [color={rgb, 255:red, 208; green, 2; blue, 27 }  ,draw opacity=1 ][line width=0.75]  (270,55) .. controls (270,32.91) and (287.91,15) .. (310,15) .. controls (332.09,15) and (350,32.91) .. (350,55) .. controls (350,77.09) and (332.09,95) .. (310,95) .. controls (287.91,95) and (270,77.09) .. (270,55) -- cycle ;
\draw  [color={rgb, 255:red, 208; green, 2; blue, 27 }  ,draw opacity=1 ][fill={rgb, 255:red, 208; green, 2; blue, 27 }  ,fill opacity=1 ] (225,20) -- (215,15) -- (225,10) -- (220,15) -- cycle ;
\draw  [color={rgb, 255:red, 208; green, 2; blue, 27 }  ,draw opacity=1 ][fill={rgb, 255:red, 208; green, 2; blue, 27 }  ,fill opacity=1 ] (305,10) -- (315,15) -- (305,20) -- (310,15) -- cycle ;
\draw  [color={rgb, 255:red, 208; green, 2; blue, 27 }  ,draw opacity=1 ][fill={rgb, 255:red, 208; green, 2; blue, 27 }  ,fill opacity=1 ] (270,160) -- (260,155) -- (270,150) -- (265,155) -- cycle ;
\draw  [color={rgb, 255:red, 208; green, 2; blue, 27 }  ,draw opacity=1 ][fill={rgb, 255:red, 208; green, 2; blue, 27 }  ,fill opacity=1 ] (270,240) -- (260,235) -- (270,230) -- (265,235) -- cycle ;
\draw   (210,190) -- (205,200) -- (200,190) -- (205,195) -- cycle ;
\draw   (330,190) -- (325,200) -- (320,190) -- (325,195) -- cycle ;
\draw [color={rgb, 255:red, 0; green, 42; blue, 255 }  ,draw opacity=1 ]   (411.3,19.26) -- (451.3,89.26)(408.7,20.74) -- (448.7,90.74) ;
\draw [color={rgb, 255:red, 3; green, 131; blue, 139 }  ,draw opacity=1 ]   (391.3,54.25) -- (411.3,88.89)(388.7,55.75) -- (408.7,90.39) ;
\draw [color={rgb, 255:red, 3; green, 131; blue, 139 }  ,draw opacity=1 ]   (451.3,19.26) -- (471.3,54.26)(448.7,20.74) -- (468.7,55.74) ;
\draw [color={rgb, 255:red, 208; green, 2; blue, 27 }  ,draw opacity=1 ][fill={rgb, 255:red, 208; green, 2; blue, 27 }  ,fill opacity=1 ][line width=0.75]    (205,235) -- (325,235) ;
\draw [color={rgb, 255:red, 208; green, 2; blue, 27 }  ,draw opacity=1 ][fill={rgb, 255:red, 208; green, 2; blue, 27 }  ,fill opacity=1 ][line width=0.75]    (205,155) -- (325,155) ;
\draw  [dash pattern={on 0.84pt off 2.51pt}] (205,155) -- (205,235) ;
\draw   [dash pattern={on 0.84pt off 2.51pt}] (325,155) -- (325,235) ;
\draw [color={rgb, 255:red, 0; green, 42; blue, 255 }  ,draw opacity=1 ]   (541.3,20.75) -- (501.3,90.39)(538.7,19.25) -- (498.7,88.89) ;
\draw [color={rgb, 255:red, 3; green, 131; blue, 139 }  ,draw opacity=1 ]   (478.7,54.26) -- (498.7,19.26)(481.3,55.74) -- (501.3,20.74) ;
\draw [color={rgb, 255:red, 3; green, 131; blue, 139 }  ,draw opacity=1 ]   (538.7,89.26) -- (558.7,54.26)(541.3,90.74) -- (561.3,55.74) ;
\draw  [color={rgb, 255:red, 208; green, 2; blue, 27 }  ,draw opacity=1 ][line width=0.75]  (390,55) .. controls (390,32.91) and (407.91,15) .. (430,15) .. controls (452.09,15) and (470,32.91) .. (470,55) .. controls (470,77.09) and (452.09,95) .. (430,95) .. controls (407.91,95) and (390,77.09) .. (390,55) -- cycle ;
\draw  [color={rgb, 255:red, 208; green, 2; blue, 27 }  ,draw opacity=1 ][line width=0.75]  (480,55) .. controls (480,32.91) and (497.91,15) .. (520,15) .. controls (542.09,15) and (560,32.91) .. (560,55) .. controls (560,77.09) and (542.09,95) .. (520,95) .. controls (497.91,95) and (480,77.09) .. (480,55) -- cycle ;
\draw [color={rgb, 255:red, 0; green, 42; blue, 255 }  ,draw opacity=1 ]   (621.3,19.26) -- (661.3,89.26)(618.7,20.74) -- (658.7,90.74) ;
\draw [color={rgb, 255:red, 3; green, 131; blue, 139 }  ,draw opacity=1 ]   (650.82,71.25) -- (620.82,90.9)(649.18,68.75) -- (619.18,88.39) ;
\draw [color={rgb, 255:red, 3; green, 131; blue, 139 }  ,draw opacity=1 ]   (649.33,68.66) -- (679.33,53.66)(650.67,71.34) -- (680.67,56.34) ;
\draw [color={rgb, 255:red, 0; green, 42; blue, 255 }  ,draw opacity=1 ]   (751.3,20.75) -- (711.3,90.39)(748.7,19.25) -- (708.7,88.89) ;
\draw [color={rgb, 255:red, 3; green, 131; blue, 139 }  ,draw opacity=1 ]   (688.7,54.26) -- (698.5,37.11) -- (708.7,19.26)(691.3,55.74) -- (701.1,38.6) -- (711.3,20.74) ;
\draw [color={rgb, 255:red, 3; green, 131; blue, 139 }  ,draw opacity=1 ]   (748.7,89.26) -- (768.7,54.26)(751.3,90.74) -- (771.3,55.74) ;
\draw  [color={rgb, 255:red, 208; green, 2; blue, 27 }  ,draw opacity=1 ][line width=0.75]  (600,55) .. controls (600,32.91) and (617.91,15) .. (640,15) .. controls (662.09,15) and (680,32.91) .. (680,55) .. controls (680,77.09) and (662.09,95) .. (640,95) .. controls (617.91,95) and (600,77.09) .. (600,55) -- cycle ;
\draw  [color={rgb, 255:red, 208; green, 2; blue, 27 }  ,draw opacity=1 ][line width=0.75]  (690,55) .. controls (690,32.91) and (707.91,15) .. (730,15) .. controls (752.09,15) and (770,32.91) .. (770,55) .. controls (770,77.09) and (752.09,95) .. (730,95) .. controls (707.91,95) and (690,77.09) .. (690,55) -- cycle ;
\draw [color={rgb, 255:red, 3; green, 131; blue, 139 }  ,draw opacity=1 ]   (660.83,21.25) -- (630.83,41.25)(659.17,18.75) -- (629.17,38.75) ;
\draw [color={rgb, 255:red, 3; green, 131; blue, 139 }  ,draw opacity=1 ]   (599.33,53.66) -- (629.33,38.66)(600.67,56.34) -- (630.67,41.34) ;
\draw  [color={rgb, 255:red, 208; green, 2; blue, 27 }  ,draw opacity=1 ][fill={rgb, 255:red, 208; green, 2; blue, 27 }  ,fill opacity=1 ] (515,10) -- (525,15) -- (515,20) -- (520,15) -- cycle ;
\draw  [color={rgb, 255:red, 208; green, 2; blue, 27 }  ,draw opacity=1 ][fill={rgb, 255:red, 208; green, 2; blue, 27 }  ,fill opacity=1 ] (725,10) -- (735,15) -- (725,20) -- (730,15) -- cycle ;
\draw  [color={rgb, 255:red, 208; green, 2; blue, 27 }  ,draw opacity=1 ][fill={rgb, 255:red, 208; green, 2; blue, 27 }  ,fill opacity=1 ] (435,20) -- (425,15) -- (435,10) -- (430,15) -- cycle ;
\draw  [color={rgb, 255:red, 208; green, 2; blue, 27 }  ,draw opacity=1 ][fill={rgb, 255:red, 208; green, 2; blue, 27 }  ,fill opacity=1 ] (645,20) -- (635,15) -- (645,10) -- (640,15) -- cycle ;
\draw  [color={rgb, 255:red, 208; green, 2; blue, 27 }  ,draw opacity=1 ][fill={rgb, 255:red, 208; green, 2; blue, 27 }  ,fill opacity=1 ] (480,160) -- (470,155) -- (480,150) -- (475,155) -- cycle ;
\draw  [color={rgb, 255:red, 208; green, 2; blue, 27 }  ,draw opacity=1 ][fill={rgb, 255:red, 208; green, 2; blue, 27 }  ,fill opacity=1 ] (480,240) -- (470,235) -- (480,230) -- (475,235) -- cycle ;
\draw   (420,190) -- (415,200) -- (410,190) -- (415,195) -- cycle ;
\draw   (540,190) -- (535,200) -- (530,190) -- (535,195) -- cycle ;
\draw [color={rgb, 255:red, 208; green, 2; blue, 27 }  ,draw opacity=1 ][fill={rgb, 255:red, 208; green, 2; blue, 27 }  ,fill opacity=1 ][line width=0.75]    (415,235) -- (535,235) ;
\draw [color={rgb, 255:red, 208; green, 2; blue, 27 }  ,draw opacity=1 ][fill={rgb, 255:red, 208; green, 2; blue, 27 }  ,fill opacity=1 ][line width=0.75]    (415,155) -- (535,155) ;
\draw  [dash pattern={on 0.84pt off 2.51pt}]  (415,155) -- (415,235) ;
\draw  [dash pattern={on 0.84pt off 2.51pt}]  (535,155) -- (535,235) ;
\draw [color={rgb, 255:red, 3; green, 131; blue, 139 }  ,draw opacity=1 ]   (466.5,155) -- (466.5,235)(463.5,155) -- (463.5,235) ;
\draw [color={rgb, 255:red, 3; green, 131; blue, 139 }  ,draw opacity=1 ]   (486.5,155) -- (486.5,235)(483.5,155) -- (483.5,235) ;
\draw [color={rgb, 255:red, 0; green, 42; blue, 255 }  ,draw opacity=1 ]   (506.5,155) -- (506.5,235)(503.5,155) -- (503.5,235) ;
\draw [color={rgb, 255:red, 3; green, 131; blue, 139 }  ,draw opacity=1 ]   (526.5,155) -- (526.5,235)(523.5,155) -- (523.5,235) ;
\draw [color={rgb, 255:red, 0; green, 42; blue, 255 }  ,draw opacity=1 ]   (446.5,155) -- (446.5,235)(443.5,155) -- (443.5,235) ;
\draw [color={rgb, 255:red, 3; green, 131; blue, 139 }  ,draw opacity=1 ]   (426.5,155) -- (426.5,235)(423.5,155) -- (423.5,235) ;
\draw  [color={rgb, 255:red, 208; green, 2; blue, 27 }  ,draw opacity=1 ][fill={rgb, 255:red, 208; green, 2; blue, 27 }  ,fill opacity=1 ] (690,160) -- (680,155) -- (690,150) -- (685,155) -- cycle ;
\draw  [color={rgb, 255:red, 208; green, 2; blue, 27 }  ,draw opacity=1 ][fill={rgb, 255:red, 208; green, 2; blue, 27 }  ,fill opacity=1 ] (690,240) -- (680,235) -- (690,230) -- (685,235) -- cycle ;
\draw   (630,190) -- (625,200) -- (620,190) -- (625,195) -- cycle ;
\draw   (750,190) -- (745,200) -- (740,190) -- (745,195) -- cycle ;
\draw [color={rgb, 255:red, 208; green, 2; blue, 27 }  ,draw opacity=1 ][fill={rgb, 255:red, 208; green, 2; blue, 27 }  ,fill opacity=1 ][line width=0.75]    (625,235) -- (745,235) ;
\draw [color={rgb, 255:red, 208; green, 2; blue, 27 }  ,draw opacity=1 ][fill={rgb, 255:red, 208; green, 2; blue, 27 }  ,fill opacity=1 ][line width=0.75]    (625,155) -- (745,155) ;
\draw   [dash pattern={on 0.84pt off 2.51pt}] (625,155) -- (625,235) ;
\draw  [dash pattern={on 0.84pt off 2.51pt}]  (745,155) -- (745,235) ;
\draw [color={rgb, 255:red, 3; green, 131; blue, 139 }  ,draw opacity=1 ]   (656.17,209.06) -- (676.17,234.06)(653.83,210.94) -- (673.83,235.94) ;
\draw [color={rgb, 255:red, 3; green, 131; blue, 139 }  ,draw opacity=1 ]   (696.5,155) -- (696.5,235)(693.5,155) -- (693.5,235) ;
\draw [color={rgb, 255:red, 0; green, 42; blue, 255 }  ,draw opacity=1 ]   (716.5,155) -- (716.5,235)(713.5,155) -- (713.5,235) ;
\draw [color={rgb, 255:red, 3; green, 131; blue, 139 }  ,draw opacity=1 ]   (736.5,155) -- (736.5,235)(733.5,155) -- (733.5,235) ;
\draw [color={rgb, 255:red, 0; green, 42; blue, 255 }  ,draw opacity=1 ]   (656.5,155) -- (656.5,235)(653.5,155) -- (653.5,235) ;
\draw [color={rgb, 255:red, 3; green, 131; blue, 139 }  ,draw opacity=1 ]   (636.17,154.06) -- (656.17,179.06)(633.83,155.94) -- (653.83,180.94) ;
\draw [color={rgb, 255:red, 3; green, 131; blue, 139 }  ,draw opacity=1 ]   (653.83,179.06) -- (673.83,154.06)(656.17,180.94) -- (676.17,155.94) ;
\draw [color={rgb, 255:red, 3; green, 131; blue, 139 }  ,draw opacity=1 ]   (656.17,210.94) -- (636.17,235.94)(653.83,209.06) -- (633.83,234.06) ;
\draw  [draw opacity=0][fill={rgb, 255:red, 221; green, 237; blue, 238 }  ,fill opacity=1 ] (40,40) .. controls (40,23.43) and (62.39,10) .. (90,10) .. controls (117.61,10) and (140,23.43) .. (140,40) -- (90,40) -- cycle ; \draw   (40,40) .. controls (40,23.43) and (62.39,10) .. (90,10) .. controls (117.61,10) and (140,23.43) .. (140,40) ;
\draw  [draw opacity=0][fill={rgb, 255:red, 221; green, 237; blue, 238 }  ,fill opacity=1 ][line width=0.75]  (140,40) .. controls (140,40) and (140,40) .. (140,40) .. controls (140,45.52) and (117.61,50) .. (90,50) .. controls (62.39,50) and (40,45.52) .. (40,40) -- (90,40) -- cycle ; \draw  [color={rgb, 255:red, 208; green, 2; blue, 27 }  ,draw opacity=1 ][line width=0.75]  (140,40) .. controls (140,40) and (140,40) .. (140,40) .. controls (140,45.52) and (117.61,50) .. (90,50) .. controls (62.39,50) and (40,45.52) .. (40,40) ;
\draw  [draw opacity=0][dash pattern={on 4.5pt off 4.5pt}][line width=0.75]  (40,40) .. controls (40,40) and (40,40) .. (40,40) .. controls (40,34.48) and (62.39,30) .. (90,30) .. controls (117.61,30) and (140,34.48) .. (140,40) -- (90,40) -- cycle ; \draw  [color={rgb, 255:red, 208; green, 2; blue, 27 }  ,draw opacity=1 ][dash pattern={on 4.5pt off 4.5pt}][line width=0.75]  (40,40) .. controls (40,40) and (40,40) .. (40,40) .. controls (40,34.48) and (62.39,30) .. (90,30) .. controls (117.61,30) and (140,34.48) .. (140,40) ;
\draw  [draw opacity=0][fill={rgb, 255:red, 221; green, 237; blue, 238 }  ,fill opacity=1 ] (140,70) .. controls (140,86.57) and (117.61,100) .. (90,100) .. controls (62.39,100) and (40,86.57) .. (40,70) -- (90,70) -- cycle ; \draw   (140,70) .. controls (140,86.57) and (117.61,100) .. (90,100) .. controls (62.39,100) and (40,86.57) .. (40,70) ;
\draw  [color={rgb, 255:red, 208; green, 2; blue, 27 }  ,draw opacity=1 ][fill={rgb, 255:red, 182; green, 214; blue, 214 }  ,fill opacity=1 ][line width=0.75]  (40,70) .. controls (40,64.48) and (62.39,60) .. (90,60) .. controls (117.61,60) and (140,64.48) .. (140,70) .. controls (140,75.52) and (117.61,80) .. (90,80) .. controls (62.39,80) and (40,75.52) .. (40,70) -- cycle ;
\draw  [color={rgb, 255:red, 208; green, 2; blue, 27 }  ,draw opacity=1 ][fill={rgb, 255:red, 208; green, 2; blue, 27 }  ,fill opacity=1 ] (83.75,45) -- (96.25,50) -- (83.75,55) -- (90,50) -- cycle ;
\draw  [color={rgb, 255:red, 208; green, 2; blue, 27 }  ,draw opacity=1 ][fill={rgb, 255:red, 208; green, 2; blue, 27 }  ,fill opacity=1 ] (83.75,75) -- (96.25,80) -- (83.75,85) -- (90,80) -- cycle ;
\draw  [draw opacity=0][fill={rgb, 255:red, 221; green, 237; blue, 238 }  ,fill opacity=1 ] (40,180) .. controls (40,163.43) and (62.39,150) .. (90,150) .. controls (117.61,150) and (140,163.43) .. (140,180) -- (90,180) -- cycle ; \draw   (40,180) .. controls (40,163.43) and (62.39,150) .. (90,150) .. controls (117.61,150) and (140,163.43) .. (140,180) ;
\draw  [draw opacity=0][fill={rgb, 255:red, 221; green, 237; blue, 238 }  ,fill opacity=1 ][line width=0.75]  (140,180) .. controls (140,180) and (140,180) .. (140,180) .. controls (140,185.52) and (117.61,190) .. (90,190) .. controls (62.39,190) and (40,185.52) .. (40,180) -- (90,180) -- cycle ; \draw  [color={rgb, 255:red, 208; green, 2; blue, 27 }  ,draw opacity=1 ][line width=0.75]  (140,180) .. controls (140,180) and (140,180) .. (140,180) .. controls (140,185.52) and (117.61,190) .. (90,190) .. controls (62.39,190) and (40,185.52) .. (40,180) ;
\draw  [draw opacity=0][dash pattern={on 4.5pt off 4.5pt}][line width=0.75]  (40,180) .. controls (40,180) and (40,180) .. (40,180) .. controls (40,174.48) and (62.39,170) .. (90,170) .. controls (117.61,170) and (140,174.48) .. (140,180) -- (90,180) -- cycle ; \draw  [color={rgb, 255:red, 208; green, 2; blue, 27 }  ,draw opacity=1 ][dash pattern={on 4.5pt off 4.5pt}][line width=0.75]  (40,180) .. controls (40,180) and (40,180) .. (40,180) .. controls (40,174.48) and (62.39,170) .. (90,170) .. controls (117.61,170) and (140,174.48) .. (140,180) ;
\draw  [draw opacity=0][fill={rgb, 255:red, 221; green, 237; blue, 238 }  ,fill opacity=1 ] (140,210) .. controls (140,226.57) and (117.61,240) .. (90,240) .. controls (62.39,240) and (40,226.57) .. (40,210) -- (90,210) -- cycle ; \draw   (140,210) .. controls (140,226.57) and (117.61,240) .. (90,240) .. controls (62.39,240) and (40,226.57) .. (40,210) ;
\draw  [draw opacity=0][fill={rgb, 255:red, 182; green, 214; blue, 214 }  ,fill opacity=1 ][line width=0.75]  (40,210) .. controls (40,204.48) and (62.39,200) .. (90,200) .. controls (117.61,200) and (140,204.48) .. (140,210) .. controls (140,215.52) and (117.61,220) .. (90,220) .. controls (62.39,220) and (40,215.52) .. (40,210) -- cycle ;
\draw  [color={rgb, 255:red, 208; green, 2; blue, 27 }  ,draw opacity=1 ][fill={rgb, 255:red, 208; green, 2; blue, 27 }  ,fill opacity=1 ] (83.75,185) -- (96.25,190) -- (83.75,195) -- (90,190) -- cycle ;
\draw  [color={rgb, 255:red, 208; green, 2; blue, 27 }  ,draw opacity=1 ][fill={rgb, 255:red, 208; green, 2; blue, 27 }  ,fill opacity=1 ] (83.75,215) -- (96.25,220) -- (83.75,225) -- (90,220) -- cycle ;
\draw  [draw opacity=0] (107.69,218.9) .. controls (106.02,212.64) and (105,204.24) .. (105,195) .. controls (105,193.06) and (105.05,191.16) .. (105.13,189.3) -- (115,195) -- cycle ; \draw   (107.69,218.9) .. controls (106.02,212.64) and (105,204.24) .. (105,195) .. controls (105,193.06) and (105.05,191.16) .. (105.13,189.3) ;
\draw  [draw opacity=0][line width=0.75]  (104.72,200.44) .. controls (125.15,201.7) and (140,205.5) .. (140,210) .. controls (140,215.52) and (117.61,220) .. (90,220) .. controls (62.39,220) and (40,215.52) .. (40,210) .. controls (40,205.5) and (54.85,201.7) .. (75.28,200.44) -- (90,210) -- cycle ; \draw  [color={rgb, 255:red, 208; green, 2; blue, 27 }  ,draw opacity=1 ][line width=0.75]  (104.72,200.44) .. controls (125.15,201.7) and (140,205.5) .. (140,210) .. controls (140,215.52) and (117.61,220) .. (90,220) .. controls (62.39,220) and (40,215.52) .. (40,210) .. controls (40,205.5) and (54.85,201.7) .. (75.28,200.44) ;
\draw  [draw opacity=0] (74.87,189.3) .. controls (74.95,191.16) and (75,193.06) .. (75,195) .. controls (75,204.24) and (73.98,212.64) .. (72.31,218.9) -- (65,195) -- cycle ; \draw   (74.87,189.3) .. controls (74.95,191.16) and (75,193.06) .. (75,195) .. controls (75,204.24) and (73.98,212.64) .. (72.31,218.9) ;
\draw  [draw opacity=0] (115,160) .. controls (115,160) and (115,160) .. (115,160) .. controls (115,160) and (115,160) .. (115,160) .. controls (115,162.76) and (103.81,165) .. (90,165) .. controls (76.19,165) and (65,162.76) .. (65,160) -- (90,160) -- cycle ; \draw   (115,160) .. controls (115,160) and (115,160) .. (115,160) .. controls (115,160) and (115,160) .. (115,160) .. controls (115,162.76) and (103.81,165) .. (90,165) .. controls (76.19,165) and (65,162.76) .. (65,160) ;
\draw  [draw opacity=0][dash pattern={on 0.84pt off 2.51pt}] (105.13,189.3) .. controls (105.91,172.68) and (110.03,160) .. (115,160) -- (115,195) -- cycle ; \draw  [dash pattern={on 0.84pt off 2.51pt}] (105.13,189.3) .. controls (105.91,172.68) and (110.03,160) .. (115,160) ;
\draw  [draw opacity=0][dash pattern={on 0.84pt off 2.51pt}] (65,160) .. controls (65,160) and (65,160) .. (65,160) .. controls (69.97,160) and (74.09,172.68) .. (74.87,189.3) -- (65,195) -- cycle ; \draw  [dash pattern={on 0.84pt off 2.51pt}] (65,160) .. controls (65,160) and (65,160) .. (65,160) .. controls (69.97,160) and (74.09,172.68) .. (74.87,189.3) ;
\draw  [draw opacity=0][dash pattern={on 0.84pt off 2.51pt}] (115,230) .. controls (115,230) and (115,230) .. (115,230) .. controls (112.12,230) and (109.52,225.73) .. (107.69,218.9) -- (115,195) -- cycle ; \draw  [dash pattern={on 0.84pt off 2.51pt}] (115,230) .. controls (115,230) and (115,230) .. (115,230) .. controls (112.12,230) and (109.52,225.73) .. (107.69,218.9) ;
\draw  [dash pattern={on 0.84pt off 2.51pt}] (65,230) .. controls (65,227.24) and (76.19,225) .. (90,225) .. controls (103.81,225) and (115,227.24) .. (115,230) .. controls (115,232.76) and (103.81,235) .. (90,235) .. controls (76.19,235) and (65,232.76) .. (65,230) -- cycle ;
\draw  [draw opacity=0][dash pattern={on 0.84pt off 2.51pt}] (72.31,218.9) .. controls (70.48,225.73) and (67.88,230) .. (65,230) -- (65,195) -- cycle ; \draw  [dash pattern={on 0.84pt off 2.51pt}] (72.31,218.9) .. controls (70.48,225.73) and (67.88,230) .. (65,230) ;
\draw  [draw opacity=0][dash pattern={on 4.5pt off 4.5pt}][line width=0.75]  (75.28,200.44) .. controls (79.93,200.15) and (84.88,200) .. (90,200) .. controls (95.12,200) and (100.07,200.15) .. (104.72,200.44) -- (90,210) -- cycle ; \draw  [color={rgb, 255:red, 208; green, 2; blue, 27 }  ,draw opacity=1 ][dash pattern={on 4.5pt off 4.5pt}][line width=0.75]  (75.28,200.44) .. controls (79.93,200.15) and (84.88,200) .. (90,200) .. controls (95.12,200) and (100.07,200.15) .. (104.72,200.44) ;

\draw (15,70) node    {$\ket{\psi }$};
\draw (15,40) node    {$\bra{\psi }$};
\draw (160,55) node    {$=$};
\draw (160,195) node    {$=$};
\draw (220,110) node    {$\bra{\psi }$};
\draw (310,110) node    {$\ket{\psi }$};
\draw (265,135) node    {$\bra{\psi }$};
\draw (265,255) node    {$\ket{\psi }$};
\draw (580,55) node    {$+$};
\draw (790,55) node    {$+$};
\draw (820,55) node    {$\cdots $};
\draw (580,195) node    {$+$};
\draw (790,195) node    {$+$};
\draw (820,195) node    {$\cdots $};
\draw (370,55) node    {$=$};
\draw (370,195) node    {$=$};
\draw (15,210) node    {$\ket{\psi }$};
\draw (15,180) node    {$\bra{\psi }$};

\end{tikzpicture}

%% file: figures/tikz/tenentropy.tex
\tikzset{every picture/.style={line width=0.75pt}} %

\begin{tikzpicture}[x=0.75pt,y=0.75pt,yscale=-1,xscale=1]

\draw  [draw opacity=0][fill={rgb, 255:red, 177; green, 250; blue, 255 }  ,fill opacity=1 ] (116,89) -- (397.4,89) -- (397.4,175.6) -- (116,175.6) -- cycle ;
\draw  [draw opacity=0][fill={rgb, 255:red, 255; green, 181; blue, 88 }  ,fill opacity=1 ] (168.6,130.4) -- (345.6,130.4) -- (345.6,174.4) -- (168.6,174.4) -- cycle ;
\draw   (177.2,143.6) -- (203.8,143.6) -- (203.8,171.4) -- (177.2,171.4) -- cycle ;
\draw   (177.2,92.4) -- (204.2,92.4) -- (204.2,120.4) -- (177.2,120.4) -- cycle ;
\draw    (191.2,120.6) -- (191.2,143.6) ;
\draw   (220.2,143.6) -- (246.8,143.6) -- (246.8,171.4) -- (220.2,171.4) -- cycle ;
\draw   (220.2,92.4) -- (247.2,92.4) -- (247.2,120.4) -- (220.2,120.4) -- cycle ;
\draw    (234.2,120.6) -- (234.2,143.6) ;
\draw    (203.8,157.9) -- (220.2,158) ;
\draw    (203.8,106.9) -- (220.2,107) ;
\draw   (264.2,143.6) -- (290.8,143.6) -- (290.8,171.4) -- (264.2,171.4) -- cycle ;
\draw   (264.2,92.4) -- (291.2,92.4) -- (291.2,120.4) -- (264.2,120.4) -- cycle ;
\draw    (278.2,120.6) -- (278.2,143.6) ;
\draw    (246.2,158) -- (264.2,158) ;
\draw    (247.8,106.9) -- (264.2,107) ;
\draw   (308.2,143.6) -- (334.8,143.6) -- (334.8,171.4) -- (308.2,171.4) -- cycle ;
\draw   (308.2,92.4) -- (335.2,92.4) -- (335.2,120.4) -- (308.2,120.4) -- cycle ;
\draw    (322.2,120.6) -- (322.2,143.6) ;
\draw    (290.8,157.9) -- (308.2,158) ;
\draw    (291.8,106.9) -- (308.2,107) ;
\draw   (352.2,143.6) -- (378.8,143.6) -- (378.8,171.4) -- (352.2,171.4) -- cycle ;
\draw   (352.2,92.4) -- (379.2,92.4) -- (379.2,120.4) -- (352.2,120.4) -- cycle ;
\draw    (366.2,120.6) -- (366.2,143.6) ;
\draw    (334.2,158) -- (352.2,158) ;
\draw    (335.8,106.9) -- (352.2,107) ;
\draw   (133.2,143.6) -- (159.8,143.6) -- (159.8,171.4) -- (133.2,171.4) -- cycle ;
\draw   (133.2,92.4) -- (160.2,92.4) -- (160.2,120.4) -- (133.2,120.4) -- cycle ;
\draw    (147.2,120.6) -- (147.2,143.6) ;
\draw    (116.8,157.9) -- (133.2,158) ;
\draw    (116.8,106.9) -- (133.2,107) ;
\draw    (159.2,158) -- (177.2,158) ;
\draw    (160.8,106.9) -- (177.2,107) ;
\draw    (378.8,157.9) -- (395.2,158) ;
\draw    (379.8,106.9) -- (396.2,107) ;
\draw [color={rgb, 255:red, 255; green, 0; blue, 0 }  ,draw opacity=1 ]   (170.1,130.4) -- (170.1,172.4)(167.1,130.4) -- (167.1,172.4) ;
\draw [color={rgb, 255:red, 255; green, 0; blue, 0 }  ,draw opacity=1 ]   (346.9,130.59) -- (347.1,174.39)(343.9,130.61) -- (344.1,174.41) ;
\draw [color={rgb, 255:red, 0; green, 4; blue, 255 }  ,draw opacity=1 ]   (168.6,128.9) -- (345.4,129.1)(168.6,131.9) -- (345.4,132.1) ;

\draw (182,149.4) node [anchor=north west][inner sep=0.75pt]    {$A$};
\draw (182,96.4) node [anchor=north west][inner sep=0.75pt]    {$A^{*}$};
\draw (225,149.4) node [anchor=north west][inner sep=0.75pt]    {$A$};
\draw (225,96.4) node [anchor=north west][inner sep=0.75pt]    {$A^{*}$};
\draw (269,149.4) node [anchor=north west][inner sep=0.75pt]    {$A$};
\draw (269,96.4) node [anchor=north west][inner sep=0.75pt]    {$A^{*}$};
\draw (313,149.4) node [anchor=north west][inner sep=0.75pt]    {$A$};
\draw (313,96.4) node [anchor=north west][inner sep=0.75pt]    {$A^{*}$};
\draw (357,149.4) node [anchor=north west][inner sep=0.75pt]    {$A$};
\draw (357,96.4) node [anchor=north west][inner sep=0.75pt]    {$A^{*}$};
\draw (138,149.4) node [anchor=north west][inner sep=0.75pt]    {$A$};
\draw (138,96.4) node [anchor=north west][inner sep=0.75pt]    {$A^{*}$};
\draw (98,120.4) node [anchor=north west][inner sep=0.75pt]    {$...$};
\draw (402,120.4) node [anchor=north west][inner sep=0.75pt]    {$...$};

\end{tikzpicture}

%% file: figures/tikz/holographic.tex
\tikzset{every picture/.style={line width=0.75pt}} %

\begin{tikzpicture}[x=0.75pt,y=0.75pt,yscale=-1,xscale=1]

\draw  [draw opacity=0][fill={rgb, 255:red, 188; green, 120; blue, 245 }  ,fill opacity=1 ] (379.93,100.69) -- (387.44,100.69) -- (387.44,204.77) -- (379.93,204.77) -- cycle ;
\draw  [draw opacity=0][fill={rgb, 255:red, 105; green, 230; blue, 242 }  ,fill opacity=1 ] (54.59,99.74) -- (158.14,99.74) -- (158.14,151.58) -- (54.59,151.58) -- cycle ;
\draw  [draw opacity=0][fill={rgb, 255:red, 221; green, 237; blue, 238 }  ,fill opacity=1 ] (54.59,151.58) -- (158.14,151.58) -- (158.14,203.42) -- (54.59,203.42) -- cycle ;
\draw [color={rgb, 255:red, 255; green, 0; blue, 31 }  ,draw opacity=1 ]   (54.59,150.08) -- (158.14,150.08)(54.59,153.08) -- (158.14,153.08) ;
\draw  [draw opacity=0][fill={rgb, 255:red, 105; green, 230; blue, 242 }  ,fill opacity=1 ] (308.41,101.03) -- (308.41,204.77) -- (256.67,204.77) -- (256.67,101.03) -- cycle ;
\draw  [draw opacity=0][fill={rgb, 255:red, 221; green, 237; blue, 238 }  ,fill opacity=1 ] (256.67,101.03) -- (256.67,204.77) -- (204.93,204.77) -- (204.93,101.03) -- cycle ;
\draw [color={rgb, 255:red, 255; green, 0; blue, 31 }  ,draw opacity=1 ]   (258.17,101.03) -- (258.17,204.77)(255.17,101.03) -- (255.17,204.77) ;
\draw  [draw opacity=0][fill={rgb, 255:red, 105; green, 230; blue, 242 }  ,fill opacity=1 ] (439.18,101.03) -- (439.18,204.77) -- (387.44,204.77) -- (387.44,101.03) -- cycle ;
\draw [color={rgb, 255:red, 255; green, 0; blue, 31 }  ,draw opacity=1 ]   (388.94,101.03) -- (388.94,204.77)(385.94,101.03) -- (385.94,204.77) ;
\draw    (564.89,139.5) -- (550.48,139.5) ;
\draw    (564.89,165.26) -- (550.48,165.26) ;
\draw    (541.52,147.05) -- (541.46,156.87) ;
\draw    (564.89,191.63) -- (550.48,191.63) ;
\draw    (541.46,172.45) -- (541.46,183.24) ;
\draw    (541.52,199.17) -- (541.46,209.6) ;
\draw    (564.89,113.14) -- (550.48,113.14) ;
\draw    (541.52,94.92) -- (541.46,104.75) ;
\draw    (541.59,122.25) -- (541.46,131.11) ;
\draw  [draw opacity=0][fill={rgb, 255:red, 105; green, 230; blue, 242 }  ,fill opacity=1 ] (599.4,102.5) -- (599.4,202.19) -- (562.31,202.19) -- (562.31,102.5) -- cycle ;
\draw    (562.31,102.5) -- (562.31,202.19) ;
\draw  [fill={rgb, 255:red, 188; green, 120; blue, 245 }  ,fill opacity=1 ] (551.07,102.85) -- (551.07,122.25) -- (533.07,122.25) -- (533.07,102.85) -- cycle ;

\draw  [fill={rgb, 255:red, 188; green, 120; blue, 245 }  ,fill opacity=1 ] (551.07,129.43) -- (551.07,148.83) -- (533.07,148.83) -- (533.07,129.43) -- cycle ;

\draw  [fill={rgb, 255:red, 188; green, 120; blue, 245 }  ,fill opacity=1 ] (551.07,156.02) -- (551.07,175.42) -- (533.07,175.42) -- (533.07,156.02) -- cycle ;

\draw  [fill={rgb, 255:red, 188; green, 120; blue, 245 }  ,fill opacity=1 ] (551.07,182.6) -- (551.07,202) -- (533.07,202) -- (533.07,182.6) -- cycle ;

\draw [color={rgb, 255:red, 255; green, 0; blue, 31 }  ,draw opacity=1 ] [dash pattern={on 0.84pt off 2.51pt}]  (558.94,99.23) -- (558.94,202.98)(555.94,99.23) -- (555.94,202.98) ;

\draw (92.97,166.95) node [anchor=north west][inner sep=0.75pt]    {$\ket{\psi }$};
\draw (79.56,116.08) node [anchor=north west][inner sep=0.75pt]    {$\bra{0_{\mathrm{CFT}}}$};
\draw (35.51,145.5) node [anchor=north west][inner sep=0.75pt]    {$\Sigma $};
\draw (172.57,143.96) node [anchor=north west][inner sep=0.75pt]    {$=$};
\draw (217.43,144.51) node [anchor=north west][inner sep=0.75pt]    {$\mathrm{BH}$};
\draw (265.05,144.89) node [anchor=north west][inner sep=0.75pt]    {$\mathrm{Bath}$};
\draw (248.65,80.82) node [anchor=north west][inner sep=0.75pt]    {$\Sigma $};
\draw (321.24,143.91) node [anchor=north west][inner sep=0.75pt]    {$=$};
\draw (347.13,144.51) node [anchor=north west][inner sep=0.75pt]    {$\mathrm{QM}$};
\draw (395.82,144.89) node [anchor=north west][inner sep=0.75pt]    {$\mathrm{Bath}$};
\draw (379.43,80.82) node [anchor=north west][inner sep=0.75pt]    {$\Sigma $};
\draw (453.58,140.81) node [anchor=north west][inner sep=0.75pt]    {$\xrightarrow{\mathrm{discretize}}$};
\draw (534.6,105.69) node [anchor=north west][inner sep=0.75pt]    {$A$};
\draw (534.6,132.28) node [anchor=north west][inner sep=0.75pt]    {$A$};
\draw (534.6,158.86) node [anchor=north west][inner sep=0.75pt]    {$A$};
\draw (534.6,185.45) node [anchor=north west][inner sep=0.75pt]    {$A$};
\draw (551.32,79.97) node [anchor=north west][inner sep=0.75pt]    {$\Sigma $};

\end{tikzpicture}

%% file: figures/tikz/tentransfer.tex
\tikzset{every picture/.style={line width=0.75pt}} %

\begin{tikzpicture}[x=0.75pt,y=0.75pt,yscale=-1,xscale=1]

\draw  [draw opacity=0][fill={rgb, 255:red, 177; green, 250; blue, 255 }  ,fill opacity=1 ] (146.46,157.25) -- (173.59,157.25) -- (173.59,203.81) -- (146.46,203.81) -- cycle ;
\draw  [draw opacity=0][fill={rgb, 255:red, 177; green, 250; blue, 255 }  ,fill opacity=1 ] (98.99,66.43) -- (133.82,66.43) -- (133.82,147.24) -- (98.99,147.24) -- cycle ;
\draw   (104.49,117.37) -- (128.87,117.37) -- (128.87,141.79) -- (104.49,141.79) -- cycle ;
\draw   (104.49,72.4) -- (129.23,72.4) -- (129.23,97) -- (104.49,97) -- cycle ;
\draw    (117.32,97.17) -- (117.32,117.37) ;
\draw   (143.9,117.37) -- (168.27,117.37) -- (168.27,141.79) -- (143.9,141.79) -- cycle ;
\draw   (143.9,72.4) -- (168.64,72.4) -- (168.64,97) -- (143.9,97) -- cycle ;
\draw    (156.73,97.17) -- (156.73,117.37) ;
\draw    (128.87,129.94) -- (143.9,130.02) ;
\draw    (128.87,85.14) -- (143.9,85.23) ;
\draw   (184.22,117.37) -- (208.59,117.37) -- (208.59,141.79) -- (184.22,141.79) -- cycle ;
\draw   (184.22,72.4) -- (208.96,72.4) -- (208.96,97) -- (184.22,97) -- cycle ;
\draw    (197.05,97.17) -- (197.05,117.37) ;
\draw    (167.72,130.02) -- (184.22,130.02) ;
\draw    (169.19,85.14) -- (184.22,85.23) ;
\draw   (224.54,117.37) -- (248.91,117.37) -- (248.91,141.79) -- (224.54,141.79) -- cycle ;
\draw   (224.54,72.4) -- (249.28,72.4) -- (249.28,97) -- (224.54,97) -- cycle ;
\draw    (237.37,97.17) -- (237.37,117.37) ;
\draw    (208.59,129.94) -- (224.54,130.02) ;
\draw    (209.51,85.14) -- (224.54,85.23) ;
\draw   (264.86,117.37) -- (289.23,117.37) -- (289.23,141.79) -- (264.86,141.79) -- cycle ;
\draw   (264.86,72.4) -- (289.6,72.4) -- (289.6,97) -- (264.86,97) -- cycle ;
\draw    (277.69,97.17) -- (277.69,117.37) ;
\draw    (248.36,130.02) -- (264.86,130.02) ;
\draw    (249.83,85.14) -- (264.86,85.23) ;
\draw   (64.17,117.37) -- (88.55,117.37) -- (88.55,141.79) -- (64.17,141.79) -- cycle ;
\draw   (64.17,72.4) -- (88.91,72.4) -- (88.91,97) -- (64.17,97) -- cycle ;
\draw    (77,97.17) -- (77,117.37) ;
\draw    (49.14,129.94) -- (64.17,130.02) ;
\draw    (49.14,85.14) -- (64.17,85.23) ;
\draw    (88,130.02) -- (104.49,130.02) ;
\draw    (89.46,85.14) -- (104.49,85.23) ;
\draw    (289.23,129.94) -- (304.26,130.02) ;
\draw    (290.15,85.14) -- (305.18,85.23) ;
\draw  [color={rgb, 255:red, 0; green, 4; blue, 255 }  ,draw opacity=1 ] (98.99,66.43) -- (133.82,66.43) -- (133.82,147.24) -- (98.99,147.24) -- cycle ;
\draw  [color={rgb, 255:red, 0; green, 4; blue, 255 }  ,draw opacity=1 ] (146.46,157.25) -- (173.59,157.25) -- (173.59,203.81) -- (146.46,203.81) -- cycle ;
\draw    (173.77,170.34) -- (188.8,170.43) ;
\draw    (173.77,189.66) -- (188.8,189.75) ;
\draw    (131.62,170.34) -- (146.65,170.43) ;
\draw    (131.62,189.66) -- (146.65,189.75) ;
\draw  [draw opacity=0][fill={rgb, 255:red, 177; green, 250; blue, 255 }  ,fill opacity=1 ] (452.71,181.85) -- (484.78,181.85) -- (484.78,228.4) -- (452.71,228.4) -- cycle ;
\draw  [draw opacity=0][fill={rgb, 255:red, 177; green, 250; blue, 255 }  ,fill opacity=1 ] (406.89,65.2) -- (441.71,65.2) -- (441.71,171.83) -- (406.89,171.83) -- cycle ;
\draw   (412.39,141.97) -- (436.77,141.97) -- (436.77,166.39) -- (412.39,166.39) -- cycle ;
\draw   (412.39,72.4) -- (437.13,72.4) -- (437.13,97) -- (412.39,97) -- cycle ;
\draw    (425.22,97.17) -- (425.22,106.13) ;
\draw   (451.79,141.97) -- (476.17,141.97) -- (476.17,166.39) -- (451.79,166.39) -- cycle ;
\draw   (451.79,72.4) -- (476.54,72.4) -- (476.54,97) -- (451.79,97) -- cycle ;
\draw    (464.62,97.17) -- (464.62,141.62) ;
\draw    (436.77,154.53) -- (451.79,154.62) ;
\draw    (436.77,85.14) -- (451.79,85.23) ;
\draw   (492.11,141.97) -- (516.49,141.97) -- (516.49,166.39) -- (492.11,166.39) -- cycle ;
\draw   (492.11,72.4) -- (516.86,72.4) -- (516.86,97) -- (492.11,97) -- cycle ;
\draw    (504.94,97.17) -- (504.94,141.62) ;
\draw    (475.62,154.62) -- (492.11,154.62) ;
\draw    (477.09,85.14) -- (492.11,85.23) ;
\draw   (532.43,141.97) -- (556.81,141.97) -- (556.81,166.39) -- (532.43,166.39) -- cycle ;
\draw   (532.43,72.4) -- (557.18,72.4) -- (557.18,97) -- (532.43,97) -- cycle ;
\draw    (545.26,97.17) -- (545.26,141.62) ;
\draw    (516.49,154.53) -- (532.43,154.62) ;
\draw    (517.41,85.14) -- (532.43,85.23) ;
\draw   (572.75,141.97) -- (597.13,141.97) -- (597.13,166.39) -- (572.75,166.39) -- cycle ;
\draw   (572.75,72.4) -- (597.5,72.4) -- (597.5,97) -- (572.75,97) -- cycle ;
\draw    (585.58,97.17) -- (585.58,141.62) ;
\draw    (556.26,154.62) -- (572.75,154.62) ;
\draw    (557.73,85.14) -- (572.75,85.23) ;
\draw   (372.07,141.97) -- (396.45,141.97) -- (396.45,166.39) -- (372.07,166.39) -- cycle ;
\draw   (372.07,72.4) -- (396.81,72.4) -- (396.81,97) -- (372.07,97) -- cycle ;
\draw    (384.9,97.17) -- (384.9,141.97) ;
\draw    (357.04,154.53) -- (372.07,154.62) ;
\draw    (357.04,85.14) -- (372.07,85.23) ;
\draw    (395.9,154.62) -- (412.39,154.62) ;
\draw    (397.36,85.14) -- (412.39,85.23) ;
\draw    (597.13,154.53) -- (612.16,154.62) ;
\draw    (598.05,85.14) -- (613.07,85.23) ;
\draw  [color={rgb, 255:red, 0; green, 4; blue, 255 }  ,draw opacity=1 ] (406.89,65.2) -- (441.71,65.2) -- (441.71,171.83) -- (406.89,171.83) -- cycle ;
\draw  [color={rgb, 255:red, 0; green, 4; blue, 255 }  ,draw opacity=1 ] (452.71,181.85) -- (484.78,181.85) -- (484.78,228.4) -- (452.71,228.4) -- cycle ;
\draw    (484.42,194.93) -- (499.45,195.02) ;
\draw    (484.42,214.26) -- (499.45,214.35) ;
\draw    (437.68,194.93) -- (452.71,195.02) ;
\draw    (437.68,214.26) -- (452.71,214.35) ;
\draw    (425.22,130.73) -- (425.22,141.62) ;
\draw   (411.93,106.22) -- (436.67,106.22) -- (436.67,130.81) -- (411.93,130.81) -- cycle ;

\draw (108.31,121.54) node [anchor=north west][inner sep=0.75pt]    {$A$};
\draw (108.01,74.87) node [anchor=north west][inner sep=0.75pt]    {$A^{*}$};
\draw (147.71,121.54) node [anchor=north west][inner sep=0.75pt]    {$A$};
\draw (147.42,74.87) node [anchor=north west][inner sep=0.75pt]    {$A^{*}$};
\draw (188.03,121.54) node [anchor=north west][inner sep=0.75pt]    {$A$};
\draw (187.74,74.87) node [anchor=north west][inner sep=0.75pt]    {$A^{*}$};
\draw (228.35,121.54) node [anchor=north west][inner sep=0.75pt]    {$A$};
\draw (228.06,74.87) node [anchor=north west][inner sep=0.75pt]    {$A^{*}$};
\draw (268.67,121.54) node [anchor=north west][inner sep=0.75pt]    {$A$};
\draw (268.38,74.87) node [anchor=north west][inner sep=0.75pt]    {$A^{*}$};
\draw (67.99,121.54) node [anchor=north west][inner sep=0.75pt]    {$A$};
\draw (67.69,74.87) node [anchor=north west][inner sep=0.75pt]    {$A^{*}$};
\draw (31.16,96.07) node [anchor=north west][inner sep=0.75pt]    {$...$};
\draw (309.74,96.07) node [anchor=north west][inner sep=0.75pt]    {$...$};
\draw (151.21,171.61) node [anchor=north west][inner sep=0.75pt]    {$M$};
\draw (105.18,170.38) node [anchor=north west][inner sep=0.75pt]  [color={rgb, 255:red, 0; green, 4; blue, 0 }  ,opacity=1 ]  {$=$};
\draw (416.2,146.14) node [anchor=north west][inner sep=0.75pt]    {$A$};
\draw (415.91,74.87) node [anchor=north west][inner sep=0.75pt]    {$A^{*}$};
\draw (455.61,146.14) node [anchor=north west][inner sep=0.75pt]    {$A$};
\draw (455.31,74.87) node [anchor=north west][inner sep=0.75pt]    {$A^{*}$};
\draw (495.93,146.14) node [anchor=north west][inner sep=0.75pt]    {$A$};
\draw (495.63,74.87) node [anchor=north west][inner sep=0.75pt]    {$A^{*}$};
\draw (536.25,146.14) node [anchor=north west][inner sep=0.75pt]    {$A$};
\draw (535.95,74.87) node [anchor=north west][inner sep=0.75pt]    {$A^{*}$};
\draw (576.57,146.14) node [anchor=north west][inner sep=0.75pt]    {$A$};
\draw (576.27,74.87) node [anchor=north west][inner sep=0.75pt]    {$A^{*}$};
\draw (375.88,146.14) node [anchor=north west][inner sep=0.75pt]    {$A$};
\draw (375.59,74.87) node [anchor=north west][inner sep=0.75pt]    {$A^{*}$};
\draw (339.06,96.07) node [anchor=north west][inner sep=0.75pt]    {$...$};
\draw (617.64,96.07) node [anchor=north west][inner sep=0.75pt]    {$...$};
\draw (455.16,196.96) node [anchor=north west][inner sep=0.75pt]    {$M_{O}$};
\draw (413.08,194.98) node [anchor=north west][inner sep=0.75pt]  [color={rgb, 255:red, 0; green, 4; blue, 0 }  ,opacity=1 ]  {$=$};
\draw (417.08,111) node [anchor=north west][inner sep=0.75pt]    {$O$};

\end{tikzpicture}

%% file: figures/tikz/cnpoint.tex
\tikzset{every picture/.style={line width=0.75pt}} %

\begin{tikzpicture}[x=0.75pt,y=0.75pt,yscale=-1,xscale=1]

\draw  [draw opacity=0][fill={rgb, 255:red, 221; green, 237; blue, 238 }  ,fill opacity=1 ] (284.08,182.39) -- (299.82,164.13) -- (307.2,170.23) -- (291.46,188.49) -- cycle ;
\draw  [draw opacity=0][fill={rgb, 255:red, 221; green, 237; blue, 238 }  ,fill opacity=1 ] (35.42,157.9) .. controls (35.42,120.22) and (95.62,89.68) .. (169.89,89.68) .. controls (244.15,89.68) and (304.35,120.22) .. (304.35,157.9) .. controls (304.35,195.57) and (244.15,226.12) .. (169.89,226.12) .. controls (95.62,226.12) and (35.42,195.57) .. (35.42,157.9) -- cycle ;
\draw  [draw opacity=0][fill={rgb, 255:red, 221; green, 237; blue, 238 }  ,fill opacity=1 ] (27.45,150.31) .. controls (27.45,116.83) and (91.22,89.68) .. (169.89,89.68) .. controls (248.56,89.68) and (312.33,116.83) .. (312.33,150.31) .. controls (312.33,183.8) and (248.56,210.95) .. (169.89,210.95) .. controls (91.22,210.95) and (27.45,183.8) .. (27.45,150.31) -- cycle ;
\draw  [draw opacity=0][fill={rgb, 255:red, 221; green, 237; blue, 238 }  ,fill opacity=1 ] (27,136.37) .. controls (27,97.6) and (90.77,66.17) .. (169.44,66.17) .. controls (248.11,66.17) and (311.88,97.6) .. (311.88,136.37) .. controls (311.88,175.13) and (248.11,206.56) .. (169.44,206.56) .. controls (90.77,206.56) and (27,175.13) .. (27,136.37) -- cycle ;
\draw  [color={rgb, 255:red, 0; green, 0; blue, 0 }  ,draw opacity=1 ][fill={rgb, 255:red, 255; green, 255; blue, 255 }  ,fill opacity=1 ] (66.99,144.47) .. controls (66.99,117.7) and (113.26,96) .. (170.34,96) .. controls (227.41,96) and (273.68,117.7) .. (273.68,144.47) .. controls (273.68,171.23) and (227.41,192.93) .. (170.34,192.93) .. controls (113.26,192.93) and (66.99,171.23) .. (66.99,144.47) -- cycle ;
\draw  [dash pattern={on 0.84pt off 2.51pt}] (27.45,143) .. controls (27.45,132.74) and (36.3,124.42) .. (47.21,124.42) .. controls (58.12,124.42) and (66.97,132.74) .. (66.97,143) .. controls (66.97,153.26) and (58.12,161.57) .. (47.21,161.57) .. controls (36.3,161.57) and (27.45,153.26) .. (27.45,143) -- cycle ;
\draw  [draw opacity=0] (27.45,143.9) .. controls (27.45,143.9) and (27.45,143.9) .. (27.45,143.9) .. controls (27.45,189.91) and (91.42,227.2) .. (170.34,227.2) .. controls (247.15,227.2) and (309.8,191.87) .. (313.09,147.57) -- (170.34,143.9) -- cycle ; \draw   (27.45,143.9) .. controls (27.45,143.9) and (27.45,143.9) .. (27.45,143.9) .. controls (27.45,189.91) and (91.42,227.2) .. (170.34,227.2) .. controls (247.15,227.2) and (309.8,191.87) .. (313.09,147.57) ;
\draw  [draw opacity=0] (27,141.17) .. controls (27,99.75) and (90.97,66.17) .. (169.89,66.17) .. controls (248.8,66.17) and (312.78,99.75) .. (312.78,141.17) .. controls (312.78,142.3) and (312.73,143.43) .. (312.63,144.56) -- (169.89,141.17) -- cycle ; \draw   (27,141.17) .. controls (27,99.75) and (90.97,66.17) .. (169.89,66.17) .. controls (248.8,66.17) and (312.78,99.75) .. (312.78,141.17) .. controls (312.78,142.3) and (312.73,143.43) .. (312.63,144.56) ;
\draw  [dash pattern={on 4.5pt off 4.5pt}]  (25.96,143.75) .. controls (28.28,121.52) and (45.28,104.61) .. (70.26,93.12) .. controls (96.84,80.9) and (132.41,74.82) .. (168.15,74.82) .. controls (168.79,74.82) and (169.43,74.82) .. (170.07,74.83) .. controls (229.19,75.19) and (288.07,92.26) .. (307.57,124.99) .. controls (311.09,130.9) and (313.35,137.31) .. (314.14,144.23)(28.94,144.06) .. controls (31.16,122.77) and (47.71,106.79) .. (71.51,95.85) .. controls (97.76,83.78) and (132.88,77.82) .. (168.15,77.82) .. controls (168.78,77.82) and (169.42,77.82) .. (170.05,77.83) .. controls (227.83,78.18) and (285.85,94.41) .. (305,126.53) .. controls (308.3,132.07) and (310.42,138.08) .. (311.16,144.57) ;
\draw    (28.88,143.44) .. controls (40.52,179.41) and (77.71,201.87) .. (121.35,211.42) .. controls (136.8,214.8) and (153.07,216.56) .. (169.34,216.72) .. controls (170.13,216.73) and (170.93,216.73) .. (171.73,216.73) .. controls (227.04,216.73) and (282.04,198.26) .. (303.84,161.5) .. controls (307.06,156.07) and (309.55,150.24) .. (311.2,144.01)(26.02,144.36) .. controls (37.95,181.22) and (75.85,204.53) .. (120.71,214.35) .. controls (136.35,217.78) and (152.83,219.56) .. (169.31,219.72) .. controls (170.11,219.73) and (170.92,219.73) .. (171.73,219.73) .. controls (228.28,219.73) and (284.19,200.51) .. (306.42,163.03) .. controls (309.78,157.36) and (312.37,151.28) .. (314.1,144.78) ;
\draw  [dash pattern={on 0.84pt off 2.51pt}] (273.68,144.47) .. controls (273.68,133.9) and (282.53,125.33) .. (293.45,125.33) .. controls (304.36,125.33) and (313.21,133.9) .. (313.21,144.47) .. controls (313.21,155.04) and (304.36,163.61) .. (293.45,163.61) .. controls (282.53,163.61) and (273.68,155.04) .. (273.68,144.47) -- cycle ;
\draw  [draw opacity=0][fill={rgb, 255:red, 221; green, 237; blue, 238 }  ,fill opacity=1 ] (83.53,109.38) .. controls (78.53,104.49) and (99.95,99.6) .. (96.38,105.88) .. controls (92.81,112.17) and (84.24,170.18) .. (85.67,177.86) .. controls (87.1,185.55) and (72.11,178.56) .. (75.68,172.97) .. controls (79.25,167.38) and (88.53,114.27) .. (83.53,109.38) -- cycle ;
\draw    (83.53,109.38) .. controls (88.53,118.46) and (80.68,163.19) .. (75.68,172.97) ;
\draw    (96.38,105.88) .. controls (91.38,116.65) and (84.96,164.59) .. (85.67,177.86) ;
\draw  [draw opacity=0][fill={rgb, 255:red, 221; green, 237; blue, 238 }  ,fill opacity=1 ] (238.42,100.99) .. controls (236.13,95.25) and (254.39,100.74) .. (249.13,104.49) .. controls (243.87,108.23) and (225.71,136.34) .. (224.03,143.02) .. controls (222.36,149.7) and (209.51,141.08) .. (214.51,137.89) .. controls (219.51,134.7) and (240.71,106.73) .. (238.42,100.99) -- cycle ;
\draw  [draw opacity=0][fill={rgb, 255:red, 221; green, 237; blue, 238 }  ,fill opacity=1 ] (212.2,141.15) .. controls (209.91,135.41) and (227.98,140.43) .. (222.72,144.18) .. controls (217.46,147.93) and (190.25,190.17) .. (188.57,196.85) .. controls (186.89,203.53) and (172.75,199.5) .. (177.75,196.31) .. controls (182.75,193.13) and (214.48,146.89) .. (212.2,141.15) -- cycle ;
\draw  [draw opacity=0][fill={rgb, 255:red, 221; green, 237; blue, 238 }  ,fill opacity=1 ] (251.27,177.86) .. controls (257.6,178.27) and (242.06,188.97) .. (240.56,182.76) .. controls (239.07,176.54) and (220.06,149.76) .. (214.38,145.46) .. controls (208.7,141.16) and (222.97,137.27) .. (224.03,143.02) .. controls (225.1,148.77) and (244.94,177.46) .. (251.27,177.86) -- cycle ;
\draw  [draw opacity=0][fill={rgb, 255:red, 221; green, 237; blue, 238 }  ,fill opacity=1 ] (220.78,135.82) .. controls (227.11,136.22) and (215.87,151.68) .. (214.38,145.46) .. controls (212.89,139.24) and (182.91,95.78) .. (177.23,91.48) .. controls (171.56,87.18) and (185.82,83.29) .. (186.89,89.04) .. controls (187.95,94.79) and (214.44,135.41) .. (220.78,135.82) -- cycle ;
\draw    (249.13,104.49) .. controls (237.35,119.16) and (224.14,137.61) .. (224.03,143.02) .. controls (223.92,148.43) and (236.99,167.38) .. (252.7,181.36) ;
\draw    (177.23,91.48) .. controls (187.03,102.67) and (212.72,135.24) .. (212.01,142.92) .. controls (211.3,150.61) and (186.46,184.99) .. (177.89,198.27) ;
\draw    (187.74,90.79) .. controls (193.45,102.67) and (214.15,134.12) .. (217.72,134.12) .. controls (221.29,134.12) and (235.57,112.03) .. (239.85,100.57) ;
\draw    (188.57,196.85) .. controls (196.31,181.36) and (214.87,149.91) .. (218.43,149.91) .. controls (222,149.91) and (239.99,178.7) .. (242.13,184.29) ;

\draw  [draw opacity=0][fill={rgb, 255:red, 221; green, 237; blue, 238 }  ,fill opacity=1 ] (420.12,114.91) .. controls (418.14,110.16) and (436.06,111.95) .. (431.51,115.05) .. controls (426.95,118.15) and (409.11,144.15) .. (407.66,149.67) .. controls (406.21,155.2) and (395.08,148.07) .. (399.41,145.43) .. controls (403.74,142.8) and (422.1,119.66) .. (420.12,114.91) -- cycle ;
\draw  [draw opacity=0][fill={rgb, 255:red, 221; green, 237; blue, 238 }  ,fill opacity=1 ] (397.41,148.13) .. controls (395.43,143.38) and (411.07,147.53) .. (406.52,150.63) .. controls (401.96,153.73) and (383.13,178.72) .. (381.67,184.25) .. controls (380.22,189.78) and (368.44,185.39) .. (372.77,182.75) .. controls (377.1,180.12) and (399.39,152.88) .. (397.41,148.13) -- cycle ;
\draw  [draw opacity=0][fill={rgb, 255:red, 221; green, 237; blue, 238 }  ,fill opacity=1 ] (434.74,181.71) .. controls (439.6,183.95) and (423.27,187.68) .. (421.97,182.54) .. controls (420.68,177.4) and (404.22,155.25) .. (399.3,151.69) .. controls (394.38,148.13) and (406.74,144.92) .. (407.66,149.67) .. controls (408.58,154.43) and (429.89,179.46) .. (434.74,181.71) -- cycle ;
\draw  [draw opacity=0][fill={rgb, 255:red, 221; green, 237; blue, 238 }  ,fill opacity=1 ] (404.84,143.72) .. controls (410.32,144.05) and (400.59,156.83) .. (399.3,151.69) .. controls (398,146.55) and (376.07,118.91) .. (371.16,115.35) .. controls (366.24,111.8) and (379.13,109.1) .. (380.05,113.85) .. controls (380.98,118.61) and (399.35,143.38) .. (404.84,143.72) -- cycle ;
\draw  [draw opacity=0][fill={rgb, 255:red, 255; green, 255; blue, 255 }  ,fill opacity=1 ] (413.29,104.35) -- (437.57,103.96) -- (437.77,114.92) -- (413.5,115.31) -- cycle ;
\draw  [draw opacity=0][fill={rgb, 255:red, 255; green, 255; blue, 255 }  ,fill opacity=1 ] (361.52,103.6) -- (385.79,103.21) -- (385.99,114.17) -- (361.72,114.56) -- cycle ;
\draw  [draw opacity=0][fill={rgb, 255:red, 255; green, 255; blue, 255 }  ,fill opacity=1 ] (364.55,181.77) -- (388.82,181.38) -- (389.03,192.34) -- (364.75,192.73) -- cycle ;
\draw  [draw opacity=0][fill={rgb, 255:red, 255; green, 255; blue, 255 }  ,fill opacity=1 ] (414.91,181.49) -- (439.18,181.1) -- (439.39,192.06) -- (415.12,192.45) -- cycle ;
\draw    (368.4,112.81) .. controls (376.89,122.06) and (398.15,143.89) .. (397.53,150.25) .. controls (396.91,156.61) and (382.16,173.47) .. (370.83,184.7) ;
\draw    (380.05,113.85) .. controls (385,123.68) and (399.1,142.31) .. (402.19,142.31) .. controls (405.28,142.31) and (417.65,124.04) .. (421.36,114.56) ;
\draw    (382.16,183.2) .. controls (388.86,170.39) and (399.72,155.37) .. (402.81,155.37) .. controls (405.9,155.37) and (420.75,177.83) .. (422.61,182.46) ;
\draw    (433.13,112.81) .. controls (422.93,124.94) and (407.76,145.2) .. (407.66,149.67) .. controls (407.56,154.15) and (422.61,173.47) .. (436.36,183.2) ;

\draw  [draw opacity=0][fill={rgb, 255:red, 221; green, 237; blue, 238 }  ,fill opacity=1 ] (513.4,113.41) .. controls (511.42,108.67) and (529.34,110.46) .. (524.79,113.56) .. controls (520.23,116.65) and (502.39,142.65) .. (500.94,148.18) .. controls (499.48,153.7) and (488.36,146.57) .. (492.69,143.93) .. controls (497.02,141.3) and (515.38,118.16) .. (513.4,113.41) -- cycle ;
\draw  [draw opacity=0][fill={rgb, 255:red, 221; green, 237; blue, 238 }  ,fill opacity=1 ] (490.69,146.63) .. controls (488.71,141.88) and (504.35,146.04) .. (499.8,149.14) .. controls (495.24,152.23) and (476.4,177.23) .. (474.95,182.75) .. controls (473.5,188.28) and (461.72,183.89) .. (466.05,181.26) .. controls (470.38,178.62) and (492.67,151.38) .. (490.69,146.63) -- cycle ;
\draw  [draw opacity=0][fill={rgb, 255:red, 221; green, 237; blue, 238 }  ,fill opacity=1 ] (528.02,180.21) .. controls (532.88,182.46) and (516.55,186.18) .. (515.25,181.04) .. controls (513.96,175.9) and (497.49,153.75) .. (492.58,150.19) .. controls (487.66,146.64) and (500.02,143.42) .. (500.94,148.18) .. controls (501.86,152.93) and (523.17,177.96) .. (528.02,180.21) -- cycle ;
\draw  [draw opacity=0][fill={rgb, 255:red, 221; green, 237; blue, 238 }  ,fill opacity=1 ] (498.12,142.22) .. controls (503.6,142.55) and (493.87,155.34) .. (492.58,150.19) .. controls (491.28,145.05) and (469.35,117.41) .. (464.43,113.85) .. controls (459.52,110.3) and (472.41,107.6) .. (473.33,112.36) .. controls (474.25,117.11) and (492.63,141.88) .. (498.12,142.22) -- cycle ;
\draw  [draw opacity=0][fill={rgb, 255:red, 255; green, 255; blue, 255 }  ,fill opacity=1 ] (506.57,102.86) -- (530.84,102.47) -- (531.05,113.42) -- (506.78,113.81) -- cycle ;
\draw  [draw opacity=0][fill={rgb, 255:red, 255; green, 255; blue, 255 }  ,fill opacity=1 ] (454.8,102.11) -- (479.07,101.72) -- (479.27,112.67) -- (455,113.06) -- cycle ;
\draw  [draw opacity=0][fill={rgb, 255:red, 255; green, 255; blue, 255 }  ,fill opacity=1 ] (459.65,180.74) -- (483.92,180.35) -- (484.13,191.31) -- (459.85,191.7) -- cycle ;
\draw  [draw opacity=0][fill={rgb, 255:red, 255; green, 255; blue, 255 }  ,fill opacity=1 ] (508.19,179.99) -- (532.46,179.6) -- (532.67,190.56) -- (508.39,190.95) -- cycle ;
\draw    (461.68,111.31) .. controls (470.17,120.56) and (491.43,142.4) .. (490.81,148.75) .. controls (490.19,155.11) and (475.44,171.97) .. (464.11,183.2) ;
\draw    (526.4,111.31) .. controls (516.2,123.45) and (501.03,143.7) .. (500.94,148.18) .. controls (500.84,152.65) and (515.89,171.97) .. (529.64,181.71) ;
\draw  [draw opacity=0][fill={rgb, 255:red, 221; green, 237; blue, 238 }  ,fill opacity=1 ] (502.94,142.31) .. controls (496.47,140.82) and (498.9,133.93) .. (506.18,135.42) .. controls (513.46,136.92) and (514.27,156.39) .. (507.8,160.14) .. controls (501.32,163.88) and (496.47,155.04) .. (502.13,154.3) .. controls (507.8,153.55) and (509.41,143.81) .. (502.94,142.31) -- cycle ;
\draw    (473.33,112.36) .. controls (478.28,122.18) and (492.38,140.81) .. (495.47,140.81) .. controls (498.56,140.81) and (510.92,122.55) .. (514.63,113.07) ;
\draw    (475.44,181.71) .. controls (482.14,168.89) and (493,153.88) .. (496.09,153.88) .. controls (499.18,153.88) and (514.03,176.33) .. (515.89,180.96) ;
\draw    (506.18,134.37) .. controls (514.27,140.82) and (511.84,157.29) .. (506.99,161.34) ;
\draw    (502.13,140.37) .. controls (508.61,144.86) and (508.61,150.85) .. (502.13,155.34) ;

\draw  [draw opacity=0][fill={rgb, 255:red, 221; green, 237; blue, 238 }  ,fill opacity=1 ] (608.4,111.41) .. controls (606.42,106.67) and (624.34,108.46) .. (619.79,111.56) .. controls (615.23,114.65) and (597.39,140.65) .. (595.94,146.18) .. controls (594.48,151.7) and (583.36,144.57) .. (587.69,141.93) .. controls (592.02,139.3) and (610.38,116.16) .. (608.4,111.41) -- cycle ;
\draw  [draw opacity=0][fill={rgb, 255:red, 221; green, 237; blue, 238 }  ,fill opacity=1 ] (585.69,144.63) .. controls (583.71,139.88) and (599.35,144.04) .. (594.8,147.14) .. controls (590.24,150.23) and (571.4,175.23) .. (569.95,180.75) .. controls (568.5,186.28) and (556.72,181.89) .. (561.05,179.26) .. controls (565.38,176.62) and (587.67,149.38) .. (585.69,144.63) -- cycle ;
\draw  [draw opacity=0][fill={rgb, 255:red, 221; green, 237; blue, 238 }  ,fill opacity=1 ] (623.02,178.21) .. controls (627.88,180.46) and (611.55,184.18) .. (610.25,179.04) .. controls (608.96,173.9) and (592.49,151.75) .. (587.58,148.19) .. controls (582.66,144.64) and (595.02,141.42) .. (595.94,146.18) .. controls (596.86,150.93) and (618.17,175.96) .. (623.02,178.21) -- cycle ;
\draw  [draw opacity=0][fill={rgb, 255:red, 221; green, 237; blue, 238 }  ,fill opacity=1 ] (593.12,140.22) .. controls (598.6,140.55) and (588.87,153.34) .. (587.58,148.19) .. controls (586.28,143.05) and (564.35,115.41) .. (559.43,111.85) .. controls (554.52,108.3) and (567.41,105.6) .. (568.33,110.36) .. controls (569.25,115.11) and (587.63,139.88) .. (593.12,140.22) -- cycle ;
\draw  [draw opacity=0][fill={rgb, 255:red, 255; green, 255; blue, 255 }  ,fill opacity=1 ] (601.57,100.86) -- (625.84,100.47) -- (626.05,111.42) -- (601.78,111.81) -- cycle ;
\draw  [draw opacity=0][fill={rgb, 255:red, 255; green, 255; blue, 255 }  ,fill opacity=1 ] (549.8,100.11) -- (574.07,99.72) -- (574.27,110.67) -- (550,111.06) -- cycle ;
\draw  [draw opacity=0][fill={rgb, 255:red, 255; green, 255; blue, 255 }  ,fill opacity=1 ] (554.65,178.74) -- (578.92,178.35) -- (579.13,189.31) -- (554.85,189.7) -- cycle ;
\draw  [draw opacity=0][fill={rgb, 255:red, 255; green, 255; blue, 255 }  ,fill opacity=1 ] (603.19,177.99) -- (627.46,177.6) -- (627.67,188.56) -- (603.39,188.95) -- cycle ;
\draw    (556.68,109.31) .. controls (565.17,118.56) and (586.43,140.4) .. (585.81,146.75) .. controls (585.19,153.11) and (570.44,169.97) .. (559.11,181.2) ;
\draw    (621.4,109.31) .. controls (611.2,121.45) and (596.03,141.7) .. (595.94,146.18) .. controls (595.84,150.65) and (610.89,169.97) .. (624.64,179.71) ;
\draw  [draw opacity=0][fill={rgb, 255:red, 221; green, 237; blue, 238 }  ,fill opacity=1 ] (597.94,140.31) .. controls (591.47,138.82) and (593.9,131.93) .. (601.18,133.42) .. controls (608.46,134.92) and (609.27,154.39) .. (602.8,158.14) .. controls (596.32,161.88) and (591.47,153.04) .. (597.13,152.3) .. controls (602.8,151.55) and (604.41,141.81) .. (597.94,140.31) -- cycle ;
\draw    (568.33,110.36) .. controls (573.28,120.18) and (587.38,138.81) .. (590.47,138.81) .. controls (593.56,138.81) and (605.92,120.55) .. (609.63,111.07) ;
\draw    (570.44,179.71) .. controls (577.14,166.89) and (588,151.88) .. (591.09,151.88) .. controls (594.18,151.88) and (609.03,174.33) .. (610.89,178.96) ;
\draw    (601.18,132.37) .. controls (609.27,138.82) and (606.84,155.29) .. (601.99,159.34) ;
\draw    (597.13,138.37) .. controls (603.61,142.86) and (603.61,148.85) .. (597.13,153.34) ;
\draw  [draw opacity=0][fill={rgb, 255:red, 221; green, 237; blue, 238 }  ,fill opacity=1 ] (581.85,140.31) .. controls (588.32,138.82) and (585.89,131.93) .. (578.61,133.42) .. controls (571.33,134.92) and (570.52,154.39) .. (576.99,158.14) .. controls (583.46,161.88) and (588.32,153.04) .. (582.65,152.3) .. controls (576.99,151.55) and (575.37,141.81) .. (581.85,140.31) -- cycle ;
\draw    (580.2,132.6) .. controls (569,137.2) and (571,157.2) .. (580,159.2) ;
\draw    (584.2,138.6) .. controls (577.2,141.6) and (576.2,149.6) .. (584.2,153.6) ;

\draw (365.59,68.39) node [anchor=north west][inner sep=0.75pt]    {$C_{ }{}_{4\text{\mbox{-}point}} =\ $};
\draw (443.06,137.27) node [anchor=north west][inner sep=0.75pt]    {$+\ $};
\draw (563.72,205.79) node [anchor=north west][inner sep=0.75pt]    {$+\ ...\ $};
\draw (539.06,135.27) node [anchor=north west][inner sep=0.75pt]    {$+\ $};

\end{tikzpicture}

%% file: figures/tikz/conpoint.tex
\tikzset{every picture/.style={line width=0.75pt}} %

\begin{tikzpicture}[x=0.75pt,y=0.75pt,yscale=-1,xscale=1]

\draw  [draw opacity=0][fill={rgb, 255:red, 221; green, 237; blue, 238 }  ,fill opacity=1 ] (294.08,194.39) -- (309.82,176.13) -- (317.2,182.23) -- (301.46,200.49) -- cycle ;
\draw  [draw opacity=0][fill={rgb, 255:red, 221; green, 237; blue, 238 }  ,fill opacity=1 ] (45.42,169.9) .. controls (45.42,132.22) and (105.62,101.68) .. (179.89,101.68) .. controls (254.15,101.68) and (314.35,132.22) .. (314.35,169.9) .. controls (314.35,207.57) and (254.15,238.12) .. (179.89,238.12) .. controls (105.62,238.12) and (45.42,207.57) .. (45.42,169.9) -- cycle ;
\draw  [draw opacity=0][fill={rgb, 255:red, 221; green, 237; blue, 238 }  ,fill opacity=1 ] (37.45,162.31) .. controls (37.45,128.83) and (101.22,101.68) .. (179.89,101.68) .. controls (258.56,101.68) and (322.33,128.83) .. (322.33,162.31) .. controls (322.33,195.8) and (258.56,222.95) .. (179.89,222.95) .. controls (101.22,222.95) and (37.45,195.8) .. (37.45,162.31) -- cycle ;
\draw  [draw opacity=0][fill={rgb, 255:red, 221; green, 237; blue, 238 }  ,fill opacity=1 ] (37,148.37) .. controls (37,109.6) and (100.77,78.17) .. (179.44,78.17) .. controls (258.11,78.17) and (321.88,109.6) .. (321.88,148.37) .. controls (321.88,187.13) and (258.11,218.56) .. (179.44,218.56) .. controls (100.77,218.56) and (37,187.13) .. (37,148.37) -- cycle ;
\draw  [color={rgb, 255:red, 0; green, 0; blue, 0 }  ,draw opacity=1 ][fill={rgb, 255:red, 255; green, 255; blue, 255 }  ,fill opacity=1 ] (76.99,156.47) .. controls (76.99,129.7) and (123.26,108) .. (180.34,108) .. controls (237.41,108) and (283.68,129.7) .. (283.68,156.47) .. controls (283.68,183.23) and (237.41,204.93) .. (180.34,204.93) .. controls (123.26,204.93) and (76.99,183.23) .. (76.99,156.47) -- cycle ;
\draw  [dash pattern={on 0.84pt off 2.51pt}] (37.45,155) .. controls (37.45,144.74) and (46.3,136.42) .. (57.21,136.42) .. controls (68.12,136.42) and (76.97,144.74) .. (76.97,155) .. controls (76.97,165.26) and (68.12,173.57) .. (57.21,173.57) .. controls (46.3,173.57) and (37.45,165.26) .. (37.45,155) -- cycle ;
\draw  [draw opacity=0] (37.45,155.9) .. controls (37.45,155.9) and (37.45,155.9) .. (37.45,155.9) .. controls (37.45,201.91) and (101.42,239.2) .. (180.34,239.2) .. controls (257.15,239.2) and (319.8,203.87) .. (323.09,159.57) -- (180.34,155.9) -- cycle ; \draw   (37.45,155.9) .. controls (37.45,155.9) and (37.45,155.9) .. (37.45,155.9) .. controls (37.45,201.91) and (101.42,239.2) .. (180.34,239.2) .. controls (257.15,239.2) and (319.8,203.87) .. (323.09,159.57) ;
\draw  [draw opacity=0] (37,153.17) .. controls (37,111.75) and (100.97,78.17) .. (179.89,78.17) .. controls (258.8,78.17) and (322.78,111.75) .. (322.78,153.17) .. controls (322.78,154.3) and (322.73,155.43) .. (322.63,156.56) -- (179.89,153.17) -- cycle ; \draw   (37,153.17) .. controls (37,111.75) and (100.97,78.17) .. (179.89,78.17) .. controls (258.8,78.17) and (322.78,111.75) .. (322.78,153.17) .. controls (322.78,154.3) and (322.73,155.43) .. (322.63,156.56) ;
\draw  [dash pattern={on 4.5pt off 4.5pt}]  (35.96,155.75) .. controls (38.28,133.52) and (55.28,116.61) .. (80.26,105.12) .. controls (106.84,92.9) and (142.41,86.82) .. (178.15,86.82) .. controls (178.79,86.82) and (179.43,86.82) .. (180.07,86.83) .. controls (239.19,87.19) and (298.07,104.26) .. (317.57,136.99) .. controls (321.09,142.9) and (323.35,149.31) .. (324.14,156.23)(38.94,156.06) .. controls (41.16,134.77) and (57.71,118.79) .. (81.51,107.85) .. controls (107.76,95.78) and (142.88,89.82) .. (178.15,89.82) .. controls (178.78,89.82) and (179.42,89.82) .. (180.05,89.83) .. controls (237.83,90.18) and (295.85,106.41) .. (315,138.53) .. controls (318.3,144.07) and (320.42,150.08) .. (321.16,156.57) ;
\draw    (38.88,155.44) .. controls (50.52,191.41) and (87.71,213.87) .. (131.35,223.42) .. controls (146.8,226.8) and (163.07,228.56) .. (179.34,228.72) .. controls (180.13,228.73) and (180.93,228.73) .. (181.73,228.73) .. controls (237.04,228.73) and (292.04,210.26) .. (313.84,173.5) .. controls (317.06,168.07) and (319.55,162.24) .. (321.2,156.01)(36.02,156.36) .. controls (47.95,193.22) and (85.85,216.53) .. (130.71,226.35) .. controls (146.35,229.78) and (162.83,231.56) .. (179.31,231.72) .. controls (180.11,231.73) and (180.92,231.73) .. (181.73,231.73) .. controls (238.28,231.73) and (294.19,212.51) .. (316.42,175.03) .. controls (319.78,169.36) and (322.37,163.28) .. (324.1,156.78) ;
\draw  [dash pattern={on 0.84pt off 2.51pt}] (283.68,156.47) .. controls (283.68,145.9) and (292.53,137.33) .. (303.45,137.33) .. controls (314.36,137.33) and (323.21,145.9) .. (323.21,156.47) .. controls (323.21,167.04) and (314.36,175.61) .. (303.45,175.61) .. controls (292.53,175.61) and (283.68,167.04) .. (283.68,156.47) -- cycle ;
\draw  [draw opacity=0][fill={rgb, 255:red, 221; green, 237; blue, 238 }  ,fill opacity=1 ] (93.53,121.38) .. controls (88.53,116.49) and (109.95,111.6) .. (106.38,117.88) .. controls (102.81,124.17) and (94.24,182.18) .. (95.67,189.86) .. controls (97.1,197.55) and (82.11,190.56) .. (85.68,184.97) .. controls (89.25,179.38) and (98.53,126.27) .. (93.53,121.38) -- cycle ;
\draw    (93.53,121.38) .. controls (98.53,130.46) and (90.68,175.19) .. (85.68,184.97) ;
\draw    (106.38,117.88) .. controls (101.38,128.65) and (94.96,176.59) .. (95.67,189.86) ;
\draw  [draw opacity=0][fill={rgb, 255:red, 221; green, 237; blue, 238 }  ,fill opacity=1 ] (248.42,112.99) .. controls (246.13,107.25) and (264.39,112.74) .. (259.13,116.49) .. controls (253.87,120.23) and (235.71,148.34) .. (234.03,155.02) .. controls (232.36,161.7) and (219.51,153.08) .. (224.51,149.89) .. controls (229.51,146.7) and (250.71,118.73) .. (248.42,112.99) -- cycle ;
\draw  [draw opacity=0][fill={rgb, 255:red, 221; green, 237; blue, 238 }  ,fill opacity=1 ] (222.2,153.15) .. controls (219.91,147.41) and (237.98,152.43) .. (232.72,156.18) .. controls (227.46,159.93) and (200.25,202.17) .. (198.57,208.85) .. controls (196.89,215.53) and (182.75,211.5) .. (187.75,208.31) .. controls (192.75,205.13) and (224.48,158.89) .. (222.2,153.15) -- cycle ;
\draw  [draw opacity=0][fill={rgb, 255:red, 221; green, 237; blue, 238 }  ,fill opacity=1 ] (261.27,189.86) .. controls (267.6,190.27) and (252.06,200.97) .. (250.56,194.76) .. controls (249.07,188.54) and (230.06,161.76) .. (224.38,157.46) .. controls (218.7,153.16) and (232.97,149.27) .. (234.03,155.02) .. controls (235.1,160.77) and (254.94,189.46) .. (261.27,189.86) -- cycle ;
\draw  [draw opacity=0][fill={rgb, 255:red, 221; green, 237; blue, 238 }  ,fill opacity=1 ] (230.78,147.82) .. controls (237.11,148.22) and (225.87,163.68) .. (224.38,157.46) .. controls (222.89,151.24) and (192.91,107.78) .. (187.23,103.48) .. controls (181.56,99.18) and (195.82,95.29) .. (196.89,101.04) .. controls (197.95,106.79) and (224.44,147.41) .. (230.78,147.82) -- cycle ;
\draw    (259.13,116.49) .. controls (247.35,131.16) and (234.14,149.61) .. (234.03,155.02) .. controls (233.92,160.43) and (246.99,179.38) .. (262.7,193.36) ;
\draw    (187.23,103.48) .. controls (197.03,114.67) and (222.72,147.24) .. (222.01,154.92) .. controls (221.3,162.61) and (196.46,196.99) .. (187.89,210.27) ;
\draw    (197.74,102.79) .. controls (203.45,114.67) and (224.15,146.12) .. (227.72,146.12) .. controls (231.29,146.12) and (245.57,124.03) .. (249.85,112.57) ;
\draw    (198.57,208.85) .. controls (206.31,193.36) and (224.87,161.91) .. (228.43,161.91) .. controls (232,161.91) and (249.99,190.7) .. (252.13,196.29) ;

\draw  [draw opacity=0][fill={rgb, 255:red, 221; green, 237; blue, 238 }  ,fill opacity=1 ] (430.12,126.91) .. controls (428.14,122.16) and (446.06,123.95) .. (441.51,127.05) .. controls (436.95,130.15) and (419.11,156.15) .. (417.66,161.67) .. controls (416.21,167.2) and (405.08,160.07) .. (409.41,157.43) .. controls (413.74,154.8) and (432.1,131.66) .. (430.12,126.91) -- cycle ;
\draw  [draw opacity=0][fill={rgb, 255:red, 221; green, 237; blue, 238 }  ,fill opacity=1 ] (407.41,160.13) .. controls (405.43,155.38) and (421.07,159.53) .. (416.52,162.63) .. controls (411.96,165.73) and (393.13,190.72) .. (391.67,196.25) .. controls (390.22,201.78) and (378.44,197.39) .. (382.77,194.75) .. controls (387.1,192.12) and (409.39,164.88) .. (407.41,160.13) -- cycle ;
\draw  [draw opacity=0][fill={rgb, 255:red, 221; green, 237; blue, 238 }  ,fill opacity=1 ] (444.74,193.71) .. controls (449.6,195.95) and (433.27,199.68) .. (431.97,194.54) .. controls (430.68,189.4) and (414.22,167.25) .. (409.3,163.69) .. controls (404.38,160.13) and (416.74,156.92) .. (417.66,161.67) .. controls (418.58,166.43) and (439.89,191.46) .. (444.74,193.71) -- cycle ;
\draw  [draw opacity=0][fill={rgb, 255:red, 221; green, 237; blue, 238 }  ,fill opacity=1 ] (414.84,155.72) .. controls (420.32,156.05) and (410.59,168.83) .. (409.3,163.69) .. controls (408,158.55) and (386.07,130.91) .. (381.16,127.35) .. controls (376.24,123.8) and (389.13,121.1) .. (390.05,125.85) .. controls (390.98,130.61) and (409.35,155.38) .. (414.84,155.72) -- cycle ;
\draw  [draw opacity=0][fill={rgb, 255:red, 255; green, 255; blue, 255 }  ,fill opacity=1 ] (423.29,116.35) -- (447.57,115.96) -- (447.77,126.92) -- (423.5,127.31) -- cycle ;
\draw  [draw opacity=0][fill={rgb, 255:red, 255; green, 255; blue, 255 }  ,fill opacity=1 ] (371.52,115.6) -- (395.79,115.21) -- (395.99,126.17) -- (371.72,126.56) -- cycle ;
\draw  [draw opacity=0][fill={rgb, 255:red, 255; green, 255; blue, 255 }  ,fill opacity=1 ] (374.55,193.77) -- (398.82,193.38) -- (399.03,204.34) -- (374.75,204.73) -- cycle ;
\draw  [draw opacity=0][fill={rgb, 255:red, 255; green, 255; blue, 255 }  ,fill opacity=1 ] (424.91,193.49) -- (449.18,193.1) -- (449.39,204.06) -- (425.12,204.45) -- cycle ;
\draw    (378.4,124.81) .. controls (386.89,134.06) and (408.15,155.89) .. (407.53,162.25) .. controls (406.91,168.61) and (392.16,185.47) .. (380.83,196.7) ;
\draw    (390.05,125.85) .. controls (395,135.68) and (409.1,154.31) .. (412.19,154.31) .. controls (415.28,154.31) and (427.65,136.04) .. (431.36,126.56) ;
\draw    (392.16,195.2) .. controls (398.86,182.39) and (409.72,167.37) .. (412.81,167.37) .. controls (415.9,167.37) and (430.75,189.83) .. (432.61,194.46) ;
\draw    (443.13,124.81) .. controls (432.93,136.94) and (417.76,157.2) .. (417.66,161.67) .. controls (417.56,166.15) and (432.61,185.47) .. (446.36,195.2) ;

\draw  [draw opacity=0][fill={rgb, 255:red, 221; green, 237; blue, 238 }  ,fill opacity=1 ] (523.4,125.41) .. controls (521.42,120.67) and (539.34,122.46) .. (534.79,125.56) .. controls (530.23,128.65) and (512.39,154.65) .. (510.94,160.18) .. controls (509.48,165.7) and (498.36,158.57) .. (502.69,155.93) .. controls (507.02,153.3) and (525.38,130.16) .. (523.4,125.41) -- cycle ;
\draw  [draw opacity=0][fill={rgb, 255:red, 221; green, 237; blue, 238 }  ,fill opacity=1 ] (500.69,158.63) .. controls (498.71,153.88) and (514.35,158.04) .. (509.8,161.14) .. controls (505.24,164.23) and (486.4,189.23) .. (484.95,194.75) .. controls (483.5,200.28) and (471.72,195.89) .. (476.05,193.26) .. controls (480.38,190.62) and (502.67,163.38) .. (500.69,158.63) -- cycle ;
\draw  [draw opacity=0][fill={rgb, 255:red, 221; green, 237; blue, 238 }  ,fill opacity=1 ] (538.02,192.21) .. controls (542.88,194.46) and (526.55,198.18) .. (525.25,193.04) .. controls (523.96,187.9) and (507.49,165.75) .. (502.58,162.19) .. controls (497.66,158.64) and (510.02,155.42) .. (510.94,160.18) .. controls (511.86,164.93) and (533.17,189.96) .. (538.02,192.21) -- cycle ;
\draw  [draw opacity=0][fill={rgb, 255:red, 221; green, 237; blue, 238 }  ,fill opacity=1 ] (508.12,154.22) .. controls (513.6,154.55) and (503.87,167.34) .. (502.58,162.19) .. controls (501.28,157.05) and (479.35,129.41) .. (474.43,125.85) .. controls (469.52,122.3) and (482.41,119.6) .. (483.33,124.36) .. controls (484.25,129.11) and (502.63,153.88) .. (508.12,154.22) -- cycle ;
\draw  [draw opacity=0][fill={rgb, 255:red, 255; green, 255; blue, 255 }  ,fill opacity=1 ] (516.57,114.86) -- (540.84,114.47) -- (541.05,125.42) -- (516.78,125.81) -- cycle ;
\draw  [draw opacity=0][fill={rgb, 255:red, 255; green, 255; blue, 255 }  ,fill opacity=1 ] (464.8,114.11) -- (489.07,113.72) -- (489.27,124.67) -- (465,125.06) -- cycle ;
\draw  [draw opacity=0][fill={rgb, 255:red, 255; green, 255; blue, 255 }  ,fill opacity=1 ] (469.65,192.74) -- (493.92,192.35) -- (494.13,203.31) -- (469.85,203.7) -- cycle ;
\draw  [draw opacity=0][fill={rgb, 255:red, 255; green, 255; blue, 255 }  ,fill opacity=1 ] (518.19,191.99) -- (542.46,191.6) -- (542.67,202.56) -- (518.39,202.95) -- cycle ;
\draw    (471.68,123.31) .. controls (480.17,132.56) and (501.43,154.4) .. (500.81,160.75) .. controls (500.19,167.11) and (485.44,183.97) .. (474.11,195.2) ;
\draw    (536.4,123.31) .. controls (526.2,135.45) and (511.03,155.7) .. (510.94,160.18) .. controls (510.84,164.65) and (525.89,183.97) .. (539.64,193.71) ;
\draw  [draw opacity=0][fill={rgb, 255:red, 221; green, 237; blue, 238 }  ,fill opacity=1 ] (512.94,154.31) .. controls (506.47,152.82) and (508.9,145.93) .. (516.18,147.42) .. controls (523.46,148.92) and (524.27,168.39) .. (517.8,172.14) .. controls (511.32,175.88) and (506.47,167.04) .. (512.13,166.3) .. controls (517.8,165.55) and (519.41,155.81) .. (512.94,154.31) -- cycle ;
\draw    (483.33,124.36) .. controls (488.28,134.18) and (502.38,152.81) .. (505.47,152.81) .. controls (508.56,152.81) and (520.92,134.55) .. (524.63,125.07) ;
\draw    (485.44,193.71) .. controls (492.14,180.89) and (503,165.88) .. (506.09,165.88) .. controls (509.18,165.88) and (524.03,188.33) .. (525.89,192.96) ;
\draw    (516.18,146.37) .. controls (524.27,152.82) and (521.84,169.29) .. (516.99,173.34) ;
\draw    (512.13,152.37) .. controls (518.61,156.86) and (518.61,162.85) .. (512.13,167.34) ;

\draw  [draw opacity=0][fill={rgb, 255:red, 221; green, 237; blue, 238 }  ,fill opacity=1 ] (618.4,123.41) .. controls (616.42,118.67) and (634.34,120.46) .. (629.79,123.56) .. controls (625.23,126.65) and (607.39,152.65) .. (605.94,158.18) .. controls (604.48,163.7) and (593.36,156.57) .. (597.69,153.93) .. controls (602.02,151.3) and (620.38,128.16) .. (618.4,123.41) -- cycle ;
\draw  [draw opacity=0][fill={rgb, 255:red, 221; green, 237; blue, 238 }  ,fill opacity=1 ] (595.69,156.63) .. controls (593.71,151.88) and (609.35,156.04) .. (604.8,159.14) .. controls (600.24,162.23) and (581.4,187.23) .. (579.95,192.75) .. controls (578.5,198.28) and (566.72,193.89) .. (571.05,191.26) .. controls (575.38,188.62) and (597.67,161.38) .. (595.69,156.63) -- cycle ;
\draw  [draw opacity=0][fill={rgb, 255:red, 221; green, 237; blue, 238 }  ,fill opacity=1 ] (633.02,190.21) .. controls (637.88,192.46) and (621.55,196.18) .. (620.25,191.04) .. controls (618.96,185.9) and (602.49,163.75) .. (597.58,160.19) .. controls (592.66,156.64) and (605.02,153.42) .. (605.94,158.18) .. controls (606.86,162.93) and (628.17,187.96) .. (633.02,190.21) -- cycle ;
\draw  [draw opacity=0][fill={rgb, 255:red, 221; green, 237; blue, 238 }  ,fill opacity=1 ] (603.12,152.22) .. controls (608.6,152.55) and (598.87,165.34) .. (597.58,160.19) .. controls (596.28,155.05) and (574.35,127.41) .. (569.43,123.85) .. controls (564.52,120.3) and (577.41,117.6) .. (578.33,122.36) .. controls (579.25,127.11) and (597.63,151.88) .. (603.12,152.22) -- cycle ;
\draw  [draw opacity=0][fill={rgb, 255:red, 255; green, 255; blue, 255 }  ,fill opacity=1 ] (611.57,112.86) -- (635.84,112.47) -- (636.05,123.42) -- (611.78,123.81) -- cycle ;
\draw  [draw opacity=0][fill={rgb, 255:red, 255; green, 255; blue, 255 }  ,fill opacity=1 ] (559.8,112.11) -- (584.07,111.72) -- (584.27,122.67) -- (560,123.06) -- cycle ;
\draw  [draw opacity=0][fill={rgb, 255:red, 255; green, 255; blue, 255 }  ,fill opacity=1 ] (564.65,190.74) -- (588.92,190.35) -- (589.13,201.31) -- (564.85,201.7) -- cycle ;
\draw  [draw opacity=0][fill={rgb, 255:red, 255; green, 255; blue, 255 }  ,fill opacity=1 ] (613.19,189.99) -- (637.46,189.6) -- (637.67,200.56) -- (613.39,200.95) -- cycle ;
\draw    (566.68,121.31) .. controls (575.17,130.56) and (596.43,152.4) .. (595.81,158.75) .. controls (595.19,165.11) and (580.44,181.97) .. (569.11,193.2) ;
\draw    (631.4,121.31) .. controls (621.2,133.45) and (606.03,153.7) .. (605.94,158.18) .. controls (605.84,162.65) and (620.89,181.97) .. (634.64,191.71) ;
\draw  [draw opacity=0][fill={rgb, 255:red, 221; green, 237; blue, 238 }  ,fill opacity=1 ] (607.94,152.31) .. controls (601.47,150.82) and (603.9,143.93) .. (611.18,145.42) .. controls (618.46,146.92) and (619.27,166.39) .. (612.8,170.14) .. controls (606.32,173.88) and (601.47,165.04) .. (607.13,164.3) .. controls (612.8,163.55) and (614.41,153.81) .. (607.94,152.31) -- cycle ;
\draw    (578.33,122.36) .. controls (583.28,132.18) and (597.38,150.81) .. (600.47,150.81) .. controls (603.56,150.81) and (615.92,132.55) .. (619.63,123.07) ;
\draw    (580.44,191.71) .. controls (587.14,178.89) and (598,163.88) .. (601.09,163.88) .. controls (604.18,163.88) and (619.03,186.33) .. (620.89,190.96) ;
\draw    (611.18,144.37) .. controls (619.27,150.82) and (616.84,167.29) .. (611.99,171.34) ;
\draw    (607.13,150.37) .. controls (613.61,154.86) and (613.61,160.85) .. (607.13,165.34) ;
\draw  [draw opacity=0][fill={rgb, 255:red, 221; green, 237; blue, 238 }  ,fill opacity=1 ] (591.85,152.31) .. controls (598.32,150.82) and (595.89,143.93) .. (588.61,145.42) .. controls (581.33,146.92) and (580.52,166.39) .. (586.99,170.14) .. controls (593.46,173.88) and (598.32,165.04) .. (592.65,164.3) .. controls (586.99,163.55) and (585.37,153.81) .. (591.85,152.31) -- cycle ;
\draw    (590.2,144.6) .. controls (579,149.2) and (581,169.2) .. (590,171.2) ;
\draw    (594.2,150.6) .. controls (587.2,153.6) and (586.2,161.6) .. (594.2,165.6) ;
\draw  [color={rgb, 255:red, 0; green, 42; blue, 255 }  ,draw opacity=1 ][fill={rgb, 255:red, 208; green, 2; blue, 27 }  ,fill opacity=1 ][line width=1.5]  (93.38,204.38) -- (93.38,204.38) -- (87.75,210) -- (93.38,215.62) -- (93.38,215.62) -- (87.75,210) -- (82.13,215.62) -- (82.13,215.62) -- (87.75,210) -- (82.13,204.38) -- (82.13,204.38) -- (87.75,210) -- cycle ;
\draw  [color={rgb, 255:red, 0; green, 42; blue, 255 }  ,draw opacity=1 ][fill={rgb, 255:red, 208; green, 2; blue, 27 }  ,fill opacity=1 ][line width=1.5]  (196.38,224.38) -- (196.38,224.38) -- (190.75,230) -- (196.38,235.62) -- (196.38,235.62) -- (190.75,230) -- (185.13,235.62) -- (185.13,235.62) -- (190.75,230) -- (185.13,224.38) -- (185.13,224.38) -- (190.75,230) -- cycle ;
\draw  [color={rgb, 255:red, 0; green, 42; blue, 255 }  ,draw opacity=1 ][fill={rgb, 255:red, 208; green, 2; blue, 27 }  ,fill opacity=1 ][line width=1.5]  (274.38,206.38) -- (274.38,206.38) -- (268.75,212) -- (274.38,217.62) -- (274.38,217.62) -- (268.75,212) -- (263.13,217.62) -- (263.13,217.62) -- (268.75,212) -- (263.13,206.38) -- (263.13,206.38) -- (268.75,212) -- cycle ;
\draw  [color={rgb, 255:red, 0; green, 42; blue, 255 }  ,draw opacity=1 ][fill={rgb, 255:red, 208; green, 2; blue, 27 }  ,fill opacity=1 ][line width=1.5]  (106.38,93.38) -- (106.38,93.38) -- (100.75,99) -- (106.38,104.62) -- (106.38,104.62) -- (100.75,99) -- (95.13,104.62) -- (95.13,104.62) -- (100.75,99) -- (95.13,93.38) -- (95.13,93.38) -- (100.75,99) -- cycle ;
\draw  [color={rgb, 255:red, 0; green, 42; blue, 255 }  ,draw opacity=1 ][fill={rgb, 255:red, 208; green, 2; blue, 27 }  ,fill opacity=1 ][line width=1.5]  (193.38,83.38) -- (193.38,83.38) -- (187.75,89) -- (193.38,94.62) -- (193.38,94.62) -- (187.75,89) -- (182.13,94.62) -- (182.13,94.62) -- (187.75,89) -- (182.13,83.38) -- (182.13,83.38) -- (187.75,89) -- cycle ;
\draw  [color={rgb, 255:red, 0; green, 42; blue, 255 }  ,draw opacity=1 ][fill={rgb, 255:red, 208; green, 2; blue, 27 }  ,fill opacity=1 ][line width=1.5]  (264.5,94.5) -- (264.5,94.5) -- (258.75,100) -- (264.25,105.75) -- (264.25,105.75) -- (258.75,100) -- (253,105.5) -- (253,105.5) -- (258.75,100) -- (253.25,94.25) -- (253.25,94.25) -- (258.75,100) -- cycle ;
\draw  [color={rgb, 255:red, 0; green, 42; blue, 255 }  ,draw opacity=1 ][fill={rgb, 255:red, 208; green, 2; blue, 27 }  ,fill opacity=1 ][line width=1.5]  (389.38,115.26) -- (389.38,115.26) -- (383.76,120.89) -- (389.38,126.51) -- (389.38,126.51) -- (383.76,120.89) -- (378.13,126.51) -- (378.13,126.51) -- (383.76,120.89) -- (378.13,115.26) -- (378.13,115.26) -- (383.76,120.89) -- cycle ;
\draw  [color={rgb, 255:red, 0; green, 42; blue, 255 }  ,draw opacity=1 ][fill={rgb, 255:red, 208; green, 2; blue, 27 }  ,fill opacity=1 ][line width=1.5]  (442.6,115.32) -- (442.6,115.32) -- (436.98,120.94) -- (442.6,126.56) -- (442.6,126.56) -- (436.98,120.94) -- (431.36,126.56) -- (431.36,126.56) -- (436.98,120.94) -- (431.36,115.32) -- (431.36,115.32) -- (436.98,120.94) -- cycle ;
\draw  [color={rgb, 255:red, 0; green, 42; blue, 255 }  ,draw opacity=1 ][fill={rgb, 255:red, 208; green, 2; blue, 27 }  ,fill opacity=1 ][line width=1.5]  (392.08,196.7) -- (392.08,196.7) -- (386.46,202.33) -- (392.08,207.95) -- (392.08,207.95) -- (386.46,202.33) -- (380.83,207.95) -- (380.83,207.95) -- (386.46,202.33) -- (380.83,196.7) -- (380.83,196.7) -- (386.46,202.33) -- cycle ;
\draw  [color={rgb, 255:red, 0; green, 42; blue, 255 }  ,draw opacity=1 ][fill={rgb, 255:red, 208; green, 2; blue, 27 }  ,fill opacity=1 ][line width=1.5]  (446.36,195.2) -- (446.36,195.2) -- (440.74,200.83) -- (446.36,206.45) -- (446.36,206.45) -- (440.74,200.83) -- (435.11,206.45) -- (435.11,206.45) -- (440.74,200.83) -- (435.11,195.2) -- (435.11,195.2) -- (440.74,200.83) -- cycle ;
\draw  [color={rgb, 255:red, 0; green, 42; blue, 255 }  ,draw opacity=1 ][fill={rgb, 255:red, 208; green, 2; blue, 27 }  ,fill opacity=1 ][line width=1.5]  (481.38,113.26) -- (481.38,113.26) -- (475.76,118.89) -- (481.38,124.51) -- (481.38,124.51) -- (475.76,118.89) -- (470.13,124.51) -- (470.13,124.51) -- (475.76,118.89) -- (470.13,113.26) -- (470.13,113.26) -- (475.76,118.89) -- cycle ;
\draw  [color={rgb, 255:red, 0; green, 42; blue, 255 }  ,draw opacity=1 ][fill={rgb, 255:red, 208; green, 2; blue, 27 }  ,fill opacity=1 ][line width=1.5]  (534.6,113.32) -- (534.6,113.32) -- (528.98,118.94) -- (534.6,124.56) -- (534.6,124.56) -- (528.98,118.94) -- (523.36,124.56) -- (523.36,124.56) -- (528.98,118.94) -- (523.36,113.32) -- (523.36,113.32) -- (528.98,118.94) -- cycle ;
\draw  [color={rgb, 255:red, 0; green, 42; blue, 255 }  ,draw opacity=1 ][fill={rgb, 255:red, 208; green, 2; blue, 27 }  ,fill opacity=1 ][line width=1.5]  (484.08,194.7) -- (484.08,194.7) -- (478.46,200.33) -- (484.08,205.95) -- (484.08,205.95) -- (478.46,200.33) -- (472.83,205.95) -- (472.83,205.95) -- (478.46,200.33) -- (472.83,194.7) -- (472.83,194.7) -- (478.46,200.33) -- cycle ;
\draw  [color={rgb, 255:red, 0; green, 42; blue, 255 }  ,draw opacity=1 ][fill={rgb, 255:red, 208; green, 2; blue, 27 }  ,fill opacity=1 ][line width=1.5]  (538.36,193.2) -- (538.36,193.2) -- (532.74,198.83) -- (538.36,204.45) -- (538.36,204.45) -- (532.74,198.83) -- (527.11,204.45) -- (527.11,204.45) -- (532.74,198.83) -- (527.11,193.2) -- (527.11,193.2) -- (532.74,198.83) -- cycle ;
\draw  [color={rgb, 255:red, 0; green, 42; blue, 255 }  ,draw opacity=1 ][fill={rgb, 255:red, 208; green, 2; blue, 27 }  ,fill opacity=1 ][line width=1.5]  (576.38,111.26) -- (576.38,111.26) -- (570.76,116.89) -- (576.38,122.51) -- (576.38,122.51) -- (570.76,116.89) -- (565.13,122.51) -- (565.13,122.51) -- (570.76,116.89) -- (565.13,111.26) -- (565.13,111.26) -- (570.76,116.89) -- cycle ;
\draw  [color={rgb, 255:red, 0; green, 42; blue, 255 }  ,draw opacity=1 ][fill={rgb, 255:red, 208; green, 2; blue, 27 }  ,fill opacity=1 ][line width=1.5]  (629.6,111.32) -- (629.6,111.32) -- (623.98,116.94) -- (629.6,122.56) -- (629.6,122.56) -- (623.98,116.94) -- (618.36,122.56) -- (618.36,122.56) -- (623.98,116.94) -- (618.36,111.32) -- (618.36,111.32) -- (623.98,116.94) -- cycle ;
\draw  [color={rgb, 255:red, 0; green, 42; blue, 255 }  ,draw opacity=1 ][fill={rgb, 255:red, 208; green, 2; blue, 27 }  ,fill opacity=1 ][line width=1.5]  (579.08,192.7) -- (579.08,192.7) -- (573.46,198.33) -- (579.08,203.95) -- (579.08,203.95) -- (573.46,198.33) -- (567.83,203.95) -- (567.83,203.95) -- (573.46,198.33) -- (567.83,192.7) -- (567.83,192.7) -- (573.46,198.33) -- cycle ;
\draw  [color={rgb, 255:red, 0; green, 42; blue, 255 }  ,draw opacity=1 ][fill={rgb, 255:red, 208; green, 2; blue, 27 }  ,fill opacity=1 ][line width=1.5]  (633.36,191.2) -- (633.36,191.2) -- (627.74,196.83) -- (633.36,202.45) -- (633.36,202.45) -- (627.74,196.83) -- (622.11,202.45) -- (622.11,202.45) -- (627.74,196.83) -- (622.11,191.2) -- (622.11,191.2) -- (627.74,196.83) -- cycle ;

\draw (375.59,80.39) node [anchor=north west][inner sep=0.75pt]    {$C_{ }{}_{O;4\text{\mbox{-}point}} =\ $};
\draw (453.06,149.27) node [anchor=north west][inner sep=0.75pt]    {$+\ $};
\draw (573.72,217.79) node [anchor=north west][inner sep=0.75pt]    {$+\ ...\ $};
\draw (549.06,147.27) node [anchor=north west][inner sep=0.75pt]    {$+\ $};
\draw (61,226.4) node [anchor=north west][inner sep=0.75pt]    {$O( x_{1})$};
\draw (172,246.4) node [anchor=north west][inner sep=0.75pt]    {$O( x_{2})$};
\draw (258,226.4) node [anchor=north west][inner sep=0.75pt]    {$O( x_{3})$};
\draw (244,67.4) node [anchor=north west][inner sep=0.75pt]    {$O( x_{4})$};
\draw (164,54.4) node [anchor=north west][inner sep=0.75pt]    {$O( x_{5})$};
\draw (70,66.4) node [anchor=north west][inner sep=0.75pt]    {$O( x_{6})$};

\end{tikzpicture}

%% file: figures/tikz/contours.tex
\tikzset{every picture/.style={line width=0.75pt}} %

\begin{tikzpicture}[x=0.75pt,y=0.75pt,yscale=-1,xscale=1]

\draw    (184.01,220.66) -- (410.38,220.66) ;
\draw [shift={(412.38,220.66)}, rotate = 180] [color={rgb, 255:red, 0; green, 0; blue, 0 }  ][line width=0.75]    (10.93,-3.29) .. controls (6.95,-1.4) and (3.31,-0.3) .. (0,0) .. controls (3.31,0.3) and (6.95,1.4) .. (10.93,3.29)   ;
\draw    (195.83,234.45) -- (195.6,68.18) ;
\draw [shift={(195.6,66.18)}, rotate = 89.92] [color={rgb, 255:red, 0; green, 0; blue, 0 }  ][line width=0.75]    (10.93,-3.29) .. controls (6.95,-1.4) and (3.31,-0.3) .. (0,0) .. controls (3.31,0.3) and (6.95,1.4) .. (10.93,3.29)   ;
\draw [color={rgb, 255:red, 35; green, 204; blue, 221 }  ,draw opacity=1 ][line width=3]    (195.6,220.66) -- (412.38,220.66) ;
\draw [color={rgb, 255:red, 65; green, 117; blue, 5 }  ,draw opacity=1 ][line width=3]    (412.27,112.28) -- (195.49,112.28) ;
\draw [draw opacity=0][fill={rgb, 255:red, 221; green, 237; blue, 238 }  ,fill opacity=1 ]   (422.92,84.61) .. controls (441.23,104.35) and (439.03,112.54) .. (439.03,136.82) .. controls (439.03,161.11) and (473.61,161.11) .. (473.61,136.82) .. controls (473.61,112.54) and (472.13,104.35) .. (487.52,84.61) ;
\draw [color={rgb, 255:red, 65; green, 117; blue, 5 }  ,draw opacity=1 ]   (439.03,136.82) .. controls (439.76,158.77) and (473.61,160.28) .. (473.61,136.82) ;
\draw [color={rgb, 255:red, 126; green, 211; blue, 33 }  ,draw opacity=1 ]   (422.92,84.61) .. controls (441.23,100.5) and (438.29,116.39) .. (439.03,136.82) ;
\draw [color={rgb, 255:red, 126; green, 211; blue, 33 }  ,draw opacity=1 ]   (487.52,84.61) .. controls (471.26,102.77) and (472.87,116.39) .. (473.61,136.82) ;
\draw  [color={rgb, 255:red, 65; green, 117; blue, 5 }  ,draw opacity=1 ][dash pattern={on 0.84pt off 2.51pt}] (456.32,140.95) .. controls (446.77,140.95) and (439.03,139.1) .. (439.03,136.82) .. controls (439.03,134.54) and (446.77,132.69) .. (456.32,132.69) .. controls (465.86,132.69) and (473.61,134.54) .. (473.61,136.82) .. controls (473.61,139.1) and (465.86,140.95) .. (456.32,140.95) -- cycle ;
\draw [color={rgb, 255:red, 35; green, 204; blue, 221 }  ,draw opacity=1 ][fill={rgb, 255:red, 221; green, 237; blue, 238 }  ,fill opacity=1 ]   (425.54,189.65) .. controls (442.41,216.25) and (439.66,231.91) .. (445.23,242.86) .. controls (450.8,253.8) and (466.41,253.03) .. (471.35,241.29) .. controls (476.29,229.55) and (473.46,216.25) .. (487.52,189.65) ;
\draw  [color={rgb, 255:red, 208; green, 2; blue, 27 }  ,draw opacity=1 ][fill={rgb, 255:red, 221; green, 237; blue, 238 }  ,fill opacity=1 ] (456.53,198.97) .. controls (439.42,198.97) and (425.54,194.79) .. (425.54,189.65) .. controls (425.54,184.5) and (439.42,180.33) .. (456.53,180.33) .. controls (473.65,180.33) and (487.52,184.5) .. (487.52,189.65) .. controls (487.52,194.79) and (473.65,198.97) .. (456.53,198.97) -- cycle ;
\draw [color={rgb, 255:red, 126; green, 211; blue, 33 }  ,draw opacity=1 ][line width=3]    (195.6,220.66) -- (195.49,112.28) ;
\draw  [draw opacity=0][fill={rgb, 255:red, 208; green, 2; blue, 27 }  ,fill opacity=1 ] (189.81,220.66) .. controls (189.81,217.38) and (192.4,214.72) .. (195.6,214.72) .. controls (198.8,214.72) and (201.4,217.38) .. (201.4,220.66) .. controls (201.4,223.94) and (198.8,226.6) .. (195.6,226.6) .. controls (192.4,226.6) and (189.81,223.94) .. (189.81,220.66) -- cycle ;
\draw  [color={rgb, 255:red, 208; green, 2; blue, 27 }  ,draw opacity=1 ][fill={rgb, 255:red, 221; green, 237; blue, 238 }  ,fill opacity=1 ] (455.22,93.63) .. controls (437.38,93.63) and (422.92,89.59) .. (422.92,84.61) .. controls (422.92,79.64) and (437.38,75.6) .. (455.22,75.6) .. controls (473.06,75.6) and (487.52,79.64) .. (487.52,84.61) .. controls (487.52,89.59) and (473.06,93.63) .. (455.22,93.63) -- cycle ;

\draw (380.65,233.12) node [anchor=north west][inner sep=0.75pt]    {$\mathrm{Re}( z)$};
\draw (171.09,46.28) node [anchor=north west][inner sep=0.75pt]    {$\mathrm{Im}( z)$};
\draw (166.42,98.82) node [anchor=north west][inner sep=0.75pt]    {$\frac{iL}{4}$};
\draw (156.85,232.55) node [anchor=north west][inner sep=0.75pt]    {$z=0\ (\mathrm{bdy})$};
\draw (447.69,137.44) node [anchor=north west][inner sep=0.75pt]    {$S^{2}$};
\draw (439.08,105.68) node [anchor=north west][inner sep=0.75pt]    {$\mathrm{LdS}_{2}$};
\draw (426.36,209.35) node [anchor=north west][inner sep=0.75pt]    {$\mathrm{-EAdS}_{2}$};
\draw (402,162.4) node [anchor=north west][inner sep=0.75pt]    {$...$};

\end{tikzpicture}

%% file: figures/tikz/discisland.tex
\tikzset{every picture/.style={line width=0.75pt}} %

\begin{tikzpicture}[x=0.75pt,y=0.75pt,yscale=-1,xscale=1]

\draw [fill={rgb, 255:red, 221; green, 237; blue, 238 }  ,fill opacity=1 ]   (217,131.92) .. controls (301.71,268.07) and (365.22,269.05) .. (442.4,131.92) ;
\draw [fill={rgb, 255:red, 221; green, 237; blue, 238 }  ,fill opacity=1 ]   (442.4,131.92) .. controls (360.71,-6.46) and (294.17,-5.22) .. (217,131.92) ;
\draw  [dash pattern={on 4.5pt off 4.5pt}]  (215.6,131.38) .. controls (219.14,122.11) and (238.09,114.9) .. (264.43,110.67) .. controls (283.67,107.58) and (306.68,105.94) .. (329.69,105.82) .. controls (330.55,105.82) and (331.41,105.82) .. (332.27,105.82) .. controls (383.49,105.82) and (433.49,113.69) .. (442.76,129.19) .. controls (443.19,129.92) and (443.54,130.66) .. (443.81,131.41)(218.4,132.45) .. controls (221.76,123.64) and (240.43,117.57) .. (264.91,113.63) .. controls (284.01,110.56) and (306.86,108.94) .. (329.71,108.82) .. controls (330.56,108.82) and (331.42,108.82) .. (332.27,108.82) .. controls (381.77,108.82) and (431.13,115.59) .. (440.18,130.73) .. controls (440.51,131.28) and (440.78,131.85) .. (440.99,132.42) ;
\draw    (218.45,131.54) .. controls (220.7,140.28) and (232.28,146.68) .. (248.51,151.32) .. controls (269.19,157.23) and (297.7,160.08) .. (326.48,160.08) .. controls (327.55,160.08) and (328.62,160.08) .. (329.69,160.07) .. controls (379.37,159.7) and (429.5,151.12) .. (439.65,134.46) .. controls (440.24,133.51) and (440.67,132.53) .. (440.96,131.52)(215.55,132.29) .. controls (217.97,141.69) and (229.88,149.12) .. (247.68,154.2) .. controls (268.59,160.18) and (297.39,163.08) .. (326.48,163.08) .. controls (327.55,163.08) and (328.63,163.08) .. (329.71,163.07) .. controls (380.99,162.69) and (431.82,153.07) .. (442.22,136.02) .. controls (442.95,134.82) and (443.49,133.59) .. (443.84,132.32) ;
\draw  [draw opacity=0][fill={rgb, 255:red, 221; green, 237; blue, 238 }  ,fill opacity=1 ] (275.86,154.59) -- (379.69,145.67) -- (381.13,162.11) -- (277.31,171.03) -- cycle ;
\draw  [draw opacity=0][fill={rgb, 255:red, 221; green, 237; blue, 238 }  ,fill opacity=1 ] (266.68,144.01) -- (300.6,147.81) -- (298.38,167.17) -- (264.46,163.38) -- cycle ;
\draw [color={rgb, 255:red, 255; green, 0; blue, 31 }  ,draw opacity=1 ]   (265.23,155.4) .. controls (284.64,157.43) and (300.57,158.75) .. (316.5,159.18) .. controls (320.35,159.29) and (324.21,159.34) .. (328.11,159.34) .. controls (343.84,159.34) and (360.37,158.47) .. (380.91,156.56)(264.92,158.38) .. controls (284.41,160.42) and (300.41,161.74) .. (316.42,162.18) .. controls (320.3,162.29) and (324.18,162.34) .. (328.11,162.34) .. controls (343.92,162.34) and (360.53,161.46) .. (381.19,159.55) ;
\draw [shift={(381.05,158.05)}, rotate = 174.7] [color={rgb, 255:red, 255; green, 0; blue, 31 }  ,draw opacity=1 ][line width=0.75]    (0,5.59) -- (0,-5.59)   ;
\draw [shift={(265.07,156.89)}, rotate = 185.97] [color={rgb, 255:red, 255; green, 0; blue, 31 }  ,draw opacity=1 ][line width=0.75]    (0,5.59) -- (0,-5.59)   ;
\draw [color={rgb, 255:red, 0; green, 42; blue, 255 }  ,draw opacity=1 ]   (264.2,170.4) .. controls (303.2,174.4) and (334.2,176.4) .. (382.2,171.4) ;
\draw [shift={(382.2,171.4)}, rotate = 174.05] [color={rgb, 255:red, 0; green, 42; blue, 255 }  ,draw opacity=1 ][line width=0.75]    (0,5.59) -- (0,-5.59)   ;
\draw [shift={(264.2,170.4)}, rotate = 185.86] [color={rgb, 255:red, 0; green, 42; blue, 255 }  ,draw opacity=1 ][line width=0.75]    (0,5.59) -- (0,-5.59)   ;
\draw [color={rgb, 255:red, 0; green, 42; blue, 255 }  ,draw opacity=1 ] [dash pattern={on 4.5pt off 4.5pt}]  (315.2,120.2) .. controls (341.25,119.37) and (369,120.6) .. (394.6,125) ;
\draw [shift={(394.6,125)}, rotate = 189.75] [color={rgb, 255:red, 0; green, 42; blue, 255 }  ,draw opacity=1 ][line width=0.75]    (0,5.59) -- (0,-5.59)   ;
\draw [shift={(315.2,120.2)}, rotate = 178.18] [color={rgb, 255:red, 0; green, 42; blue, 255 }  ,draw opacity=1 ][line width=0.75]    (0,5.59) -- (0,-5.59)   ;
\draw  [draw opacity=0][fill={rgb, 255:red, 221; green, 237; blue, 238 }  ,fill opacity=1 ] (312.8,100.67) -- (368.09,96.21) -- (369.08,108.25) -- (313.8,112.7) -- cycle ;
\draw  [draw opacity=0][fill={rgb, 255:red, 221; green, 237; blue, 238 }  ,fill opacity=1 ] (343.21,97.64) -- (398.3,106.52) -- (396.18,119.38) -- (341.09,110.5) -- cycle ;
\draw [color={rgb, 255:red, 255; green, 0; blue, 31 }  ,draw opacity=1 ] [dash pattern={on 4.5pt off 4.5pt}]  (314.61,105.44) .. controls (321.63,105.22) and (328.17,105.1) .. (334.41,105.1) .. controls (345.47,105.1) and (355.57,105.48) .. (365.67,106.34) .. controls (375.88,107.22) and (386.08,108.6) .. (397.28,110.61)(314.7,108.44) .. controls (321.69,108.22) and (328.2,108.1) .. (334.41,108.1) .. controls (345.38,108.1) and (355.4,108.47) .. (365.42,109.33) .. controls (375.53,110.2) and (385.65,111.57) .. (396.75,113.56) ;
\draw [shift={(397.02,112.08)}, rotate = 190.14] [color={rgb, 255:red, 255; green, 0; blue, 31 }  ,draw opacity=1 ][line width=0.75]    (0,5.59) -- (0,-5.59)   ;
\draw [shift={(314.66,106.94)}, rotate = 178.21] [color={rgb, 255:red, 255; green, 0; blue, 31 }  ,draw opacity=1 ][line width=0.75]    (0,5.59) -- (0,-5.59)   ;

\draw (309.09,176.34) node [anchor=north west][inner sep=0.75pt]    {$I_{1}$};
\draw (349.43,122.4) node [anchor=north west][inner sep=0.75pt]    {$I_{2}$};
\draw (309.53,140.66) node [anchor=north west][inner sep=0.75pt]    {$R_{1}$};
\draw (352.39,86.72) node [anchor=north west][inner sep=0.75pt]    {$R_{2}$};

\end{tikzpicture}

%% file: figures/tikz/holointer.tex
\tikzset{every picture/.style={line width=0.75pt}} %

\begin{tikzpicture}[x=0.75pt,y=0.75pt,yscale=-1,xscale=1]

\draw [color={rgb, 255:red, 61; green, 199; blue, 208 }  ,draw opacity=1 ][fill={rgb, 255:red, 221; green, 237; blue, 238 }  ,fill opacity=1 ]   (150.46,102.77) .. controls (98.6,119.13) and (75.53,120.7) .. (75.2,141.9) .. controls (74.86,163.09) and (99.5,166.41) .. (150.46,179.66) ;
\draw  [draw opacity=0][fill={rgb, 255:red, 255; green, 255; blue, 255 }  ,fill opacity=1 ] (140.21,141.77) .. controls (140.21,120.53) and (145.01,103.32) .. (150.93,103.32) .. controls (156.84,103.32) and (161.64,120.53) .. (161.64,141.77) .. controls (161.64,163) and (156.84,180.21) .. (150.93,180.21) .. controls (145.01,180.21) and (140.21,163) .. (140.21,141.77) -- cycle ;
\draw  [color={rgb, 255:red, 0; green, 0; blue, 0 }  ,draw opacity=0 ][fill={rgb, 255:red, 221; green, 237; blue, 238 }  ,fill opacity=1 ] (398.96,38.45) -- (515.8,38.45) -- (515.8,245.8) -- (398.96,245.8) -- cycle ;
\draw [color={rgb, 255:red, 208; green, 2; blue, 27 }  ,draw opacity=1 ]   (399.06,85.34) -- (398.88,198.92) ;
\draw [color={rgb, 255:red, 208; green, 2; blue, 27 }  ,draw opacity=1 ]   (515.88,85.34) -- (515.7,198.92) ;
\draw  [dash pattern={on 0.84pt off 2.51pt}]  (399.06,85.34) -- (515.7,198.92) ;
\draw  [dash pattern={on 0.84pt off 2.51pt}]  (515.88,85.34) -- (398.88,198.92) ;
\draw    (340.56,142.13) -- (398.88,198.92) ;
\draw    (574.2,142.13) -- (515.7,198.92) ;
\draw    (515.88,85.34) -- (574.2,142.13) ;
\draw    (399.06,85.34) -- (340.56,142.13) ;
\draw  [draw opacity=0][fill={rgb, 255:red, 255; green, 255; blue, 255 }  ,fill opacity=1 ] (150.46,102.77) -- (246.21,102.77) -- (246.21,179.54) -- (150.46,179.54) -- cycle ;
\draw  [fill={rgb, 255:red, 255; green, 255; blue, 255 }  ,fill opacity=1 ][dash pattern={on 0.84pt off 2.51pt}] (235.49,141.21) .. controls (235.49,119.98) and (240.29,102.77) .. (246.21,102.77) .. controls (252.13,102.77) and (256.93,119.98) .. (256.93,141.21) .. controls (256.93,162.44) and (252.13,179.66) .. (246.21,179.66) .. controls (240.29,179.66) and (235.49,162.44) .. (235.49,141.21) -- cycle ;
\draw    (150.46,102.77) -- (246.21,102.77) ;
\draw    (150.46,179.66) -- (246.21,179.66) ;
\draw  [draw opacity=0][fill={rgb, 255:red, 221; green, 237; blue, 238 }  ,fill opacity=0.62 ] (140.21,141.77) .. controls (140.21,120.53) and (145.01,103.32) .. (150.93,103.32) .. controls (156.84,103.32) and (161.64,120.53) .. (161.64,141.77) .. controls (161.64,163) and (156.84,180.21) .. (150.93,180.21) .. controls (145.01,180.21) and (140.21,163) .. (140.21,141.77) -- cycle ;
\draw  [draw opacity=0] (150.91,180.21) .. controls (150.76,180.23) and (150.61,180.25) .. (150.46,180.25) .. controls (144.54,180.25) and (139.74,162.9) .. (139.74,141.51) .. controls (139.74,120.11) and (144.54,102.77) .. (150.46,102.77) .. controls (150.55,102.77) and (150.64,102.77) .. (150.73,102.78) -- (150.46,141.51) -- cycle ; \draw  [color={rgb, 255:red, 208; green, 2; blue, 27 }  ,draw opacity=1 ] (150.91,180.21) .. controls (150.76,180.23) and (150.61,180.25) .. (150.46,180.25) .. controls (144.54,180.25) and (139.74,162.9) .. (139.74,141.51) .. controls (139.74,120.11) and (144.54,102.77) .. (150.46,102.77) .. controls (150.55,102.77) and (150.64,102.77) .. (150.73,102.78) ;
\draw  [draw opacity=0][dash pattern={on 0.84pt off 2.51pt}] (150.01,102.8) .. controls (150.16,102.78) and (150.31,102.77) .. (150.46,102.77) .. controls (156.38,102.77) and (161.18,120.11) .. (161.18,141.51) .. controls (161.18,162.9) and (156.38,180.25) .. (150.46,180.25) .. controls (150.37,180.25) and (150.28,180.24) .. (150.19,180.23) -- (150.46,141.51) -- cycle ; \draw  [color={rgb, 255:red, 208; green, 2; blue, 27 }  ,draw opacity=1 ][dash pattern={on 0.84pt off 2.51pt}] (150.01,102.8) .. controls (150.16,102.78) and (150.31,102.77) .. (150.46,102.77) .. controls (156.38,102.77) and (161.18,120.11) .. (161.18,141.51) .. controls (161.18,162.9) and (156.38,180.25) .. (150.46,180.25) .. controls (150.37,180.25) and (150.28,180.24) .. (150.19,180.23) ;

\draw (94.32,132.02) node [anchor=north west][inner sep=0.75pt]    {$\mathrm{QG}$};
\draw (183.93,131.14) node [anchor=north west][inner sep=0.75pt]    {$\mathrm{Bath}$};
\draw (142.77,83.86) node [anchor=north west][inner sep=0.75pt]    {$\Sigma $};
\draw (354.94,132.9) node [anchor=north west][inner sep=0.75pt]    {$\mathrm{Bath}$};
\draw (522.35,132.9) node [anchor=north west][inner sep=0.75pt]    {$\mathrm{Bath}$};
\draw (443.64,100.5) node [anchor=north west][inner sep=0.75pt]    {$\mathrm{BH}$};
\draw (263.13,130.27) node [anchor=north west][inner sep=0.75pt]    {$...$};

\end{tikzpicture}

%% file: biblio.bib
@article{Geng:2025efs,
    doi = "10.1007/JHEP07(2026)262",
    author = "Geng, Hao and Hung, Ling-Yan and Jiang, Yikun",
    title = "{It from ETH: multi-interval entanglement and replica wormholes from large-$c$ BCFT ensemble}",
    eprint = "2505.20385",
    archivePrefix = "arXiv",
    primaryClass = "hep-th",
    month = "5",
    year = "2026",
    journal = "JHEP",
    volume = "07",
    pages = "262"
}

@article{Geng:2024xpj,
    author = "Geng, Hao",
    title = "{Replica wormholes and entanglement islands in the Karch-Randall braneworld}",
    eprint = "2405.14872",
    archivePrefix = "arXiv",
    primaryClass = "hep-th",
    doi = "10.1007/JHEP01(2025)063",
    journal = "JHEP",
    volume = "01",
    pages = "063",
    year = "2025"
}

@article{Bao:2025plr,
    author = "Bao, Ning and Geng, Hao and Jiang, Yikun",
    title = "{Ryu-Takayanagi formula for multi-boundary black holes from 2D large-c CFT ensemble}",
    eprint = "2504.12388",
    archivePrefix = "arXiv",
    primaryClass = "hep-th",
    doi = "10.1007/JHEP10(2025)042",
    journal = "JHEP",
    volume = "10",
    pages = "042",
    year = "2025"
}

@article{dsgps_new,
    author = "Jung, Sunghoon and Kum, Minju and Lee, Junghwan",
    title = "{Two-point correlators in de Sitter-prepared states with bra-ket wormholes}",
    eprint = "2512.13646",
    archivePrefix = "arXiv",
    primaryClass = "gr-qc",
    month = "12",
    year = "2025"
}

@article{Witten:2023qsv,
    author = "Witten, Edward",
    title = "{Algebras, regions, and observers}",
    eprint = "2303.02837",
    archivePrefix = "arXiv",
    primaryClass = "hep-th",
    doi = "10.1090/pspum/107/01954",
    journal = "Proc. Symp. Pure Math.",
    volume = "107",
    pages = "247--276",
    year = "2024"
}

@article{Engelhardt:2024lnd,
    doi = "10.1007/JHEP05(2026)107",
    author = "Engelhardt, Netta and Folkestad, {\r{A}}smund and Levine, Adam and Verheijden, Evita and Yang, Lisa",
    title = "{Spoofing entanglement in holography}",
    eprint = "2407.14589",
    archivePrefix = "arXiv",
    primaryClass = "hep-th",
    month = "7",
    year = "2026",
    journal = "JHEP",
    volume = "05",
    pages = "107"
}

@article{Giddings:1988wv,
    author = "Giddings, Steven B. and Strominger, Andrew",
    title = "{Baby Universes, Third Quantization and the Cosmological Constant}",
    reportNumber = "HUTP-88/A036",
    doi = "10.1016/0550-3213(89)90353-2",
    journal = "Nucl. Phys. B",
    volume = "321",
    pages = "481--508",
    year = "1989"
}

@article{Hawking:1988ae,
    author = "Hawking, S. W.",
    editor = "Gibbons, G. W. and Hawking, S. W.",
    title = "{Wormholes in Space-Time}",
    doi = "10.1103/PhysRevD.37.904",
    journal = "Phys. Rev. D",
    volume = "37",
    pages = "904--910",
    year = "1988"
}

@article{mollabashi_pseudo_2021,
author = "Mollabashi, Ali and Shiba, Noburo and Takayanagi, Tadashi and Tamaoka, Kotaro and Wei, Zixia",
    title = "{Pseudo Entropy in Free Quantum Field Theories}",
    eprint = "2011.09648",
    archivePrefix = "arXiv",
    primaryClass = "hep-th",
    reportNumber = "YITP-20-148; IPMU20-120; MPP-2020-203, ?YITP-20-148; IPMU20-120; MPP-2020-203?",
    doi = "10.1103/PhysRevLett.126.081601",
    journal = "Phys. Rev. Lett.",
    volume = "126",
    number = "8",
    pages = "081601",
    year = "2021"
}

@article{nakata_holographic_2021,
    author = "Nakata, Yoshifumi and Takayanagi, Tadashi and Taki, Yusuke and Tamaoka, Kotaro and Wei, Zixia",
    title = "{New holographic generalization of entanglement entropy}",
    eprint = "2005.13801",
    archivePrefix = "arXiv",
    primaryClass = "hep-th",
    reportNumber = "YITP-20-71, IPMU20-0060",
    doi = "10.1103/PhysRevD.103.026005",
    journal = "Phys. Rev. D",
    volume = "103",
    number = "2",
    pages = "026005",
    year = "2021"
}

@article{Shaghoulian_2022,
    author = "Shaghoulian, Edgar",
    title = "{The central dogma and cosmological horizons}",
    eprint = "2110.13210",
    archivePrefix = "arXiv",
    primaryClass = "hep-th",
    doi = "10.1007/JHEP01(2022)132",
    journal = "JHEP",
    volume = "01",
    pages = "132",
    year = "2022"
}

@book{birrell_quantum_1982,
    author = "Birrell, N. D. and Davies, P. C. W.",
    title = "{Quantum Fields in Curved Space}",
    doi = "10.1017/CBO9780511622632",
    isbn = "978-0-511-62263-2, 978-0-521-27858-4",
    publisher = "Cambridge University Press",
    address = "Cambridge, UK",
    year = "1982"
}

@article{caraglio_entanglement_2008,
    author = "Caraglio, Michele and Gliozzi, Ferdinando",
    title = "{Entanglement Entropy and Twist Fields}",
    eprint = "0808.4094",
    archivePrefix = "arXiv",
    primaryClass = "hep-th",
    doi = "10.1088/1126-6708/2008/11/076",
    journal = "JHEP",
    volume = "11",
    pages = "076",
    year = "2008"
}

@article{calabrese_entanglement_2004,
    author = "Calabrese, Pasquale and Cardy, John L.",
    title = "{Entanglement entropy and quantum field theory}",
    eprint = "hep-th/0405152",
    archivePrefix = "arXiv",
    doi = "10.1088/1742-5468/2004/06/P06002",
    journal = "J. Stat. Mech.",
    volume = "0406",
    pages = "P06002",
    year = "2004"
}

@article{casini_entanglement_2009,
    author = "Casini, H. and Huerta, M.",
    title = "{Entanglement entropy in free quantum field theory}",
    eprint = "0905.2562",
    archivePrefix = "arXiv",
    primaryClass = "hep-th",
    doi = "10.1088/1751-8113/42/50/504007",
    journal = "J. Phys. A",
    volume = "42",
    pages = "504007",
    year = "2009"
}

@article{casini_entanglement_2005-1,
    author = "Casini, H. and Fosco, C. D. and Huerta, M.",
    title = "{Entanglement and alpha entropies for a massive Dirac field in two dimensions}",
    eprint = "cond-mat/0505563",
    archivePrefix = "arXiv",
    doi = "10.1088/1742-5468/2005/07/P07007",
    journal = "J. Stat. Mech.",
    volume = "0507",
    pages = "P07007",
    year = "2005"
}

@article{Perez-Garcia:2006nqo,
    author = "Perez-Garcia, David and Verstraete, Frank and Wolf, Michael M. and Cirac, J. Ignacio",
    title = "{Matrix product state representations}",
    eprint = "quant-ph/0608197",
    archivePrefix = "arXiv",
    doi = "10.26421/QIC7.5-6-1",
    journal = "Quant. Inf. Comput.",
    volume = "7",
    number = "5-6",
    pages = "401--430",
    year = "2007"
}

@article{Bridgeman:2016dhh,
    author = "Bridgeman, Jacob C. and Chubb, Christopher T.",
    title = "{Hand-waving and Interpretive Dance: An Introductory Course on Tensor Networks}",
    eprint = "1603.03039",
    archivePrefix = "arXiv",
    primaryClass = "quant-ph",
    doi = "10.1088/1751-8121/aa6dc3",
    journal = "J. Phys. A",
    volume = "50",
    pages = "223001",
    year = "2017"
}

@article{Sekino_2008,
    author = "Sekino, Yasuhiro and Susskind, Leonard",
    title = "{Fast Scramblers}",
    eprint = "0808.2096",
    archivePrefix = "arXiv",
    primaryClass = "hep-th",
    reportNumber = "SU-ITP-08-18, OIQP-08-08, SU-ITP-08/18, OIQP-08-08",
    doi = "10.1088/1126-6708/2008/10/065",
    journal = "JHEP",
    volume = "10",
    pages = "065",
    year = "2008"
}

@article{Stanford:2020wkf,
    author = "Stanford, Douglas",
    title = "{More quantum noise from wormholes}",
    eprint = "2008.08570",
    archivePrefix = "arXiv",
    primaryClass = "hep-th",
    month = "8",
    year = "2020"
}

@article{Orus:2013kga,
    author = "Orus, Roman",
    title = "{A Practical Introduction to Tensor Networks: Matrix Product States and Projected Entangled Pair States}",
    eprint = "1306.2164",
    archivePrefix = "arXiv",
    primaryClass = "cond-mat.str-el",
    doi = "10.1016/j.aop.2014.06.013",
    journal = "Annals Phys.",
    volume = "349",
    pages = "117--158",
    year = "2014"
}

@article{almheiri_page_2020,
    author = "Almheiri, Ahmed and Mahajan, Raghu and Maldacena, Juan and Zhao, Ying",
    title = "{The Page curve of Hawking radiation from semiclassical geometry}",
    eprint = "1908.10996",
    archivePrefix = "arXiv",
    primaryClass = "hep-th",
    doi = "10.1007/JHEP03(2020)149",
    journal = "JHEP",
    volume = "03",
    pages = "149",
    year = "2020"
}

@article{gibbons_cosmological_1977,
    author = "Gibbons, G. W. and Hawking, S. W.",
    title = "{Cosmological Event Horizons, Thermodynamics, and Particle Creation}",
    doi = "10.1103/PhysRevD.15.2738",
    journal = "Phys. Rev. D",
    volume = "15",
    pages = "2738--2751",
    year = "1977"
}

@article{hartman_islands_2020,
    author = "Hartman, Thomas and Jiang, Yikun and Shaghoulian, Edgar",
    title = "{Islands in cosmology}",
    eprint = "2008.01022",
    archivePrefix = "arXiv",
    primaryClass = "hep-th",
    doi = "10.1007/JHEP11(2020)111",
    journal = "JHEP",
    volume = "11",
    pages = "111",
    year = "2020"
}

@article{engelsoy_investigation_2016,
    author = {Engels{\"o}y, Julius and Mertens, Thomas G. and Verlinde, Herman},
    title = "{An investigation of AdS$_{2}$ backreaction and holography}",
    eprint = "1606.03438",
    archivePrefix = "arXiv",
    primaryClass = "hep-th",
    doi = "10.1007/JHEP07(2016)139",
    journal = "JHEP",
    volume = "07",
    pages = "139",
    year = "2016"
}

@article{guth_inflationary_1981,
    author = "Guth, Alan H.",
    editor = "Fang, Li-Zhi and Ruffini, R.",
    title = "{The Inflationary Universe: A Possible Solution to the Horizon and Flatness Problems}",
    reportNumber = "SLAC-PUB-2576",
    doi = "10.1103/PhysRevD.23.347",
    journal = "Phys. Rev. D",
    volume = "23",
    pages = "347--356",
    year = "1981"
}

@article{kontsevich_wick_2021,
    author = "Kontsevich, Maxim and Segal, Graeme",
    title = "{Wick Rotation and the Positivity of Energy in Quantum Field Theory}",
    eprint = "2105.10161",
    archivePrefix = "arXiv",
    primaryClass = "hep-th",
    doi = "10.1093/qmath/haab027",
    journal = "Quart. J. Math. Oxford Ser.",
    volume = "72",
    number = "1-2",
    pages = "673--699",
    year = "2021"
}

@article{teresi_islands_2022,
    author = "Teresi, Daniele",
    title = "{Islands and the de Sitter entropy bound}",
    eprint = "2112.03922",
    archivePrefix = "arXiv",
    primaryClass = "hep-th",
    reportNumber = "CERN-TH-2021-213",
    doi = "10.1007/JHEP10(2022)179",
    journal = "JHEP",
    volume = "10",
    pages = "179",
    year = "2022"
}

@article{chen_bra-ket_2021,
    author = "Chen, Yiming and Gorbenko, Victor and Maldacena, Juan",
    title = "{Bra-ket wormholes in gravitationally prepared states}",
    eprint = "2007.16091",
    archivePrefix = "arXiv",
    primaryClass = "hep-th",
    doi = "10.1007/JHEP02(2021)009",
    journal = "JHEP",
    volume = "02",
    pages = "009",
    year = "2021"
}

@article{mollabashi_aspects_2021,
    author = "Mollabashi, Ali and Shiba, Noburo and Takayanagi, Tadashi and Tamaoka, Kotaro and Wei, Zixia",
    title = "{Aspects of pseudoentropy in field theories}",
    eprint = "2106.03118",
    archivePrefix = "arXiv",
    primaryClass = "hep-th",
    reportNumber = "YITP-21-52; IPMU21-0033; MPP-2021-101, YITP-21-52, IPMU21-0033",
    doi = "10.1103/PhysRevResearch.3.033254",
    journal = "Phys. Rev. Res.",
    volume = "3",
    number = "3",
    pages = "033254",
    year = "2021"
}

@article{witten_note_2022,
    author = "Witten, Edward",
    title = "{A Note On Complex Spacetime Metrics}",
    eprint = "2111.06514",
    archivePrefix = "arXiv",
    primaryClass = "hep-th",
    month = "11",
    year = "2021"
}

@article{garnerone_typicality_2010,
    author = "Garnerone, Silvano and de Oliveira, Thiago R. and Zanardi, Paolo",
    title = "{Typicality in random matrix product states}",
    eprint = "0908.3877",
    archivePrefix = "arXiv",
    primaryClass = "quant-ph",
    doi = "10.1103/PhysRevA.81.032336",
    journal = "Phys. Rev. A",
    volume = "81",
    pages = "032336",
    year = "2010"
}

@article{cotler_black_2016,
    author = "Cotler, Jordan S. and Gur-Ari, Guy and Hanada, Masanori and Polchinski, Joseph and Saad, Phil and Shenker, Stephen H. and Stanford, Douglas and Streicher, Alexandre and Tezuka, Masaki",
    title = "{Black Holes and Random Matrices}",
    eprint = "1611.04650",
    archivePrefix = "arXiv",
    primaryClass = "hep-th",
    reportNumber = "SU-ITP-16-19, SU-ITP-16/19, YITP-16-124",
    doi = "10.1007/JHEP05(2017)118",
    journal = "JHEP",
    volume = "05",
    pages = "118",
    year = "2017",
    note = "[Erratum ibid. 09 (2018) 002]"
}

@article{tilloy_relativistic_2021,
    author = "Tilloy, Antoine",
    title = "{Relativistic continuous matrix product states for quantum fields without cutoff}",
    eprint = "2102.07741",
    archivePrefix = "arXiv",
    primaryClass = "quant-ph",
    doi = "10.1103/PhysRevD.104.096007",
    journal = "Phys. Rev. D",
    volume = "104",
    number = "9",
    pages = "096007",
    year = "2021"
}

@article{dong_gravity_2016,
    author = "Dong, Xi",
    title = "{The Gravity Dual of Renyi Entropy}",
    eprint = "1601.06788",
    archivePrefix = "arXiv",
    primaryClass = "hep-th",
    reportNumber = "SU-ITP-16/01, SU-ITP-16-01",
    doi = "10.1038/ncomms12472",
    journal = "Nature Commun.",
    volume = "7",
    pages = "12472",
    year = "2016"
}

@article{saad_jt_2019,
    author = "Saad, Phil and Shenker, Stephen H. and Stanford, Douglas",
    title = "{JT gravity as a matrix integral}",
    eprint = "1903.11115",
    archivePrefix = "arXiv",
    primaryClass = "hep-th",
    month = "3",
    year = "2019"
}

@article{milekhin_bra-ket_2022,
    author = "Milekhin, Alexey and Tajdini, Amirhossein",
    title = "{Bra-ket wormholes and Casimir entropy}",
    eprint = "2212.08246",
    archivePrefix = "arXiv",
    primaryClass = "hep-th",
    month = "12",
    year = "2022"
}

@article{arkani-hamed_measure_2007,
    author = "Arkani-Hamed, Nima and Dubovsky, Sergei and Nicolis, Alberto and Trincherini, Enrico and Villadoro, Giovanni",
    title = "{A measure of de Sitter entropy and eternal inflation}",
    eprint = "0704.1814",
    archivePrefix = "arXiv",
    primaryClass = "hep-th",
    doi = "10.1088/1126-6708/2007/05/055",
    journal = "JHEP",
    volume = "05",
    pages = "055",
    year = "2007"
}

@article{borot_asymptotic_2013,
    author = "Borot, G. and Guionnet, A.",
    title = "{Asymptotic expansion of beta matrix models in the one-cut regime}",
    eprint = "1107.1167",
    archivePrefix = "arXiv",
    primaryClass = "math.PR",
    doi = "10.1007/s00220-012-1619-4",
    journal = "Commun. Math. Phys.",
    volume = "317",
    pages = "447--483",
    year = "2013"
}

@article{ganahl_continuous_2017,
    author = "Ganahl, Martin and Rincon, Julian and Vidal, Guifre",
    title = "{Continuous Matrix Product States for Quantum Fields: an Energy Minimization Algorithm}",
    eprint = "1611.03779",
    archivePrefix = "arXiv",
    primaryClass = "cond-mat.str-el",
    doi = "10.1103/PhysRevLett.118.220402",
    journal = "Phys. Rev. Lett.",
    volume = "118",
    number = "22",
    pages = "220402",
    year = "2017"
}

@article{ryu_holographic_2006,
    author = "Ryu, Shinsei and Takayanagi, Tadashi",
    title = "{Holographic derivation of entanglement entropy from AdS/CFT}",
    eprint = "hep-th/0603001",
    archivePrefix = "arXiv",
    reportNumber = "NSF-KITP-06-11, NSF-KITP-06-11",
    doi = "10.1103/PhysRevLett.96.181602",
    journal = "Phys. Rev. Lett.",
    volume = "96",
    pages = "181602",
    year = "2006"
}

@article{hubeny_covariant_2007,
    author = "Hubeny, Veronika E. and Rangamani, Mukund and Takayanagi, Tadashi",
    title = "{A covariant holographic entanglement entropy proposal}",
    eprint = "0705.0016",
    archivePrefix = "arXiv",
    primaryClass = "hep-th",
    reportNumber = "DCPT-07-13, KUNS-2069",
    doi = "10.1088/1126-6708/2007/07/062",
    journal = "JHEP",
    volume = "07",
    pages = "062",
    year = "2007"
}

@article{faulkner_quantum_2013,
    author = "Faulkner, Thomas and Lewkowycz, Aitor and Maldacena, Juan",
    title = "{Quantum corrections to holographic entanglement entropy}",
    eprint = "1307.2892",
    archivePrefix = "arXiv",
    primaryClass = "hep-th",
    doi = "10.1007/JHEP11(2013)074",
    journal = "JHEP",
    volume = "11",
    pages = "074",
    year = "2013"
}

@article{engelhardt_quantum_2015,
    author = "Engelhardt, Netta and Wall, Aron C.",
    title = "{Quantum Extremal Surfaces: Holographic Entanglement Entropy beyond the Classical Regime}",
    eprint = "1408.3203",
    archivePrefix = "arXiv",
    primaryClass = "hep-th",
    doi = "10.1007/JHEP01(2015)073",
    journal = "JHEP",
    volume = "01",
    pages = "073",
    year = "2015"
}

@article{penington_entanglement_2020,
    author = "Penington, Geoffrey",
    title = "{Entanglement Wedge Reconstruction and the Information Paradox}",
    eprint = "1905.08255",
    archivePrefix = "arXiv",
    primaryClass = "hep-th",
    doi = "10.1007/JHEP09(2020)002",
    journal = "JHEP",
    volume = "09",
    pages = "002",
    year = "2020"
}

@article{almheiri_entropy_2019,
    author = "Almheiri, Ahmed and Engelhardt, Netta and Marolf, Donald and Maxfield, Henry",
    title = "{The entropy of bulk quantum fields and the entanglement wedge of an evaporating black hole}",
    eprint = "1905.08762",
    archivePrefix = "arXiv",
    primaryClass = "hep-th",
    doi = "10.1007/JHEP12(2019)063",
    journal = "JHEP",
    volume = "12",
    pages = "063",
    year = "2019"
}

@article{almheiri_replica_2020,
    author = "Almheiri, Ahmed and Hartman, Thomas and Maldacena, Juan and Shaghoulian, Edgar and Tajdini, Amirhossein",
    title = "{Replica Wormholes and the Entropy of Hawking Radiation}",
    eprint = "1911.12333",
    archivePrefix = "arXiv",
    primaryClass = "hep-th",
    doi = "10.1007/JHEP05(2020)013",
    journal = "JHEP",
    volume = "05",
    pages = "013",
    year = "2020"
}

@article{penington_replica_2020,
    author = "Penington, Geoff and Shenker, Stephen H. and Stanford, Douglas and Yang, Zhenbin",
    title = "{Replica wormholes and the black hole interior}",
    eprint = "1911.11977",
    archivePrefix = "arXiv",
    primaryClass = "hep-th",
    doi = "10.1007/JHEP03(2022)205",
    journal = "JHEP",
    volume = "03",
    pages = "205",
    year = "2022"
}

@article{hartle_wave_1983,
    author = "Hartle, J. B. and Hawking, S. W.",
    editor = "Fang, Li-Zhi and Ruffini, R.",
    title = "{Wave Function of the Universe}",
    reportNumber = "PRINT-83-0937 (CAMBRIDGE)",
    doi = "10.1103/PhysRevD.28.2960",
    journal = "Phys. Rev. D",
    volume = "28",
    pages = "2960--2975",
    year = "1983"
}

@article{jackiw_lower_1985,
    author = "Jackiw, R.",
    editor = "Baier, R. and Satz, H.",
    title = "{Lower Dimensional Gravity}",
    reportNumber = "MIT-CTP-1203",
    doi = "10.1016/0550-3213(85)90448-1",
    journal = "Nucl. Phys. B",
    volume = "252",
    pages = "343--356",
    year = "1985"
}

@article{teitelboim_gravitation_1983,
    author = "Teitelboim, C.",
    title = "{Gravitation and Hamiltonian Structure in Two Space-Time Dimensions}",
    doi = "10.1016/0370-2693(83)90012-6",
    journal = "Phys. Lett. B",
    volume = "126",
    pages = "41--45",
    year = "1983"
}

@article{maldacena_two_2021,
    author = "Maldacena, Juan and Turiaci, Gustavo J. and Yang, Zhenbin",
    title = "{Two dimensional Nearly de Sitter gravity}",
    eprint = "1904.01911",
    archivePrefix = "arXiv",
    primaryClass = "hep-th",
    doi = "10.1007/JHEP01(2021)139",
    journal = "JHEP",
    volume = "01",
    pages = "139",
    year = "2021"
}

@article{maldacena_eternal_2018,
    author = "Maldacena, Juan and Qi, Xiao-Liang",
    title = "{Eternal traversable wormhole}",
    eprint = "1804.00491",
    archivePrefix = "arXiv",
    primaryClass = "hep-th",
    month = "4",
    year = "2018"
}

@article{fumagalli_sitter_2025,
    author = "Fumagalli, Alessandro and Gorbenko, Victor and Kames-King, Joshua",
    title = "{De Sitter Bra-Ket wormholes}",
    eprint = "2408.08351",
    archivePrefix = "arXiv",
    primaryClass = "hep-th",
    doi = "10.1007/JHEP05(2025)074",
    journal = "JHEP",
    volume = "05",
    pages = "074",
    year = "2025"
}

@article{shenker_black_2014,
    author = "Shenker, Stephen H. and Stanford, Douglas",
    title = "{Black holes and the butterfly effect}",
    eprint = "1306.0622",
    archivePrefix = "arXiv",
    primaryClass = "hep-th",
    reportNumber = "SU-ITP-13-08",
    doi = "10.1007/JHEP03(2014)067",
    journal = "JHEP",
    volume = "03",
    pages = "067",
    year = "2014"
}

@article{hooft_planar_1974,
    author = "'t Hooft, Gerard",
    editor = "Taylor, J. C.",
    title = "{A Planar Diagram Theory for Strong Interactions}",
    reportNumber = "CERN-TH-1786",
    doi = "10.1016/0550-3213(74)90154-0",
    journal = "Nucl. Phys. B",
    volume = "72",
    pages = "461",
    year = "1974"
}

@article{saad_semiclassical_2019,
    author = "Saad, Phil and Shenker, Stephen H. and Stanford, Douglas",
    title = "{A semiclassical ramp in SYK and in gravity}",
    eprint = "1806.06840",
    archivePrefix = "arXiv",
    primaryClass = "hep-th",
    month = "6",
    year = "2018"
}

@article{chen_comments_2023,
    author = "Chen, Yiming and Ivo, Victor and Maldacena, Juan",
    title = "{Comments on the double cone wormhole}",
    eprint = "2310.11617",
    archivePrefix = "arXiv",
    primaryClass = "hep-th",
    doi = "10.1007/JHEP04(2024)124",
    journal = "JHEP",
    volume = "04",
    pages = "124",
    year = "2024"
}

@article{witten_anti_1998,
    author = "Witten, Edward",
    title = "{Anti de Sitter space and holography}",
    eprint = "hep-th/9802150",
    archivePrefix = "arXiv",
    reportNumber = "IASSNS-HEP-98-15",
    doi = "10.4310/ATMP.1998.v2.n2.a2",
    journal = "Adv. Theor. Math. Phys.",
    volume = "2",
    pages = "253--291",
    year = "1998"
}

@article{maldacena_large_1999,
    author = "Maldacena, Juan Martin",
    title = "{The large $N$ limit of superconformal field theories and supergravity}",
    eprint = "hep-th/9711200",
    archivePrefix = "arXiv",
    reportNumber = "HUTP-97-A097, HUTP-98-A097",
    doi = "10.4310/ATMP.1998.v2.n2.a1",
    journal = "Adv. Theor. Math. Phys.",
    volume = "2",
    pages = "231--252",
    year = "1998"
}

@article{harlow_wormholes_2016,
    author = "Harlow, Daniel",
    title = "{Wormholes, Emergent Gauge Fields, and the Weak Gravity Conjecture}",
    eprint = "1510.07911",
    archivePrefix = "arXiv",
    primaryClass = "hep-th",
    doi = "10.1007/JHEP01(2016)122",
    journal = "JHEP",
    volume = "01",
    pages = "122",
    year = "2016"
}

@article{guica_construction_2017,
    author = "Guica, Monica and Jafferis, Daniel L.",
    title = "{On the construction of charged operators inside an eternal black hole}",
    eprint = "1511.05627",
    archivePrefix = "arXiv",
    primaryClass = "hep-th",
    doi = "10.21468/SciPostPhys.3.2.016",
    journal = "SciPost Phys.",
    volume = "3",
    number = "2",
    pages = "016",
    year = "2017"
}

@article{gibbons_action_1977,
    author = "Gibbons, G. W. and Hawking, S. W.",
    title = "{Action Integrals and Partition Functions in Quantum Gravity}",
    reportNumber = "PRINT-76-0995 (CAMBRIDGE)",
    doi = "10.1103/PhysRevD.15.2752",
    journal = "Phys. Rev. D",
    volume = "15",
    pages = "2752--2756",
    year = "1977"
}

@article{susskind_world_1995,
    author = "Susskind, Leonard",
    title = "{The World as a hologram}",
    eprint = "hep-th/9409089",
    archivePrefix = "arXiv",
    reportNumber = "SU-ITP-94-33",
    doi = "10.1063/1.531249",
    journal = "J. Math. Phys.",
    volume = "36",
    pages = "6377--6396",
    year = "1995"
}

@article{Garnerone:2010lqy,
    author = "Garnerone, Silvano and de Oliveira, Thiago R. and Haas, Stephan and Zanardi, Paolo",
    title = "{Statistical properties of random matrix product states}",
    eprint = "1003.5253",
    archivePrefix = "arXiv",
    primaryClass = "quant-ph",
    doi = "10.1103/PhysRevA.82.052312",
    journal = "Phys. Rev. A",
    volume = "82",
    pages = "052312",
    year = "2010"
}

@article{cotler_non-perturbative_2025,
    author = "Cotler, Jordan and Jensen, Kristan",
    title = "{Non-perturbative de Sitter Jackiw-Teitelboim gravity}",
    eprint = "2401.01925",
    archivePrefix = "arXiv",
    primaryClass = "hep-th",
    doi = "10.1007/JHEP12(2024)016",
    journal = "JHEP",
    volume = "12",
    pages = "016",
    year = "2024"
}

@article{cotler_isometric_2023,
    author = "Cotler, Jordan and Jensen, Kristan",
    title = "{Isometric Evolution in de Sitter Quantum Gravity}",
    eprint = "2302.06603",
    archivePrefix = "arXiv",
    primaryClass = "hep-th",
    doi = "10.1103/PhysRevLett.131.211601",
    journal = "Phys. Rev. Lett.",
    volume = "131",
    number = "21",
    pages = "211601",
    year = "2023"
}

@article{cotler_emergent_2025,
    author = "Cotler, Jordan and Jensen, Kristan",
    title = "{Emergent unitarity in de Sitter from matrix integrals}",
    eprint = "1911.12358",
    archivePrefix = "arXiv",
    primaryClass = "hep-th",
    doi = "10.1007/JHEP12(2021)089",
    journal = "JHEP",
    volume = "12",
    pages = "089",
    year = "2021"
}

@article{cotler_chaos_2017,
    author = "Cotler, Jordan and Hunter-Jones, Nicholas and Liu, Junyu and Yoshida, Beni",
    title = "{Chaos, Complexity, and Random Matrices}",
    eprint = "1706.05400",
    archivePrefix = "arXiv",
    primaryClass = "hep-th",
    doi = "10.1007/JHEP11(2017)048",
    journal = "JHEP",
    volume = "11",
    pages = "048",
    year = "2017"
}

@article{jafferis_jt_2023,
    author = "Jafferis, Daniel Louis and Kolchmeyer, David K. and Mukhametzhanov, Baur and Sonner, Julian",
    title = "{Jackiw-Teitelboim gravity with matter, generalized eigenstate thermalization hypothesis, and random matrices}",
    eprint = "2209.02131",
    archivePrefix = "arXiv",
    primaryClass = "hep-th",
    doi = "10.1103/PhysRevD.108.066015",
    journal = "Phys. Rev. D",
    volume = "108",
    number = "6",
    pages = "066015",
    year = "2023"
}

@article{witten_matrix_2020,
    author = "Witten, Edward",
    title = "{Matrix Models and Deformations of JT Gravity}",
    eprint = "2006.13414",
    archivePrefix = "arXiv",
    primaryClass = "hep-th",
    doi = "10.1098/rspa.2020.0582",
    journal = "Proc. Roy. Soc. Lond. A",
    volume = "476",
    number = "2244",
    pages = "20200582",
    year = "2020"
}

@article{maxfield_path_2021,
    author = "Maxfield, Henry and Turiaci, Gustavo J.",
    title = "{The path integral of 3D gravity near extremality; or, JT gravity with defects as a matrix integral}",
    eprint = "2006.11317",
    archivePrefix = "arXiv",
    primaryClass = "hep-th",
    doi = "10.1007/JHEP01(2021)118",
    journal = "JHEP",
    volume = "01",
    pages = "118",
    year = "2021"
}

@article{verstraete_matrix_2008,
	author = "Verstraete, F. and Murg, V. and Cirac, J. I.",
    title = "{Matrix product states, projected entangled pair states, and variational renormalization group methods for quantum spin systems}",
    eprint = "0907.2796",
    archivePrefix = "arXiv",
    primaryClass = "quant-ph",
    doi = "10.1080/14789940801912366",
    journal = "Adv. Phys.",
    volume = "57",
    number = "2",
    pages = "143--224",
    year = "2008"
}

@article{stanford_fermionic_2017,
    author = "Stanford, Douglas and Witten, Edward",
    title = "{Fermionic Localization of the Schwarzian Theory}",
    eprint = "1703.04612",
    archivePrefix = "arXiv",
    primaryClass = "hep-th",
    doi = "10.1007/JHEP10(2017)008",
    journal = "JHEP",
    volume = "10",
    pages = "008",
    year = "2017"
}

@article{hernandez-cuenca_wormholes_2024,
    author = "Hern{\'a}ndez-Cuenca, Sergio",
    title = "{Wormholes and factorization in exact effective theory}",
    eprint = "2404.10035",
    archivePrefix = "arXiv",
    primaryClass = "hep-th",
    reportNumber = "MIT-CTP/5708",
    doi = "10.1007/JHEP05(2025)024",
    journal = "JHEP",
    volume = "05",
    pages = "024",
    year = "2025"
}

@article{hawking_quantum_1987,
    author = "Hawking, S. W.",
    title = "{Quantum Coherence Down the Wormhole}",
    reportNumber = "Print-87-0842 (CAMBRIDGE)",
    doi = "10.1016/0370-2693(87)90028-1",
    journal = "Phys. Lett. B",
    volume = "195",
    pages = "337",
    year = "1987"
}

@article{harlow_quantum_2025,
    doi = "10.1007/JHEP02(2026)108",
    author = "Harlow, Daniel and Usatyuk, Mykhaylo and Zhao, Ying",
    title = "{Quantum mechanics and observers for gravity in a closed universe}",
    eprint = "2501.02359",
    archivePrefix = "arXiv",
    primaryClass = "hep-th",
    reportNumber = "MIT-CTP/5824",
    month = "1",
    year = "2026",
    journal = "JHEP",
    volume = "02",
    pages = "108"
}

@article{kolchmeyer_chaos_2024,
    author = "Kolchmeyer, David K. and Liu, Hong",
    title = "{Chaos and the Emergence of the Cosmological Horizon}",
    eprint = "2411.08090",
    archivePrefix = "arXiv",
    primaryClass = "hep-th",
    reportNumber = "MIT-CTP/5805",
    month = "11",
    year = "2024"
}

@article{chandrasekaran_algebra_2023,
    author = "Chandrasekaran, Venkatesa and Longo, Roberto and Penington, Geoff and Witten, Edward",
    title = "{An algebra of observables for de Sitter space}",
    eprint = "2206.10780",
    archivePrefix = "arXiv",
    primaryClass = "hep-th",
    doi = "10.1007/JHEP02(2023)082",
    journal = "JHEP",
    volume = "02",
    pages = "082",
    year = "2023"
}

@article{bunch_quantum_1997,
    author = "Bunch, T. S. and Davies, P. C. W.",
    title = "{Quantum Field Theory in de Sitter Space: Renormalization by Point Splitting}",
    doi = "10.1098/rspa.1978.0060",
    journal = "Proc. Roy. Soc. Lond. A",
    volume = "360",
    pages = "117--134",
    year = "1978"
}

@article{hertog_holographic_2012,
   author = "Hertog, Thomas and Hartle, James",
    title = "{Holographic No-Boundary Measure}",
    eprint = "1111.6090",
    archivePrefix = "arXiv",
    primaryClass = "hep-th",
    doi = "10.1007/JHEP05(2012)095",
    journal = "JHEP",
    volume = "05",
    pages = "095",
    year = "2012"
}

@article{susskind_entanglement_2021,
    author = "Susskind, Leonard",
    title = "{Entanglement and Chaos in De Sitter Space Holography: An SYK Example}",
    eprint = "2109.14104",
    archivePrefix = "arXiv",
    primaryClass = "hep-th",
    doi = "10.22128/jhap.2021.455.1005",
    journal = "JHAP",
    volume = "1",
    number = "1",
    pages = "1--22",
    year = "2021"
}

@article{gao_traversable_2017,
    author = "Gao, Ping and Jafferis, Daniel Louis and Wall, Aron C.",
    title = "{Traversable Wormholes via a Double Trace Deformation}",
    eprint = "1608.05687",
    archivePrefix = "arXiv",
    primaryClass = "hep-th",
    doi = "10.1007/JHEP12(2017)151",
    journal = "JHEP",
    volume = "12",
    pages = "151",
    year = "2017"
}

@article{maldacena_conformal_2016,
    author = "Maldacena, Juan and Stanford, Douglas and Yang, Zhenbin",
    title = "{Conformal symmetry and its breaking in two dimensional Nearly Anti-de-Sitter space}",
    eprint = "1606.01857",
    archivePrefix = "arXiv",
    primaryClass = "hep-th",
    doi = "10.1093/ptep/ptw124",
    journal = "PTEP",
    volume = "2016",
    number = "12",
    pages = "12C104",
    year = "2016"
}

@article{linde_eternally_1986,
    author = "Linde, Andrei D.",
    title = "{Eternally Existing Selfreproducing Chaotic Inflationary Universe}",
    reportNumber = "Print-86-0418 (LEBEDEV INST), LEBEDEV-86-106",
    doi = "10.1016/0370-2693(86)90611-8",
    journal = "Phys. Lett. B",
    volume = "175",
    pages = "395--400",
    year = "1986"
}

@article{metropolis_equation_1953,
    author = "Metropolis, N. and Rosenbluth, A. W. and Rosenbluth, M. N. and Teller, A. H. and Teller, E.",
    title = "{Equation of state calculations by fast computing machines}",
    doi = "10.1063/1.1699114",
    journal = "J. Chem. Phys.",
    volume = "21",
    pages = "1087--1092",
    year = "1953"
}

@article{hastings_monte_1970,
    author = "Hastings, W. K.",
    title = "{Monte Carlo Sampling Methods Using Markov Chains and Their Applications}",
    doi = "10.1093/biomet/57.1.97",
    journal = "Biometrika",
    volume = "57",
    pages = "97--109",
    year = "1970"
}
